\pdfoutput=1
\newcommand*{\ATLASLATEXPATH}{latex/}
\documentclass[cernpreprint,UKenglish,texlive=2016,txfonts]{\ATLASLATEXPATH atlasdoc}

\usepackage[number-unit-product=~]{siunitx}
\usepackage{multirow}
\usepackage{xstring}
\usepackage[eVkern=true]{\ATLASLATEXPATH atlaspackage}
\usepackage{\ATLASLATEXPATH atlasbiblatex}

\usepackage{\ATLASLATEXPATH atlascontribute}
\usepackage{\ATLASLATEXPATH atlascover}

\usepackage[process,unit=false]{\ATLASLATEXPATH atlasphysics}

\graphicspath{{logos/}{figures/}}

\usepackage{TriggerPerfPaper2015-defs}

\AtlasTitle{Performance of the ATLAS Trigger System in 2015}

\author{The ATLAS Collaboration}

\AtlasRefCode{TRIG-2016-01}

\PreprintIdNumber{CERN-EP-2016-241}

\arXivId{1611.09661}

\AtlasJournal{EPJC}
\AtlasJournalRef{Eur. Phys. J. C 77 (2017) 317}
\AtlasDOI{10.1140/epjc/s10052-017-4852-3}

\AtlasAbstract{%
During 2015 the ATLAS experiment recorded $3.8~\mathrm{fb}^{-1}$ of
proton--proton collision data at a centre-of-mass energy of $13~\mathrm{TeV}$. 
The ATLAS trigger system is a crucial component of the experiment, responsible
for selecting events of interest at a recording rate of
approximately 1~kHz from up to 40~MHz of collisions.
This paper presents a short overview of the changes to the
trigger and data acquisition systems during the first long shutdown of the LHC
and shows the performance of the trigger system and its components based on the
2015 proton--proton collision data.}

\AtlasCoverCommentsDeadline{26 Sep 2016}

\AtlasCoverAnalysisTeam{Imma~Riu, Frank~Winklmeier (contact editors)\\\textit{and all the people that have contributed to the Trigger activity through hardware and software development, operational work and performance assessment.}}

\AtlasCoverEdBoardMember{Andrew~Lankford~(chair)}
\AtlasCoverEdBoardMember{John~Baines}
\AtlasCoverEdBoardMember{Manuela~Venturi}

\AtlasCoverEgroupEditors{atlas-trig-2016-01-editors@cern.ch}

\AtlasCoverEgroupEdBoard{atlas-trig-2016-01-editorial-board@cern.ch}

\hypersetup{pdftitle={ATLAS},pdfauthor={The ATLAS Collaboration}}

\begin{document}

\maketitle

\tableofcontents



\section{Introduction}
\label{sec:intro}
The trigger system is an essential component of any collider experiment
as it is responsible for deciding whether or not to keep an event from a given bunch-crossing interaction
for later study. During \runi (2009 to early 2013) of the Large Hadron Collider (LHC), the
trigger system~\cite{PERF-2011-02,ATLAS-CONF-2012-048,TRIG-2012-03,PERF-2013-06,TRIG-2012-01} 
of the ATLAS experiment~\cite{PERF-2007-01} operated efficiently at instantaneous
luminosities of up to $8\times\lumi{e33}$ and primarily at centre-of-mass energies, $\sqrt{s}$, of
\SI{7}{\TeV} and \SI{8}{\TeV}.
In \runii (since 2015) the increased centre-of-mass energy of \SI{13}{\TeV},
higher luminosity and increased number of proton--proton interactions per
bunch-crossing (pile-up) meant that, without upgrades of the trigger system, the trigger rates
would have exceeded the maximum allowed rates 
when running with the trigger thresholds needed to satisfy the
physics programme of the experiment. For this reason, the first long shutdown (LS1) between LHC
\runi and \runii operations was used to improve the trigger system
with almost no component left untouched.

After a brief introduction of the ATLAS detector in Section~\ref{sec:detector}, Section~\ref{sec:run2changes}
summarises the changes to the trigger and data acquisition during
LS1. Section~\ref{sec:menu} gives an overview of the trigger menu used during
2015 followed by an introduction to the reconstruction algorithms used at the
high-level trigger in Section~\ref{sec:hltrec}. 
The performance of the different trigger signatures is shown in Section~\ref{sec:sigperf} for
the data taken with \SI{25}{ns} bunch-spacing in 2015 at a peak luminosity of $5\times\lumi{e33}$
with comparison to Monte Carlo (MC) simulation.



\section{ATLAS detector}
\label{sec:detector}

ATLAS is a general-purpose detector with a
forward-backward symmetry, which provides almost full solid
angle coverage around the interaction point.\footnote{ATLAS uses a
  right-handed coordinate system with its origin at the nominal
  interaction point (IP) in the centre of the detector and the
  $z$-axis along the beam pipe. The $x$-axis points from the IP to the
  centre of the LHC ring, and the $y$-axis points upward. Cylindrical
  coordinates $(r,\phi)$ are used in the transverse plane, $\phi$
  being the azimuthal angle around the $z$-axis. The pseudorapidity is
  defined in terms of the polar angle $\theta$ as
  $\eta=-\ln\tan(\theta/2)$.}
The main components of ATLAS are an inner detector (ID), which is
surrounded by a superconducting solenoid providing a \SI{2}{\tesla} axial
magnetic field, a calorimeter system, and a muon spectrometer (MS)
in a magnetic field generated by three large
superconducting toroids with eight coils each. The ID provides track reconstruction within
$|\eta| < 2.5$, employing a pixel detector (Pixel) close to the beam pipe,
a silicon microstrip detector (SCT) at intermediate radii, and a transition
radiation tracker (TRT) at outer radii. A new innermost pixel-detector layer,
the insertable B-layer (IBL), was added during LS1  at a radius of \SI{33}{\mm}
around a new and thinner beam pipe~\cite{B-layerRef}. 
The calorimeter system covers the region $|\eta| < 4.9$, the forward 
region ($3.2 < |\eta| < 4.9$) being instrumented
with a liquid-argon (LAr) calorimeter for electromagnetic and
hadronic measurements. In the central region, a lead/LAr
electromagnetic calorimeter covers $|\eta| < 3.2$, while the hadronic
calorimeter uses two different detector technologies, with
steel/scintillator tiles ($|\eta| < 1.7$) or lead/LAr ($1.5 < |\eta| < 3.2$) as 
absorber/active material.
The MS consists of one barrel ($|\eta|<1.05$) and two end-cap sections
($1.05<|\eta|<2.7$).  Resistive plate chambers (RPC, three doublet layers for
$|\eta|<1.05$) and thin gap chambers (TGC, one triplet layer followed by two doublets
for $1.0<|\eta|<2.4$) provide triggering capability as well as $(\eta,\phi)$ position
measurements. A precise momentum measurement for muons with $|\eta|$ up to 2.7 is provided
by three layers of monitored drift tubes (MDT), with each chamber providing six to eight $\eta$
measurements along the muon trajectory. For $|\eta|>2$, the inner layer is instrumented with 
cathode strip chambers (CSC), consisting of four sensitive layers each, instead of MDTs.

\begin{figure}[htbp]
\centering
\includegraphics[width=0.85\textwidth]{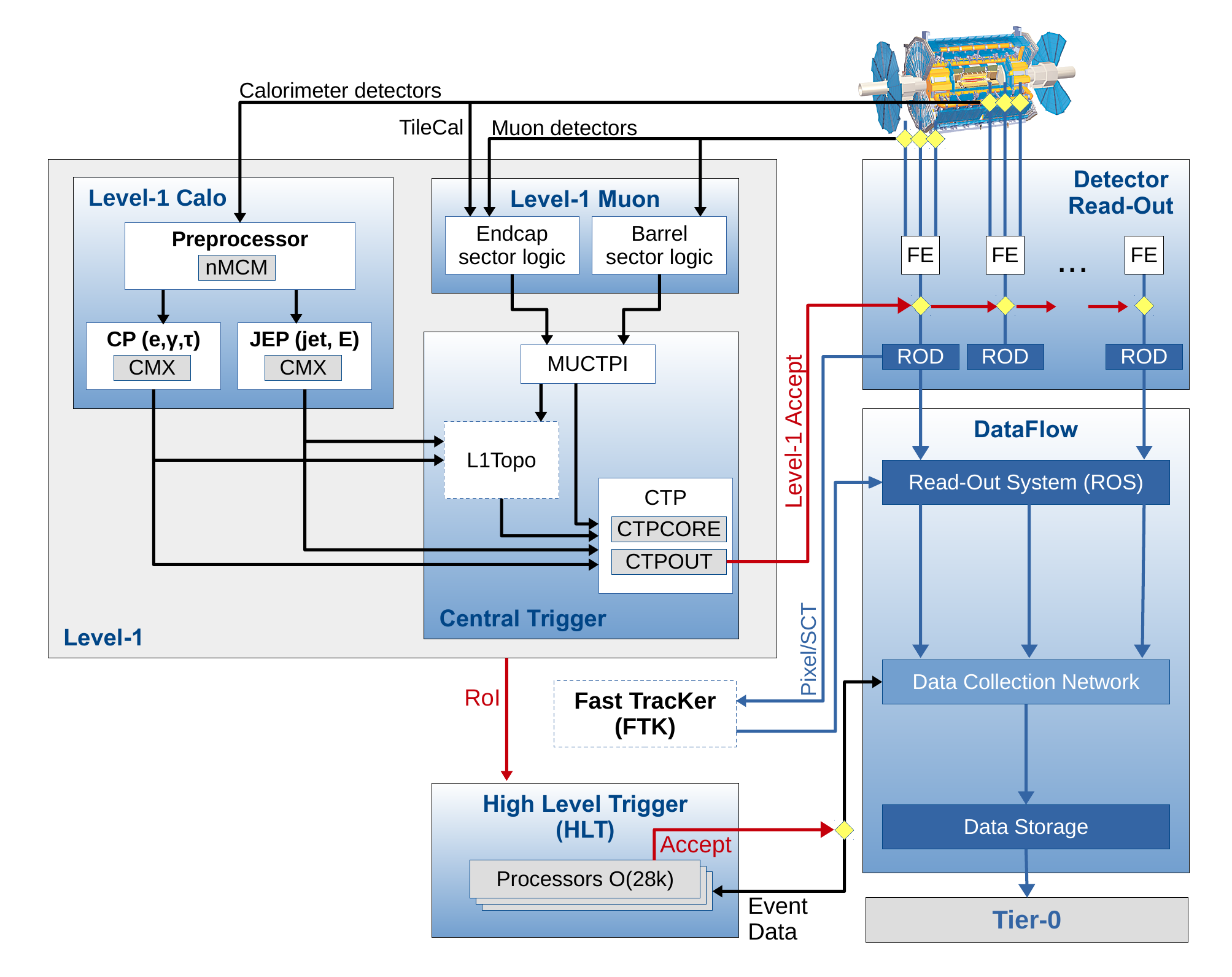}
\caption{The ATLAS TDAQ system in \runii with emphasis on the components
  relevant for triggering. L1Topo and FTK were being commissioned during 2015 and not used for the results shown here.}
\label{fig:tdaq-schema}
\end{figure}

The Trigger and Data Acquisition (TDAQ) system shown in
Figure~\ref{fig:tdaq-schema} consists of a hardware-based first-level trigger
(L1) and a software-based high-level trigger (HLT). The L1 trigger decision is
formed by the Central Trigger Processor (CTP), which receives inputs from the L1
calorimeter (L1Calo) and L1 muon (L1Muon) triggers as well as several other
subsystems such as the Minimum Bias Trigger Scintillators (MBTS), the LUCID
Cherenkov counter and the Zero-Degree Calorimeter (ZDC). The CTP is also
responsible for applying preventive dead-time. It limits the minimum time
between two consecutive L1 accepts (\emph{simple dead-time}) to avoid
overlapping readout windows, and restricts the number of L1 accepts allowed in a
given number of bunch-crossings (\emph{complex dead-time}) to avoid front-end
buffers from overflowing. In 2015 running, the simple dead-time was set to 4
bunch-crossings (\SI{100}{\ns}). A more detailed description of the L1 trigger
system can be found in Ref.~\cite{PERF-2011-02}.
After the L1 trigger acceptance, the events are buffered in the Read-Out System
(ROS) and processed by the HLT. The HLT receives Region-of-Interest (RoI)
information from L1, which can be used for regional reconstruction in the
trigger algorithms.
After the events are accepted by the HLT, they are transferred to local
storage at the experimental site and exported to the Tier-0 facility at
CERN's computing centre for offline reconstruction.

Several Monte Carlo simulated datasets were used to assess the performance of the trigger.
Fully simulated photon+jet and dijet events generated with \textsc{Pythia8}~\cite{PYTHIA8} using
the NNPDF2.3LO~\cite{NNPDF} parton distribution
function (PDF) set were used to study the photon and jet triggers. To study tau and $b$-jet triggers,
$Z\rightarrow\tau\tau$ and $t\bar{t}$ samples generated with \textsc{Powheg-Box} 2.0~\cite{POWHEG-BOX1,POWHEG-BOX2,POWHEG-BOX3}
with the CT10~\cite{CT10} PDF set and interfaced to \textsc{Pythia8} or \textsc{Pythia6}~\cite{PYTHIA6} with the
CTEQ6L1~\cite{CTEQ6L1} PDF set were used.


\section{Changes to the Trigger/DAQ system for Run 2}
\label{sec:run2changes}

%
The TDAQ system used during \runi is described
in detail in Refs. \cite{PERF-2011-02} and \cite{DAQRun1Paper}. 
Compared to \runi, the LHC has increased its centre-of-mass energy from \num{8} to
\SI{13}{\TeV}, and the nominal bunch-spacing has decreased from \num{50} to \SI{25}{\ns}. Due to the
larger transverse beam size at the interaction point ($\beta^{*}=\SI{80}{\cm}$ compared to \SI{60}{\cm} in 2012)
and a lower bunch population ($1.15\times 10^{11}$ instead of $1.6\times 10^{11}$ protons per bunch)
the peak luminosity reached in 2015 ($5.0\times\lumi{e33}$) was lower than in
\runi\ ($7.7\times\lumi{e33}$). However, due to the increase in energy, trigger rates
are on average 2.0 to 2.5 times larger for the same luminosity and with the same trigger criteria (individual
trigger rates, e.g.\ jets, can have even larger increases). The decrease in bunch-spacing also
increases certain trigger rates (e.g.\ muons) due to additional interactions from neighbouring 
bunch-crossings (out-of-time pile-up).
In order to prepare for the expected higher rates in \runii, several upgrades and additions
were implemented during LS1. The main changes relevant to the
trigger system are briefly described below.

In the L1 Central Trigger, a new topological trigger (L1Topo) consisting of two
FPGA-based (Field-Programmable Gate Arrays) processor modules was added.  The
modules are identical hardware-wise and each is programmed to perform selections
based on geometric or kinematic association between trigger objects received
from the L1Calo or L1Muon systems. This includes the refined calculation of
global event quantities such as missing transverse momentum (with magnitude
\met).  The system was fully installed and commissioned during 2016, i.e.\ it
was not used for the data described in this paper. Details of the hardware
implementation can be found in Ref.~\cite{ATLAS-TDR-023}. The Muon-to-CTP
interface (MUCPTI) and the CTP were upgraded to provide inputs to and receive
inputs from L1Topo, respectively. In order to better address sub-detector
specific requirements, the CTP now supports up to four independent complex
dead-time settings operating simultaneously. In addition, the number of L1
trigger selections (512) and bunch-group selections (16), defined later, were
doubled compared to \runi.  The changes to the L1Calo and L1Muon trigger systems
are described in separate sections below.

In \runi the HLT consisted of separate Level-2 (L2) and Event Filter (EF) farms.
While L2 requested partial event data over the network, the EF operated on full
event information assembled by separate farm nodes dedicated to Event Building
(EB). For \runii, the L2 and EF farms were merged into a single homogeneous farm
allowing better resource sharing and an overall simplification of both the
hardware and software. RoI-based reconstruction continues to be employed by
time-critical algorithms. The functionality of the EB nodes was also integrated
into the HLT farm. To achieve higher readout and output rates, the ROS, the data
collection network and data storage system were upgraded. The on-detector
front-end (FE) electronics and detector-specific readout drivers (ROD) were not
changed in any significant way.

A new Fast TracKer (FTK) system~\cite{ATLAS-TDR-021} will provide global ID track
reconstruction at the L1 trigger rate using lookup tables stored in custom
associative memory chips for the pattern recognition. Instead of a
computationally intensive helix fit, the FPGA-based track fitter performs a fast
linear fit and the tracks are made available to the HLT. This system will allow
the use of tracks at much higher event rates in the HLT than is currently affordable using
CPU systems. This system is currently being installed and expected to be
fully commissioned during 2017.


\subsection{Level-1 calorimeter trigger}
\label{sec:l1calo}

\begin{figure}[htbp]
  \centering
  \includegraphics[width=0.4\textwidth]{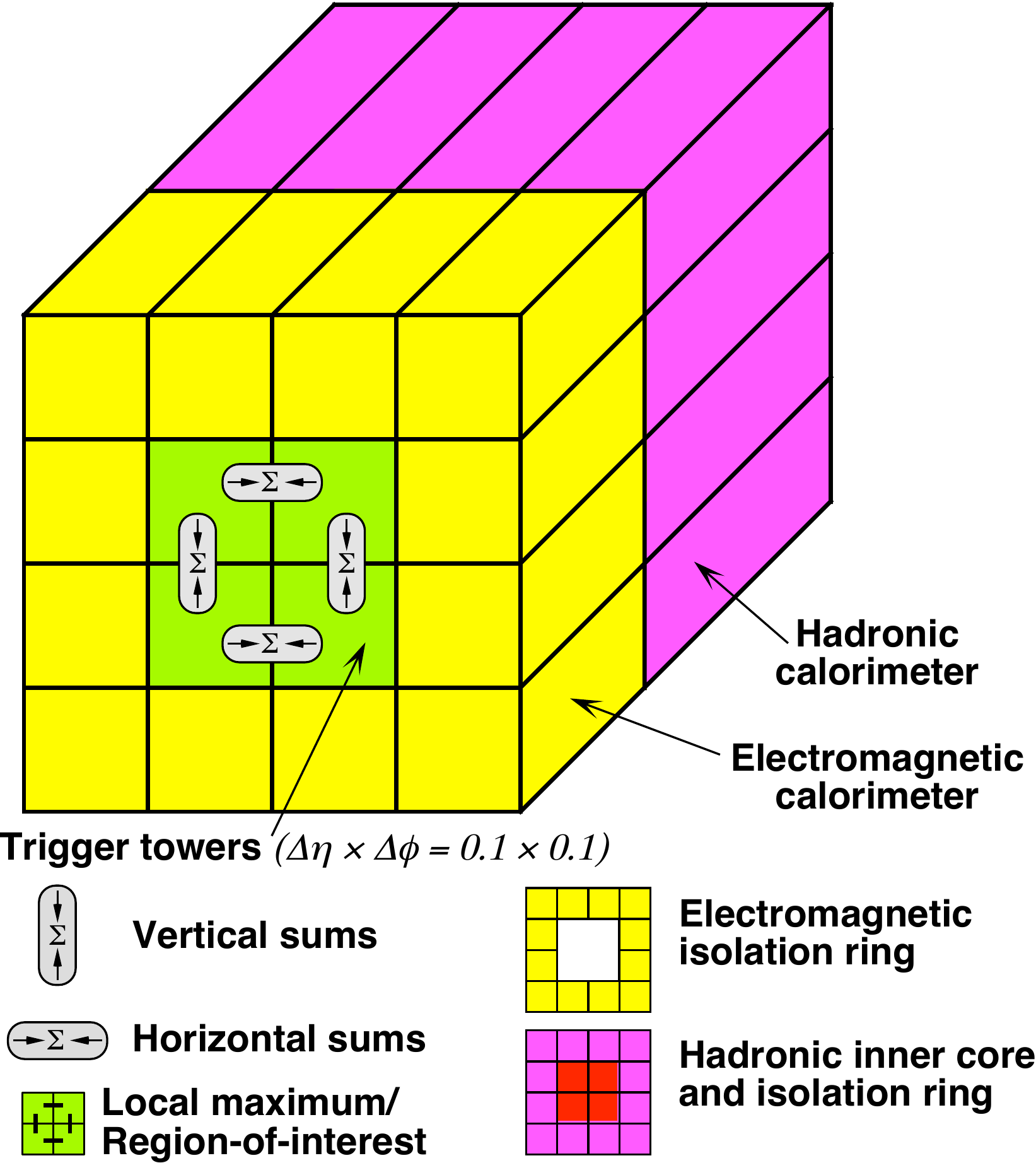}
  \caption{Schematic view of the trigger towers used as input to the L1Calo trigger algorithms.}
  \label{fig:L1CaloTT}
\end{figure}

The details of the L1Calo trigger algorithms can be found in Ref.~\cite{L1CaloPaper},
and only the basic elements are described here.
The electron/photon and tau trigger algorithm (Figure~\ref{fig:L1CaloTT}) identifies an RoI as a $2\times2$
trigger tower cluster in the electromagnetic calorimeter for which the
sum of the transverse energy from at least one of the four possible pairs of nearest
neighbour towers ($1\times2$ or $2\times1$) exceeds a predefined
threshold. Isolation-veto thresholds can be set for the electromagnetic (EM) isolation 
ring in the electromagnetic calorimeter, as well as for hadronic tower sums in a
central $2\times2$ core behind the EM cluster and in the 12-tower hadronic ring around
it. The $\ET$ threshold can be set differently for different $\eta$ regions at
a granularity of 0.1 in $\eta$ in order to correct for varying detector energy
responses. The energy of the trigger towers is calibrated at the electromagnetic energy scale (EM scale).
The EM scale correctly reconstructs the energy deposited by particles in an electromagnetic shower 
in the calorimeter but underestimates the energy deposited by hadrons.
Jet RoIs are defined as $4\times4$ or $8\times8$ trigger tower
windows for which the summed electromagnetic and hadronic transverse energy
exceeds predefined thresholds and which surround a $2\times2$ trigger tower
core that is a local maximum.  The location of this local maximum also defines
the coordinates of the jet RoI.

In preparation for \runii, due to the expected increase in luminosity and consequent 
increase in the number of pile-up events, a major upgrade of several central components of
the L1Calo electronics was undertaken to reduce the trigger rates.

\begin{figure}[htbp]
  \centering
  \includegraphics[width=0.5\textwidth]{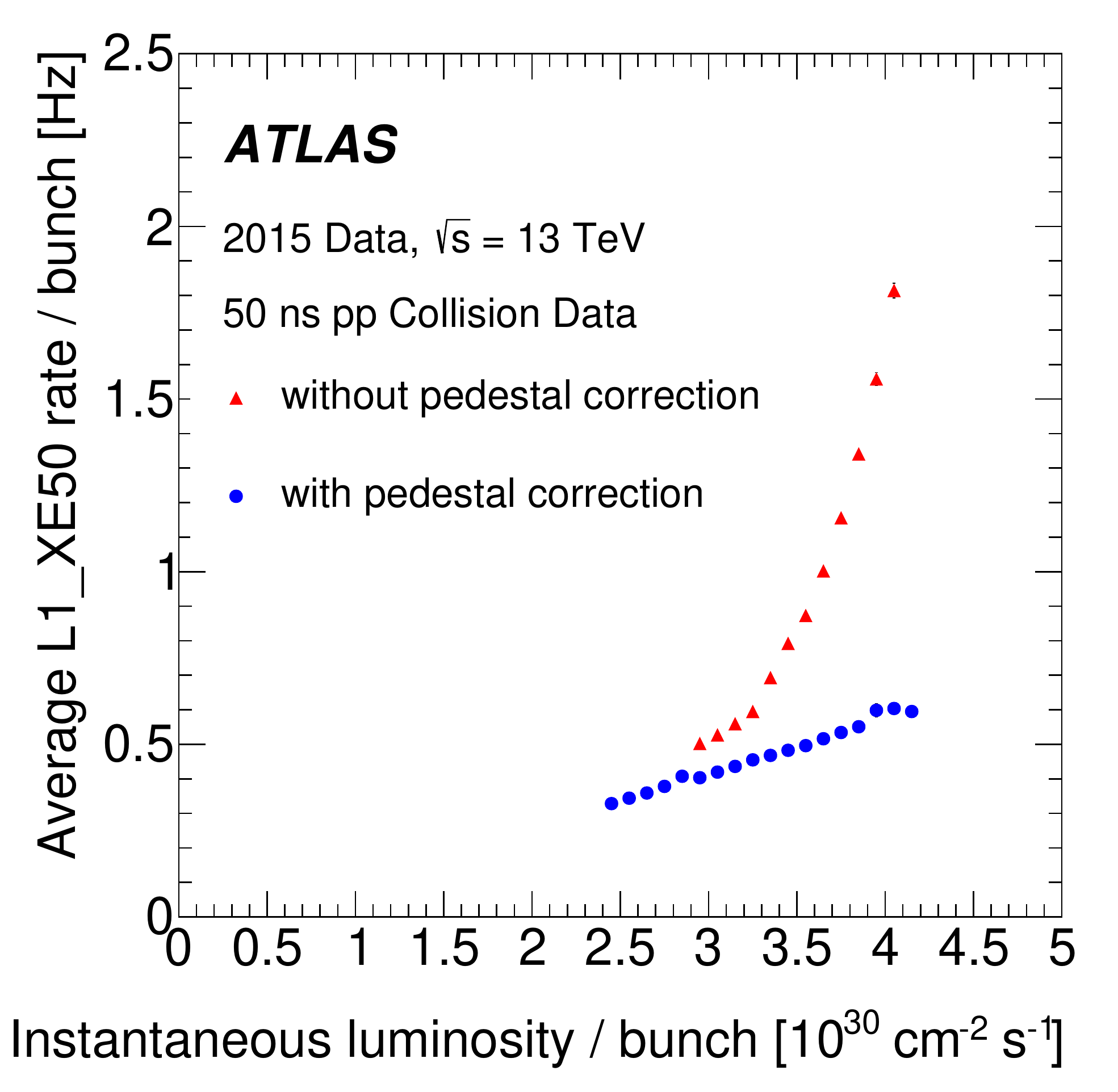}
  \caption{The per-bunch trigger rate for the L1 missing transverse momentum trigger with a
    threshold of \SI{50}{\GeV} (\trig{L1_XE50}) as a function of
    the instantaneous luminosity per bunch. The rates are shown 
    with and without pedestal correction applied.}
  \label{fig:XErates}
\end{figure}

For the preprocessor system~\cite{L1Calo-PPMpaper}, which digitises and calibrates
the analogue signals (consisting of $\sim$7000 trigger towers at a granularity of $0.1\times 0.1$ in $\eta\times\phi$)
from the calorimeter detectors, a new FPGA-based multi-chip 
module (nMCM) was developed~\cite{L1Calo-nMCMproc} and about 3000 chips (including spares) were 
produced. They replace the old ASIC-based MCMs used during
\runi. The new modules provide additional flexibility and new functionality
with respect to the old system. In particular, the nMCMs support the use of
digital autocorrelation Finite Impulse Response (FIR)
filters and the implementation of a dynamic, bunch-by-bunch pedestal correction, both 
introduced for \runii. 
These improvements lead to a significant rate reduction of the L1 jet and L1 \met\ triggers.
The bunch-by-bunch pedestal subtraction compensates for the increased trigger
rates at the beginning of a bunch train caused by the interplay of in-time and
out-of-time pile-up coupled with the LAr pulse shape~\cite{LARG-2009-02}, and
linearises the L1 trigger rate as a function of the instantaneous luminosity, as shown in
Figure~\ref{fig:XErates} for the L1 \met\ trigger.
The autocorrelation FIR filters substantially improve the bunch-crossing 
identification (BCID) efficiencies, in particular for low energy deposits. However, the use 
of this new filtering scheme initially led to an early trigger signal (and incomplete events) for a
small fraction of very high energy events. These events were saved into a stream dedicated
to mistimed events and treated separately in the relevant physics analyses. The source of
the problem was fixed in firmware by adapting the BCID decision logic for saturated pulses 
and was deployed at the start of the 2016 data-taking period.

The preprocessor outputs are then transmitted to both the Cluster
Processor (CP) and Jet/Energy-sum Processor (JEP) subsystems in parallel.
The CP subsystem identifies electron/photon and tau lepton candidates with \ET
above a programmable threshold and satisfying, if required, certain isolation 
criteria. The JEP receives jet trigger elements, which are $0.2\times 0.2$ sums in $\eta\times\phi$, and uses these
to identify jets and to produce global sums of scalar and missing transverse momentum.
Both the CP and JEP firmware were upgraded to allow an increase of the data
transmission rate over the custom-made backplanes from \SI{40}{Mbps} to
\SI{160}{Mbps}, allowing the transmission of up to four jet or five EM/tau
trigger objects per module. A trigger object contains the \ET sum, $\eta-\phi$ coordinates,
and isolation thresholds where relevant. While the JEP firmware 
changes were only minor, substantial extra selectivity was added to the CP by 
implementing energy-dependent L1 electromagnetic 
isolation criteria instead of fixed threshold cuts. This feature was added 
to the trigger menu (defined in Section~\ref{sec:menu}) at
the beginning of \runii. In 2015 
it was used to effectively select events with specific signatures, e.g.\ EM 
isolation was required for taus but not for electrons.

Finally, new extended cluster merger modules (CMX) were developed to replace
the L1Calo merger modules (CMMs) used during \runi. The new CMX modules
transmit the location and the energy of identified trigger objects to the new L1Topo modules
instead of only the threshold multiplicities as done by the CMMs. This
transmission happens with a bandwidth of \SI{6.4}{Gbps} per channel, while
the total output bandwidth amounts to above \SI{2}{Tbps}. 
Moreover, for most L1 triggers, twice as many trigger selections and isolation
thresholds can be processed with the new CMX modules compared to \runi,
considerably increasing the selectivity of the L1Calo system.


\subsection{Level-1 muon trigger}
\label{sec:l1muon}
%
%
The muon barrel trigger was not significantly changed with respect to \runi,
apart from the regions close to the feet that support the ATLAS detector,
where the presence of support structures reduces trigger coverage.
To recover trigger acceptance, a fourth layer of RPC trigger chambers was installed
before \runi in the projective region of the acceptance holes. These chambers
were not operational during \runi.
During LS1, these RPC layers were equipped with trigger electronics.
Commissioning started during 2015 and they are fully operational in 2016.  
Additional chambers were installed during LS1 to cover the acceptance holes
corresponding to two elevator shafts at the bottom of the muon spectrometer but are
not yet operational.
At the end of the commissioning phase, the new feet and elevator chambers
are expected to increase the overall barrel trigger acceptance by 2.8 and 0.8 
percentage points, respectively.

%
%
\begin{figure}[htbp]
\centering
\includegraphics[width=0.55\textwidth]{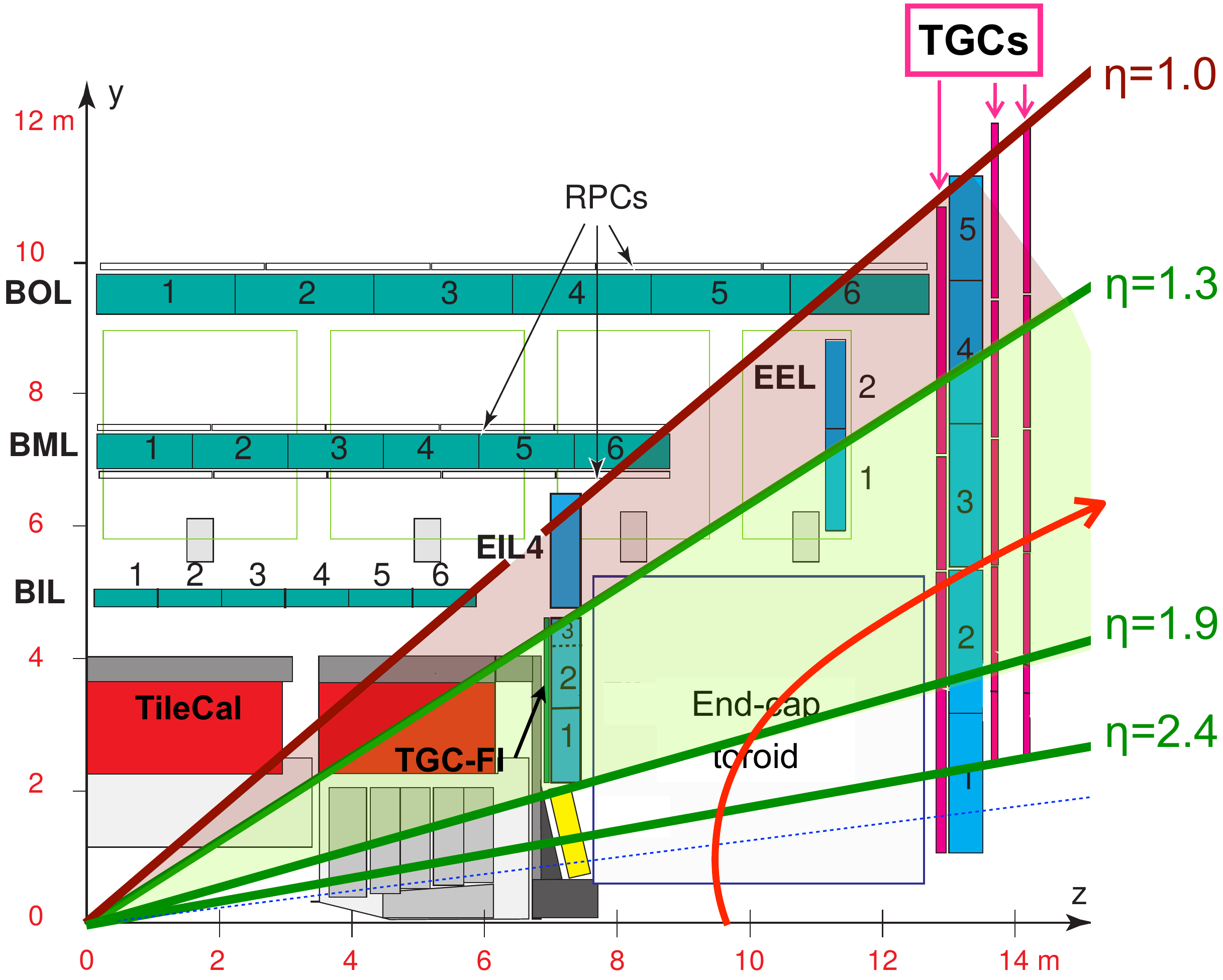}
\caption{A schematic view of the muon spectrometer with lines indicating
  various pseudorapidity regions. The curved arrow shows an example of a trajectory
  from slow particles generated at the beam pipe around $z\sim\SI{10}{\m}$.
  Triggers due to events of this type are mitigated by requiring an additional
  coincidence with the TGC-FI chambers in the region $1.3 < |\eta| < 1.9$.}
\label{fig:l1muon:FakeTrigger}
\end{figure}

\begin{figure}[htbp]
\centering
\subfloat[]{
  \includegraphics[width=0.5\textwidth]{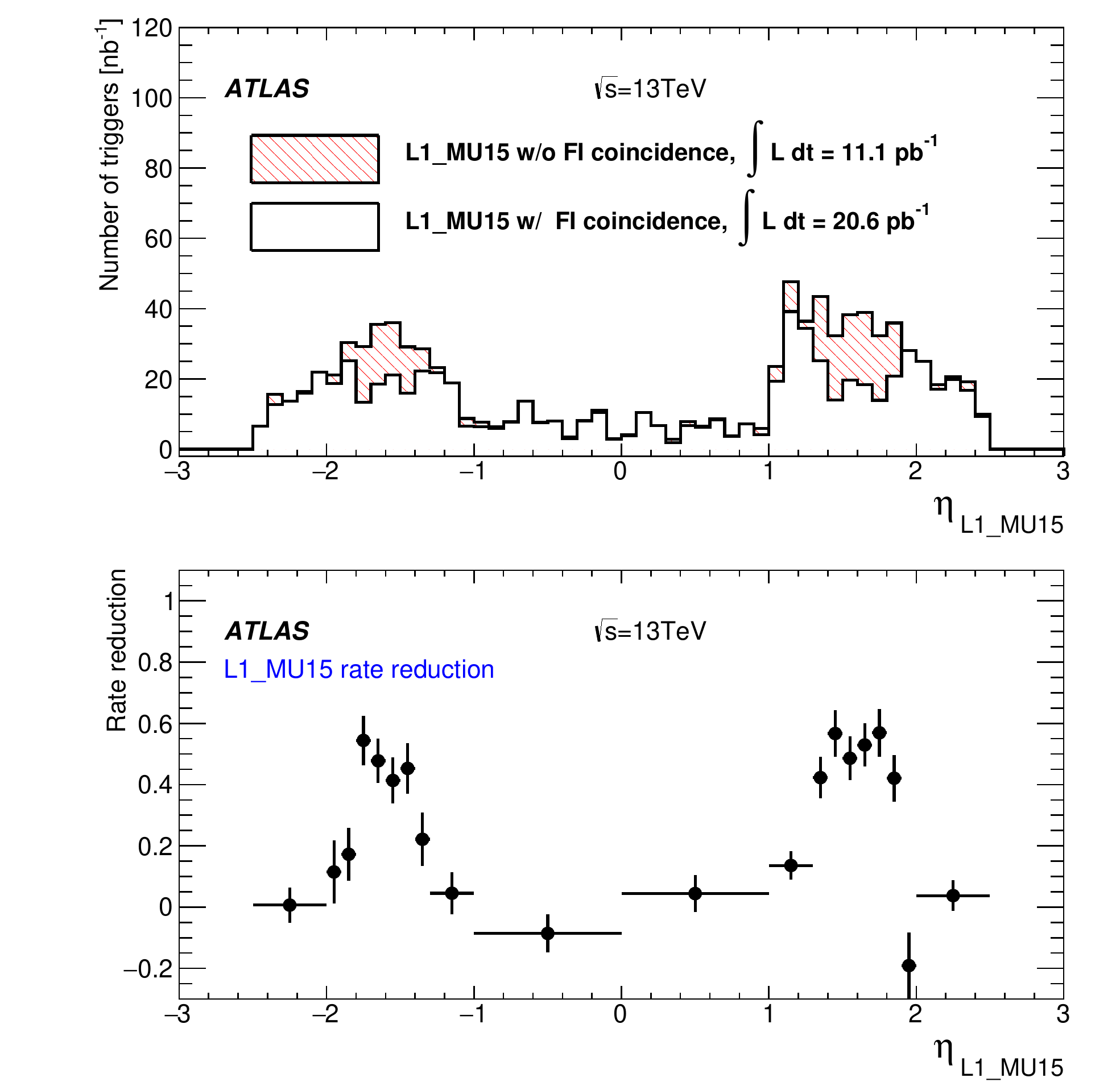}
  \label{fig:l1muon:MU15reduction:a}
}
\subfloat[]{
  \includegraphics[width=0.5\textwidth]{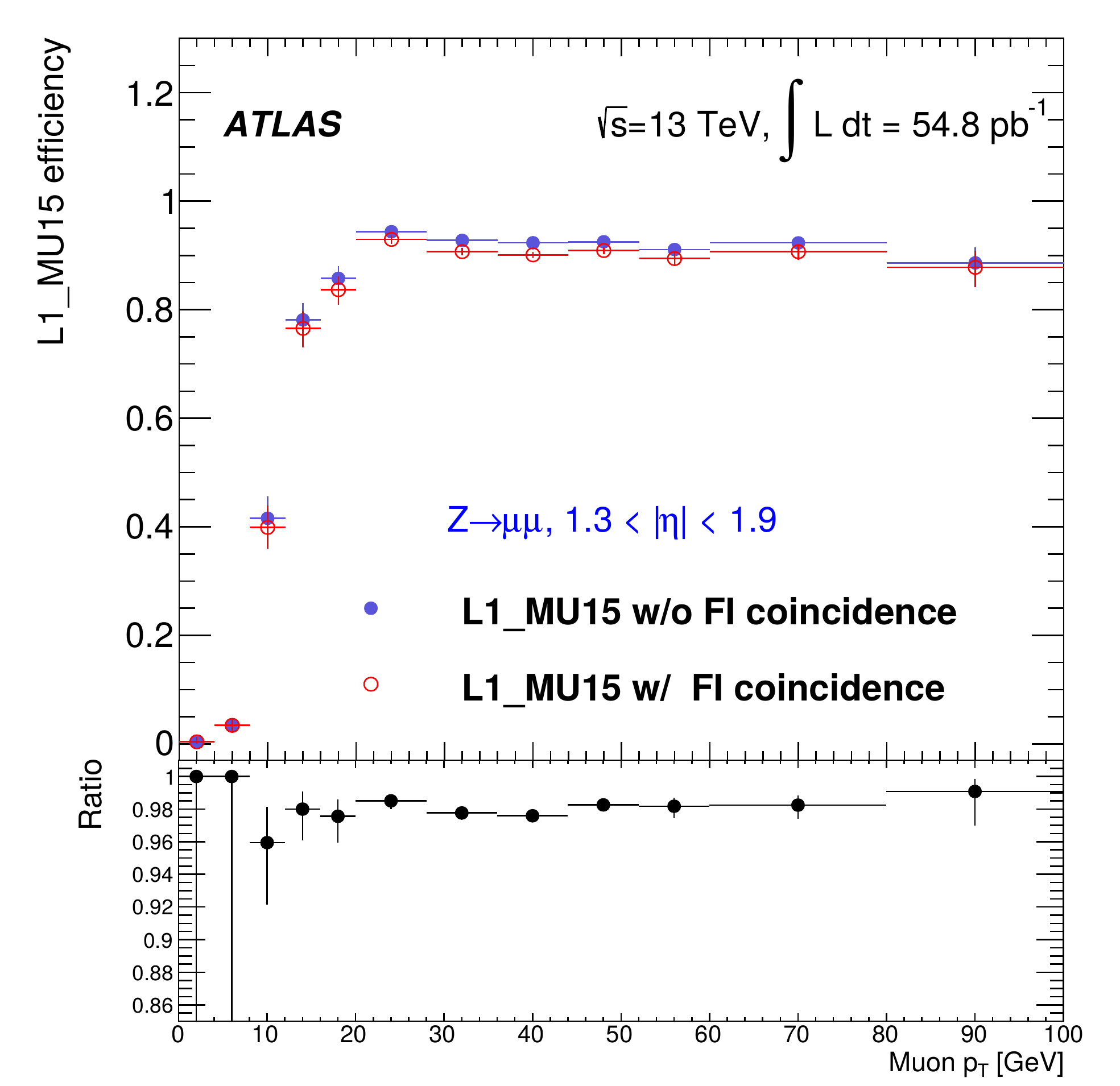}
  \label{fig:l1muon:MU15reduction:b}
}
\caption{(a) Number of events with an L1 muon trigger with transverse momentum (\pt) above \SI{15}{\GeV}
  (\trig{L1\_MU15}) as a function of the muon
  trigger $\eta$ coordinate, requiring a coincidence with
  the TGC-FI chambers (open histogram) and not requiring it (cross-hatched histogram), together with
  the fractional event rate reduction in the bottom plot. The event rate reduction
  in the regions with no TGC-FI chambers is consistent with zero within 
  the uncertainty. (b) Efficiency of \trig{L1\_MU15} in the
  end-cap region, as a function of the \pt\ of the offline muon measured via a
  tag-and-probe method (see Section~\ref{sec:sigperf}) using $Z\to \mu\mu$ events with (open dots) and without
  (filled dots) the TGC-FI coincidence, together with the ratio in the bottom panel.}
\end{figure}

During \runi, a significant fraction of the trigger rate from the end-cap region
was found to be due to particles not originating from the interaction point,
as illustrated  in Figure~\ref{fig:l1muon:FakeTrigger}. 
To reject these interactions, new trigger logic was introduced in \runii. An 
additional TGC coincidence requirement was deployed in 2015 covering the region
 $1.3 < |\eta| < 1.9$ (TGC-FI). Further coincidence logic in the region $1.0 < |\eta| < 1.3$
is being commissioned by requiring coincidence with the inner TGC chambers (EIL4) or
the Tile hadronic calorimeter.
Figure~\ref{fig:l1muon:MU15reduction:a} shows the muon trigger rate as a function
of the muon trigger pseudorapidity with and without the TGC-FI coincidence in separate
data-taking runs. The asymmetry as a function of $\eta$ is a result of the magnetic field direction
and the background particles being mostly positively charged.
In the region where this additional coincidence is applied, the trigger rate is
reduced by up to 60\% while only about 2\% of offline reconstructed muons are lost in
this region, as seen in Figure~\ref{fig:l1muon:MU15reduction:b}.

\section{Trigger menu}
\label{sec:menu}
The trigger menu defines the list of L1 and HLT triggers and consists of:
\begin{itemize}
\item \emph{primary} triggers, which are used for physics analyses and are typically unprescaled;
\item \emph{support} triggers, which are used for efficiency and performance measurements or for monitoring,
and are typically operated at a small rate (of the order of \SI{0.5}{\Hz} each) using prescale factors;
\item \emph{alternative} triggers, using alternative (sometimes experimental or new) reconstruction algorithms 
compared to the primary or support selections, and often heavily overlapping with the primary triggers;
\item \emph{backup} triggers, with tighter selections and lower expected rate;
\item \emph{calibration} triggers, which are used for detector calibration and are often operated at high rate
but storing very small events with only the relevant information needed for calibration.
\end{itemize}

The primary triggers cover all signatures relevant to the ATLAS physics
programme including electrons, photons, muons, tau leptons, ($b$-)jets and \met\,
which are used for Standard Model (SM) precision measurements including decays of the Higgs,
$W$ and $Z$ bosons, and searches for
physics beyond the SM such as heavy particles, supersymmetry or exotic
particles. A set of low transverse momentum (\pt) dimuon triggers is used to collect $B$-meson
decays, which are essential for the $B$-physics programme of ATLAS.

The trigger menu composition and trigger thresholds are
optimised for several luminosity ranges in order to maximise
the physics output of the experiment and to fit within the rate and bandwidth
constraints of the ATLAS detector, TDAQ system and offline computing. For
\runii the most relevant constraints are the maximum L1 rate of \SI{100}{\kHz} (\SI{75}{\kHz}
in \runi) defined by the ATLAS detector readout capability and an average HLT physics output rate of
\SI{1000}{\Hz} (\SI{400}{\Hz} in \runi) defined by the offline computing model. 
To ensure an optimal trigger menu within the rate constraints for a given LHC 
luminosity, prescale factors can be applied to
L1 and HLT triggers and changed during data-taking in such a way that triggers may
be disabled or only a certain fraction of events may be accepted by them.
Supporting triggers may be running at a constant rate or certain triggers enabled
later in the LHC fill when the luminosity and pile-up has reduced and the required resources are available.
Further flexibility is provided by \emph{bunch groups}, which allow triggers
to include specific requirements on the LHC proton bunches colliding in
ATLAS. These requirements include paired (colliding) bunch-crossings for physics
triggers, empty or unpaired crossings for background studies or search for long-lived particle decays, 
and dedicated bunch groups for detector calibration.

Trigger names used throughout this paper consist of the trigger level (\trig{L1}
or \trig{HLT}, the latter often omitted for brevity), multiplicity, particle
type (e.g.\  \trig{g} for photon, \trig{j} for jet, \trig{xe} for \MET, \trig{te} for $\sum$\ET~triggers) and \pt\ threshold
value in \si{\GeV} (e.g.\ \trig{L1_2MU4} requires at least two muons with
$\pt>\SI{4}{\GeV}$ at L1, \trig{HLT_mu40} requires at least one muon with
$\pt>\SI{40}{\GeV}$ at the HLT). L1 and HLT trigger items are written in upper case and
lower case letters, respectively. Each HLT trigger is configured with an L1 trigger as its seed. 
The L1 seed is not explicitly part of the trigger name except when an HLT trigger is seeded by 
more than one L1 trigger, in which case the L1 seed is denoted in the suffix of the
alternative trigger (e.g.\ \trig{HLT_mu20} and \trig{HLT_mu20_L1MU15} with the first one using \trig{L1_MU20} as its seed).
Further selection criteria (type of identification, isolation, reconstruction algorithm,
geometrical region) are suffixed to the trigger name
(e.g.\ \trig{HLT_g120_loose}).

\subsection{Physics trigger menu for 2015 data-taking}

The main goal of the trigger menu design was to maintain the unprescaled
single-electron and single-muon trigger \pt\ thresholds around \SI{25}{\GeV} despite
the expected higher trigger rates in \runii (see Section~\ref{sec:run2changes}).
This strategy ensures the collection of the majority
of the events with leptonic $W$ and $Z$ boson decays, which are the main source of events
for the study of electroweak processes. 
In addition, compared to using a large number of analysis-specific triggers, this
trigger strategy is simpler and more robust at the cost of slightly higher trigger output rates. 
Dedicated (multi-object) triggers were added for specific analyses not covered by the above.
Table~\ref{T:AtlasTriggerMenu} shows a comparison of selected primary trigger
thresholds for L1 and the HLT used during \runi and 2015 together with the typical
thresholds for offline reconstructed objects used in analyses (the latter are usually
defined as the \pt\ value at which the trigger efficiency reached the plateau). 
Trigger thresholds at
L1 were either kept the same as during \runi or slightly increased to fit within the allowed
maximum L1 rate of \SI{100}{\kHz}. At
the HLT, several selections were loosened compared to \runi or thresholds lowered thanks to the use
of more sophisticated HLT algorithms (e.g.\ multivariate analysis techniques for
electrons and taus).

\begin{table}[htbp]
\begin{center}
  \caption{Comparison of selected primary trigger thresholds (in \si{\GeV}) at the end of \runi
    and during 2015 together with typical offline requirements applied in analyses
    (the 2012 offline thresholds are not listed but have a similar relationship to the 2012 HLT thresholds).
    Electron and tau identification are assumed to fulfil the `medium' criteria
    unless otherwise stated. Photon and $b$-jet identification (`b') are
    assumed to fulfil the `loose' criteria. Trigger isolation is denoted by `i'. 
    The details of these selections are described in Section~\ref{sec:sigperf}.}
\begin{tabular}{l|cc|ccc}\hline\hline
\label{T:AtlasTriggerMenu}
Year            & \multicolumn{2}{c|}{2012}                  & \multicolumn{3}{c}{2015}                                        \\
$\sqrt{s}\,$    & \multicolumn{2}{c|}{\SI{8}{\TeV}}                & \multicolumn{3}{c}{\SI{13}{\TeV}}                                     \\
Peak luminosity & \multicolumn{2}{c|}{$7.7\times\lumi{e33}$} & \multicolumn{3}{c}{$5.0\times\lumi{e33}$}                       \\
\hline
                & \multicolumn{5}{c}{\pt~threshold$\,$[\si{\GeV}], criteria}                                                       \\ 
Category        & L1                                         & HLT               & L1            & HLT               & Offline \\
\hline
Single electron & 18                                         & 24i               & 20            & 24                & 25      \\
Single muon     & 15                                         & 24i               & 15            & 20i               & 21      \\
Single photon   & 20                                         & 120               & 22i           & 120               & 125     \\ 
Single tau      & 40                                         & 115               & 60            & 80                & 90      \\  
Single jet      & 75                                         & 360               & 100           & 360               & 400     \\
Single $b$-jet  & n/a                                        & n/a               & 100           & 225               & 235     \\
\MET            & 40                                         & 80                & 50            & 70                & 180     \\
\hline
Dielectron     & 2$\times$10                                & 2$\times$12,loose & 2$\times$10   & 2$\times$12,loose & 15      \\
Dimuon         & 2$\times$10                                & 2$\times$13       & 2$\times$10   & 2$\times$10       & 11      \\
Electron, muon  & 10, 6                                      & 12, 8             & 15, 10        & 17, 14            & 19, 15  \\
Diphoton       & 16, 12                                     & 35, 25            & 2$\times$15   & 35, 25            & 40, 30  \\
Ditau          & 15i, 11i                                   & 27, 18            & 20i, 12i       & 35, 25           & 40, 30  \\
Tau, electron   & 11i, 14                                    & 28i, 18           & 12i(+jets), 15 & 25, 17i          & 30, 19 \\
Tau, muon       & 8, 10                                      & 20, 15            & 12i(+jets), 10 & 25, 14           & 30, 15 \\
Tau, \MET       & 20, 35                                     & 38, 40            & 20, 45(+jets)  & 35, 70           & 40, 180 \\
\hline
Four jets       & 4$\times$15                                & 4$\times$80       & 3$\times$40   & 4$\times$85       & 95      \\ 
Six jets        & 4$\times$15                                & 6$\times$45       & 4$\times$15   & 6$\times$45       & 55      \\
Two $b$-jets    & 75                                & 35b,145b       & 100   & 50b,150b       & 60      \\
Four(Two) ($b$-)jets   & 4$\times$15                                & 2$\times$35b, 2$\times$35       & 3$\times$25   &  2$\times$35b, 2$\times$35      & 45      \\
\hline
$B$-physics (Dimuon)     & 6, 4                                       & 6, 4              & 6, 4          & 6, 4              & 6, 4    \\
\hline\hline
\end{tabular}
\end{center}
\end{table}

\begin{figure}[htbp]
\centering
\subfloat[]{
  \includegraphics[width=0.5\textwidth]{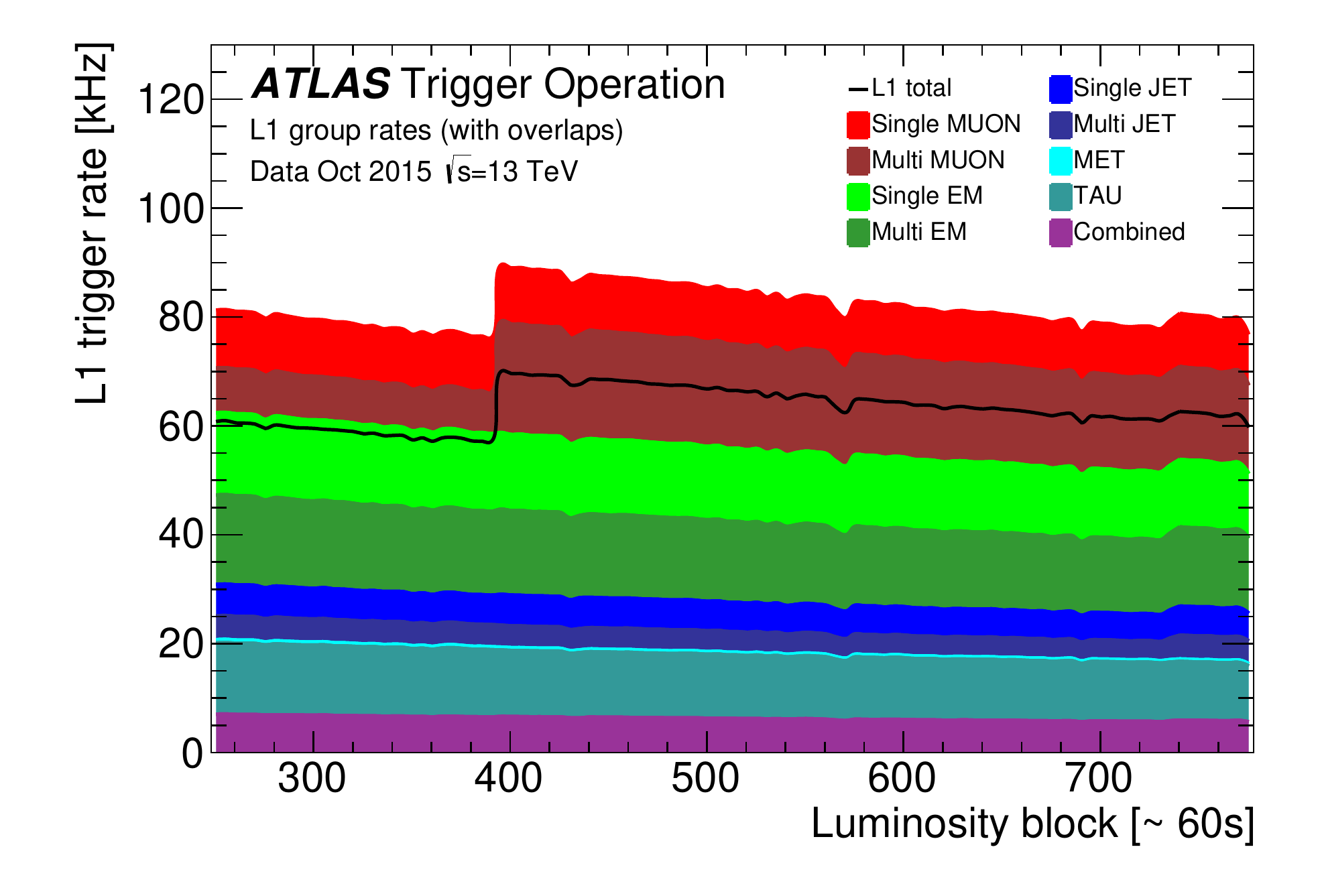}
  \label{fig:l1rates}
}
\subfloat[]{
  \includegraphics[width=0.5\textwidth]{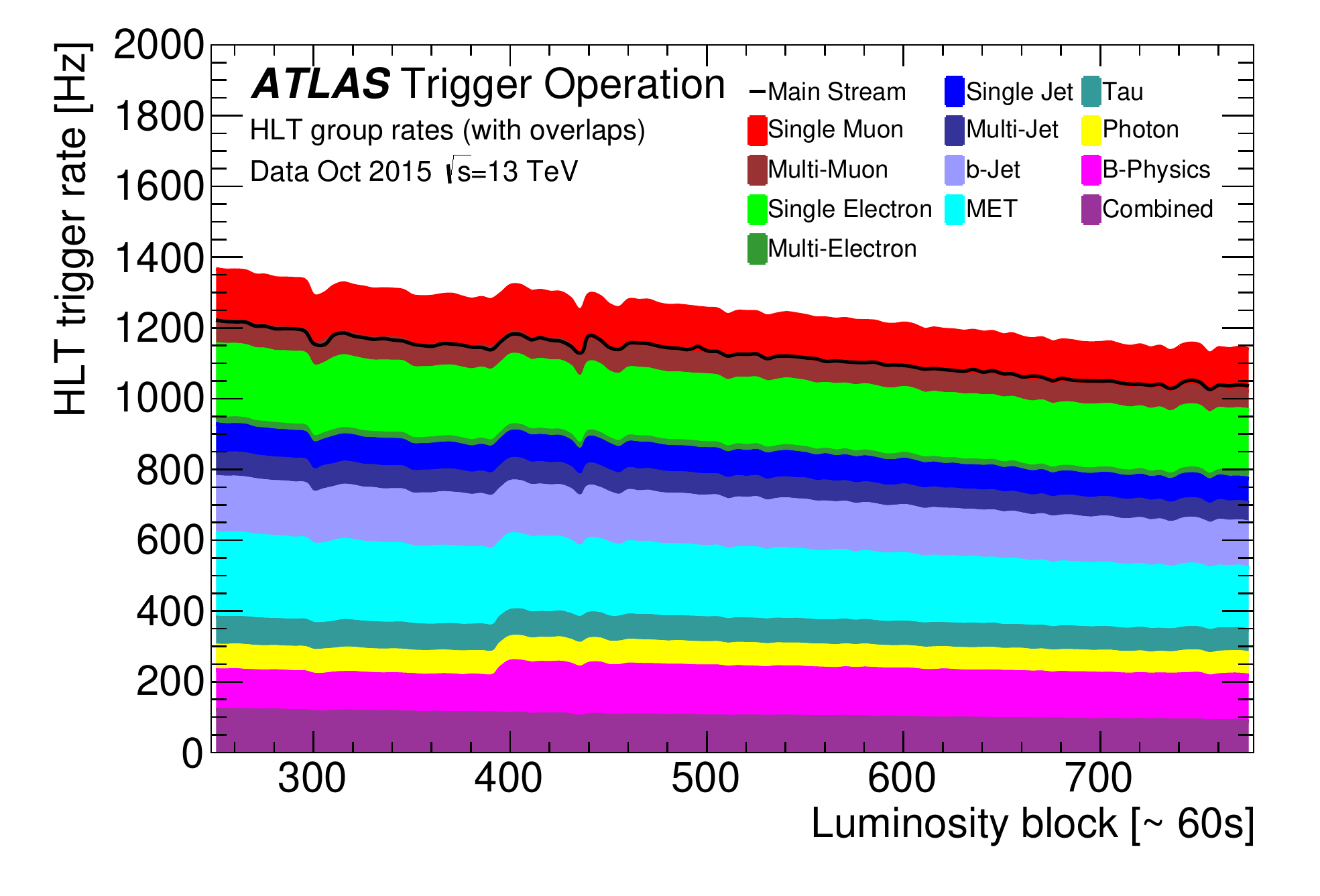}
  \label{fig:hltrates}
}
\caption{(a) L1 and (b) HLT trigger rates grouped by trigger signature 
  during an LHC fill in October 2015 with a peak luminosity of $4.5\times\lumi{e33}$.
  Due to overlaps the sum of the individual groups is higher than the 
  (a) L1 total rate and (b) \emph{Main} physics stream rate, which are shown as black lines.
  Multi-object triggers are included in the $b$-jets and tau groups.
  The rate increase around luminosity block 400 is due to the removal of prescaling of the $B$-physics triggers.
  The combined group includes multiple triggers combining
  different trigger signatures such as electrons with muons, taus, jets or \met.}
\label{fig:l1hltrates}
\end{figure}

Figures~\ref{fig:l1rates} and~\ref{fig:hltrates} show the L1 and HLT trigger
rates grouped by signatures during an LHC fill with a peak luminosity of
$4.5\times\lumi{e33}$. The preventive dead-time\footnote{The four complex
  dead-time settings were 15/370, 42/381, 9/351 and 7/350, where the first
  number specifies the number of triggers and the second number specifies the
  number of bunch-crossings, e.g. 7 triggers in 350 bunch-crossings.}
The single-electron and single-muon triggers
contribute a large fraction to the total rate. While running at these relatively
low luminosities it was possible to dedicate a large fraction of the bandwidth to
the $B$-physics triggers. Support triggers contribute about 20$\%$ of the total
rate. Since the time for trigger commissioning in 2015 was limited due to the
fast rise of the LHC luminosity (compared to \runi), several backup triggers, which
contribute additional rate, were implemented in the menu in addition to the primary
physics triggers. This is the case for electron, $b$-jet and \MET~triggers,
which are discussed in later sections of the paper. 

\subsection{Event streaming}

\begin{figure}[htbp]
\centering
\subfloat[]{\includegraphics[width=0.5\textwidth]{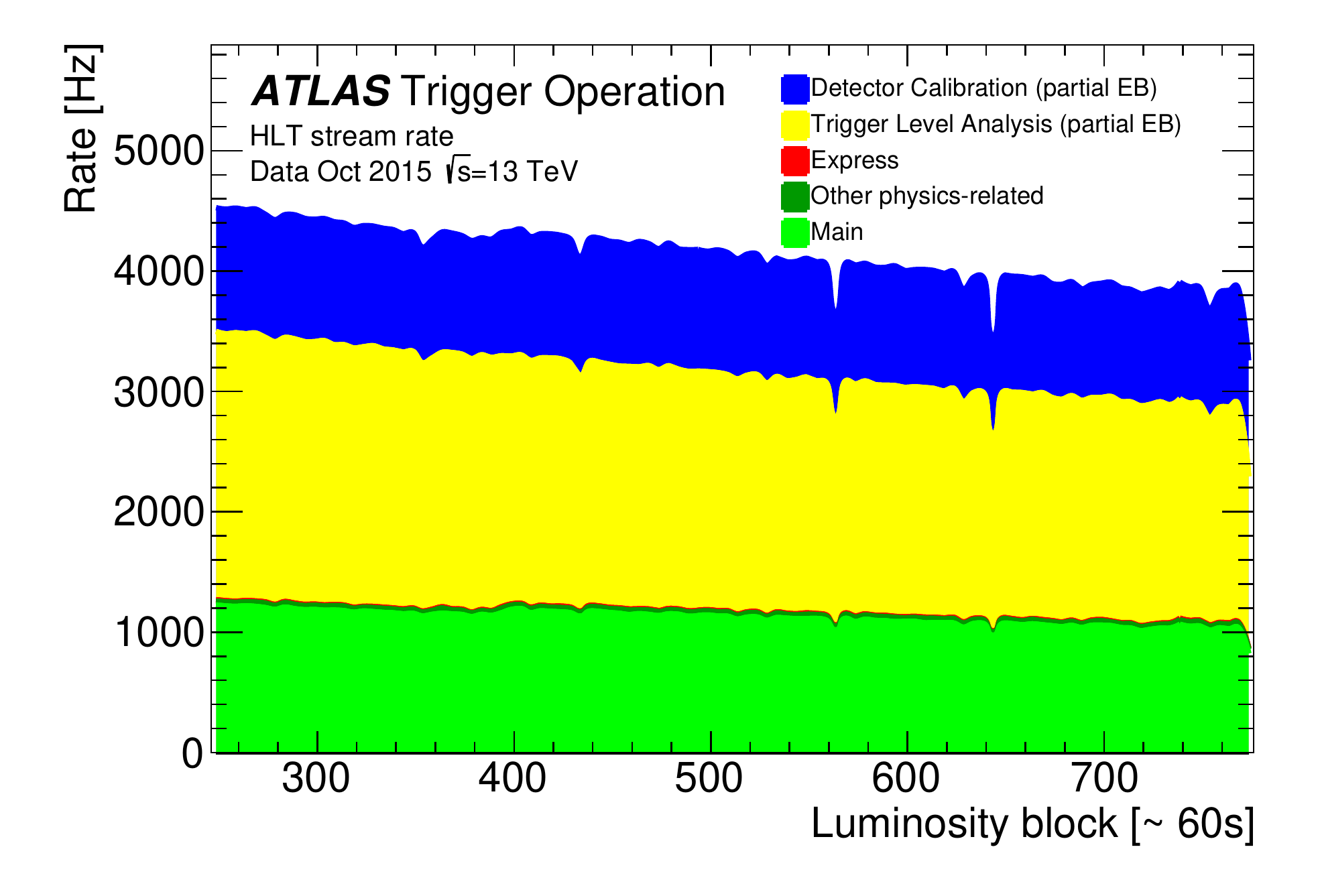}}
\subfloat[]{\includegraphics[width=0.5\textwidth]{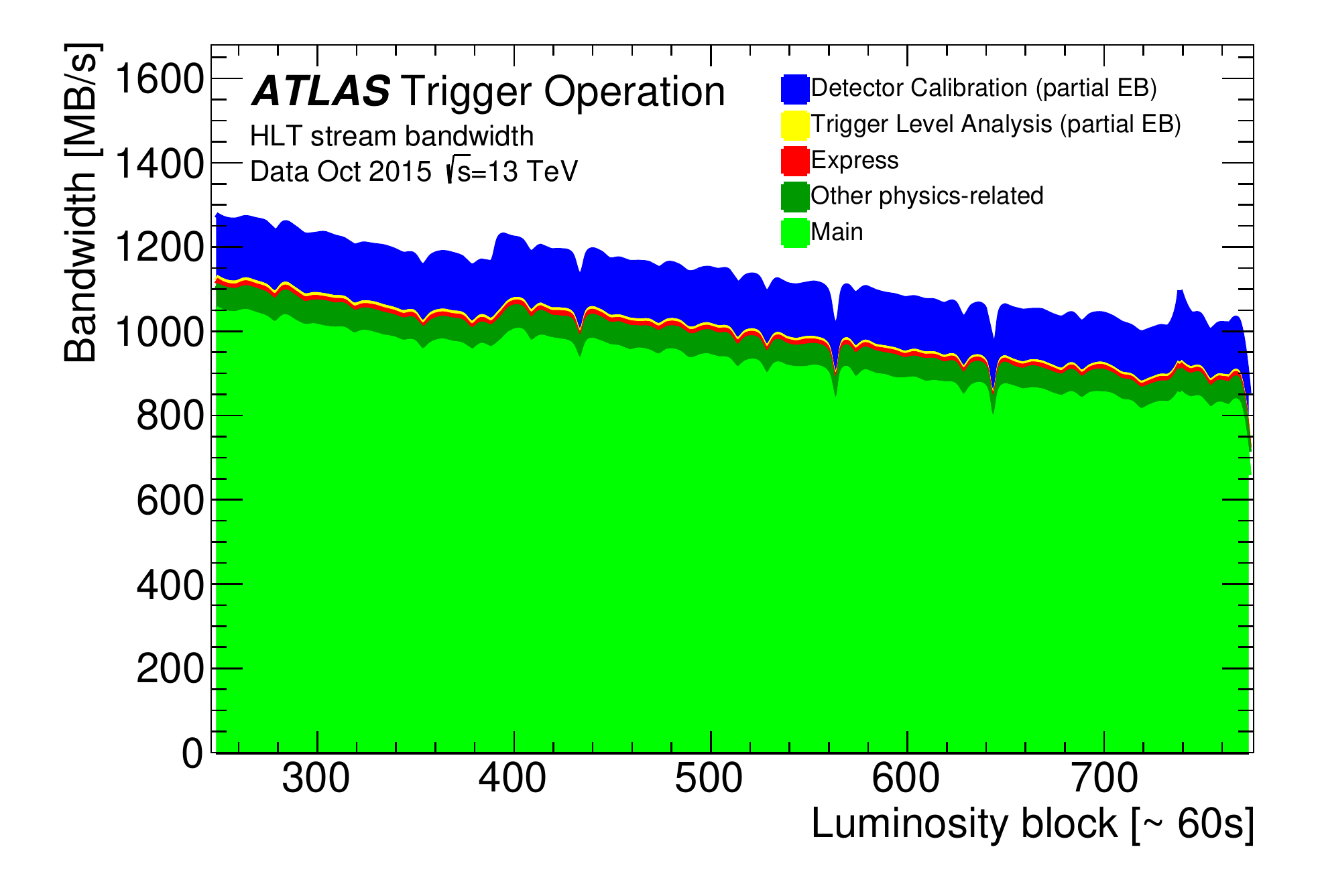}}
\caption{(a) HLT stream rates and (b) bandwidth during 
  an LHC fill in October 2015 with a peak luminosity of $4.5\times\lumi{e33}$.
  Partial Event Building (partial EB) streams only store relevant
  subdetector data and thus have smaller event sizes. The other physics-related streams
  contain events with special readout settings and are used to overlay with MC events to simulate pile-up.}
\label{fig:streamrates}
\end{figure}

Events accepted by the HLT are written into separate data \emph{streams}. Events for
physics analyses are sent to a single \emph{Main} stream replacing the three
separate physics streams (\emph{Egamma, Muons, JetTauEtMiss}) used in \runi. This change
reduces event duplication, thus reducing storage and CPU resources required for
reconstruction by roughly 10\%. A small fraction of these events at 
a rate of \SIrange{10}{20}{\Hz} are also written
to an \emph{Express} stream that is reconstructed promptly offline and used to provide
calibration and data quality information prior to the reconstruction of the
full \emph{Main} stream, which typically happens 36 hours after the data are taken.
In addition, there are about twenty additional streams for
calibration, monitoring and detector performance studies.
To reduce event size, some of these streams
use partial event building (partial EB), which writes only a predefined subset of the
ATLAS detector data per event.
For \runii, events that contain only HLT reconstructed objects, but no ATLAS detector data, can
be recorded to a new type of stream. These events are of very
small size, allowing recording at high rate.
These streams are used for calibration purposes and \emph{Trigger-Level Analysis}
as described in Section~\ref{sec:TLA}.
Figure~\ref{fig:streamrates} shows typical HLT stream rates and bandwidth during an LHC fill.

Events that cannot be properly processed at the HLT or have other DAQ-related problems
are written to dedicated \emph{debug} streams. These events are reprocessed offline with the same HLT 
configuration as used during data-taking and accepted events are stored into separate data sets
for use in physics analyses. In 2015, approximately \num{339000} events were written to debug streams.
The majority of them ($\sim 90\%$) are due to online processing timeouts that occur when the event cannot
be processed within 2--3 minutes. Long processing times are mainly due to muon
algorithms processing events with a large number of tracks in the muon spectrometer
(e.g.\ due to jets not contained in the calorimeter). 
During the debug stream reprocessing, \num{330000} events were successfully processed by the HLT
of which about 85\% were accepted.
The remaining \num{9000} events could not be processed due to data integrity issues.

\subsection{HLT processing time}

\begin{figure}[htbp]
\centering
\captionsetup[subfloat]{}
\subfloat[]{
  \includegraphics[width=0.5\textwidth]{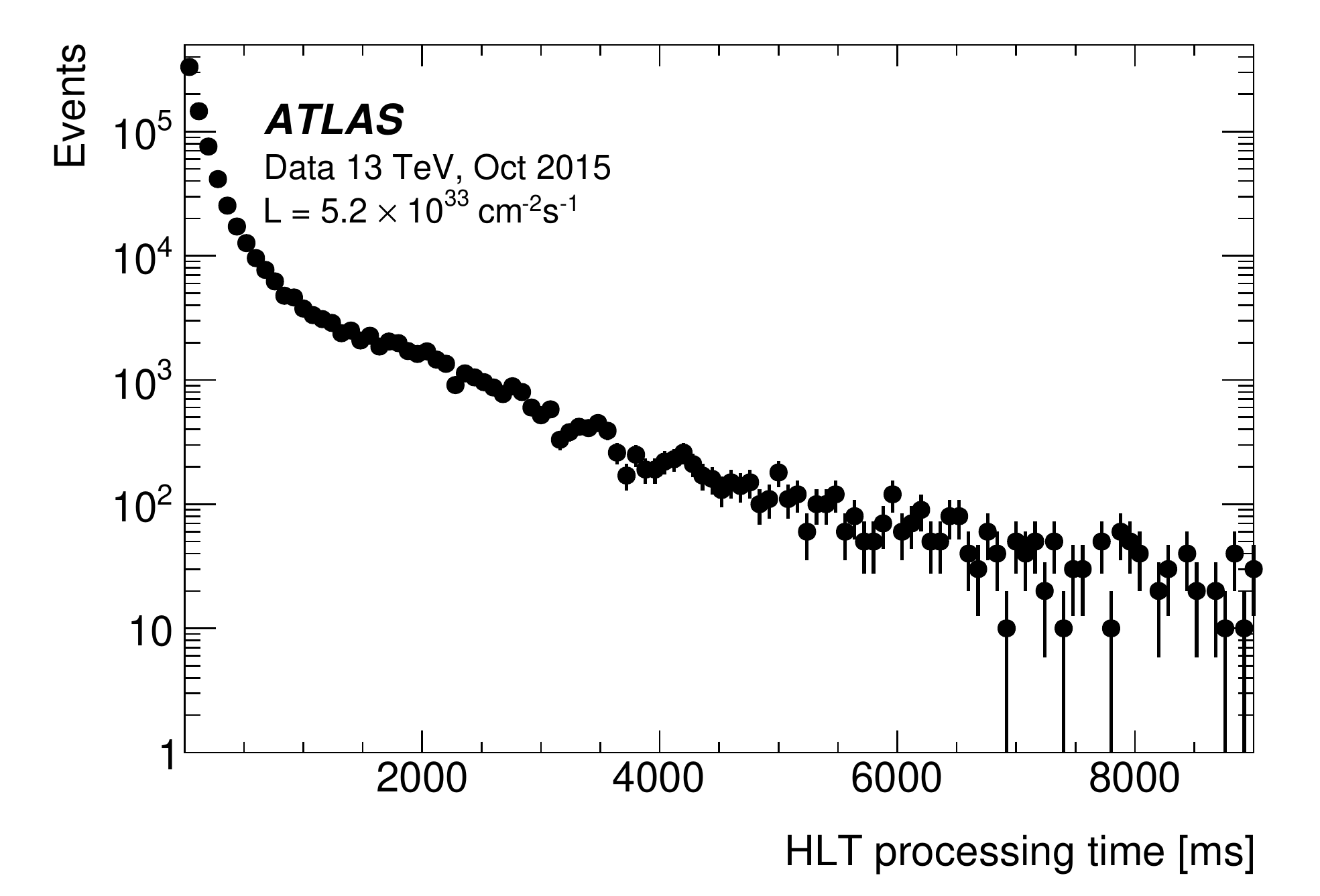}
  \label{fig:proctime_a}
}
\subfloat[]{
  \includegraphics[width=0.5\textwidth]{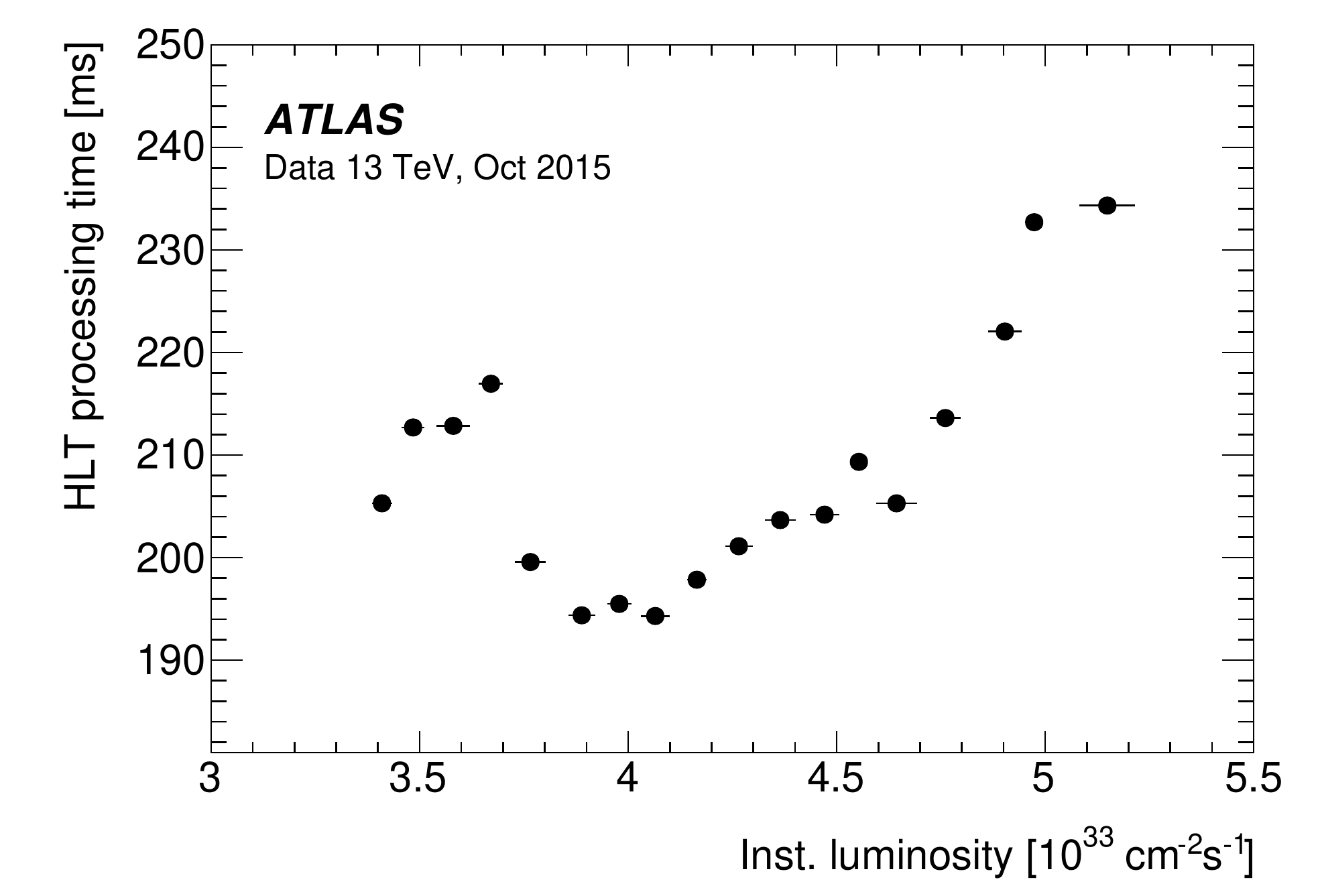}
  \label{fig:proctime_b}
}
\caption{(a) HLT processing time distribution per event for an instantaneous
  luminosity of $5.2\times\lumi{e33}$ and average pile-up $\langle\mu\rangle=15$
  and (b) mean HLT processing time as a function of the instantaneous
  luminosity.}
\label{fig:proctime}
\end{figure}

The HLT processing time per event is mainly determined by the trigger menu and
the number of pile-up interactions. The HLT farm CPU utilisation depends on the L1
trigger rate and the average HLT processing time. Figure~\ref{fig:proctime} shows (a) the HLT
processing time distribution for the highest luminosity run in 2015 with a peak
luminosity of $5.2\times\lumi{e33}$ and (b) the average HLT processing time 
as a function of the instantaneous luminosity. At the highest luminosity point the average event processing
time was approximately \SI{235}{\ms}. An L1 rate of \SI{80}{\kHz} corresponds to an average utilisation
of 67\% of a farm with \num{28000} available CPU cores. About 40\%, 35\% and 15\% of the
processing time are spent on inner detector tracking, muon spectrometer
reconstruction and calorimeter reconstruction, respectively.  The muon
reconstruction time is dominated by the large rate of low-\pt $B$-physics triggers.
The increased processing time at low luminosities observed in
Figure~\ref{fig:proctime_b} is due to additional triggers being enabled
towards the end of an LHC fill to take advantage of the available CPU and
bandwidth resources. Moreover, trigger prescale changes are made throughout the run
giving rise to some of the observed features in the curve. The clearly visible
scaling with luminosity is due to the pileup dependence of the processing time.
It is also worth noting that the processing time cannot naively be scaled
to higher luminosities as the trigger menu changes significantly in order to
keep the L1 rate below or at \SI{100}{\kHz}.

\subsection{Trigger menu for special data-taking conditions}

Special trigger menus are used for particular data-taking conditions and can
either be required for collecting a set of events for dedicated measurements or
due to specific LHC bunch configurations. In the following, three examples of dedicated
menus are given: menu for low number of bunches in the LHC, menu for collecting 
enhanced minimum-bias data for trigger rate predictions and menu during beam separation scans for luminosity
calibration (van der Meer scans).

\begin{figure}[htbp]
\centering
\includegraphics[trim=50 50 40 90,clip,width=0.5\textwidth]{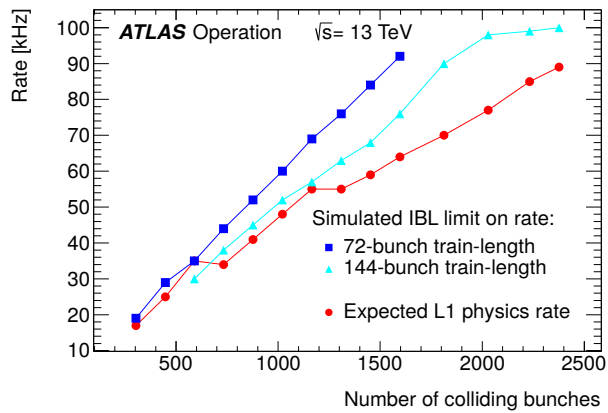}
\caption{Simulated limits on the L1 trigger rate due to the IBL fixed frequency veto for two
different filling schemes and the expected maximum L1 rate from rate
predictions. The steps in the latter indicate a change in the prescale
strategy. The simulated rate limit is confirmed with experimental tests. The rate
limit is higher for the 72-bunch train configuration since the bunches are more equally spread
across the LHC ring. 
The rate limitation was only crucial for the low luminosity phase, where the
required physics L1 rate was higher than the limit imposed by the IBL
veto. The maximum number of colliding bunches in 2015 was 2232.}
\label{fig:iblrates}
\end{figure}

When the LHC contains a low number of bunches (and thus few bunch trains), care is
needed not to trigger at resonant frequencies that could damage the wire bonds
of the IBL or SCT detectors, which reside in the magnetic field. The dangerous resonant
frequencies are between 9 and \SI{25}{\kHz} for the IBL and above \SI{100}{\kHz} for the SCT
detector. To avoid this risk, both detectors have implemented
in the readout firmware a so-called fixed frequency veto that 
prevents triggers falling within a dangerous frequency range~\cite{Barber:2005ct}.
The IBL veto poses the most stringent limit on the acceptable L1 rate in this
LHC configuration.
In order to provide trigger menus appropriate to each LHC configuration during
the startup phase, the trigger rate has been estimated after simulating the effect of the IBL veto.
Figure~\ref{fig:iblrates} shows the simulated IBL rate
limit for two different bunch configurations and the expected L1 trigger
rate of the nominal physics trigger menu.
At a low number of bunches the expected L1 trigger rate exceeds slightly the
allowed L1 rate imposed by the IBL veto. In order not to veto important physics triggers,
the required rate reduction was achieved by reducing the rate of supporting triggers.

Certain applications such as trigger algorithm development, rate predictions and
validation require a data set that is minimally biased by the triggers used to
select it. This special data set is collected using the enhanced minimum-bias
trigger menu, which consists of all primary lowest-\pt\ L1 triggers with increasing \pT\
threshold and a random trigger for very high cross-section processes. This trigger menu can
be enabled in addition to the regular physics menu and records events at \SI{300}{\Hz} for a
period of approximately one hour to obtain a data set of around one~million
events. Since the correlations between triggers are preserved, per-event weights can be
calculated and used to convert the sample into a zero-bias sample,
which is used for trigger rate predictions during the development of new
triggers~\cite{ATL-DAQ-PUB-2016-002}. This approach requires a much smaller total
number of events than a true zero-bias data set.

During van der Meer scans~\cite{vanderMeer:296752}, which are performed by
the LHC to allow the experiments to calibrate their luminosity measurements, a
dedicated trigger menu is used. ATLAS uses several luminosity algorithms (see
Ref.~\cite{DAPR-2013-01}) amongst which one relies on counting tracks
in the ID. Since the different LHC bunches do not have the exact same proton
density, it is beneficial to sample a few bunches at the maximum possible rate.
For this purpose, a minimum-bias trigger selects events for specific LHC bunches
and uses partial event building to read out only the ID data
at about \SI{5}{\kHz} for five different LHC bunches.



\FloatBarrier
\section{High-level trigger reconstruction}
\label{sec:hltrec}
After L1 trigger acceptance, the events are processed by the HLT using finer-granularity 
calorimeter information, precision measurements from the MS and tracking information
from the ID, which are not available at L1.
As needed, the HLT reconstruction can either be executed
within RoIs identified at L1 or for the full detector. In both cases the data is
retrieved on demand from the readout system. As in \runi, in order to reduce the 
processing time, most HLT triggers
use a two-stage approach with a fast first-pass reconstruction to reject the
majority of events and a slower precision reconstruction for the remaining
events. However, with the merging of the previously separate L2 and EF farms, there
is no longer a fixed bandwidth or rate limitation between the two steps. The
following sections describe the main reconstruction algorithms used in the HLT
for inner detector, calorimeter and muon reconstruction.


\subsection{Inner detector tracking}
\label{sec:id}

\newcommand{\ptoff}{\ensuremath{p_{\text{T,offline}}}\xspace}

For \runi the ID tracking in the trigger consisted of custom tracking algorithms
at L2 and offline tracking algorithms adapted for running in the EF. The ID
trigger was redesigned for \runii to take advantage of the merged HLT and
include information from the IBL. The latter significantly 
improves the tracking performance and in particular the impact parameter resolution~\cite{B-layerRef}.
In addition, provision was made for the inclusion of FTK tracks once that 
system becomes available later in \runii.

\subsubsection{Inner detector tracking algorithms}

The tracking trigger is subdivided into \emph{fast tracking} and \emph{precision tracking}
stages. The fast tracking consists of trigger-specific pattern recognition algorithms very similar to those
used at L2 during \runi, whereas the precision stage relies heavily on offline tracking algorithms. 
Despite similar naming the fast tracking as described here is not related to the FTK hardware tracking that 
will only become available during 2017.
The tracking algorithms are typically configured to run within an RoI identified by L1.
The offline tracking was reimplemented in LS1 to run three times faster than in
\runi, making it more suitable to use in the HLT.
To reduce CPU usage even further, the offline track-finding is seeded by
tracks and space-points identified by the fast tracking stage.

\subsubsection{Inner detector tracking performance}

\begin{figure}[htbp]
\centering
\subfloat[]{\includegraphics[width=0.5\textwidth]{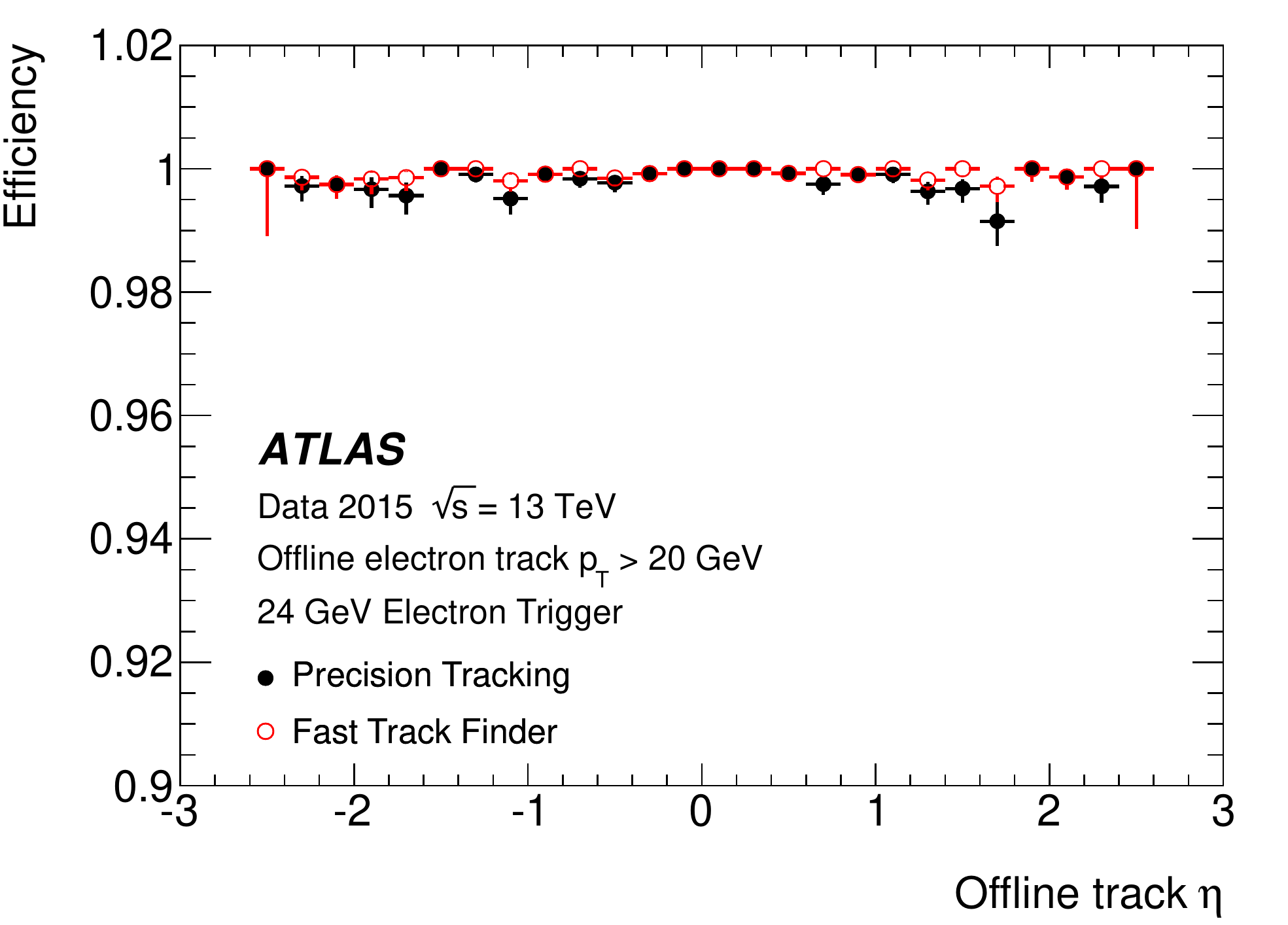}}
\subfloat[]{\includegraphics[width=0.5\textwidth]{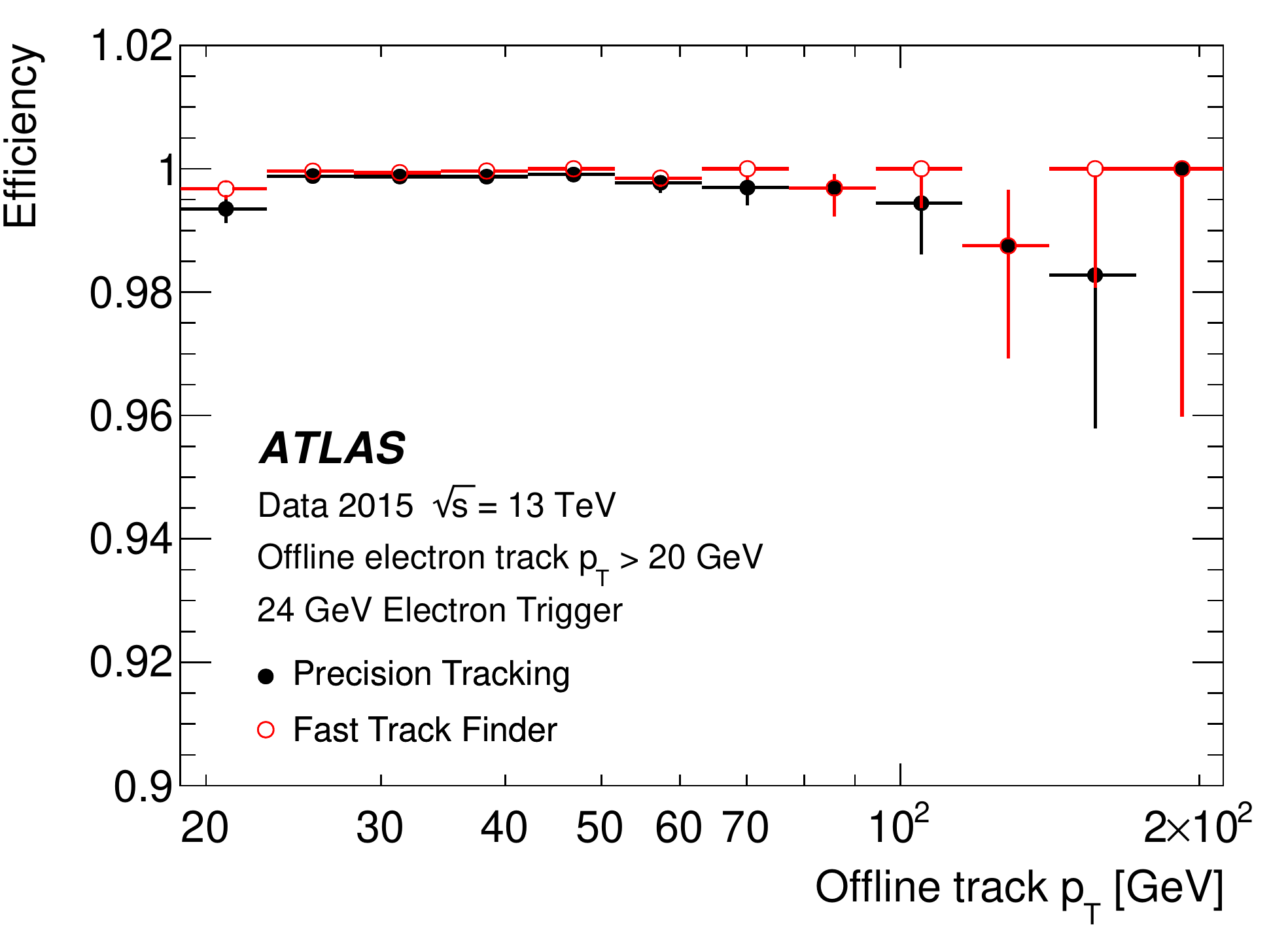}}
\caption{The ID tracking efficiency for the \SI{24}{\GeV} electron trigger is
  shown as a function of the (a) $\eta$ and (b) $\pt$ of the track of the offline electron candidate.
  Uncertainties based on Bayesian statistics are shown.}
\label{fig:idtriggerefficiecny}
\end{figure}

The tracking efficiency with respect to offline tracks has been determined for
electrons and muons. The reconstructed tracks are required to have at least two
(six) pixel (SCT) clusters and lie in the region $|\eta|<2.5$. The closest
trigger track within a cone of size
$\Delta R = \sqrt{(\Delta\eta)^2+(\Delta\phi)^2} = 0.05$ of the offline
reconstructed track is selected as the matching trigger track.

Figure~\ref{fig:idtriggerefficiecny} shows the tracking efficiency for the
\SI{24}{\GeV} medium electron trigger (see Section~\ref{sec:egamma}) as a
function of the $\eta$ and of the \pt of the offline track.  The tracking
efficiency is measured with respect to offline tracks with $\pt>\SI{20}{\GeV}$
for tight offline electron candidates from the \SI{24}{\GeV} electron support
trigger, which does not use the trigger tracks in the selection, but is
otherwise identical to the physics trigger. The efficiencies of the fast track
finder and precision tracking exceed 99\% for all pseudorapidities.
There is a small efficiency loss at low \pt\ due to bremsstrahlung energy loss
by electrons.

\begin{figure}[htbp]
\centering
\subfloat[]{
  \includegraphics[width=0.5\textwidth]{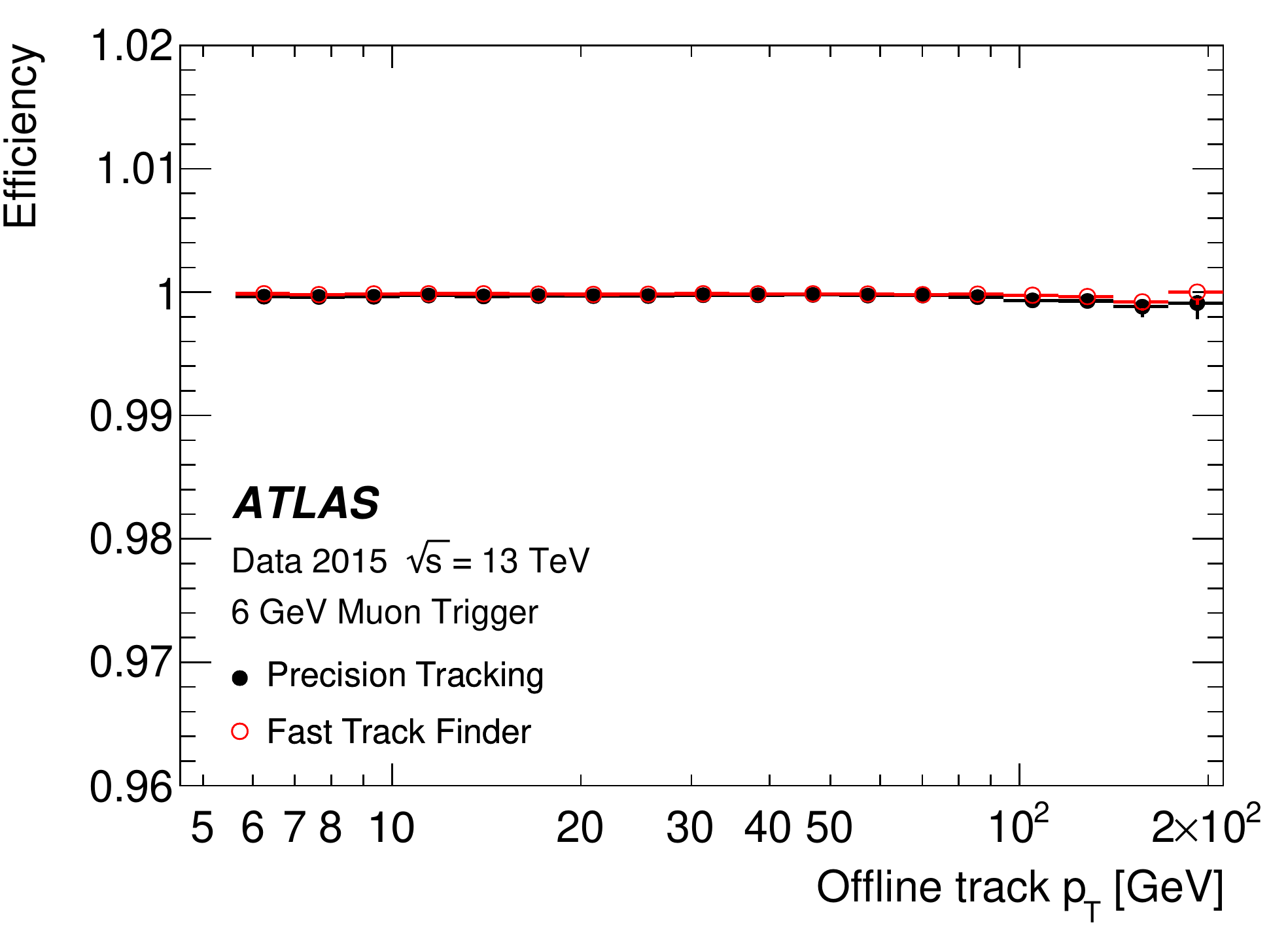}
  \label{fig:idmuontriggera}
}
\subfloat[]{
  \includegraphics[width=0.5\textwidth]{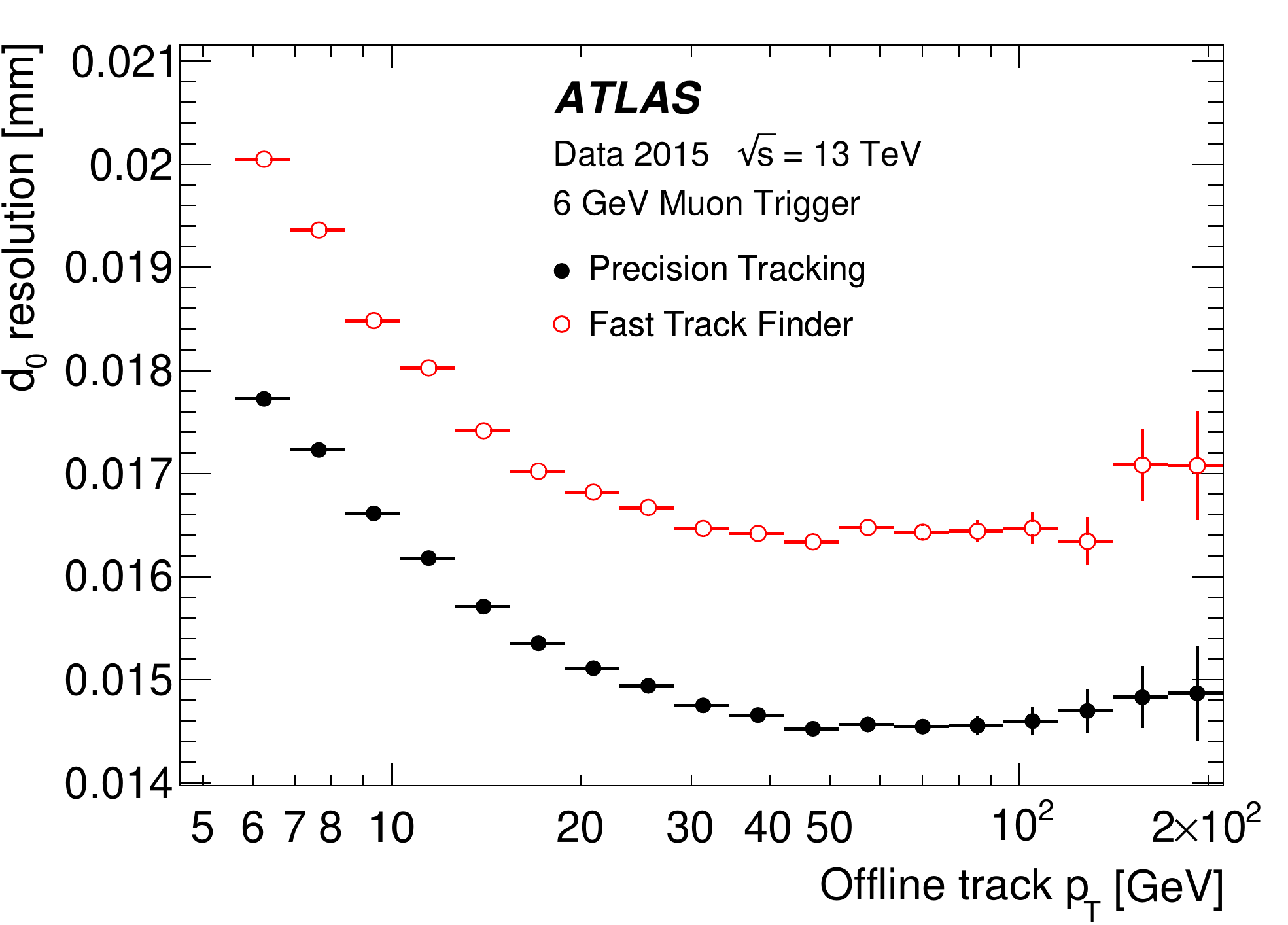}
  \label{fig:idmuontriggerb}
}
\caption{ 
The ID tracking performance for the \SI{6}{\GeV} muon trigger;
(a) efficiency as a function of the offline reconstructed muon \pt,  
(b) the resolution of the transverse impact parameter, $d_{0}$  as a function of the offline reconstructed muon \pt. Uncertainties based on Bayesian statistics are shown.}
\label{fig:idmuontrigger}
\end{figure}

Figure~\ref{fig:idmuontriggera} shows the tracking performance of the ID trigger 
for muons with respect to loose offline muon candidates with $\pt>\SI{6}{\GeV}$
selected by the \SI{6}{\GeV} muon support trigger as a function of the offline muon
transverse momentum. The efficiency is significantly better than 99\% for all
\pt\ for both the fast and precision tracking. Shown in
Figure~\ref{fig:idmuontriggerb} is the resolution of the transverse track impact
parameter with respect to offline as a function of the offline muon \pt. The
resolution in the fast (precision) tracking is better than \SI{17}{\um}
(\SI{15}{\um}) for muon candidates with offline $\pt > \SI{20}{\GeV}$.

\subsubsection{Multiple stage tracking}
\label{sec:MultipleStageTracking}

For the hadronic tau and $b$-jet triggers, tracking is run in a larger RoI than
for electrons or muons. To limit CPU usage, multiple stage track reconstruction
was implemented.

\begin{figure}[htbp]
\centering
  \includegraphics[width=0.65\textwidth]{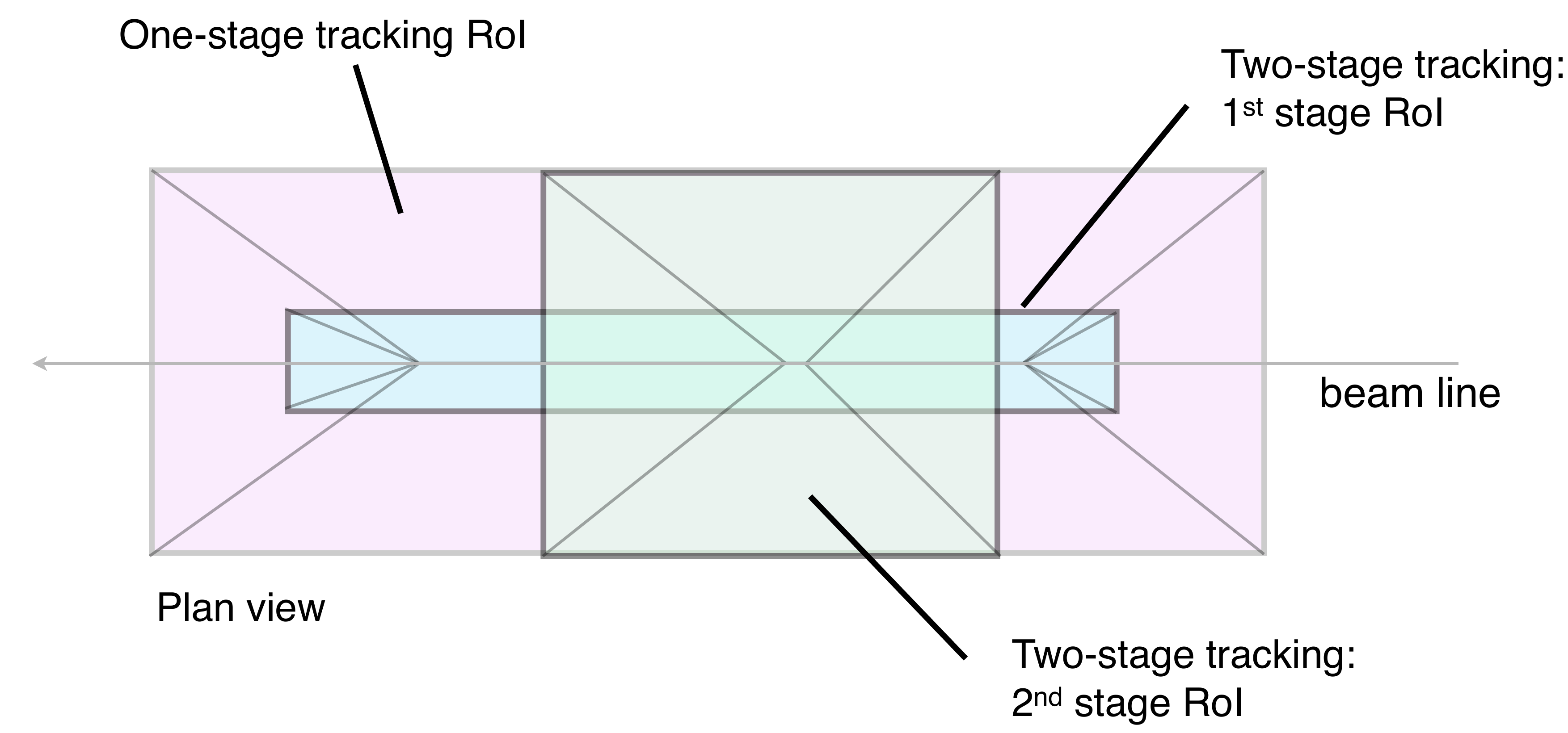}
  \includegraphics[width=0.3\textwidth]{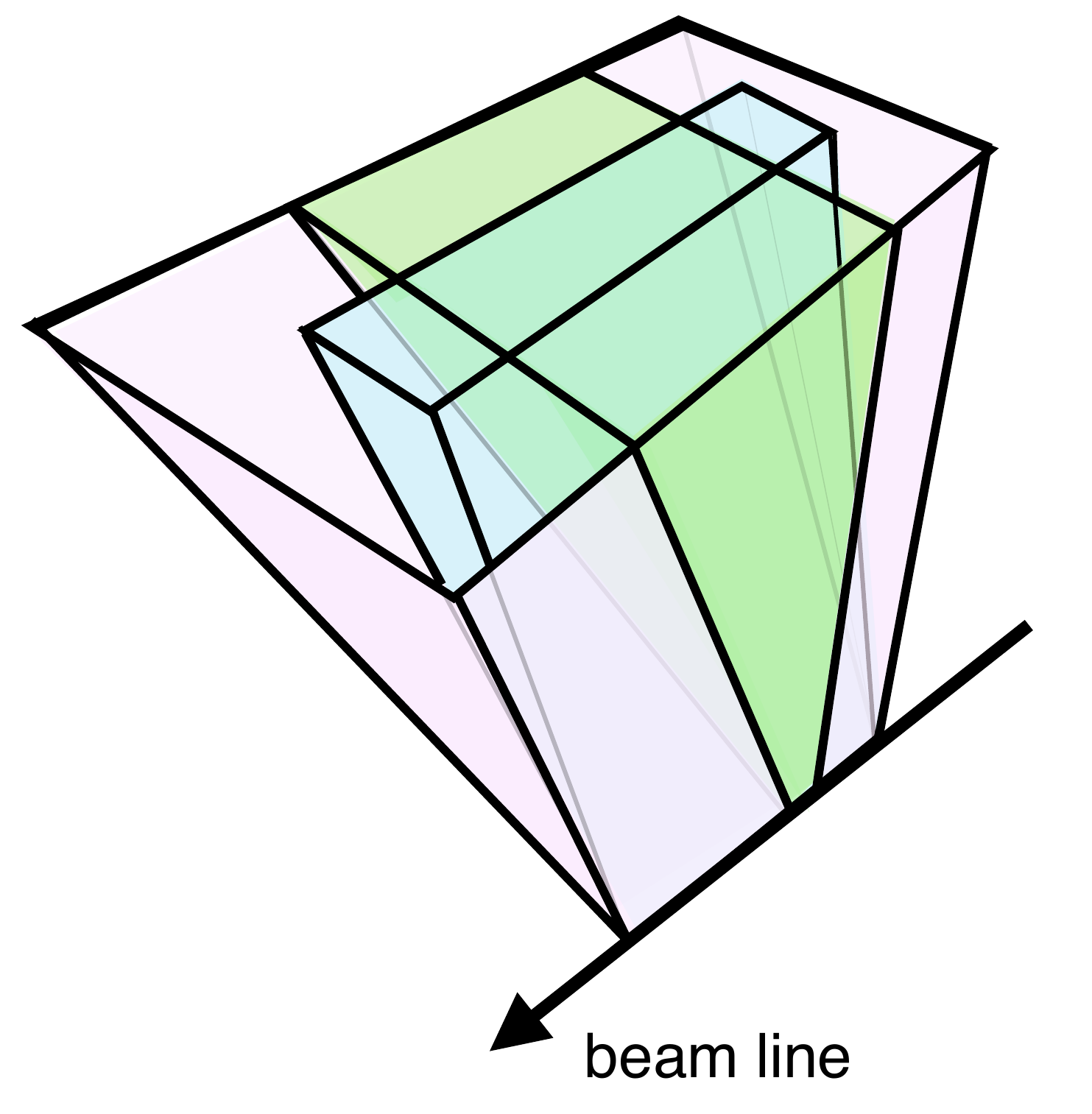}
  \caption{A schematic illustrating the RoIs from the single-stage and
    two-stage tau lepton trigger tracking, shown in plan view (x-z plane) along
    the transverse direction and in perspective view. The z-axis is along the
    beam line. The combined tracking volume of the $1^{\mathrm{st}}$ and
    $2^{\mathrm{nd}}$ stage RoI in the two-stage tracking approach is
    significantly smaller than the RoI in the one-stage tracking scheme.}
\label{fig:idroi}
\end{figure}

A two-stage processing approach was implemented for the hadronic tau
trigger. First, the leading track and its position along the beamline are
determined by executing fast tracking in an RoI that is fully extended along the
beamline ($|z|<\SI{225}{\mm}$) but narrow (0.1) in both $\eta$ and $\phi$. (See the
blue-shaded region in Figure~\ref{fig:idroi}.) Using this position along the beamline, the
second stage reconstructs all tracks in an RoI that is larger (0.4) in both $\eta$
and $\phi$ but limited to  $|\Delta z|<\SI{10}{\mm}$ with respect to the leading
track. (See the green shaded region in Figure 12.) At this second stage, fast
tracking is followed by precision tracking. For evaluation purposes, the tau
lepton signatures can also be executed in a single-stage mode, running the fast
track finder followed by the precision tracking in an RoI of the full extent
along the beam line and in eta and phi.

\begin{figure}[htbp]
\centering
\subfloat[]{
  \includegraphics[width=0.5\textwidth]{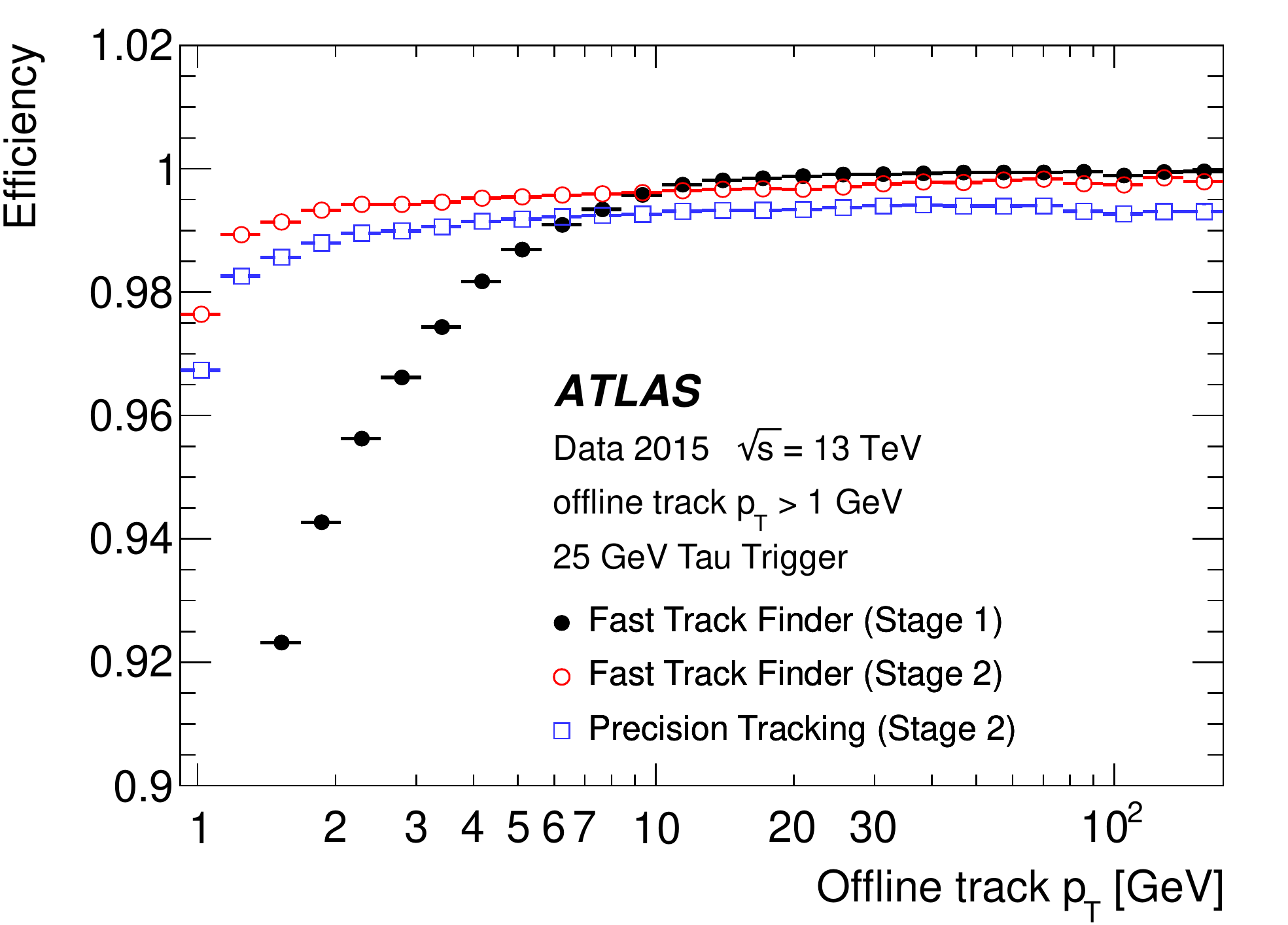}
  \label{fig:tautriggera}
}
\subfloat[]{
  \includegraphics[width=0.5\textwidth]{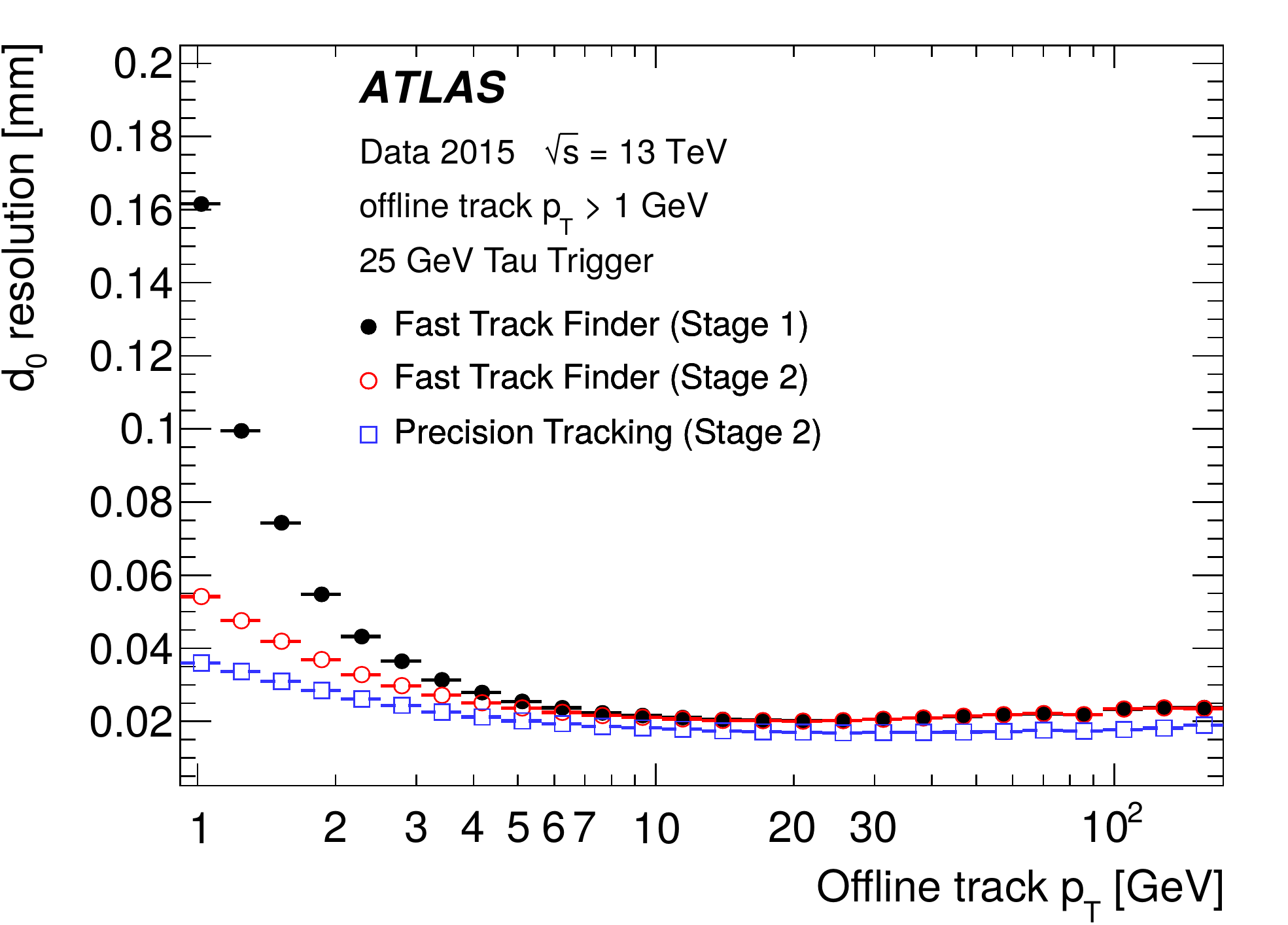}
  \label{fig:tautriggerb}
}
\caption{The ID trigger tau tracking performance with respect to
  offline tracks from very loose tau candidates with $\pt>\SI{1}{\GeV}$ from the
  \SI{25}{\GeV} tau trigger; (a) the efficiency as a function of the offline reconstructed
  tau track \pt, (b) the resolution of the transverse impact parameter, $d_{0}$
  as a function of the offline reconstructed tau track \pt. The offline reconstructed tau
  daughter tracks are required to have $\pt>\SI{1}{\GeV}$, lie in the region
  $|\eta|<2.5$ and have at least two pixel clusters and at least six SCT
  clusters. The closest matching trigger track within a cone of size
  $\Delta R = 0.05$ of the offline track is selected as the matching trigger
  track.}
\label{fig:tautrigger}
\end{figure}

Figure~\ref{fig:tautrigger} shows the performance of the tau two-stage tracking with respect to the 
offline tau tracking for tracks with $\pt>\SI{1}{\GeV}$ originating from decays of offline tau lepton candidates 
with $\pt > \SI{25}{\GeV}$, but with very loose track matching in $\Delta R$ to the offline tau candidate.
Figure~\ref{fig:tautriggera} shows the efficiency of the fast tracking from the first and second stages, 
together with the efficiency of the precision tracking for the second stage. The second-stage tracking 
efficiency is higher than 96\% everywhere, and improves to better than 99\% for tracks with $\pt>\SI{2}{\GeV}$.
The efficiency of the first-stage fast tracking has a slower turn-on, rising from 94\% at \SI{2}{\GeV} to 
better than 99\% for $\pt>\SI{5}{\GeV}$. This slow turn-on arises due to the narrow width ($\Delta\phi<0.1$) 
of the first-stage RoI and the loose tau selection that results in a larger fraction of low-\pt\ tracks
from tau candidates that bend out of the RoI (and are not reconstructed) compared to a wider RoI.
The transverse impact parameter resolution with respect to offline for loosely matched tracks is seen in 
Figure~\ref{fig:tautriggerb} and is around \SI{20}{\um} for tracks with $\pt>\SI{10}{\GeV}$ reconstructed by the
precision tracking. The tau selection algorithms based on this two-stage tracking are presented in Section~\ref{sec:tau:rec}.


\begin{figure}[htbp]
\centering
\subfloat[]{
  \includegraphics[width=0.5\textwidth]{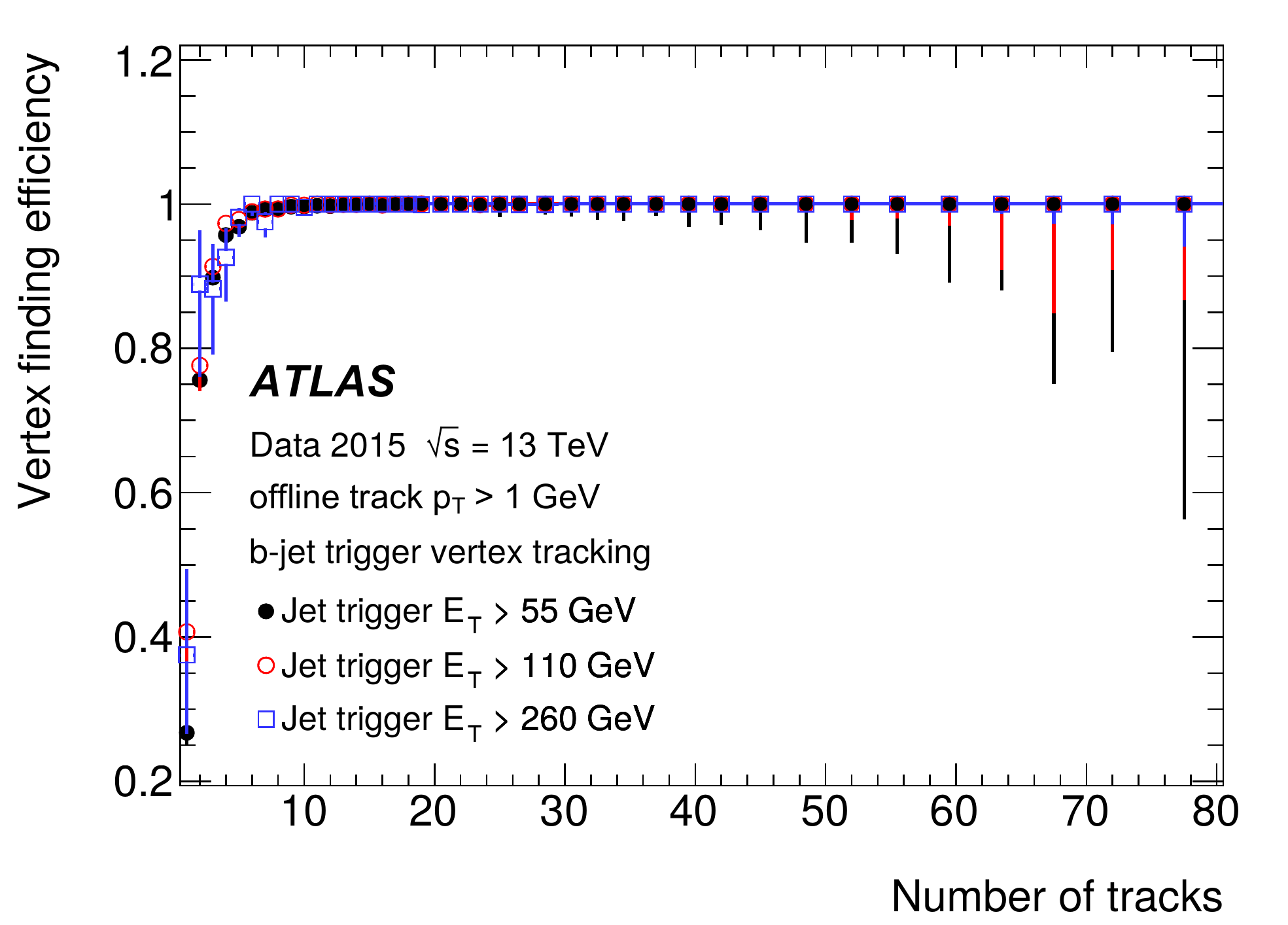}
  \label{fig:bjetvertexa}
}
\subfloat[]{
  \includegraphics[width=0.5\textwidth]{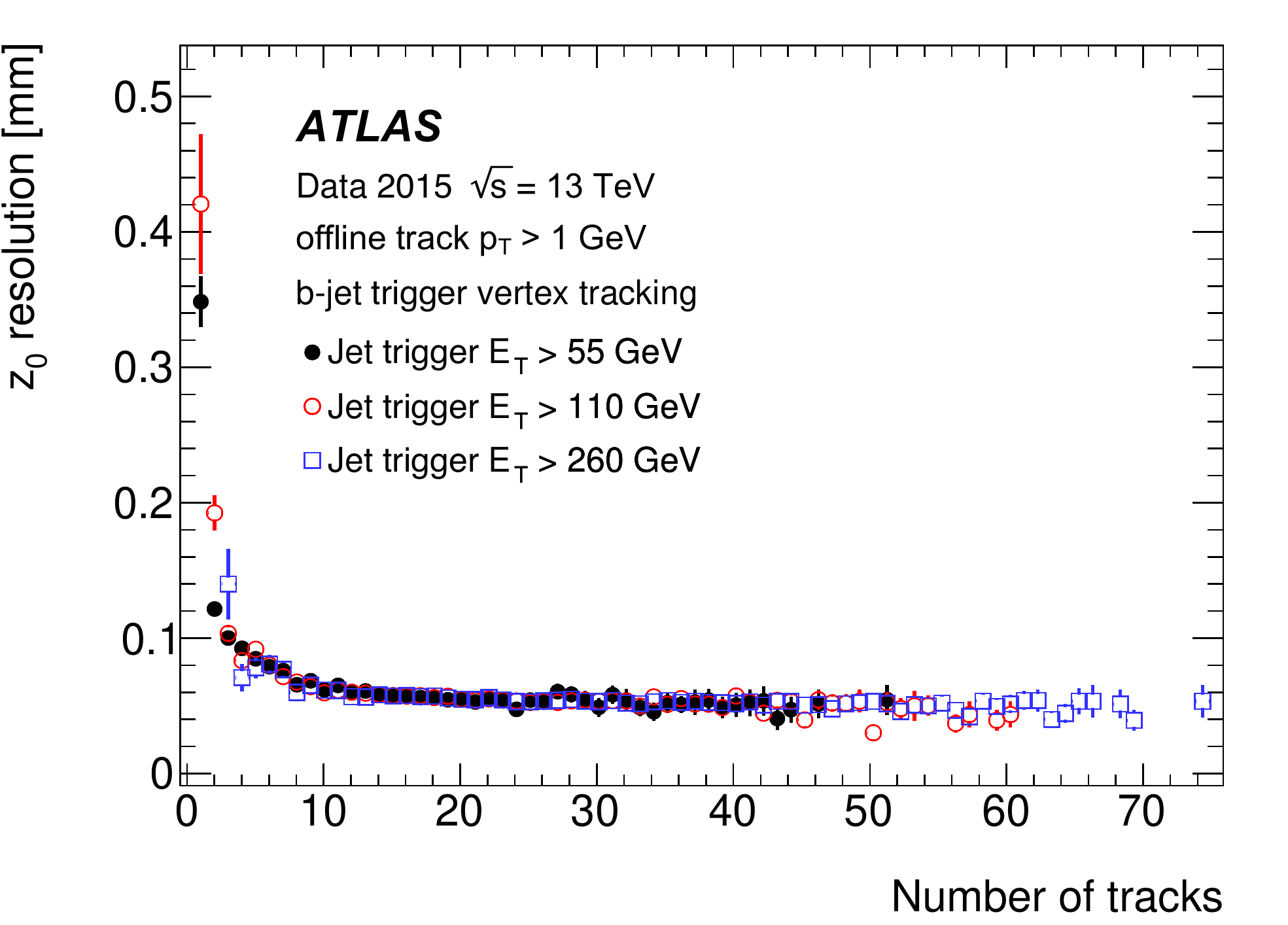}
  \label{fig:bjetvertexb}
}
\caption{The trigger performance for primary vertices in the $b$-jet signatures for 55, 110 and \SI{260}{\GeV}
  jet triggers; 
  (a) the vertexing efficiency as a function of the number of offline tracks within the
  jets used for the 
  vertex tracking, (b) the resolution in $z$ of the vertex with respect to the offline vertex position
  as a function of the 
number of offline tracks from the offline vertex.}
\label{fig:bjetvertex}
\end{figure}

For $b$-jet tracking a similar multi-stage tracking strategy was adopted.
However, in this case the first-stage vertex tracking takes all jets identified by the jet trigger 
with $\et>\SI{30}{\GeV}$ and reconstructs tracks with the fast track finder   
in a narrow region in $\eta$ and $\phi$ around the jet axis for each jet, but with $|z|<\SI{225}{\mm}$ along 
the beam line.
Following this step, the primary vertex reconstruction~\cite{ATLAS-CONF-2010-069} is performed using the tracks from the fast tracking stage.
This vertex is used to define wider RoIs around
the jet axes, with $|\Delta\eta|<0.4$ and $|\Delta\phi|<0.4$ but with $|\Delta z|<\SI{20}{\mm}$ relative
to the primary vertex $z$ position. These RoIs are then used for the second-stage 
reconstruction that runs the fast track finder in the wider $\eta$ 
and $\phi$ regions followed by the precision 
tracking, secondary vertexing and $b$-tagging algorithms.

The performance of the primary vertexing in the $b$-jet vertex tracking can be seen in Figure~\ref{fig:bjetvertexa},
which shows the vertex finding efficiency with respect to 
offline vertices in jet events with at least one jet with transverse energy above
55, 110, or \SI{260}{\GeV} and with no additional $b$-tagging requirement. The efficiency is shown 
as a function of the number of offline tracks with $\pt>\SI{1}{\GeV}$ that 
lie within the boundary of the wider RoI (defined above) from the selected jets. The efficiency rises sharply and 
is above 90\% for vertices with three or more tracks, and rises to more 
than 99.5\% for vertices with five or more tracks. The resolution in $z$ with respect to the
offline $z$ position as shown in Figure~\ref{fig:bjetvertexb} is better than \SI{100}{\um} for 
vertices with two or more offline tracks and improves to \SI{60}{\um} for 
vertices with ten or more offline tracks.

\subsubsection{Inner detector tracking timing}

\begin{figure}[htbp]
\centering
\includegraphics[width=0.5\textwidth]{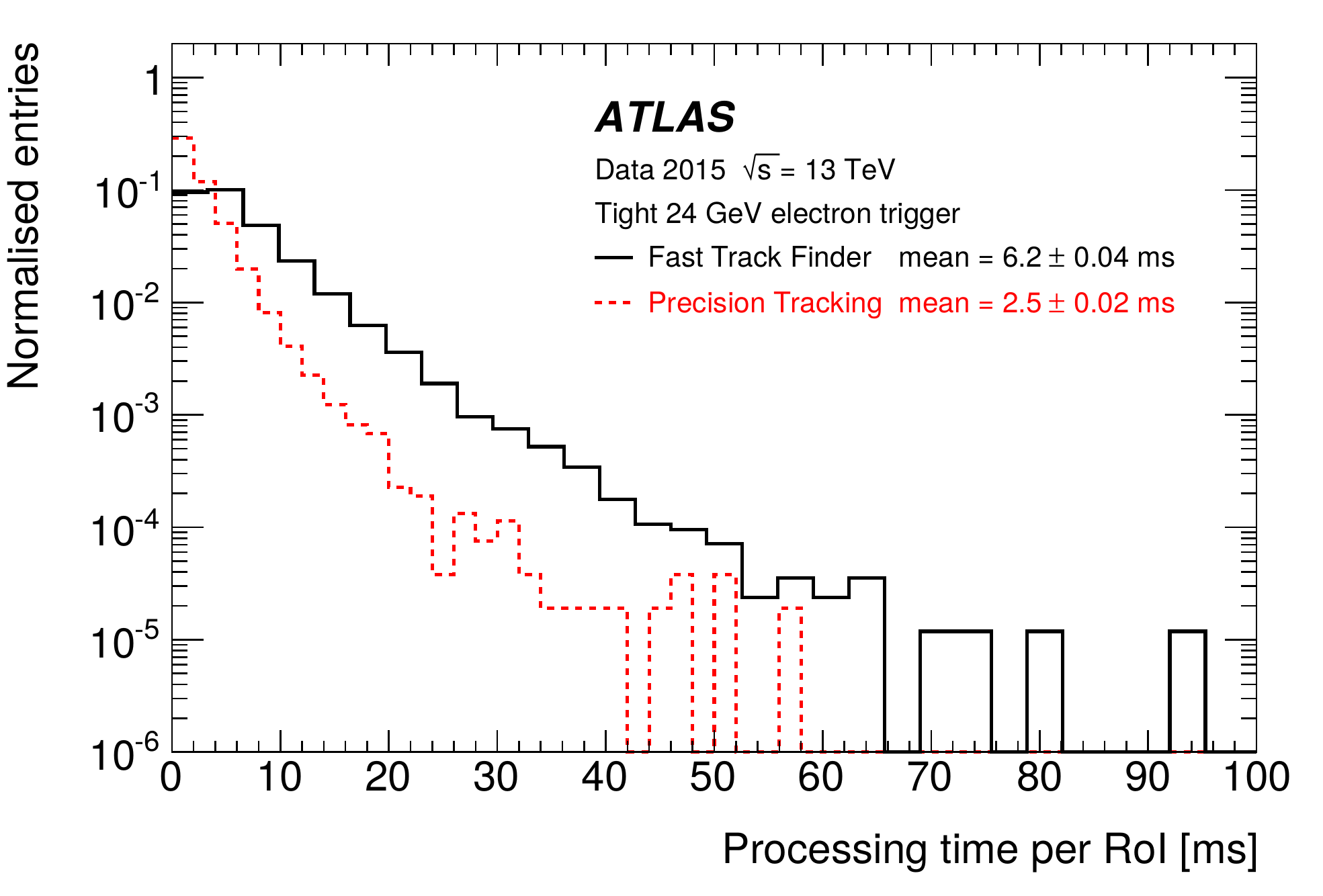}
\caption{
The CPU processing time for the fast and precision tracking per electron RoI for the \SI{24}{\GeV} electron 
trigger. The precision tracking is seeded by the tracks found in the fast tracking stage and hence requires less CPU time.}
\label{fig:idelectronlatency}
\end{figure}

The timing of the fast tracking and precision tracking stages of the electron trigger executed per RoI 
can be seen in Figure~\ref{fig:idelectronlatency} for events passing the \SI{24}{\GeV} electron trigger. 
The fast tracking takes on average \SI{6.2}{\ms} per RoI with a tail at the per-mille level at around \SI{60}{\ms}.
The precision tracking execution time has a mean of \SI{2.5}{\ms} and a tail at the per-mille level of
around \SI{20}{\ms}. The precision tracking is seeded by the tracks found in the fast tracking stage and hence requires less CPU time.

\begin{figure}[htbp]
\centering
\subfloat[]{
  \includegraphics[width=0.5\textwidth]{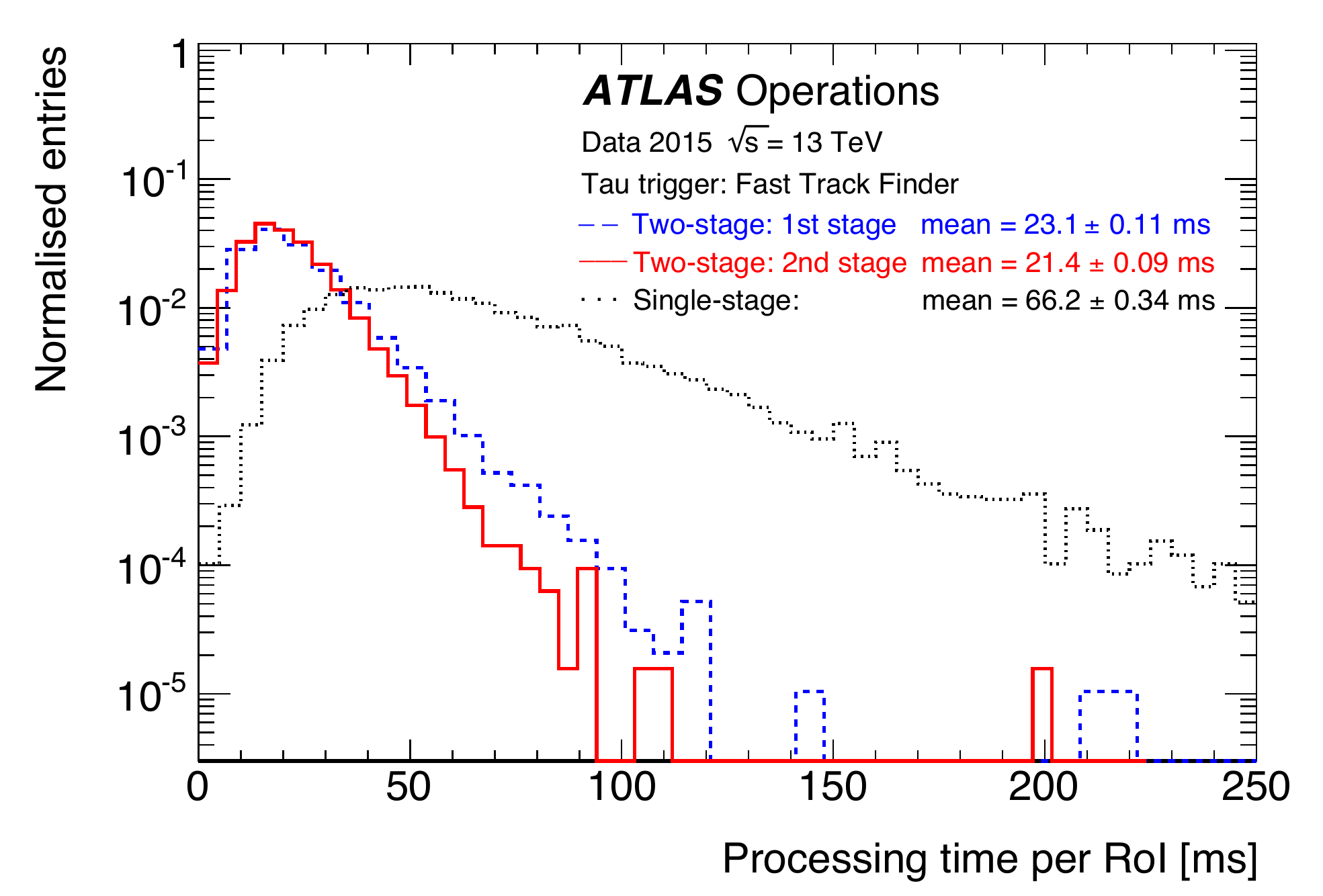}
  \label{fig:idtriggerlatencya}
}
\subfloat[]{
  \includegraphics[width=0.5\textwidth]{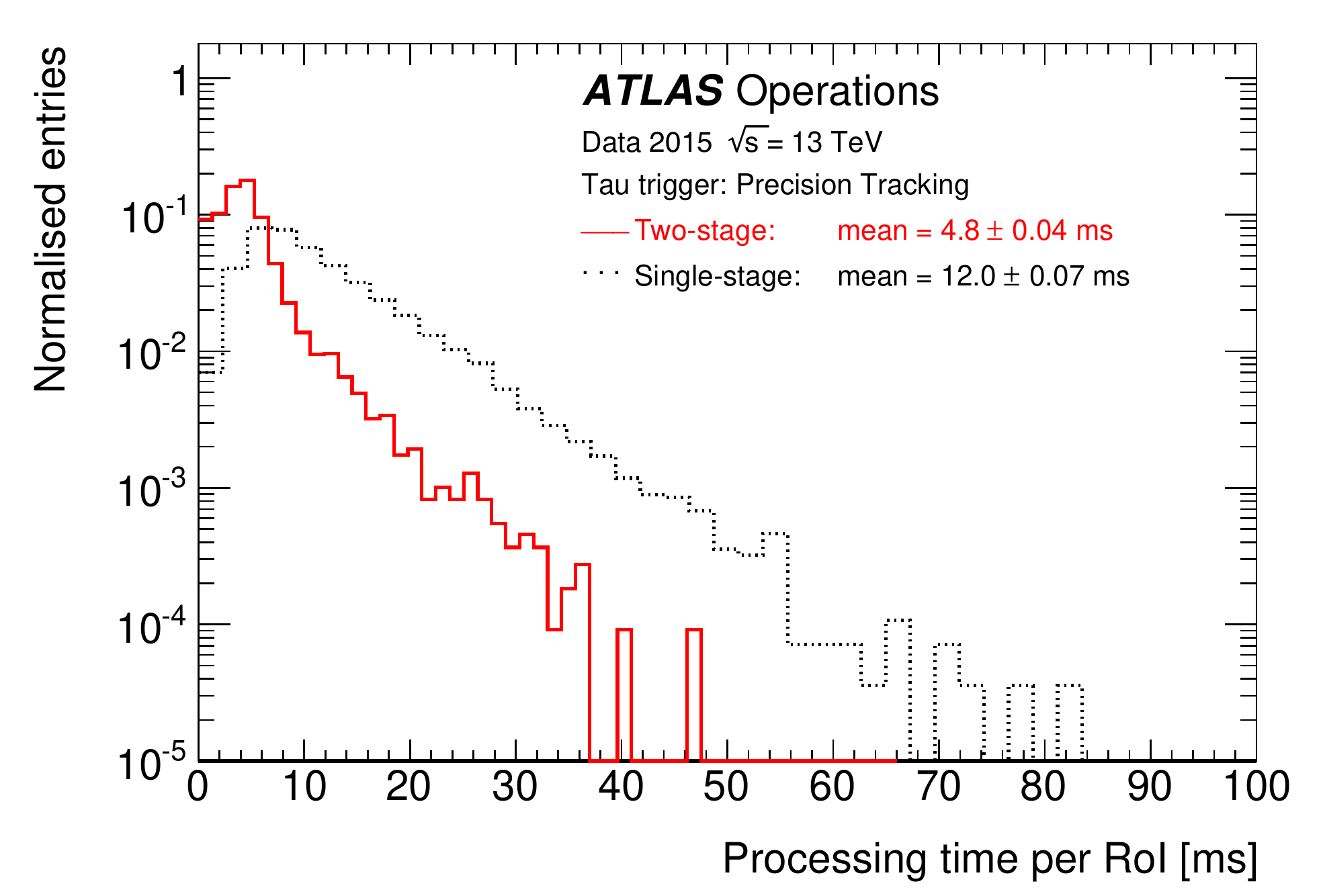}
  \label{fig:idtriggerlatencyb}
}
\caption{The ID trigger tau tracking processing time for (a) the fast track
  finder and (b) the precision tracking comparing the single-stage and two-stage
  tracking approach.}
\label{fig:idtriggerlatency}
\end{figure}

The time taken by the tau tracking in both the single-stage and two-stage
variants is shown in
Figure~\ref{fig:idtriggerlatency}. Figure~\ref{fig:idtriggerlatencya} shows the
processing times per RoI for fast tracking stages: individually for the first and
second stages of the two-stage tracking, and separately for the single-stage
tracking with the wider RoI in $\eta$, $\phi$ and $z$. The fast tracking in the
single-stage tracking has a mean execution time of approximately \SI{66}{\ms}, with a
very long tail. In contrast, the first-stage tracking with an RoI that is wide
only in the $z$ direction has a mean execution time of \SI{23}{\ms}, driven predominantly by the narrower
RoI width in $\phi$. The second-stage tracking, although wider in $\eta$
and $\phi$, takes only \SI{21}{\ms} on average because of the significant reduction in
the RoI $z$-width along the beam line.  Figure~\ref{fig:idtriggerlatencyb} shows
a comparison of the processing time per RoI for the precision tracking. The
two-stage tracking executes faster, with a mean of \SI{4.8}{\ms} compared to \SI{12}{\ms} for
the single-stage tracking. Again, this is due to the reduction in the number of
tracks to be processed from the tighter selection in $z$ along the beam line.

\subsection{Calorimeter reconstruction}
\label{sec:calo}

A series of reconstruction algorithms are used to convert signals from the calorimeter
readout into objects, specifically cells and clusters, that then serve as input to the reconstruction of
electron, photon, tau, and jet candidates and the reconstruction of \met. These cells
and clusters are also used in the determination of the shower shapes and the isolation properties of
candidate particles (including muons), both of which are later used as discriminants for particle
identification and the rejection of backgrounds. The reconstruction algorithms used in the HLT have
access to full detector granularity and thus allow improved accuracy and precision in energy and
position measurements with respect to L1.

\subsubsection{Calorimeter algorithms}
\label{sec:calo:algo}
The first stage in the reconstruction involves unpacking
the data from the calorimeter. The unpacking can be done in two different ways: either by
unpacking only the data from within the RoIs identified at L1 or by unpacking the
data from the full calorimeter. The RoI-based approach is used for well-separated objects
(e.g.\ electron, photon, muon, tau), whereas the full calorimeter reconstruction is used for jets and global
event quantities (e.g.\ \met). In both cases the raw unpacked data is then
converted into a collection of cells. 
Two different clustering algorithms are used to reconstruct the clusters of energy deposited
in the calorimeter, the sliding-window
and the topo-clustering algorithms~\cite{PERF-2014-07}. While the latter provides performance
closer to the offline reconstruction, it is also significantly slower (see Section~\ref{sec:calo:timing}).

The sliding-window algorithm operates on a grid in which the cells are
divided into projective towers. The algorithm scans this grid and positions the
window in such a way that the transverse energy contained within the window is the
local maximum. If this local maximum is above a given threshold, a cluster
is formed by summing the cells within a rectangular clustering window. 
For each layer the barycentre of the cells within that layer is determined, and then all cells within a
fixed window around that position are included in the cluster. Although
the size of the clustering window is fixed, the central position of the window
may vary slightly at each calorimeter layer, depending on how the cell energies
are distributed within them.

The topo-clustering algorithm begins with a seed cell and iteratively adds neighbouring cells to the
cluster if their energies are above a given energy threshold that is a function of the expected root-mean-square (RMS)
noise ($\sigma$). The seed cells are first identified as those cells that have energies greater
than 4$\sigma$. All neighbouring cells with energies greater than 2$\sigma$ are then added to the
cluster and, finally, all the remaining neighbours to these cells are also added.
Unlike the sliding-window clusters, the topo-clusters have no predefined shape, and consequently their
size can vary from cluster to cluster.

The reconstruction of candidate electrons and photons uses the sliding-window algorithm
with rectangular clustering windows of
size $\Delta\eta\times\Delta\phi$ = 0.075\,$\times$\,0.175 in the barrel and 0.125\,$\times$\,0.125
in the end-caps. Since the magnetic field bends the electron trajectory in the $\phi$ direction, 
the size of the window is larger in that coordinate in order to contain most of the energy.
The reconstruction of candidate taus and jets and the reconstruction of
\met all use the topo-clustering algorithm. For taus the topo-clustering uses
a window of 0.8\,$\times$\,0.8 around each of the tau RoIs identified at L1.
For jets and \met, the topo-clustering is done for the full calorimeter. In addition, the \met is
also determined based on the cell energies across the full calorimeter (see Section~\ref{sec:met}). 

\subsubsection{Calorimeter algorithm performance}

\begin{figure}[htbp]
\centering
\subfloat[]{
  \includegraphics[width=0.5\textwidth]{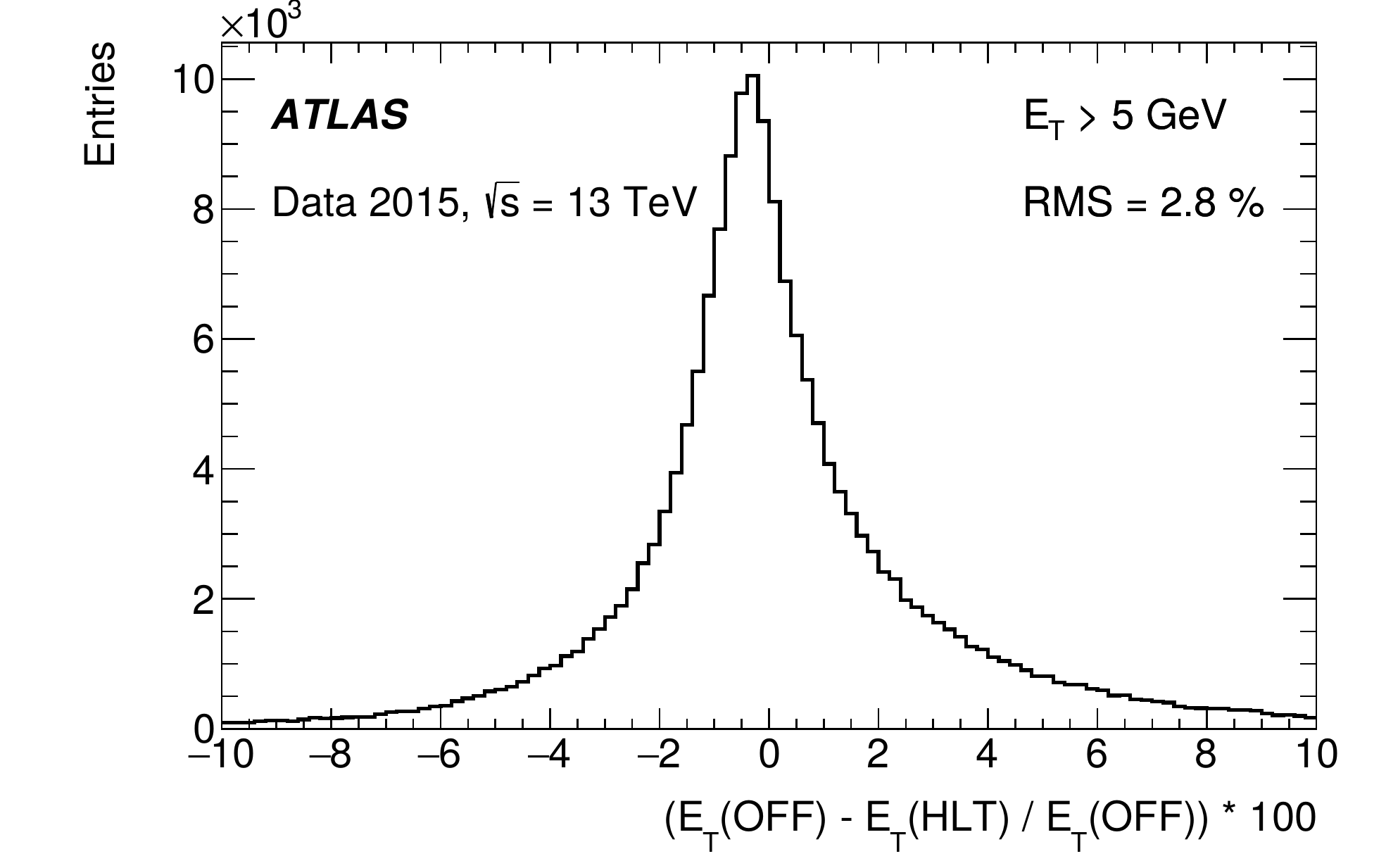}
  \label{fig:calo_perf_slw}
}
\subfloat[]{
  \includegraphics[width=0.5\textwidth]{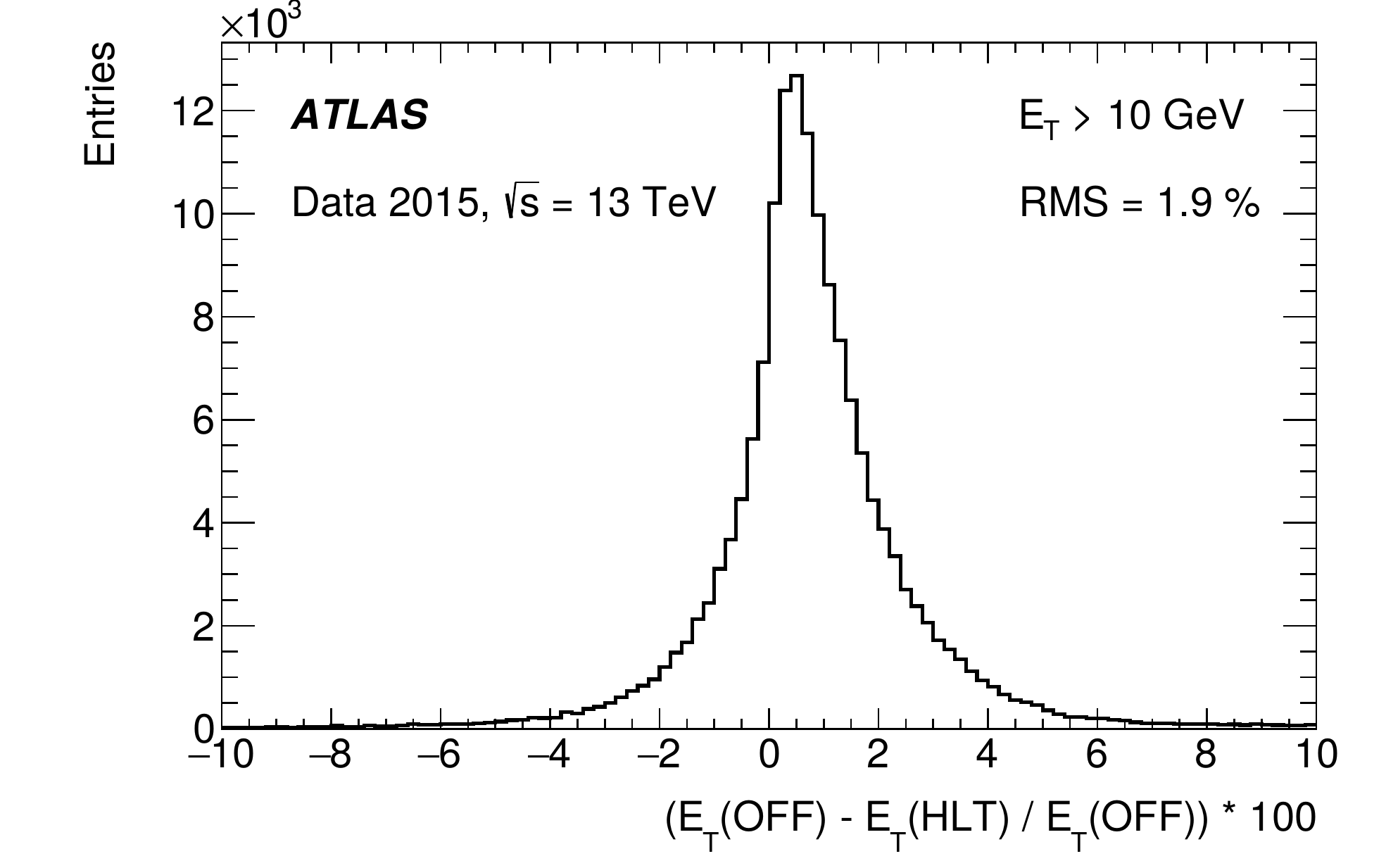}
  \label{fig:calo_perf_topo}
}
\caption{The relative differences between the online and offline \ET for (a) sliding-window clusters
and (b) topo-clusters. Online and offline clusters are matched within $\Delta R$ < 0.001.
The distribution for the topo-clusters was obtained from the RoI-based topo-clustering algorithm
that is used for online tau reconstruction.}
\label{fig:calo_perf}
\end{figure}

The harmonisation between the online and offline algorithms in \runii means that the online 
calorimeter performance is now much closer to the offline performance. The \ET resolutions of the
sliding-window clusters and the topo-clusters with respect to their offline counterparts are 
shown in Figure~\ref{fig:calo_perf}. The \ET resolution of the sliding-window clusters is 3\% 
for clusters above \SI{5}{\GeV}, while the \ET resolution of the topo-clustering algorithm is 2\% for 
clusters above \SI{10}{\GeV}.
The slight shift in cell energies between the HLT and offline is due to the fact
that out-of-time pile-up effects were not corrected in the online
reconstruction, resulting in slightly higher reconstructed cell energies in the
HLT (this was changed for 2016). In addition, the topo-cluster based
reconstruction shown in Figure~\ref{fig:calo_perf_topo} suffered from a mismatch of
some calibration constants between online and offline during most of 2015,
resulting in a shift towards lower HLT cell energies.

\subsubsection{Calorimeter algorithm timing}
\label{sec:calo:timing}

\begin{figure}[htbp]
\centering
\subfloat[]{
  \includegraphics[width=0.5\textwidth]{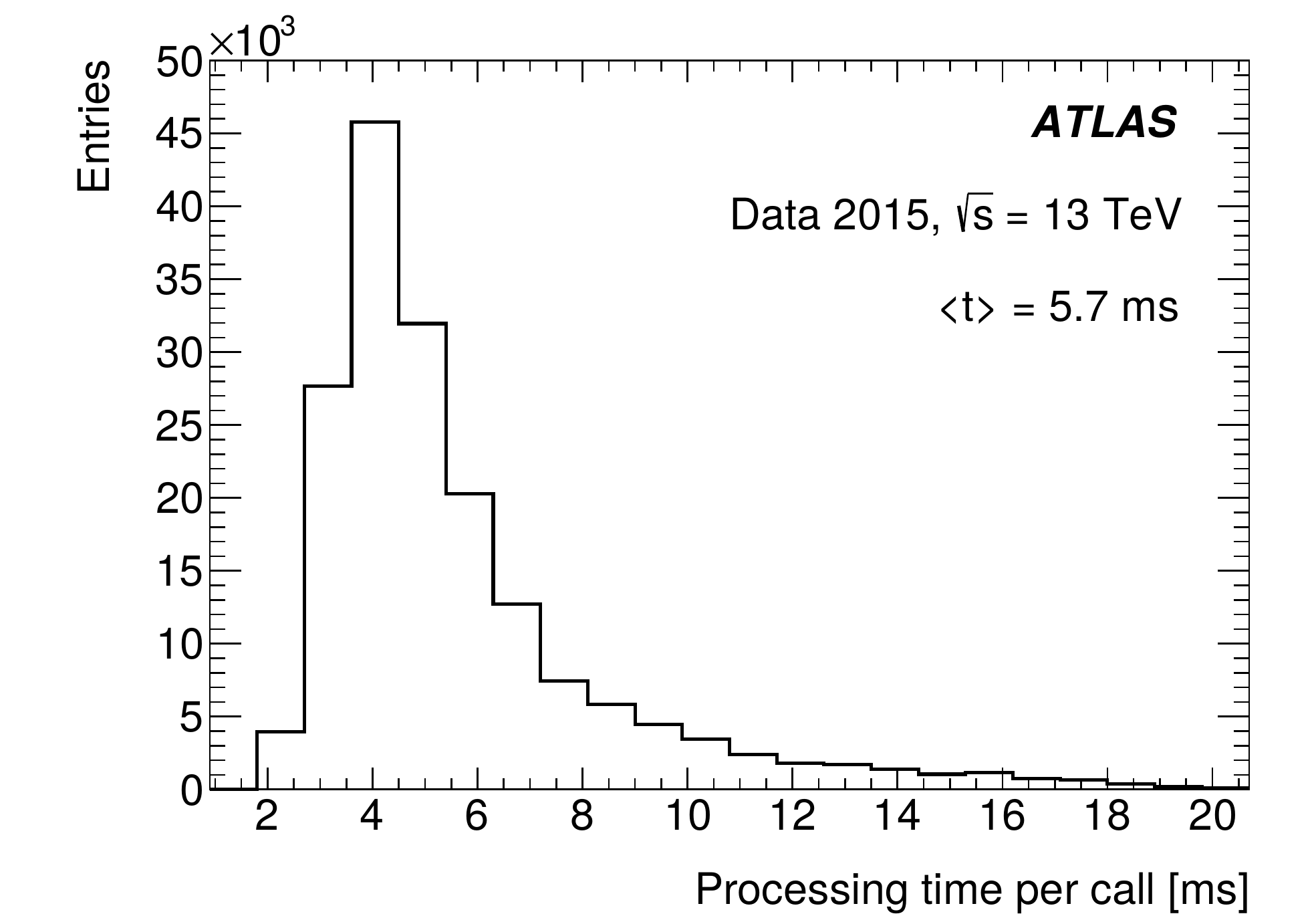}
  \label{fig:calo_time_a}
}
\subfloat[]{
  \includegraphics[width=0.5\textwidth]{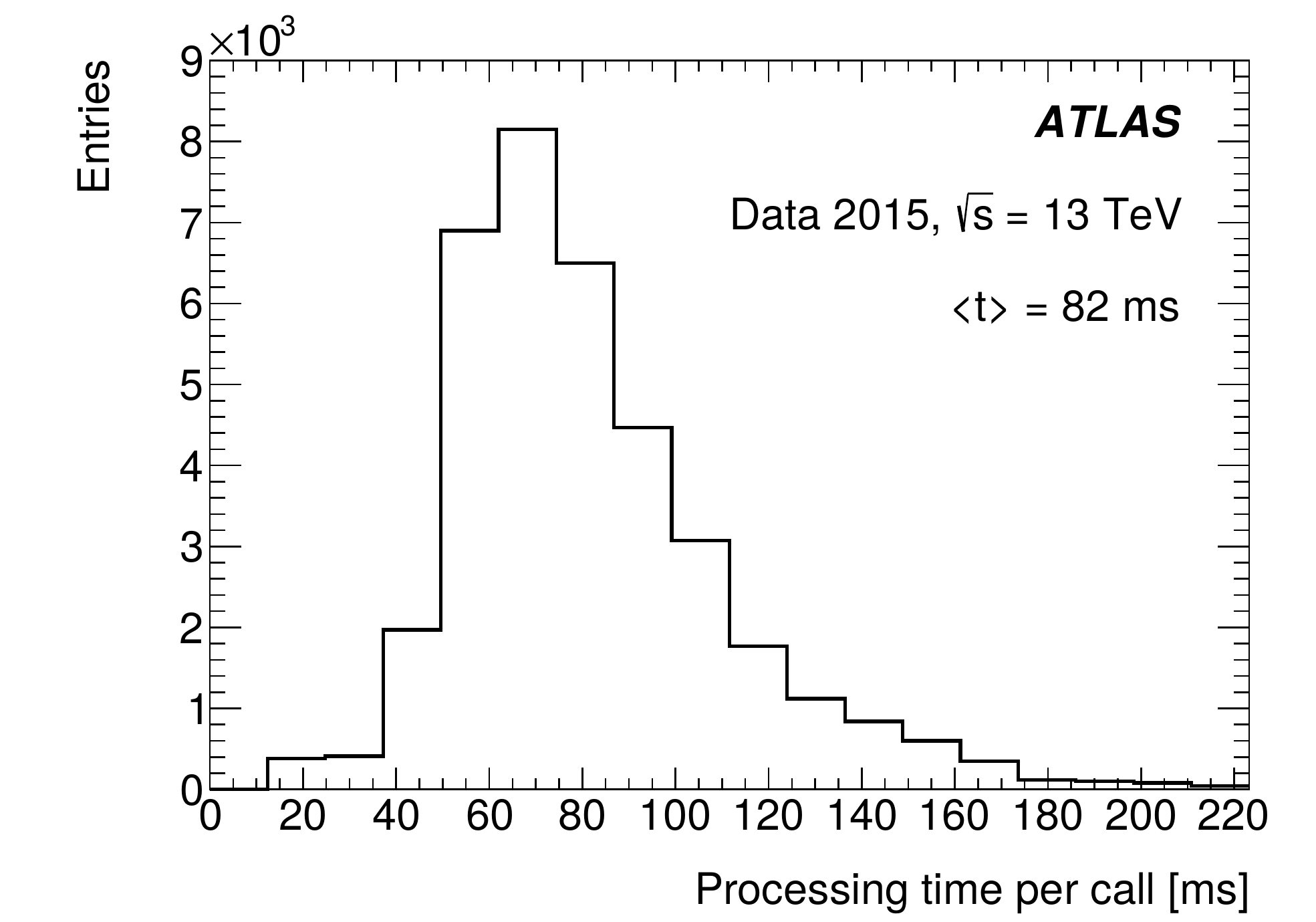}
  \label{fig:calo_time_b}
}
\caption{The distributions of processing times for the topo-clustering algorithm executed (a) within an RoI
  and (b) on the full calorimeter. The processing times within an RoI are obtained from
  tau RoIs with a size of $\Delta\eta\times\Delta\phi=0.8\times0.8$.}
\label{fig:calo_time}
\end{figure}

Due to the optimisation of the offline clustering algorithms during LS1, offline clustering algorithms
can be used in the HLT directly after the L1 selection.
At the data preparation stage, a specially optimised
infrastructure with a memory caching mechanism allows very fast unpacking of data, even
from the full calorimeter, which comprises approximately \num{187000} cells. 
The mean processing time for the data preparation stage is \SI{2}{\ms} per RoI and \SI{20}{\ms} 
for the full calorimeter, and both are roughly independent of pile-up.
The topo-clustering, however, requires a fixed estimate of the expected pile-up noise (cell energy
contributions from pile-up interactions) in order
to determine the cluster-building thresholds and, when there is a discrepancy between the
expected pile-up noise and the actual pile-up noise, the processing time can show some dependence
on the pile-up conditions. The mean processing time for the topo-clustering is \SI{6}{\ms} per
RoI and \SI{82}{\ms} for the full calorimeter. The distributions of the topo-clustering processing
times are shown in Figure~\ref{fig:calo_time_a} for an RoI and Figure~\ref{fig:calo_time_b} for the full calorimeter.
The RoI-based topo-clustering can run multiple times if there is more than one
RoI per event. The topo-clustering over the full calorimeter runs at most once per event,
even if the event satisfied both jet and \met\ selections at L1.
The mean processing time of the sliding
window clustering algorithm is not shown but is typically less than \SI{2.5}{\ms} per RoI.

\subsection{Tracking in the muon spectrometer}
\label{sec:muonrec}

Muons are identified at the L1 trigger by the spatial and temporal coincidence
of hits either in the RPC or TGC chambers within the rapidity range of
$|\eta|<2.4$. The degree of deviation from the hit pattern expected for a muon
with infinite momentum is used to estimate the \pt\ of the muon with six
possible thresholds. The HLT receives this information together with the RoI position and
makes use of the precision MDT and CSC chambers to further refine the L1 muon candidates.

\subsubsection{Muon tracking algorithms}
The HLT muon reconstruction is split into \emph{fast} (trigger specific) and
\emph{precision} (close to offline) reconstruction stages, which were used during
\runi at L2 and EF, respectively.

In the fast reconstruction stage, each L1 muon candidate is refined by
including the precision data from the MDT chambers in the RoI defined by the L1
candidate. A track fit is performed using the MDT drift times and positions, and
a \pt\ measurement is assigned using lookup tables, creating \emph{MS-only}
muon candidates. The MS-only muon track is
back-extrapolated to the interaction point using the offline track extrapolator
(based on a detailed detector description instead of the lookup-table-based
approach used in \runi) and combined with tracks reconstructed in the ID to form
a \emph{combined} muon candidate with refined track parameter resolution.

In the precision reconstruction stage, the muon reconstruction starts
from the refined RoIs identified by the fast stage, reconstructing segments and
tracks using information from the trigger and precision chambers. As in the fast
stage, muon candidates are first formed by using the muon detectors
(MS-only) and are subsequently combined with ID tracks leading to
combined muons. If no matching ID track can be found, 
combined muon candidates are searched for by extrapolating ID tracks to the MS.
This latter \emph{inside-out} approach is slower and hence only used if the \emph{outside-in}
search fails. It recovers about 1-5\% of the muons, most of them at low \pt.

The combined muon candidates are used for the majority of the muon
triggers. However, MS-only candidates are used for specialised triggers that
cannot rely on the existence of an ID track, e.g.\ triggers for long-lived
particles that decay within the ID volume.

\subsubsection{Muon tracking performance}
\begin{figure}[htbp]
\centering
\includegraphics[width=0.5\textwidth]{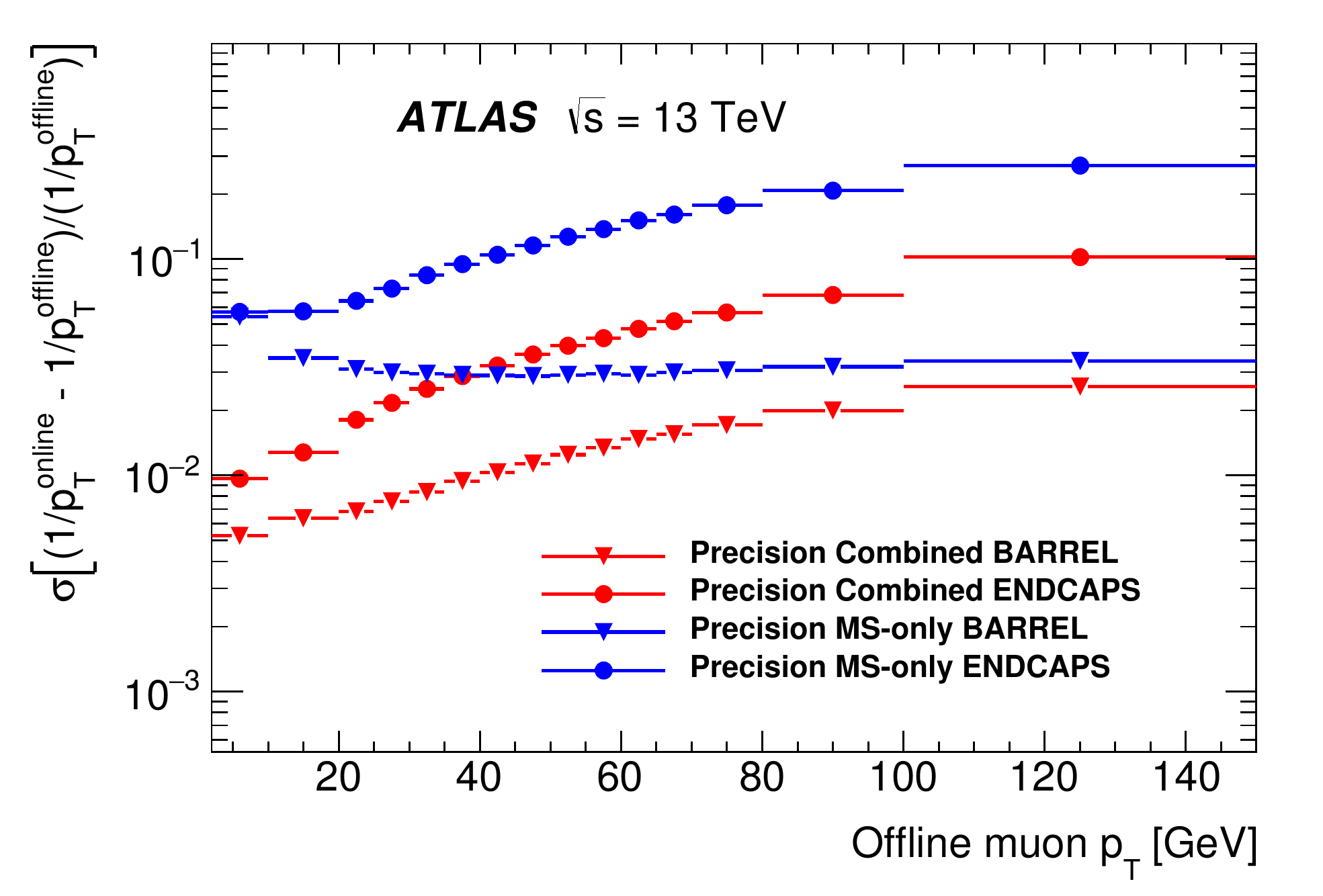}
\caption{Width of the residuals for inverse-\pT\ as a function of offline muon
  \pT\ for the precision MS-only and combined algorithms in the barrel
  ($|\eta|<1.05$) and end-caps ($1.0<|\eta|<2.4$).}
\label{fig:mu:ptreso}
\end{figure}

\begin{figure}[htbp]
\centering
\subfloat[]{\includegraphics[width=0.5\textwidth]{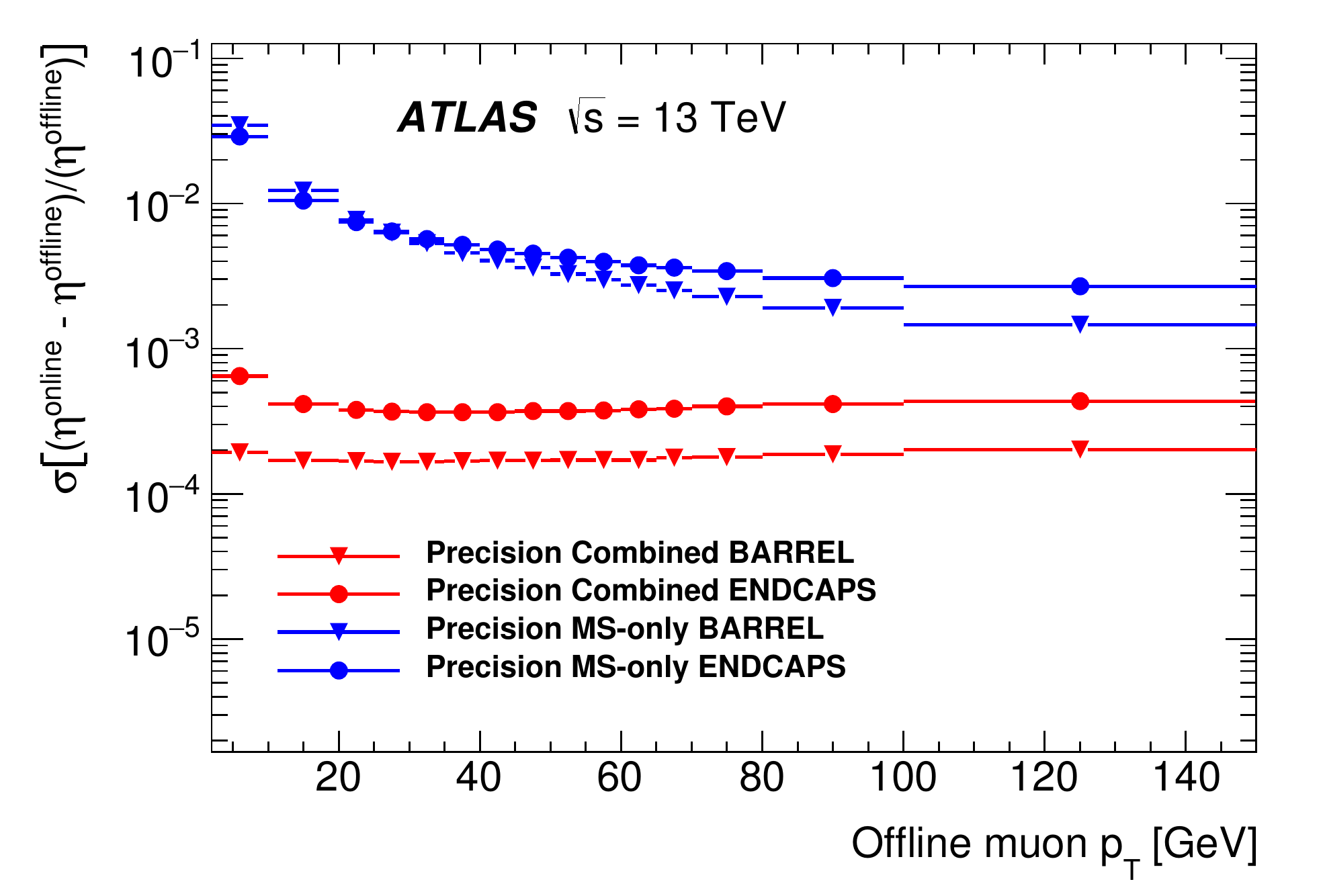}}
\subfloat[]{\includegraphics[width=0.5\textwidth]{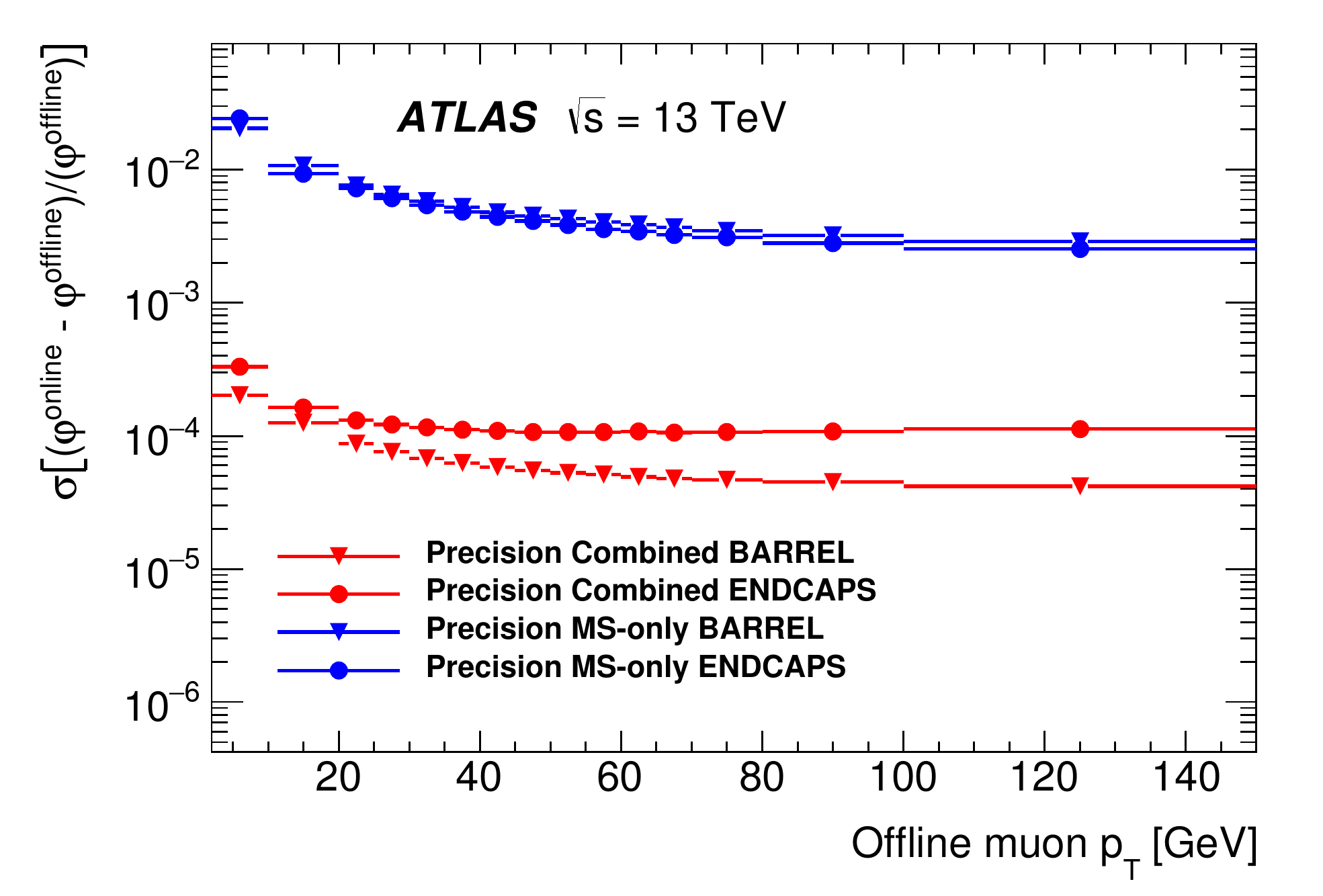}}
\caption{Width of the residuals as a function of the offline muon \pT\ for (a)
  $\eta$ and (b) $\phi$ for the precision MS-only and combined algorithms in the
  barrel ($|\eta|<1.05$) and end-caps ($1.0<|\eta|<2.4$).}
\label{fig:mu:etaphireso}
\end{figure}

Comparisons between online and offline muon track parameters using $Z\rightarrow\mu\mu$ candidate events
are presented in this section while muon trigger efficiencies are described in
Section~\ref{sec:muon}. Distributions of the residuals between online and
offline track parameters ($1/\pT$, $\eta$ and $\phi$) are constructed in
bins of \pT\ and two subsequent Gaussian fits are performed on the core of the
distribution to extract the widths, $\sigma$, of the residual distributions as a
function of $\pT$. The inverse-\pT\ residual widths,
$\sigma((1/\pT^{\mathrm{online}}-1/\pt^{\mathrm{offline}})/(1/\pt^{\mathrm{offline}}))$,
are shown in Figure~\ref{fig:mu:ptreso} as a function of the offline muon \pT\ for
the precision MS-only and precision combined reconstruction.
The resolution for combined muons is better than the resolution for MS-only
muons due to the higher precision of the ID track measurements, especially at
low \pT. As the tracks become closer to straight lines at high \pT, it
becomes more difficult to precisely measure the \pT\ of both the MS and ID tracks, and
hence the resolution degrades. The \pT resolution for low-\pT\ MS-only muons is degraded when
muons in the barrel are bent out of the detector before traversing the entire
muon spectrometer. The resolution is generally better in the barrel than in
the end-caps due to the difference in detector granularity.
The $\eta$ residual widths,
$\sigma(\eta^{\mathrm{online}}-\eta^{\mathrm{offline}})$, and $\phi$ residual
widths, $\sigma(\phi^{\mathrm{online}}-\phi^{\mathrm{offline}})$, are shown as a
function of \pT\ in Figure~\ref{fig:mu:etaphireso} for both the MS-only and combined
algorithms. As the trajectories are straighter at high $\pT$, the precision of their
position improves and so the spatial resolution decreases with \pT. Good
agreement between track parameters calculated online and offline is observed.
 
\subsubsection{Muon tracking timing}

\begin{figure}[htbp]
\centering
\subfloat[]{
  \includegraphics[width=0.5\textwidth]{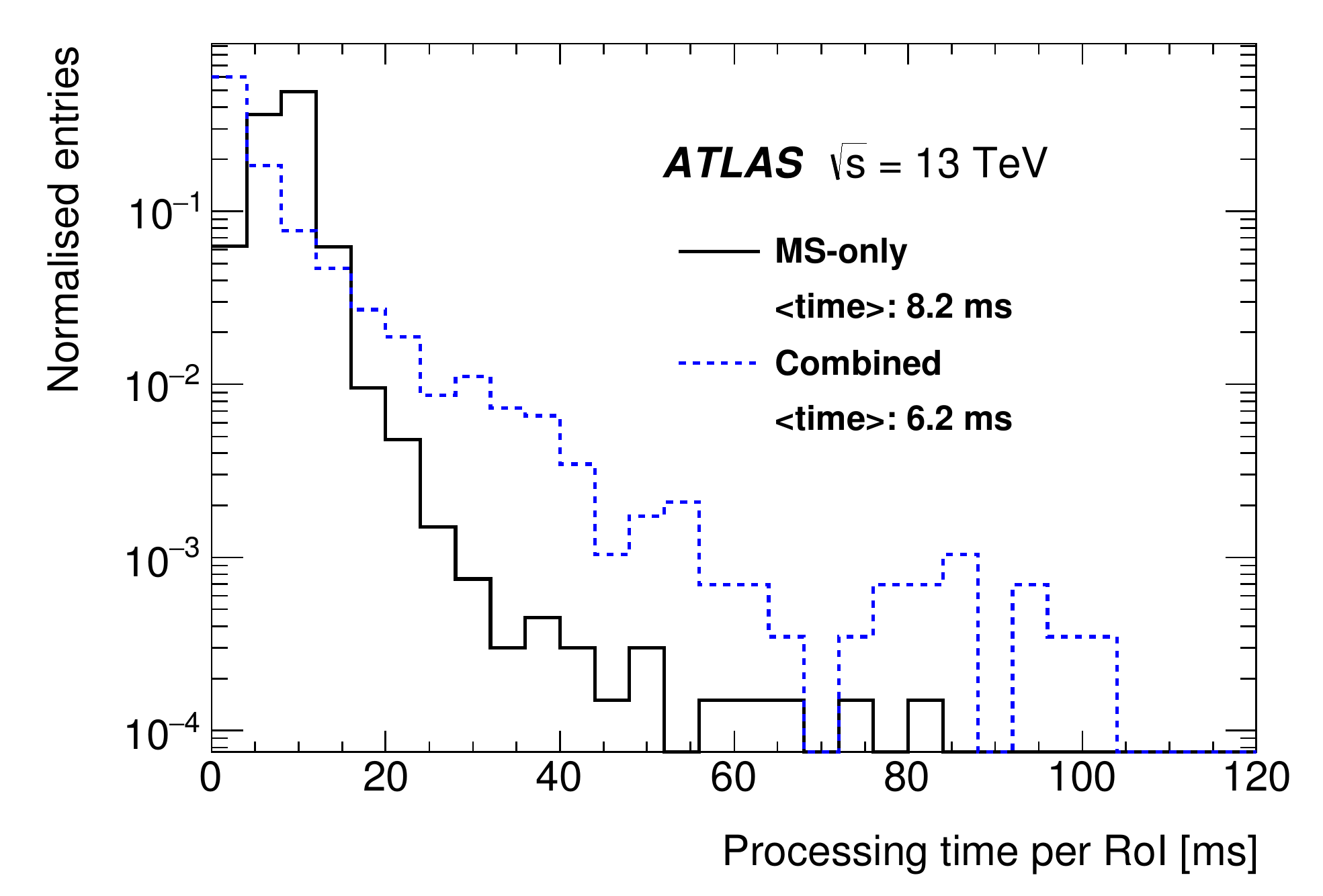}
  \label{fig:mu:timinga}
}
\subfloat[]{
  \includegraphics[width=0.5\textwidth]{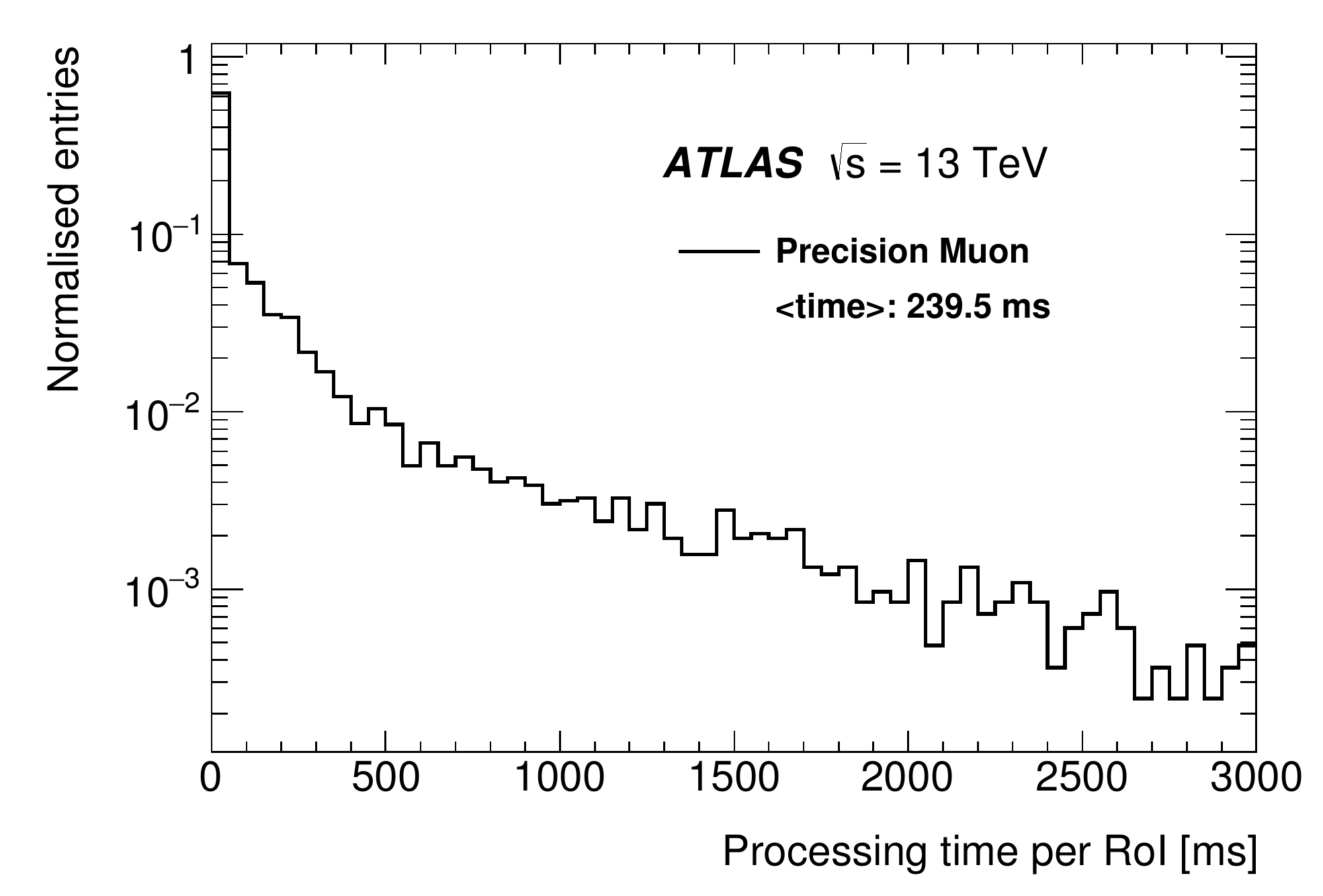}
  \label{fig:mu:timingb}
}
\caption{Processing times per RoI for the (a) fast MS-only and fast combined algorithms
  and (b) precision muon-finding algorithm. The time for the combined algorithm includes only
  the time for the ID--MS combination and not the tracking itself.
  The mean time of each algorithm is indicated in the legend. 
  The large number of entries in the first bin in (b) is due to algorithm caching~\cite{HLTSteering:2008}.}
\label{fig:mu:timing}
\end{figure}

Figure~\ref{fig:mu:timing} shows the processing times per RoI for the (a) fast
MS-only and fast combined algorithms and (b) precision muon algorithm. The large
time difference between the fast and precision algorithms, with the precision reconstruction
using too much time to be run by itself at the full L1 muon trigger rate, motivates
the need for a two-stage reconstruction.

\FloatBarrier
\section{Trigger signature performance}
\label{sec:sigperf}
The following sections describe the different selection criteria placed upon the
reconstructed objects described in Section~\ref{sec:hltrec} in order to form individual trigger
signatures that identify leptons, hadrons, and global event quantities such as
\met. For each case the primary triggers used during 2015
are listed together with their output rate and performance. Where possible the
trigger efficiency measured in data is compared with MC simulation.  The following
methods are used to derive an unbiased measurement of the trigger efficiency:

\begin{itemize}
\item \emph{Tag-and-probe method}, which uses a sample of offline-selected events
that contain a pair of related objects reconstructed offline, such as electrons
from a \Zee decay, where one has triggered the event and the other one is used
to measure the trigger efficiency;
\item \emph{Bootstrap method}, where the efficiency of a higher trigger threshold
is determined using events triggered by a lower threshold.
\end{itemize}

Trigger efficiencies are computed with respect to an offline-selected data sample.
The ratio of the measured trigger efficiency to the simulated one is used as a correction factor
in physics analyses. Unless otherwise specified, performance studies use good-quality
data corresponding to
an integrated luminosity of \SI{3.2}{\ifb} collected during 2015 with a bunch-spacing of \SI{25}{\ns}.
Trigger rates shown in the following sections are usually extracted from multiple data-taking
runs to cover the maximum range in instantaneous luminosity. Due to different beam and detector
conditions between runs, this can result in slightly different trigger rates for nearby luminosity
values.


\subsection{Minimum-bias and forward triggers}
\label{sec:minbias}
Studies of the total cross-section, hadronisation, diffraction, hadrons containing strange quarks and
other non-perturbative properties of $pp$ interactions require the use of a
high-efficiency trigger for selecting all inelastic 
interactions that result in particle production within the detector. 
The MBTS minimum-bias trigger is highly efficient, even for events containing only two charged particles 
with $\pt~>\SI{100}{\MeV}$ and $|\eta| < 2.5$.

The primary minimum-bias and high-multiplicity data set at $\sqrt{s}=\SI{13}{\TeV}$ was recorded in June 2015. The average pile-up
$\left<\mu\right>$ varied between 0.003 and 0.03, and the interaction rate
had a maximum of about \SI{15}{\kHz}. More than 200 million interactions were recorded
during a one-week data-taking period.
Most of the readout bandwidth was dedicated to the loosest
\trig{L1_MBTS_1} trigger (described below) recording events at \SIrange{1.0}{1.5}{\kHz} on average.
 
\subsubsection{Reconstruction and selection}
The MBTS are used as the primary L1 hardware
triggers for recording inelastic events with minimum bias, as reported in 
Refs.~\cite{STDM-2015-02,STDM-2015-05}. The plastic scintillation
counters composing the system were replaced during LS1 and consist of two
planes of twelve counters, each plane formed of an inner ring of eight counters
and an outer ring of four counters. These rings are sensitive to charged
particles in the interval $2.07~<~|\eta|~<~3.86$. Each counter is connected to a
photomultiplier tube and provides a fast trigger via a constant fraction
discriminator and is read out through the Tile calorimeter data acquisition
system. 

The MBTS triggers require a certain multiplicity of counters to be above threshold
in a bunch-crossing with colliding beams. The \trig{L1_MBTS_1} and
\trig{L1_MBTS_2} triggers require any one or two of the 24 counters to be above
threshold, respectively. The coincidence of two hits in the latter suppresses
beam-induced backgrounds from low-energy neutrons and photons. The
\trig{L1_MBTS_1_1} trigger requires at least one counter to be above threshold
in both the $+z$ and $-z$ hemispheres of the detector and is used to seed the
high-multiplicity HLT triggers. The same trigger selections are also applied to empty
(no beam present) and unpaired (one beam present) beam-crossings to
investigate beam-induced backgrounds. No additional HLT selection is applied to \trig{L1_MBTS_1}
and \trig{L1_MBTS_2} triggered events.

The \trig{mb_sptrk} trigger is used to determine the efficiency of the MBTS. It
is seeded using a random trigger on filled bunches and requires at least two
reconstructed space-points in the Pixel system and three in the SCT, along with
at least one reconstructed track with $\pt > \SI{200}{\MeV}$. Studies using MC
simulation and a fully unbiased data sample have demonstrated that this control
trigger is unbiased with respect to the offline selection.

The primary high-multiplicity trigger (e.g.\ used in the measurement of
two-particle correlations~\cite{HION-2015-09}) is
\trig{mb_sp900_trk60_hmt_L1MBTS_1_1} and requires at least 900 reconstructed
space-points in the SCT and at least 60 reconstructed tracks with $\pt > \SI{400}{\MeV}$.
This higher \pt\ requirement for the high-multiplicity trigger is compatible 
with the \pt\ cut used for physics analysis and reduces the computational complexity of the
track-finding algorithms in the HLT to an acceptable level. 

\subsubsection{Trigger efficiencies}

\begin{figure}[htbp]
\centering
\subfloat[]{\includegraphics[width=0.5\textwidth]{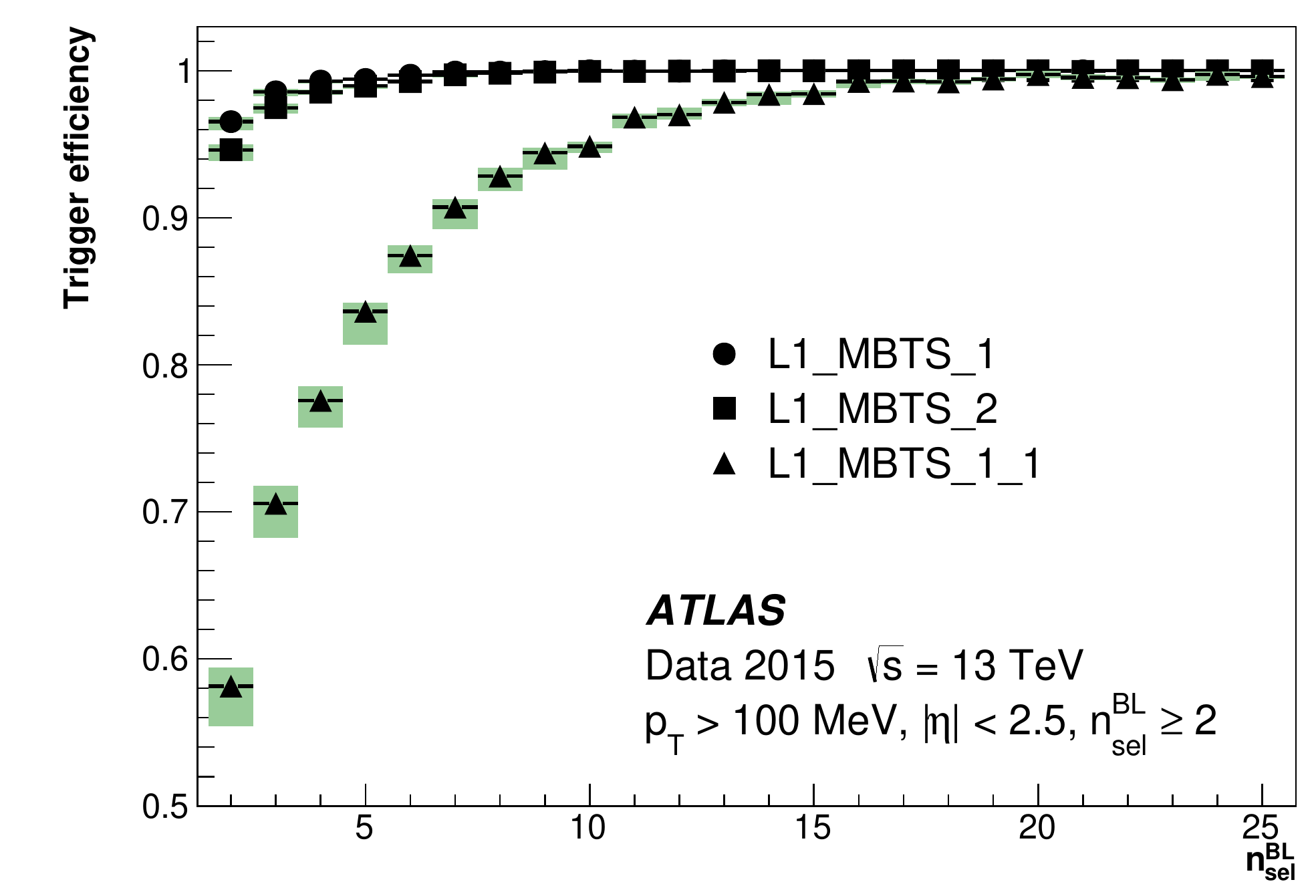}}
\subfloat[]{\includegraphics[width=0.5\textwidth]{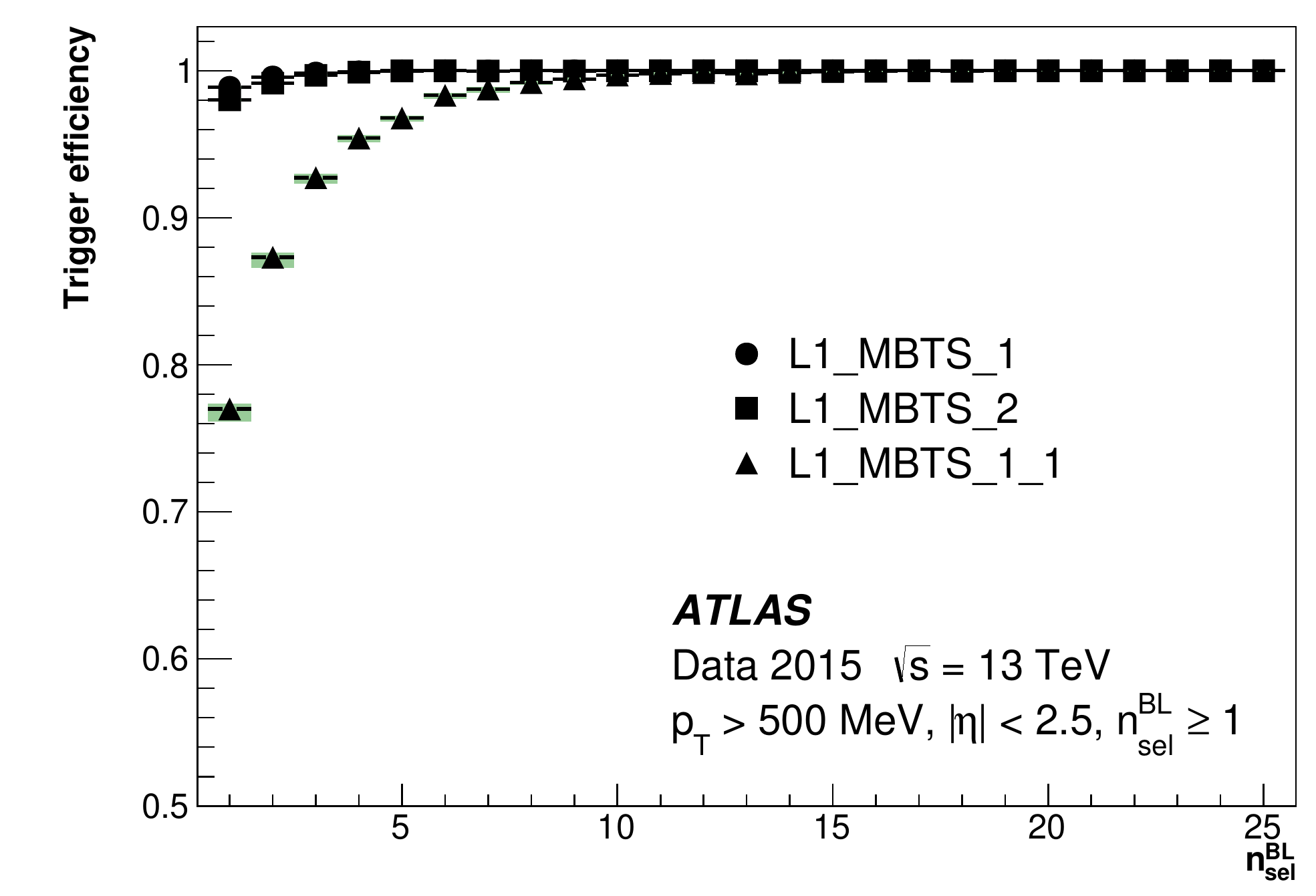}}
\caption{Efficiency of \trig{L1_MBTS_1}, \trig{L1_MBTS_2} and \trig{L1_MBTS_1_1}
  triggers as a function of the number tracks compatible with the beam line for
  two different transverse momentum requirements (a) $\pT>\SI{100}{\MeV}$ and (b)
  $\pT>\SI{500}{\MeV}$. The bands denote the total uncertainty. 
}
\label{fig:minbias_efficiency_mbts}
\end{figure}

The MBTS trigger efficiency is defined as the ratio of events passing  MBTS trigger, the control
trigger (\trig{mb_sptrk}) and offline selection to events passing the control
trigger and offline selection.
The efficiency is shown in Figure~\ref{fig:minbias_efficiency_mbts} for two
offline selections as a function of the number of selected tracks 
compatible in transverse impact parameter ($|d_0|<\SI{1.5}{\mm}$) with the beam line
($n_\mathrm{sel}^\mathrm{BL}$) for (a) $\pt > \SI{100}{\MeV}$
and (b) $\pt > \SI{500}{\MeV}$.
The efficiency is close to 95\% in the first bin, quickly rising to
100\% for \trig{L1_MBTS_1} and \trig{L1_MBTS_2}. The \trig{L1_MBTS_1_1} trigger, 
which requires at least one hit on both sides of the detector,
only approaches 100\% efficiency for events with around 15 tracks.
The primary reason for the lower efficiency of the \trig{L1_MBTS_1_1} trigger
compared to \trig{L1_MBTS_1}  or \trig{L1_MBTS_2} is
that at low multiplicities about 30\% of the inelastic events are due to diffractive
interactions where usually one proton stays intact and thus particles
from the interactions are only produced on one side of the detector.
Systematic uncertainties in the trigger efficiency are evaluated by removing the
cut on the transverse impact parameter with respect to the beam line from the track
selection and applying a longitudinal impact parameter cut with respect to the
primary vertex (for events where a primary vertex is reconstructed). This
results in a less than 0.1\% shift. The difference in response between the two
hemispheres is additionally evaluated to be at most 0.12\%. 

\begin{figure}[htbp]
\centering
\subfloat[]{\includegraphics[width=0.51\textwidth]{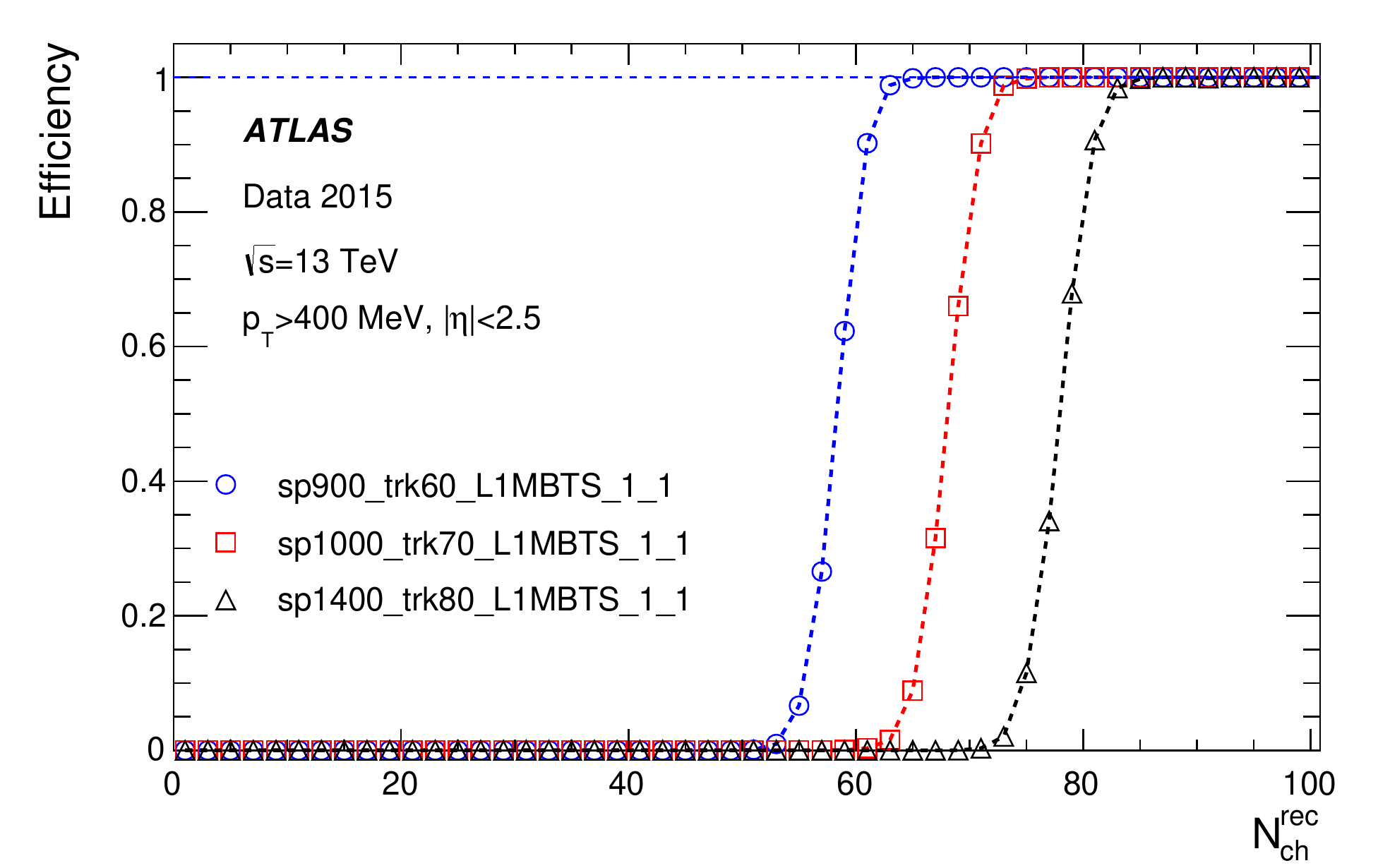}\label{fig:minbias_efficiency_mht400}}
\subfloat[]{\includegraphics[width=0.51\textwidth]{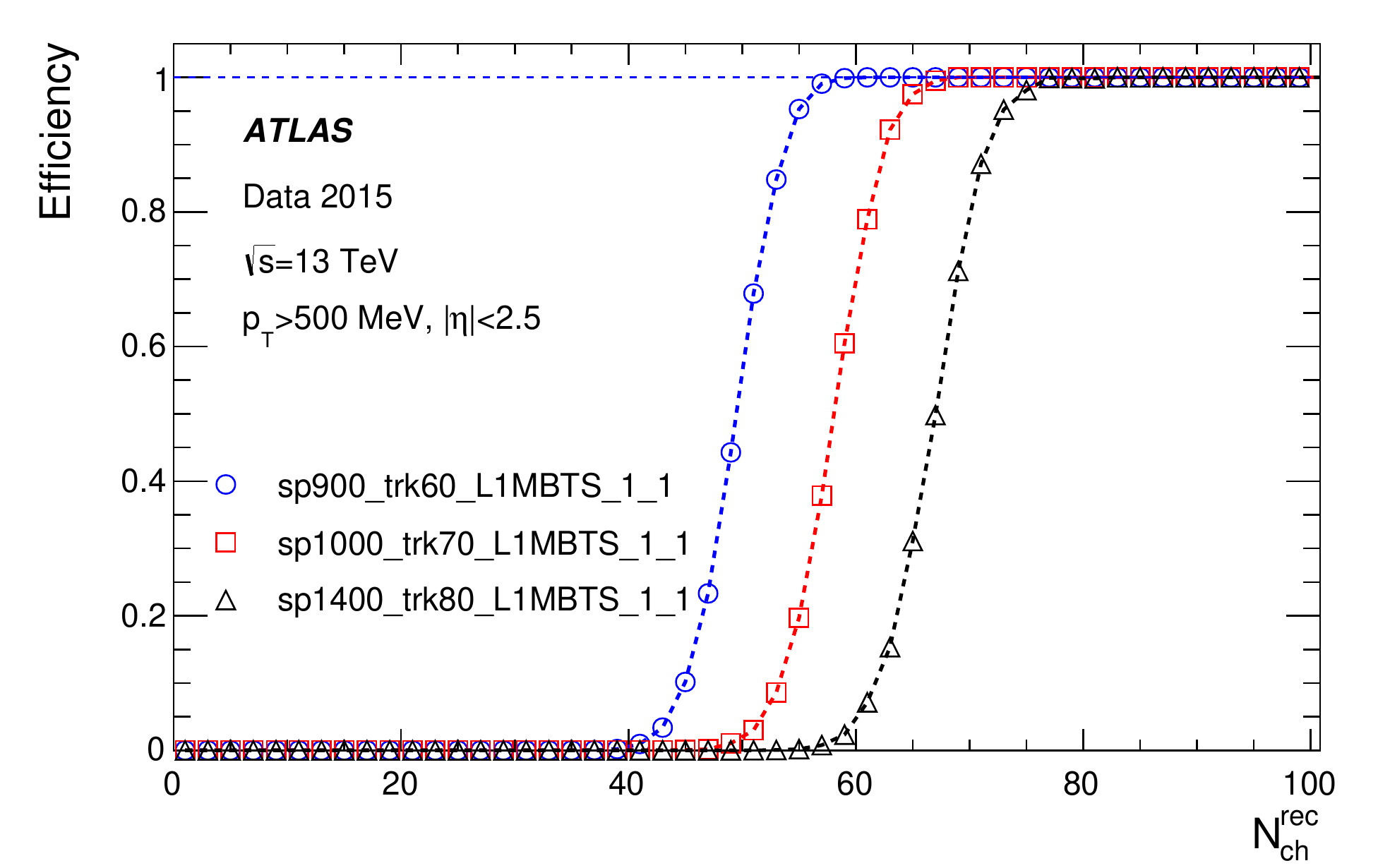}\label{fig:minbias_efficiency_mht500}}
\caption{Efficiency of high-multiplicity triggers as a function of the number of
  tracks compatible with the primary vertex for two different offline transverse
  momentum requirements (a) $\pT>\SI{400}{\MeV}$ and (b) $\pT>\SI{500}{\MeV}$. The curves
  represent three different selections on the minimum number of SCT space-points
  and reconstructed tracks (900/60, 1000/70 and 1400/80).}
\label{fig:minbias_efficiency_mht}
\end{figure}

The \trig{L1_MBTS_1} trigger is used as the control trigger for the determination of the
efficiency turn-on curves for the high-multiplicity data set. The efficiency is
parameterised as a function of the number of offline tracks associated with the
primary vertex. Figure~\ref{fig:minbias_efficiency_mht} shows the efficiency for three
different selections of the minimum number of SCT space-points and reconstructed tracks
and for two selections of the offline track $\pt$ requirement (above \SIlist{400;500}{\MeV}).
In the case of matching offline and trigger \pt\ selections ($\pt>\SI{400}{\MeV}$) shown in 
Figure~\ref{fig:minbias_efficiency_mht400}, the triggers are 100\% efficient for a value of five tracks above the
offline threshold (e.g.\ \trig{trk60} becomes fully efficient for 65 offline tracks). If the offline requirement is raised to \SI{500}{\MeV} as shown in Figure~\ref{fig:minbias_efficiency_mht500}, the trigger
is 100\% efficient for the required number of tracks.

\FloatBarrier
\subsection{Electrons and photons}
\label{sec:egamma}

Events with electrons and photons in the final state are important signatures for many ATLAS physics
analyses, from SM precision physics, such as Higgs boson, top quark, $W$ and $Z$ boson properties and production
rate measurements, to searches for
new physics. Various triggers cover the energy range between a few GeV and several TeV. Low-$\ET$ triggers
are used to collect data for measuring the properties of $J/\psi\rightarrow ee$, diphoton or low mass Drell-Yan
production. Single-electron triggers with $\ET$ above \SI{24}{\GeV}, dielectron triggers with lower thresholds
and diphoton triggers are used for the signal selection in a wide variety of ATLAS physics analyses such as
studies of the Higgs boson.

\subsubsection{Electron and photon reconstruction and selection}
\label{sec:egammasubsec}

At L1 the electron and photon triggers use the algorithms described in Section~\ref{sec:l1calo}.
The isolation and hadronic leakage veto cuts are not required for EM clusters with
transverse energy above \SI{50}{\GeV}.

At the HLT, electron and photon candidates are reconstructed and selected 
in several steps in order to reject events as fast as possible, thus allowing
algorithms which reproduce closely the offline algorithms and require more CPU time
to run at a reduced rate later in the trigger sequence. 
At first, fast calorimeter algorithms build clusters from the 
calorimeter cells (covering $0.025\times0.025$ in $\eta\times\phi$ space)
within the RoI ($\Delta\eta\times\Delta\phi=0.4\times0.4$)
identified by L1. Since electrons and photons deposit most of their energy
in the second layer of the EM calorimeter, this layer is used
to find the cell with the largest deposited transverse energy in the RoI.
EM calorimeter clusters of size $3\times7$ in the barrel ($|\eta|<1.4$) and
$5\times5$ in the end-cap ($1.4<|\eta|<2.47$) are used to reconstruct electrons
and photons. The identification of electrons and photons is based on the cluster \et\
as well as cluster shape parameters such as $R_{\text{had}}$, $R_\eta$ and
$E_{\text{ratio}}$,\footnote{$R_{\text{had}} = E^{\text{had}}_{\mathrm{T}} / E^{\text{EM}}_{\mathrm{T}}$ is the ratio
of the cluster transverse energy in the hadronic calorimeter to that in the
EM calorimeter. $R_\eta$ is based on the cluster shape in the second layer of the EM
calorimeter and defined as the ratio of transverse energy in a core region
of $3\times7$ cells in $\eta \times \phi$ to that in a $7\times7$ region, expanded
in $\eta$ from the $3\times7$ core. $E_{\text{ratio}}$ is defined as the ratio of the energy
difference between the largest and second-largest energy deposits in the cluster over the
sum of these energies in the front layer of the EM calorimeter.} the latter being used for 
electron candidates and a few tight photon triggers.
Electron candidates are required to have tracks from the fast tracking stage
with $\pt > \SI{1}{\GeV}$ and to match clusters within $\Delta\eta<$ 0.2.

The second step relies on precise offline-like algorithms. The energy of the
clusters is calibrated for electron and photon triggers separately using a multivariate
technique where the response of the calorimeter layers is corrected in data and
simulation~\cite{ATL-PHYS-PUB-2016-015}. Precision tracks extrapolated to the 
second layer of the EM calorimeter are required to match to clusters within
$\Delta\eta$ of 0.05 and $\Delta\phi$ of 0.05.
Electron identification relies on a multivariate technique using a likelihood (LH)
discriminant with three operating points named \emph{loose LH}, \textit{medium LH}
and \textit{tight LH}. An additional working point named \emph{very loose LH} is
used for supporting triggers. The LH-based identification
makes use of variables similar to the cut-based identification employed during \runi~\cite{ATLAS-CONF-2012-048}
but has better background rejection for the same signal efficiency.
The discriminating variables used offline are also used by the trigger,
exploiting the characteristic features of energy deposits in the EM calorimeters (longitudinal and 
lateral shower shapes), track quality, track-cluster matching, and 
particle identification by the TRT.
All variables are described in Refs.~\cite{ATLAS-CONF-2016-024,ATL-PHYS-PUB-2016-014}.
The composition of the likelihood is the same as in the offline reconstruction with the exception of momentum loss
due to bremsstrahlung, $\Delta p/p$, which is not accounted for in the online environment.
The photon identification relies only on the cluster shower-shape variables and three working
points are also defined: \emph{loose}, \textit{medium} and \textit{tight}.

Not applied during 2015 but foreseen for higher luminosities during \runii is an additional
requirement on isolation for the lowest-threshold unprescaled single-electron
trigger. The isolation parameter is calculated as the sum of the $\pt$ values of all tracks
in a cone of size $\Delta R=0.2$
around the electron for tracks with $\pt > \SI{1}{\GeV}$ and $|\Delta z_{0} \sin\theta| <$ 0.3, where
$\Delta z_0$ is the distance along $z$ between the longitudinal impact parameter of the 
track and the leading track in the RoI.
The ratio of this quantity to the EM cluster $\ET$, namely $\sum{\pt}/\ET$, is used
to estimate the energy deposited by other particles.

\subsubsection{Electron and photon trigger menu and rates}

The primary L1 and HLT electron and photon triggers used in 2015 
are listed in Table~\ref{T:AtlasTriggerMenu}.
The lowest-threshold single-electron trigger (\trig{e24_lhmedium_L1EM20VH}) applies a \SI{24}{\GeV}
transverse energy thresh\-old and requires the electron to pass medium LH identification
requirements. The trigger is seeded by
\trig{L1_EM20VH}, which requires $\ET>\SI{20}{\GeV}$, and applies
an \ET-dependent veto against energy deposited in the hadronic
calorimeter behind the electromagnetic cluster of the electron candidate (hadronic veto, denoted by \trig{H} in 
the trigger name). The \ET\ threshold varies
slightly as a function of $\eta$ to compensate for passive material in front of the calorimeter (denoted by
\trig{V} in the trigger name).
To recover efficiency in the high transverse energy regime, this trigger is complemented
by a trigger requiring a transverse energy above \SI{120}{\GeV} with loose LH
identification (\trig{e120_lhloose}). With a maximum instantaneous luminosity of 
$5.2\times\lumi{e33}$ reached during the 2015 data-taking, the
rates of electron triggers could be sustained without the use of additional
electromagnetic or track isolation requirements at L1 or HLT. The
lowest-threshold dielectron trigger (\trig{2e12_lhloose_L12EM10VH}) applies a \SI{12}{\GeV}
transverse energy threshold and requires the two electrons to pass loose LH
identification requirements. The trigger is seeded by
\trig{L1_2EM10VH}, which requires two electrons with \ET\ above \SI{10}{\GeV} and a hadronic energy veto.

The primary single-photon trigger used in 2015 is \trig{g120_loose}. It requires
a transverse energy above \SI{120}{\GeV} and applies loose photon identification
criteria. It is seeded by \trig{L1_EM22VHI}, which requires an isolated electromagnetic cluster
(denoted by \trig{I} in the trigger name) with \ET\ above \SI{22}{\GeV} and applies a hadronic
veto and $\eta$-dependent \ET\ thresholds as described above. As mentioned
earlier, the electromagnetic isolation and hadronic veto requirements are not applied for
\ET\ above \SI{50}{\GeV}. The two main diphoton triggers are \trig{g35_loose_g25_loose},
which requires two photons above \SIlist{35;25}{\GeV} thresholds
and loose photon identification requirements, and \trig{2g20_tight},
which requires two photons with \ET\ above \SI{20}{\GeV} and tight identification. Both
triggers are seeded by \trig{L1_2EM15VH}, which requires two electromagnetic clusters
with \ET\ above \SI{15}{\GeV} and a hadronic veto.

\begin{figure}[htbp]
\centering
\includegraphics[width=0.5\textwidth]{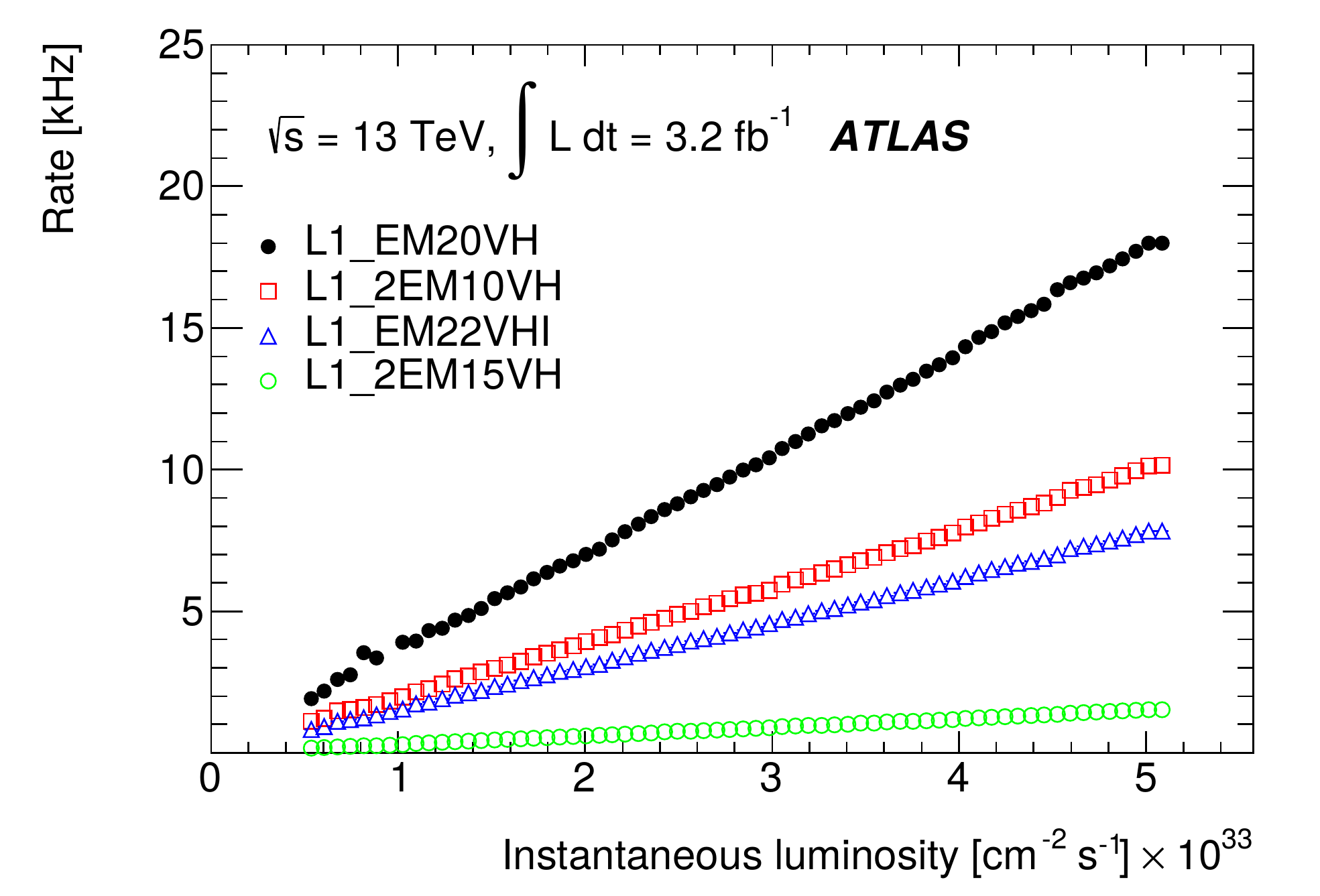}
\caption{L1 trigger rates as a function of the
instantaneous luminosity for selected single- and multi-object triggers.}
\label{fig:electron:l1rate}
\end{figure}

\begin{figure}[htbp]
\centering
\subfloat[]{\includegraphics[width=0.5\textwidth]{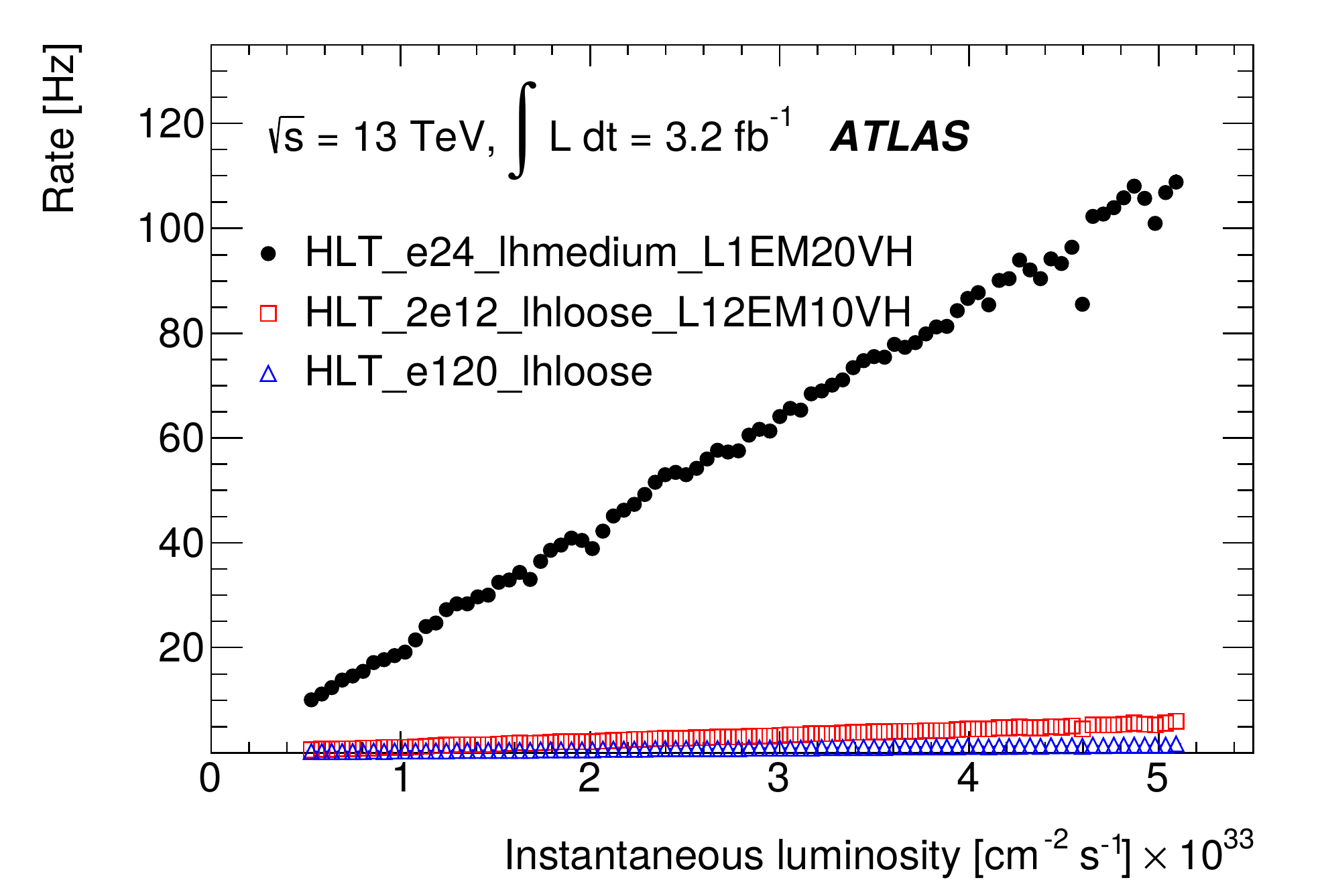}}
\subfloat[]{\includegraphics[width=0.5\textwidth]{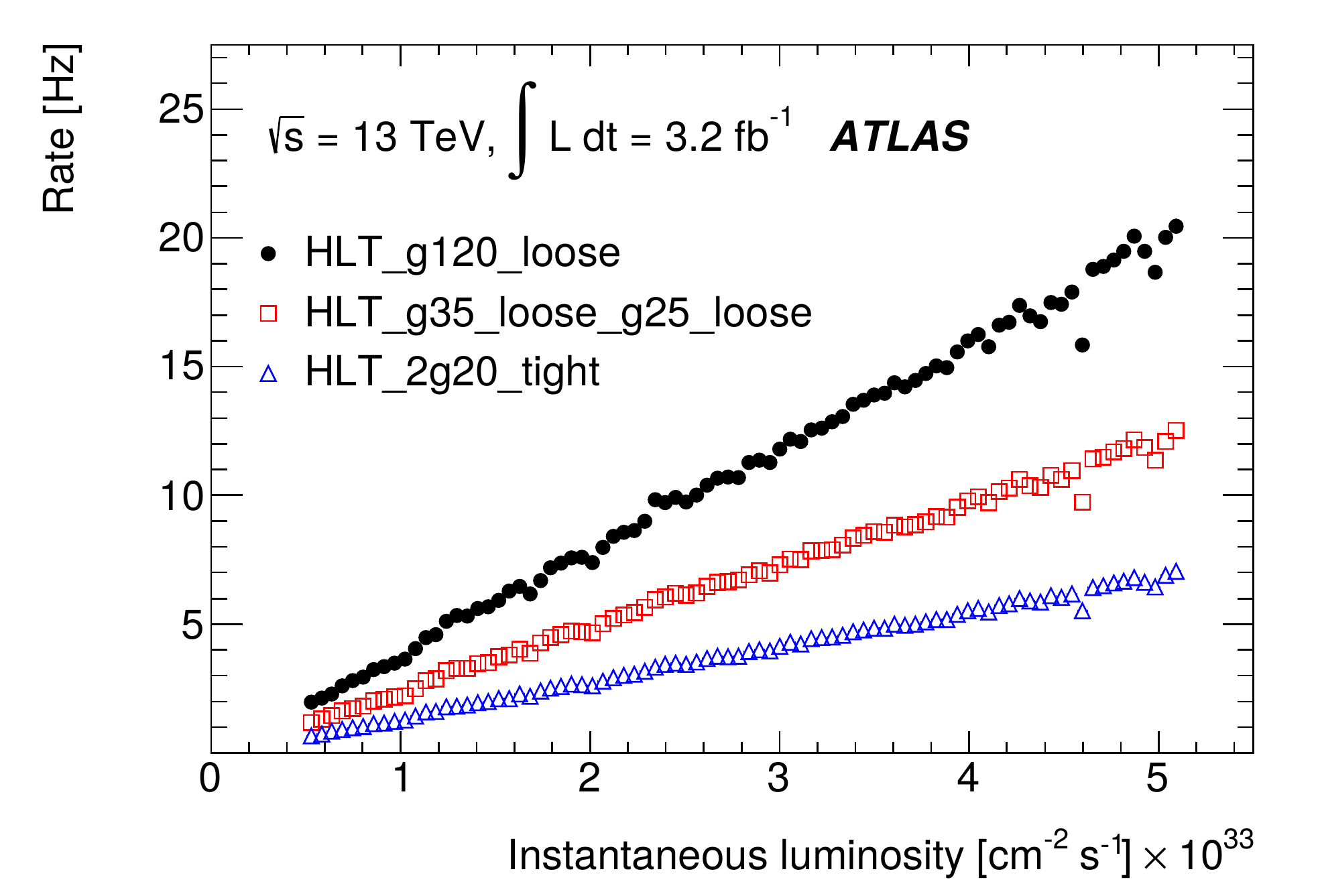}}
\caption{HLT trigger rates for (a) electron and (b) photon triggers as a function of the
instantaneous luminosity for selected single- and multi-object triggers.}
\label{fig:electron:hltrate}
\end{figure}

Figures~\ref{fig:electron:l1rate} and~\ref{fig:electron:hltrate} show the rates of the electron and photon triggers as a
function of the instantaneous luminosity. These trigger rates scale linearly with the instantaneous luminosity.

\subsubsection{Electron and photon trigger efficiencies}

\begin{figure}[htbp]
\centering
\subfloat[]{\includegraphics[width=0.5\textwidth]{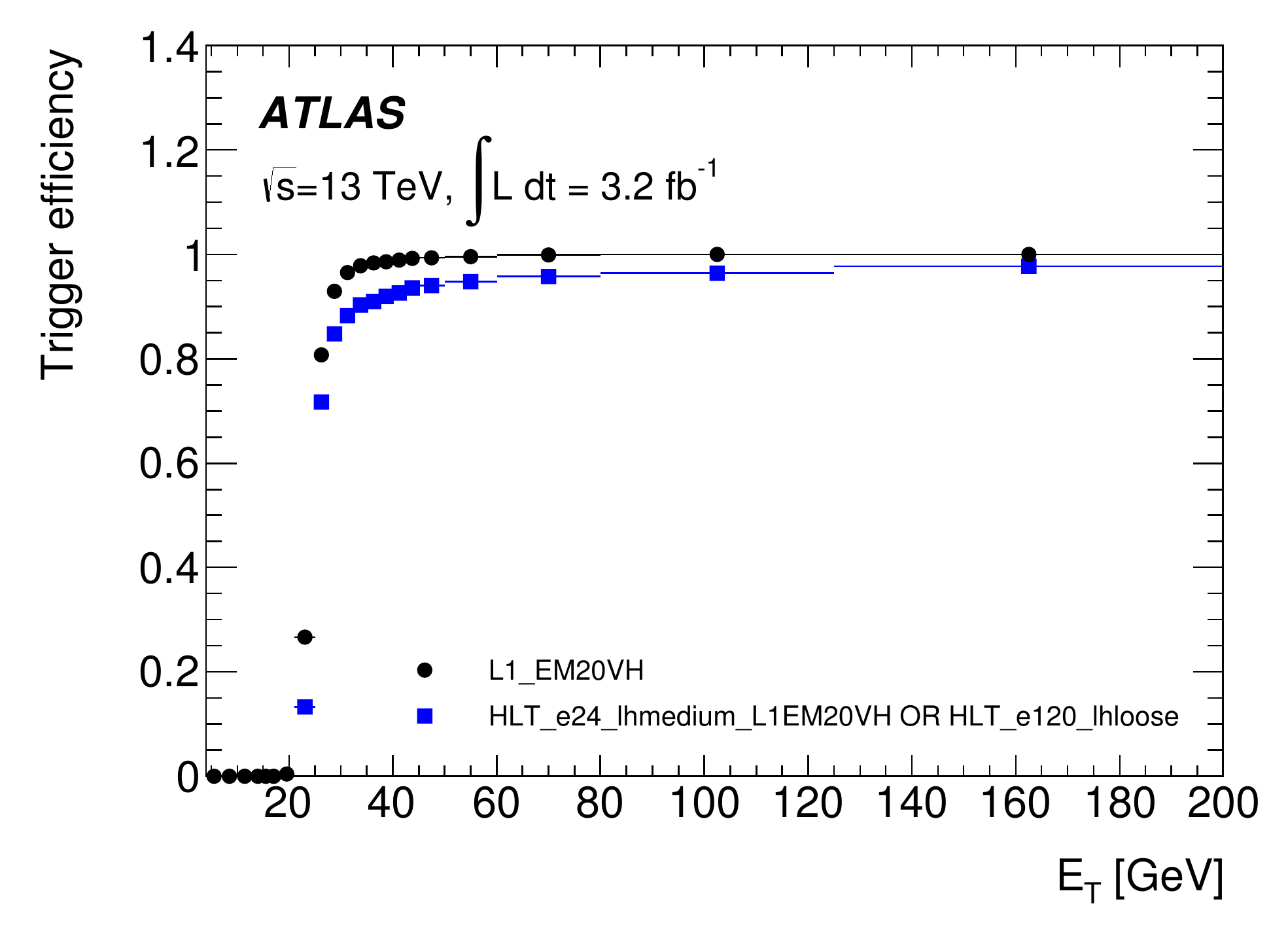}}
\subfloat[]{\includegraphics[width=0.5\textwidth]{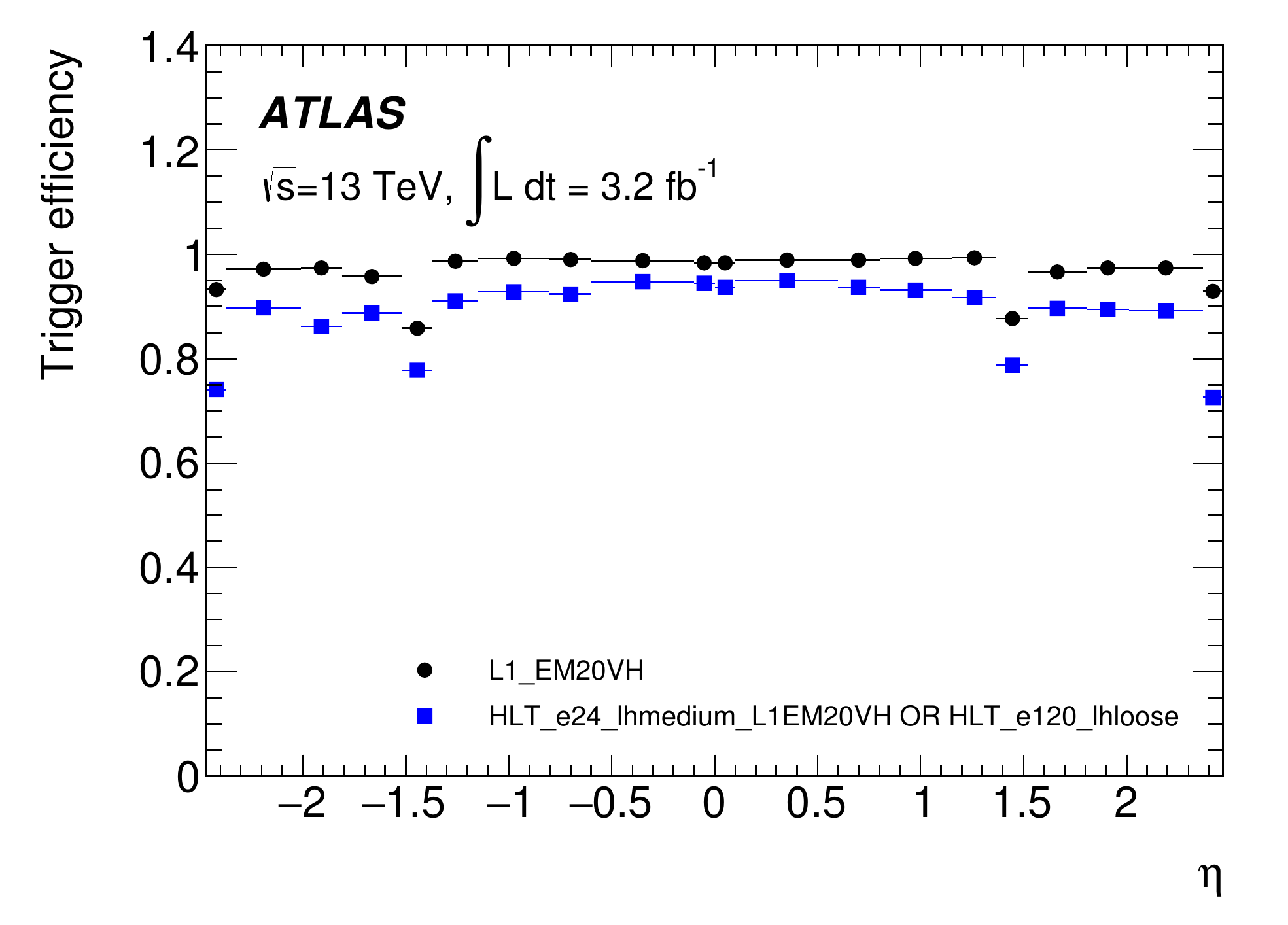}}
\caption{Efficiency of the \trig{L1_EM20VH} trigger and the logical `or' of
  the \trig{e24_lhmedium_L1EM20VH} and \trig{e120_lhloose} triggers 
  as a function of (a) the probe electron transverse energy \ET and (b) pseudorapidity $\eta$. 
  The offline reconstructed electron candidate
  is required to have an \ET value at least \SI{1}{\GeV} above the trigger threshold.}
\label{fig:electron:eff}
\end{figure}

The performance of electron triggers is studied using a sample of \Zee 
events. The tag-and-probe method utilises events triggered
by a single-electron trigger and requires two offline reconstructed electrons
with an invariant mass between \num{80} and \SI{100}{\GeV}.
After identifying the electron that triggered the event 
(tag electron), the other electron (probe electron) is unbiased by the trigger
selection, thus allowing its use to measure the electron trigger efficiency.
HLT electrons (L1 EM objects) are matched to the probe electron if
their separation is $\Delta R < 0.07 (0.15)$.
The trigger efficiency is calculated as the ratio of the number of
probe electrons passing the trigger selection to the number of probe electrons.
The efficiency of the combination of the lowest unprescaled single-electron trigger
\trig{e24_lhmedium_L1EM20VH} and the high transverse momentum electron
trigger \trig{e120_lhloose} with respect to the offline objects
is shown in Figure~\ref{fig:electron:eff} as a
function of the offline reconstructed electron transverse energy and 
pseudorapidity. The figure also shows the efficiency of the L1 trigger
(\trig{L1_EM20VH}) seeding the lowest unprescaled single-electron trigger.
A sharp turn-on can be observed for both the L1 and overall (L1 and HLT)
efficiency, and the HLT inefficiency with respect to L1 is small.
Inefficiencies observed around pseudorapidities of $-1.4$ and $1.4$ are due
to the transition region between the barrel and end-cap calorimeter.

\begin{figure}[htbp]
\centering
\subfloat[]{\includegraphics[width=0.5\textwidth]{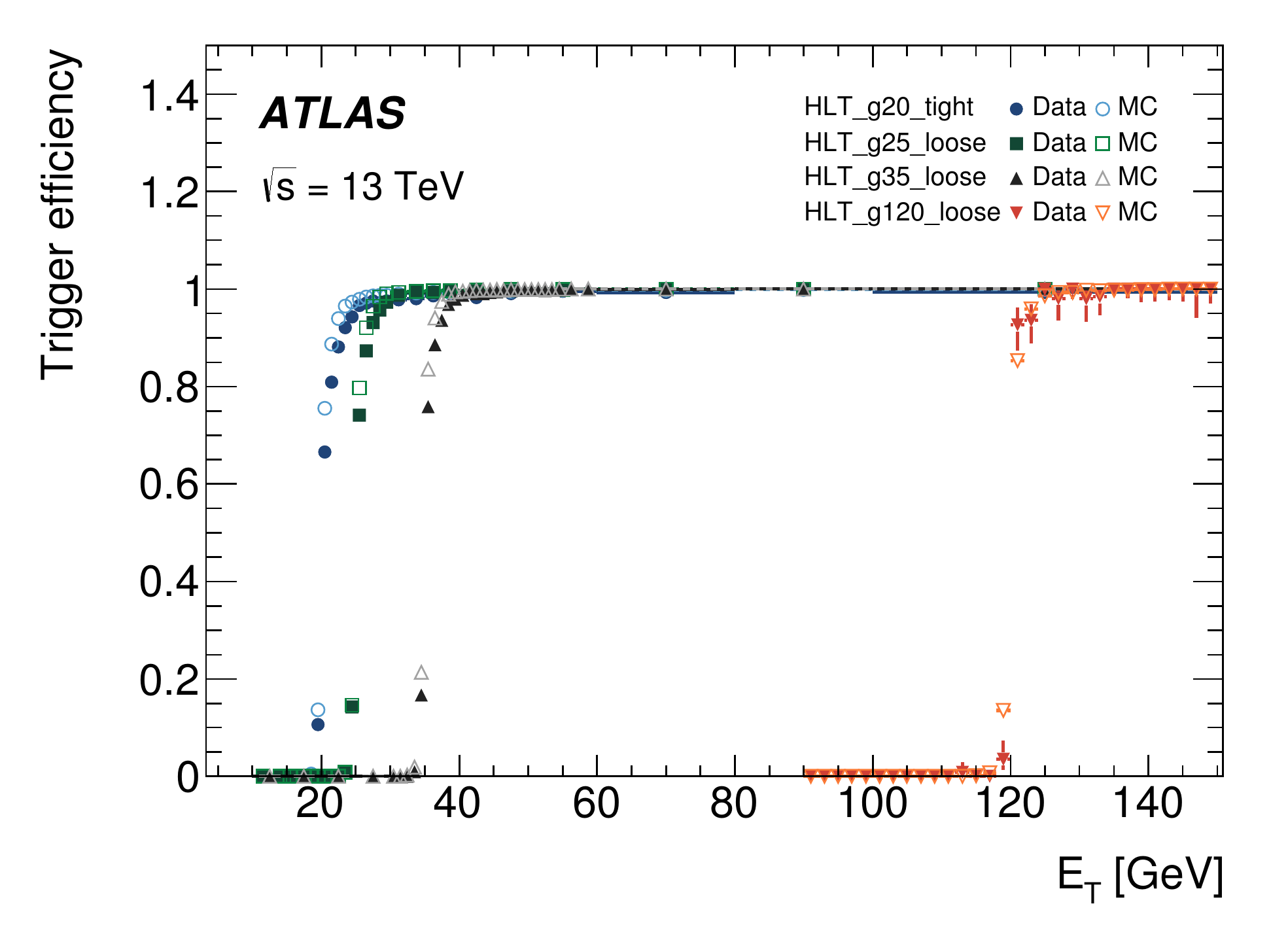}}
\subfloat[]{\includegraphics[width=0.5\textwidth]{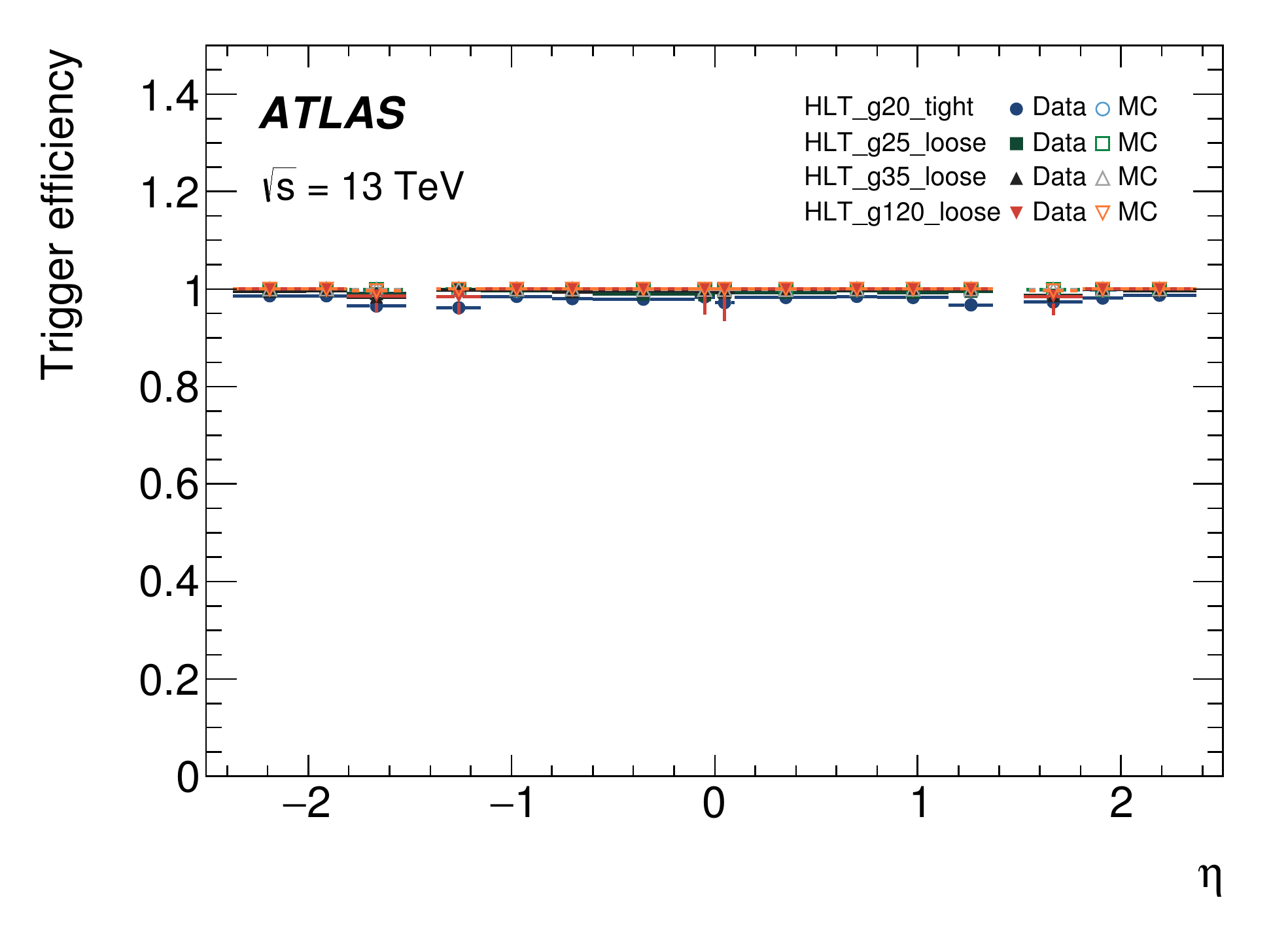}}
\caption{Efficiency of HLT photon triggers \trig{g20_tight}, \trig{g25_loose}, \trig{g35_loose},
  and \trig{g120_loose} relative to a looser HLT photon trigger as a function of (a) the transverse
  energy \ET and (b) pseudorapidity $\eta$ of the photon candidates reconstructed offline and
  satisfying the tight identification and isolation requirements. The offline reconstructed photon candidate
  is required to have an \ET value at least \SI{5}{\GeV} above the trigger threshold.
  The transition region between the barrel and end-cap calorimeter ($1.37<|\eta|<1.52$) is excluded.}
\label{fig:photon:eff}
\end{figure}

The photon trigger efficiency is computed using the bootstrap method
as the efficiency of the HLT trigger relative to a trigger with a lower \ET threshold.
Figure~\ref{fig:photon:eff} shows the efficiency of the main single-photon trigger
and the photons of the main diphoton trigger as a function of
the offline reconstructed photon transverse energy and pseudorapidity for data and MC simulation.
Very good agreement is observed between data and simulation.

\FloatBarrier
\subsection{Muons}
\label{sec:muon}

Muons are produced in many final states of interest to the ATLAS physics programme,
from SM precision physics to searches for new physics.  Muons are identified with high
purity compared to other signatures and cover a wide transverse momentum range, from
a few \si{\GeV} to several \si{\TeV}. Muon trigger thresholds in the \pT\ range from \SIrange{4}{10}{\GeV}
are used to collect data for measurements of processes such as $J/\psi\rightarrow\mu\mu$,
low-\pT\ dimuons, and $Z\rightarrow\tau\tau$~\cite{BPHY-2012-06,STDM-2011-18}.
Higher \pT\ thresholds are used to collect data for new-physics searches as well as measuring 
the properties and production rates of SM particles such as the Higgs, \Wboson and \Zboson\ bosons,
and top quarks~\cite{HIGG-2014-06,STDM-2015-03,TOPQ-2015-15}.

\subsubsection{Muon reconstruction and selection}

The trigger reconstruction algorithms for muons at L1 and the HLT are described in
Sections~\ref{sec:l1muon} and~\ref{sec:muonrec}, respectively. The selection criteria
depend on the algorithm used for reconstruction.
The MS-only algorithm selects solely on the \pT\ of the muon candidate measured
by the muon spectrometer; the combined algorithm makes selections based on the match
between the ID and MS tracks and their combined \pT; and the
isolated muon algorithm applies selection criteria based on the amount of energy
in the isolation cones.

\subsubsection{Muon trigger menu and rates}
 
The lowest-threshold single-muon trigger (\trig{mu20_iloose_L1MU15}) requires
a minimum transverse momentum of \SI{20}{\GeV} for combined muon candidates in addition to
a loose isolation: the scalar sum of the track \pT\ values in a cone of size $\Delta R = 0.2$
around the muon candidate is required to be smaller than 12\% of the
muon transverse momentum. The isolation requirement reduces the rate
by a factor of approximately 2.5 with a negligible efficiency loss.
The trigger is seeded by \trig{L1_MU15}, which requires a transverse momentum above \SI{15}{\GeV}.
At a transverse momentum above \SI{50}{\GeV} this trigger is complemented by a trigger not
requiring isolation (\trig{mu50}), to recover a small efficiency loss in the high
transverse momentum region.

\begin{figure}[htbp]
\centering
\subfloat[]{\includegraphics[width=0.5\textwidth]{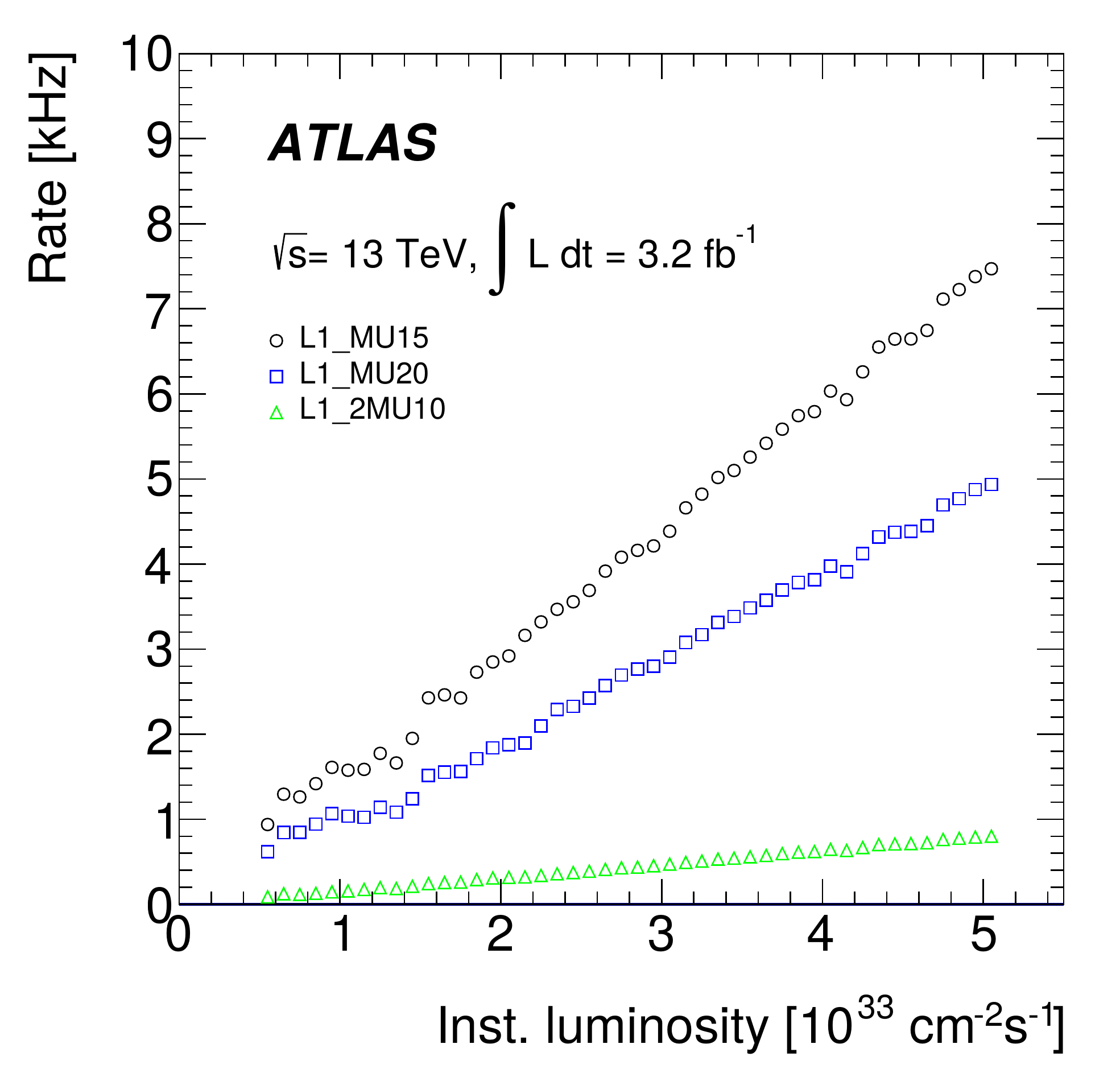}}
\subfloat[]{\includegraphics[width=0.5\textwidth]{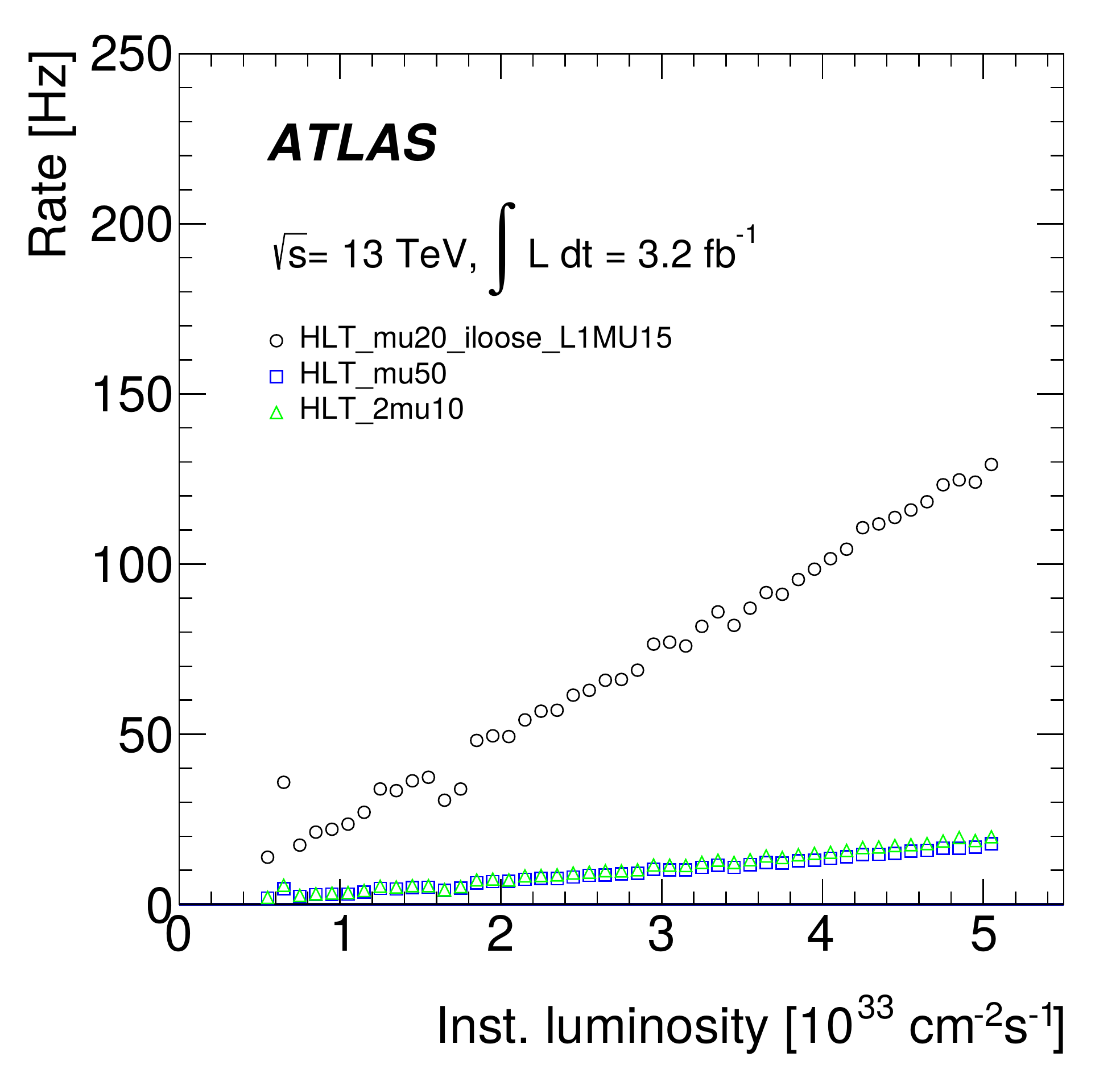}}
\caption{(a) L1 and (b) HLT muon trigger rates as a function of the
instantaneous luminosity for primary single and dimuon
triggers.}
\label{fig:mu:rate}
\end{figure}

The lowest-threshold unprescaled dimuon trigger (\trig{2mu10}) requires a minimum transverse
momentum of \SI{10}{\GeV} for combined muon candidates. The trigger is seeded by
\trig{L1_2MU10}, which requires two muons with transverse
momentum above \SI{10}{\GeV}. 
Figure~\ref{fig:mu:rate} shows the rates of these triggers as a
function of the instantaneous luminosity. The trigger rates scale linearly with the
instantaneous luminosity.
Dimuon triggers with lower \pt\ thresholds and further selections
(e.g.\ on the dimuon invariant mass) were also active and are discussed in
Section~\ref{sec:bphys}.
Additionally, an asymmetric dimuon trigger (\trig{mu18_mu8noL1}) is included, where 
\trig{mu18} is seeded by \trig{L1_MU15} and \trig{mu8noL1}
performs a search for a muon in the full detector at the HLT. 
By requiring only one muon at L1, the dimuon trigger does not suffer a loss of efficiency
that would otherwise have if two muons were required at L1.
This trigger is typically used by physics
searches involving two relatively high-\pT\ muons to
improve the acceptance with respect to the standard dimuon triggers.

\subsubsection{Muon trigger efficiencies}

The L1 and HLT muon efficiencies are determined using a tag-and-probe method
with $Z\rightarrow\mu\mu$ candidate events.
Events are required to contain a pair of reference muons with opposite charge and an
invariant mass within \SI{10}{\GeV} of the \Zboson mass. 
Reference muons reconstructed offline using both ID and MS information are required to be inside
the fiducial volume of the muon triggers ($|\eta|<2.4$) and pass the 
\emph{medium} identification requirements~\cite{PERF-2014-05,PERF-2015-10}.

\begin{figure}[htbp]
\centering
\subfloat[]{\includegraphics[width=0.5\textwidth]{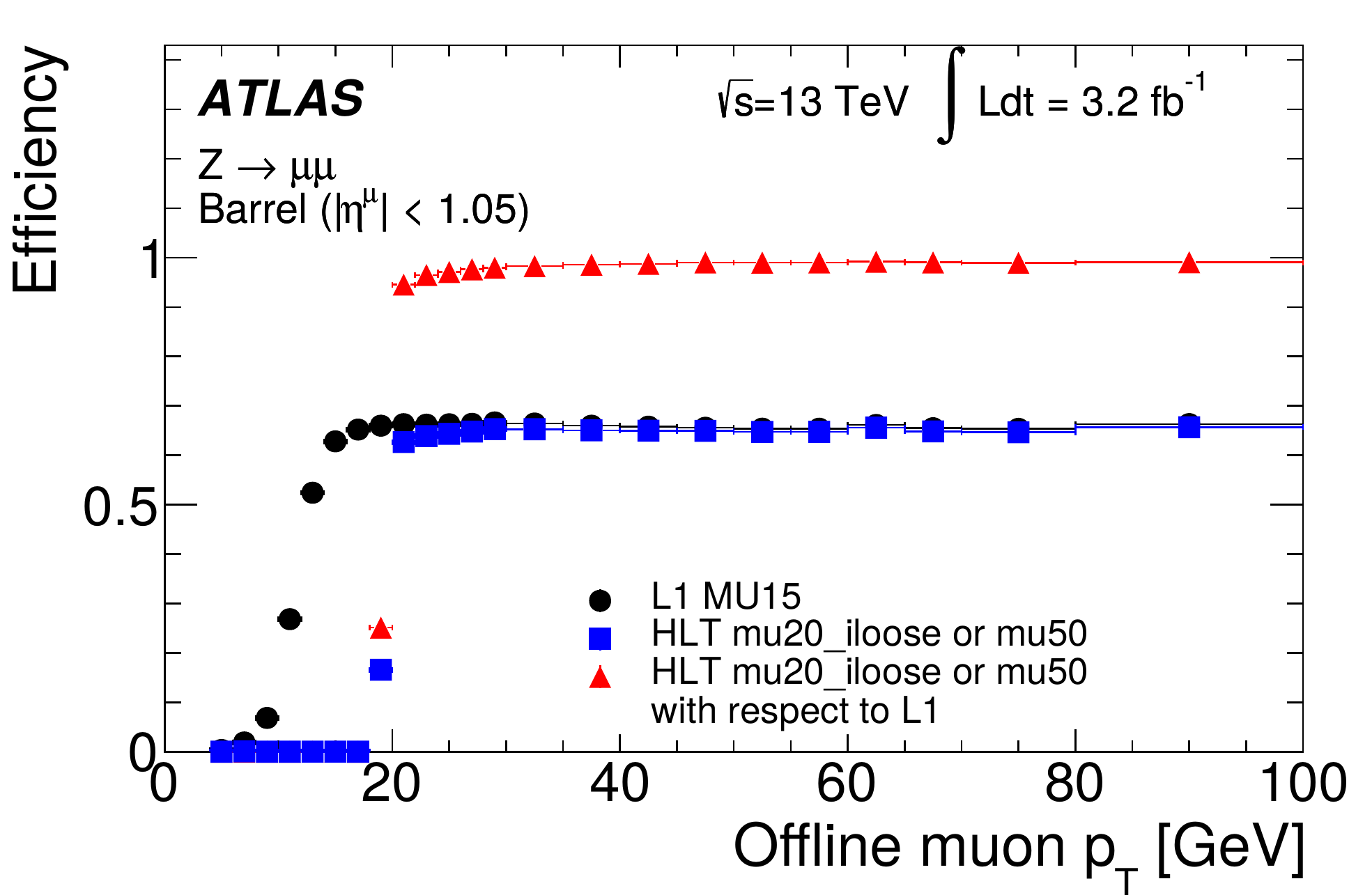}}
\subfloat[]{\includegraphics[width=0.5\textwidth]{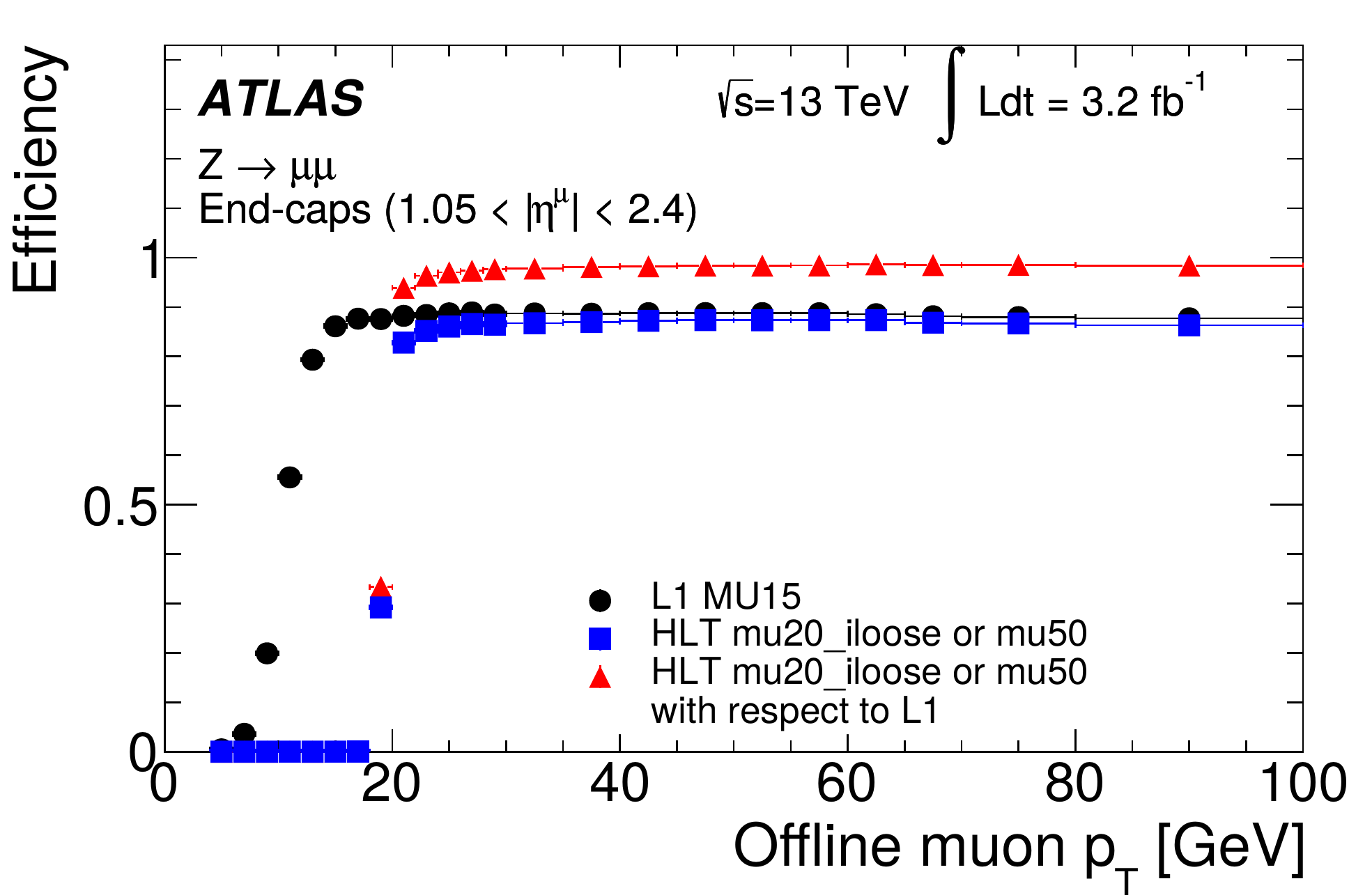}}
\caption{Efficiency of the L1 muon trigger \trig{L1_MU15} and the combination of the HLT muon triggers
  \trig{mu20_iloose_L1MU15} and \trig{mu50} as a function of the probe muon
  $p_\mathrm{T}$, separately for (a) the barrel and (b) the end-cap regions.}
\label{fig:mu:eff:pt}
\end{figure}

\begin{figure}[htbp]
\centering
\subfloat[]{
  \includegraphics[width=0.5\textwidth]{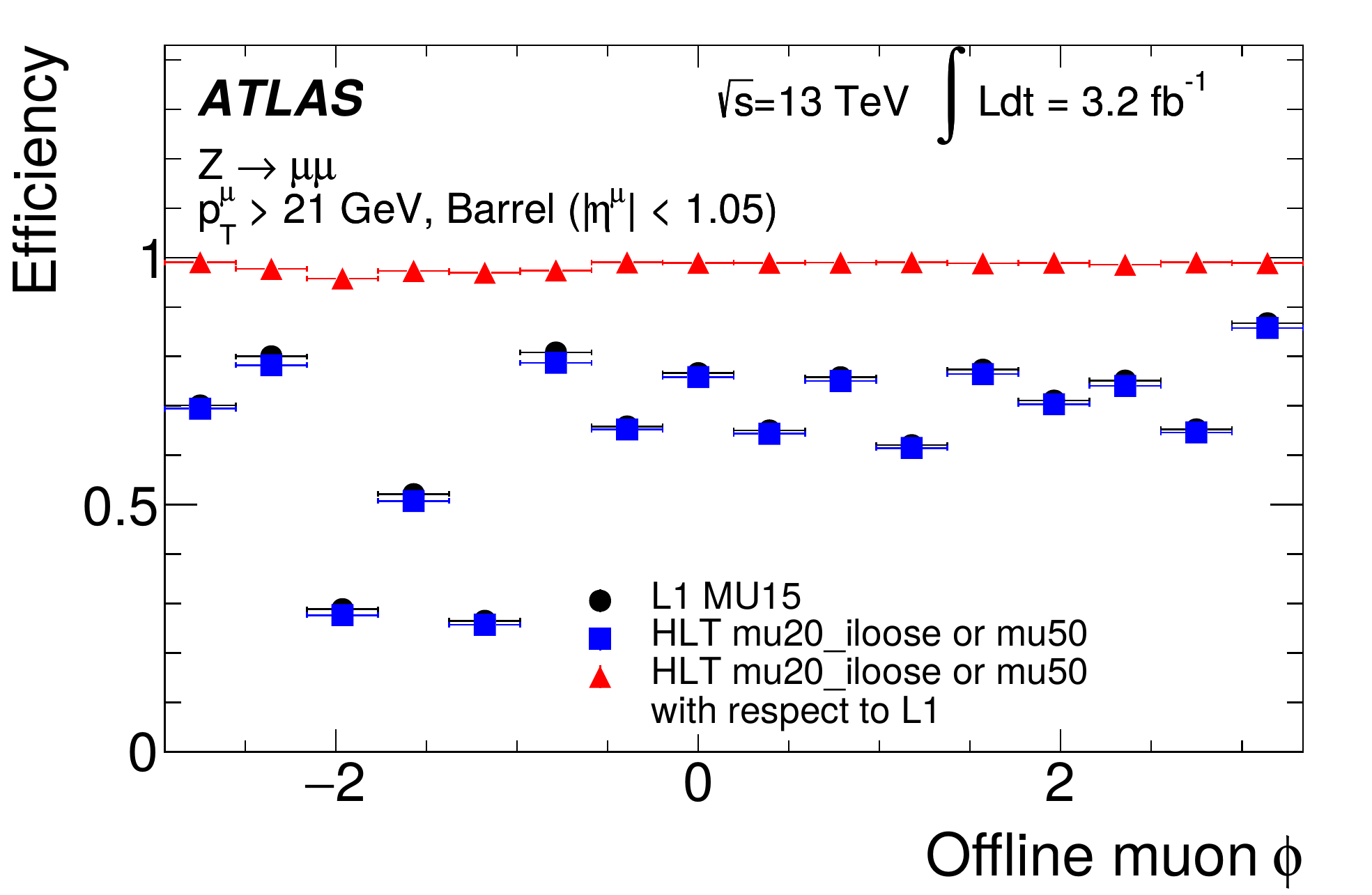}
  \label{fig:mu:eff:phi:barrel}
}
\subfloat[]{
  \includegraphics[width=0.5\textwidth]{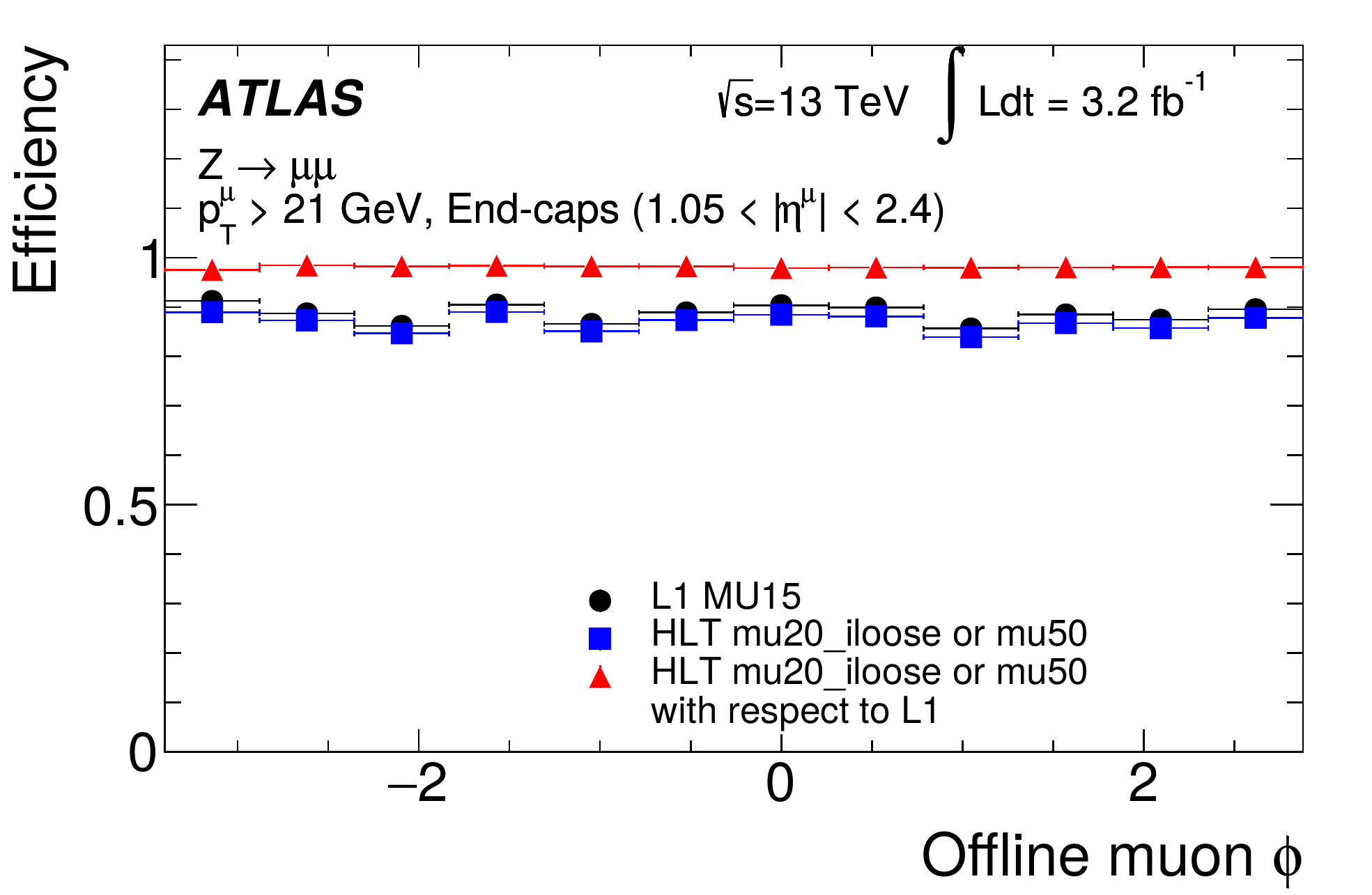}
  \label{fig:mu:eff:phi:endcap}
}
\caption{Efficiency of the L1 muon trigger \trig{L1_MU15} and the combination of the HLT muon triggers
  \trig{mu20_iloose_L1MU15} and \trig{mu50} as a function of the probe muon $\phi$,
  separately for (a) the barrel and (b) the end-cap regions.}
\label{fig:mu:eff:phi}
\end{figure}

The absolute efficiency of the \trig{L1_MU15} trigger and the absolute and relative
efficiencies of the logical `or' of \trig{mu20_iloose} and \trig{mu50} as
a function of the \pt\ of the offline muon track are shown in Figure~\ref{fig:mu:eff:pt}.
The L1 muon trigger efficiency is close to 70\% in the barrel and 90\% in the
end-caps. The different efficiencies are due to the different geometrical
acceptance of the barrel and end-cap trigger systems and local detector
inefficiencies. The HLT efficiency relative to L1 is close to 100\% both in the
barrel and in the end-caps.
Figure~\ref{fig:mu:eff:phi} shows the muon trigger efficiency as a function of the azimuthal angle $\phi$
of the offline muon track for (a) the barrel and (b) the end-cap regions. The reduced barrel
acceptance can be seen in the eight bins corresponding to the sectors
containing the toroid coils and in the two feet sectors around $\phi\approx-1.6$ and $\phi\approx-2.0$, respectively.

\FloatBarrier
\subsection{Jets}
\label{sec:jet}

Jet triggers are used for signal selection in a wide variety of physics measurements and
detector performance studies. Precision measurements of inclusive jet, dijet and multi-jet topologies
rely on the events selected with the single-jet and multi-jet triggers. Events selected by the
single-jet triggers are also used for the calibration of the calorimeter jet energy scale and
resolution. All-hadronic decays of \ttbar\ events can be studied using multi-jet signatures
and the all-hadronic decay of the weak bosons, Higgs bosons and top quarks can be selected in high
transverse momentum (`boosted') topologies using large-radius jets. Searches for physics beyond
the SM, such as high-mass dijet resonances, supersymmetry or large extra dimensions,
often utilise single-jet and multi-jet unprescaled triggers with a high transverse momentum threshold. 

\subsubsection{Jet reconstruction}
\label{sec:jet_reco}

A detailed description of the jet triggers used during \runi can be found in Ref.~\cite{TRIG-2012-01}.
Jets are reconstructed in the HLT using the anti-$k_t$ jet algorithm~\cite{antikT} with a radius
parameter of $R=0.4$ or $R=1.0$. The inputs to the algorithm are calorimeter
topo-clusters that are reconstructed from the full set of calorimeter
cell information calibrated by default at the EM scale. The jets are
calibrated in a procedure similar to that adopted for offline physics analyses~\cite{PERF-2012-01}.
First, contributions to the jet energy from pile-up
collisions are subtracted on an event-by-event basis using the calculated area
of each jet and the measured energy density within $|\eta|<2$. Second, the
response of the calorimeter is corrected using a series of \pt-
and $\eta$-dependent calibration factors derived from simulation.

The jet reconstruction in the HLT is highly flexible and some triggers use
non-standard inputs or a calibration procedure that differs from the default
outlined above. For example, the clusters can be reconstructed using
cells from a restricted region in the calorimeter defined using the RoIs
identified by the L1 trigger. The clusters can also be calibrated using
local calibration weights that are applied after classifying each cluster as
electromagnetic or hadronic in origin. Furthermore, the jet calibration can be
applied in four ways: no jet calibration, pile-up subtraction only, jet response
correction only, or both pile-up subtraction and jet response corrections (default). 
Finally, the jet reconstruction can be run twice to produce \emph{reclustered} 
jets~\cite{Nachman:2014kla}, in which the input to the second jet-finding is the output from the 
first, e.g.\ to build large-$R$ jets from small-$R$ jets.

\subsubsection{Jet trigger menu and rates}
The jet trigger menu consists of
\emph{single-jet} triggers, which require at least one jet above a given transverse energy threshold, 
\emph{multi-jet} triggers, which require at least $N$ jets above a given transverse energy threshold,
\HT\ triggers, which require the scalar sum of the transverse energy of all jets in the
event, \HT, above a given threshold,
and \emph{analysis-specific} triggers for specific topologies of interest. 
The jet triggers use at L1 either a random trigger (on colliding bunches) 
or an L1 jet algorithm.
The random trigger is typically used for triggers that select events with
offline jet $\pt<\SI{45}{\GeV}$ to avoid bias due to inefficiencies
of the L1 jet algorithm for low-\pt jets. In the following, only the
most commonly used jet triggers are discussed.

The lowest-threshold unprescaled single-jet trigger for standard jets ($R=0.4$) selects
events that contain a jet at L1 with transverse energy above \SI{100}{\GeV} (\trig{L1_J100})
and a jet in the HLT with transverse energy above \SI{360}{\GeV} (\trig{j360}). This trigger
has a rate of \SI{18}{\Hz} at a luminosity of $5\times\lumi{e33}$.
The lowest-threshold unprescaled multi-jet triggers are \trig{3j175}, \trig{4j85},
\trig{5j60} and \trig{6j45}, which have rates of \SIlist{6;20;15;12}{\Hz}, respectively.
The lowest-threshold unprescaled \HT\ trigger
used in 2015 is \trig{ht850} with a rate of \SI{12}{\Hz} where one jet with transverse
energy above \SI{100}{\GeV} is required at L1 and 
\HT\ is required to be above \SI{850}{\GeV} at HLT.

In addition to the unprescaled triggers, a set of lower-threshold triggers
select events that contain jets with lower transverse momentum and are typically
prescaled to give an event rate of \SI{1}{\Hz} each. The lowest-threshold
single-jet trigger
in 2015 is \trig{j15}, which uses a random trigger at L1. Multiple thresholds
for single jets exist between \trig{j15} and \trig{j360} to cover the
entire \pt\ spectrum.

\subsubsection{Jet trigger efficiencies}

\begin{figure}[htbp]
\captionsetup[subfloat]{captionskip=1mm}
\centering
\subfloat[]{
  \includegraphics[width=0.5\textwidth]{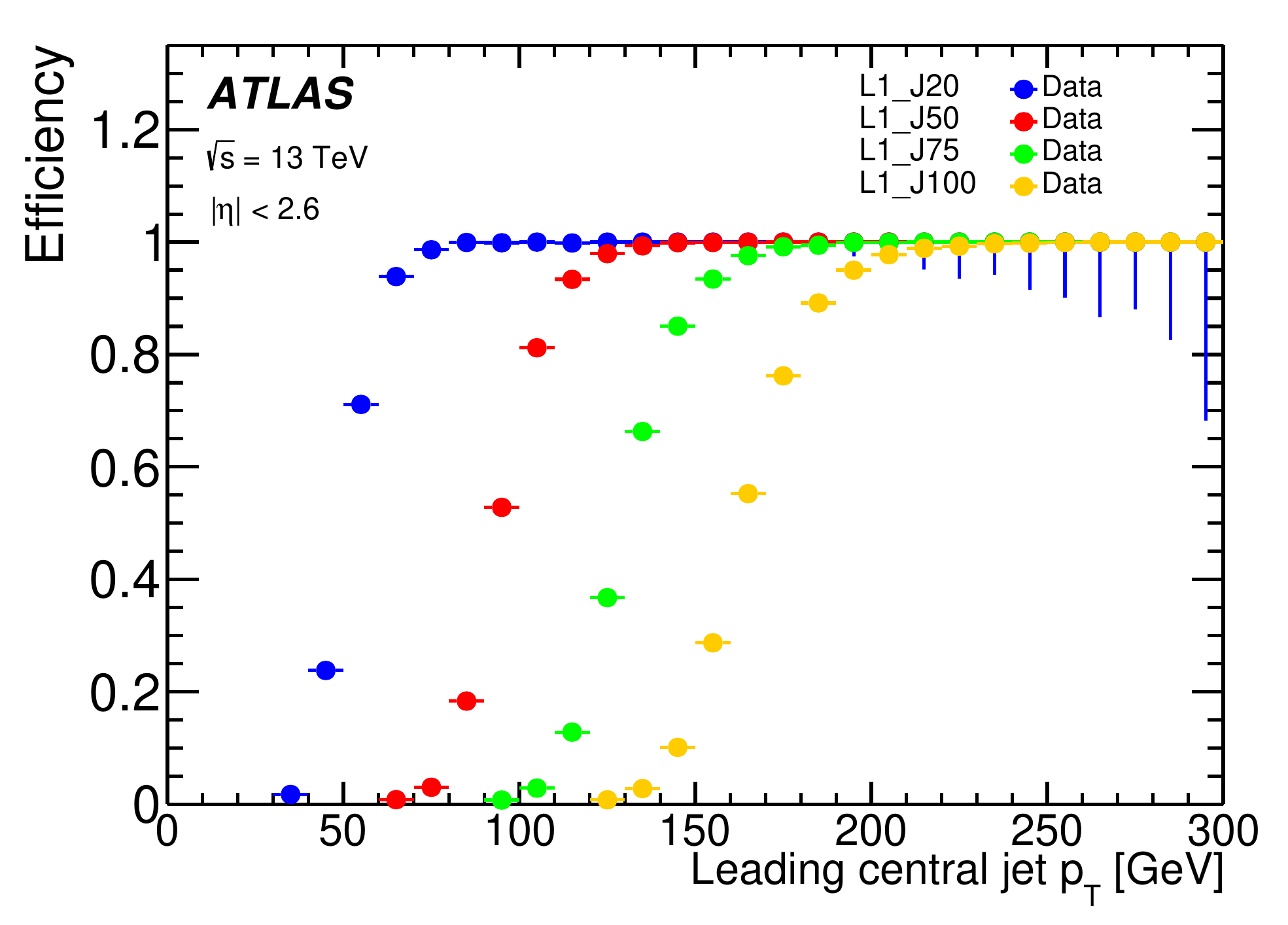}
  \label{fig:jet_efficiency_singlejet_L1}
}
\subfloat[]{
  \includegraphics[width=0.5\textwidth]{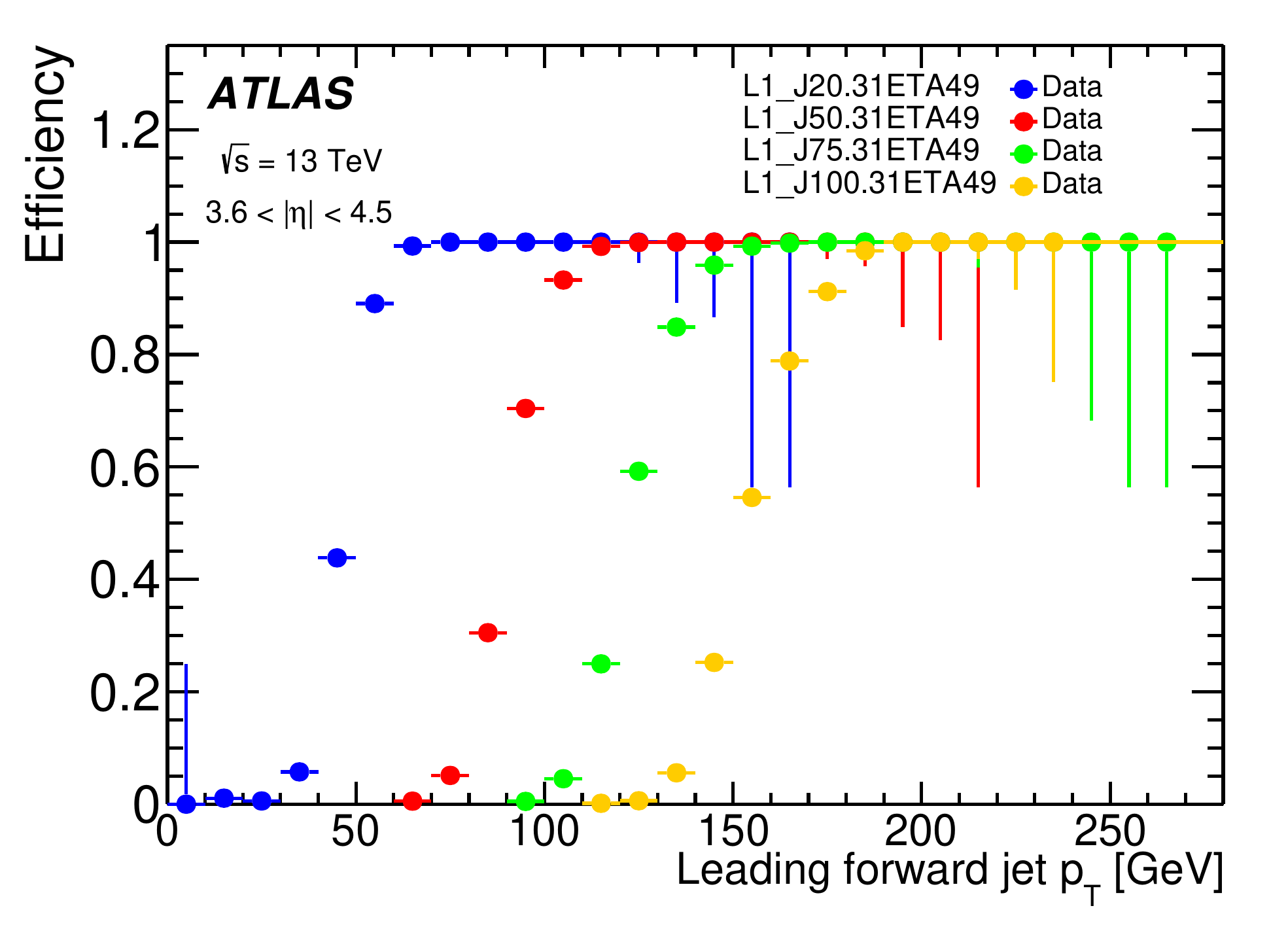}
  \label{fig:jet_efficiency_singlejet_L1fw}
}
\\
\subfloat[]{
  \includegraphics[width=0.5\textwidth]{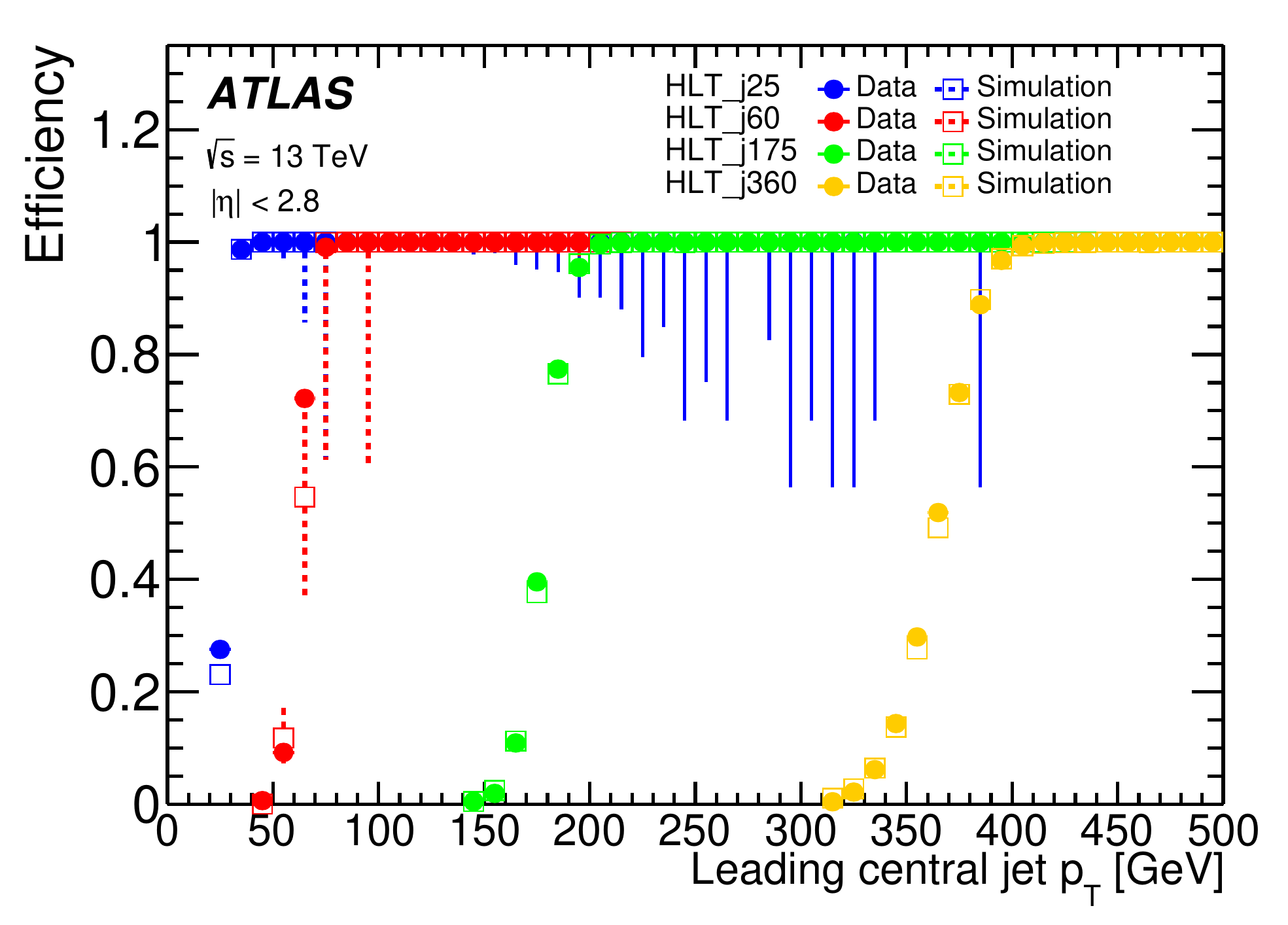}
  \label{fig:jet_efficiency_singlejet_HLT}
}
\subfloat[]{
  \includegraphics[width=0.5\textwidth]{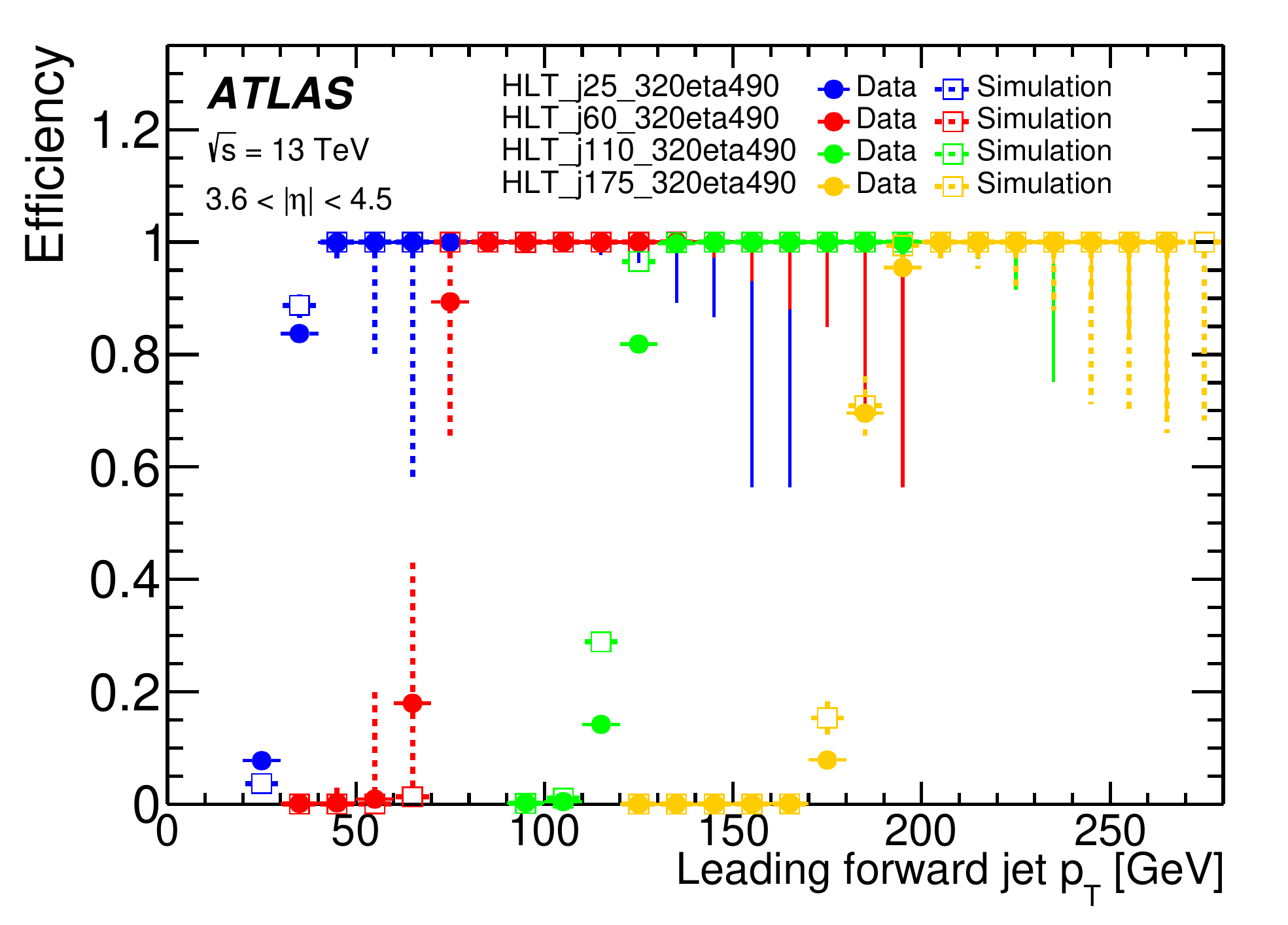}
  \label{fig:jet_efficiency_singlejet_HLTfw}
}
\caption{Efficiency of single-jet triggers as a function of offline jet \pT{} for (a) L1 in
  the central region, (b) L1 in the forward region, (c) HLT in the central region, and (d) HLT
  in the forward region. }
\label{fig:jet_efficiency_singlejet}
\end{figure}

\begin{figure}[htbp]
\centering
\includegraphics[width=0.5\textwidth]{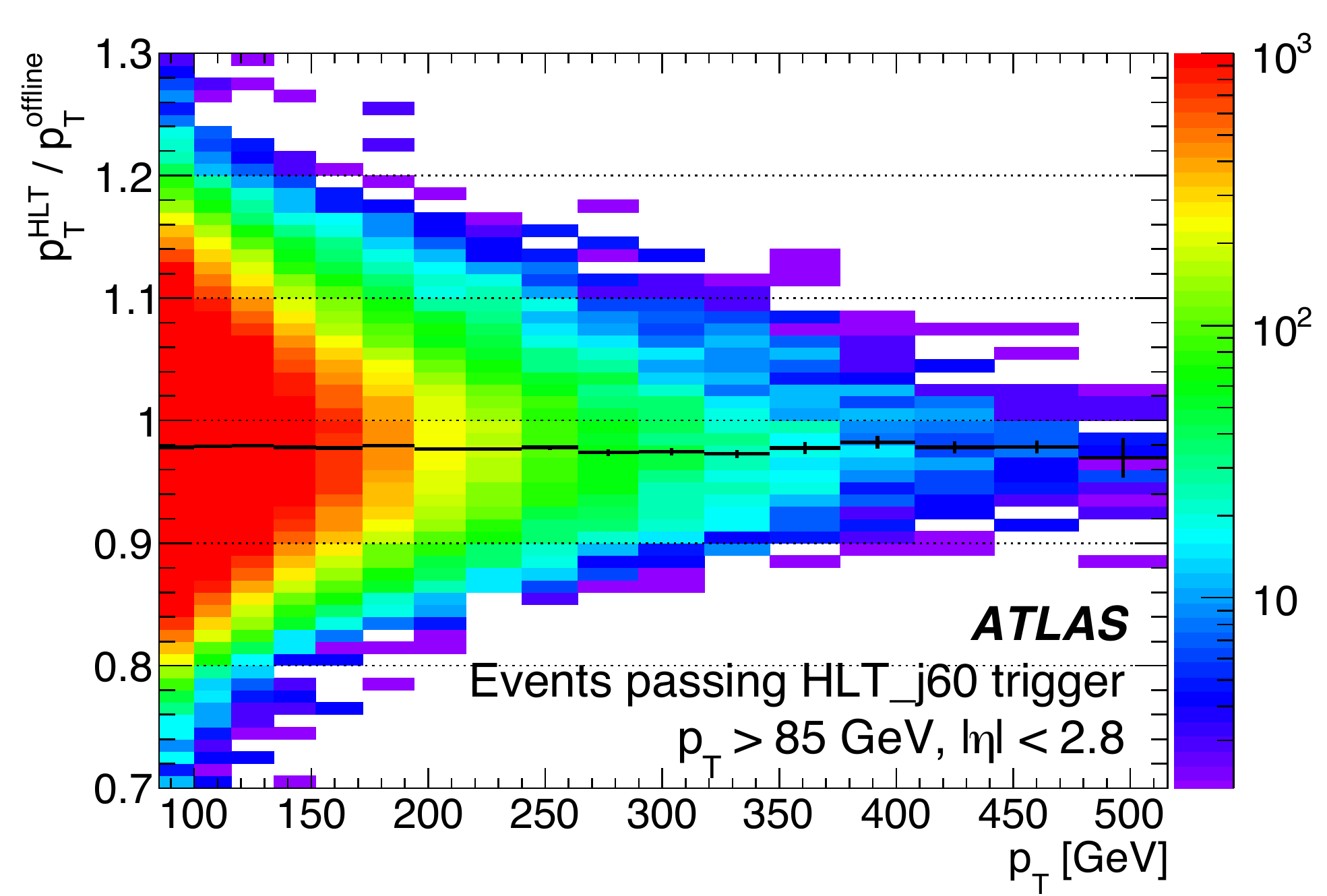}
\caption{Comparison between the jet energy scales of trigger and offline jets. The black points represent the
  mean of the distribution at a given \pt\ value. The 2\% shift is due to differences in the jet calibration
  applied online and offline.}
\label{fig:jet_performance_comparison}
\end{figure}

Jet trigger efficiencies are determined using the bootstrap method with respect to the \pt\ of the jet.
The single-jet trigger efficiencies for L1 and the HLT are shown in Figure~\ref{fig:jet_efficiency_singlejet}
for both the central and forward regions of the calorimeter. The ranges in $|\eta|$ are chosen to ensure
that the probe jet is fully contained within the $|\eta|$ region of study. 
Good agreement is observed between simulation and data. The sharp HLT efficiency
turn-on curves in Figure~\ref{fig:jet_efficiency_singlejet} are due to good agreement between the energy scale
of jets in the HLT and offline, as shown in Figure~\ref{fig:jet_performance_comparison}.

\begin{figure}[htbp]
\centering
\subfloat[]{\includegraphics[width=0.5\textwidth]{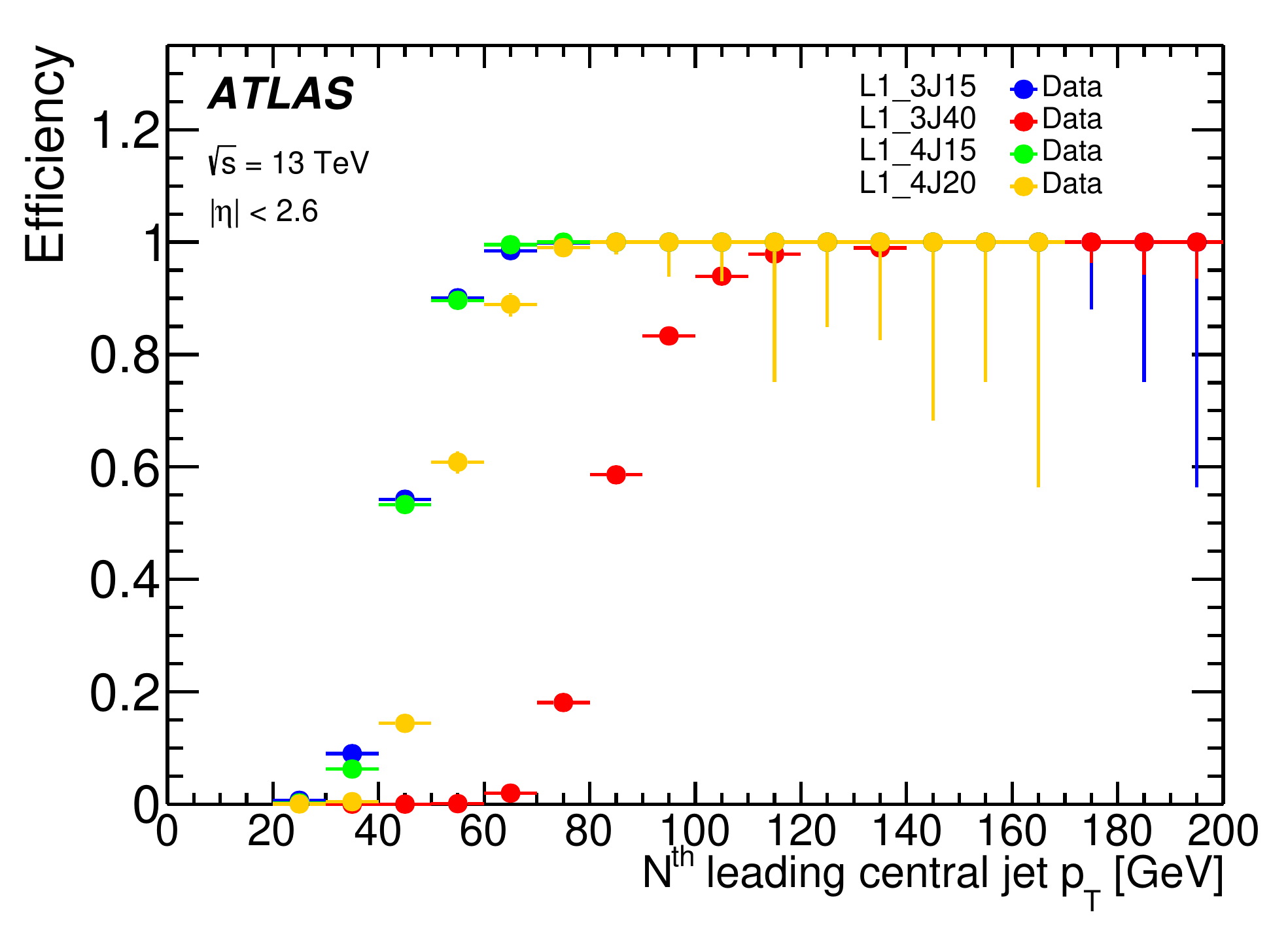}}
\subfloat[]{\includegraphics[width=0.5\textwidth]{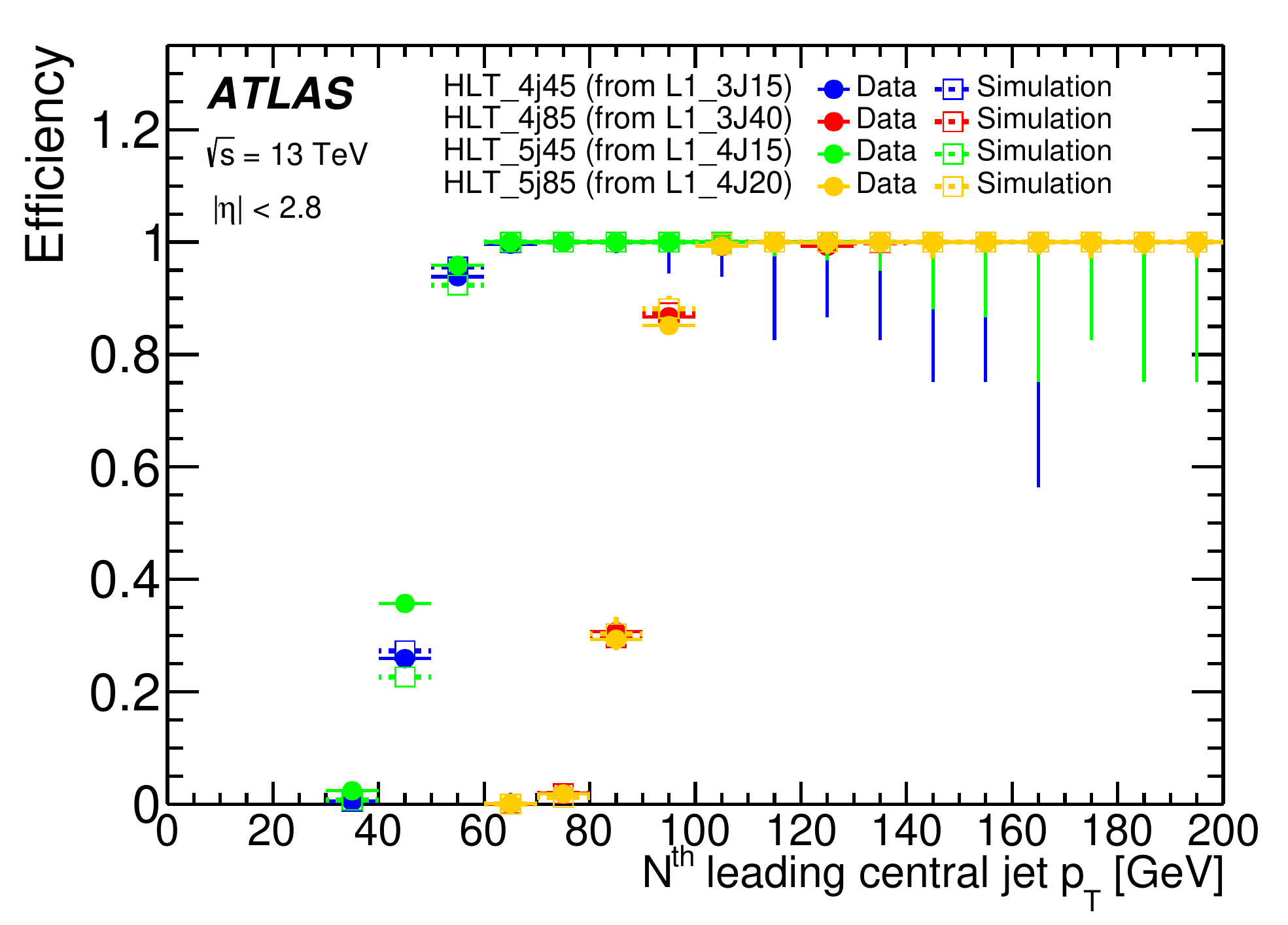}}
\caption{Efficiency of multi-jet (a) L1 and (b) HLT triggers as a function of offline jet \pT{}.}
\label{fig:jet_efficiency_multijet}
\end{figure}

The multi-jet trigger efficiencies are dominated by the trigger efficiency of the $N$th leading jet
and are shown in Figure~\ref{fig:jet_efficiency_multijet} for (a) L1 and (b) HLT as a function of
the $N$th leading jet transverse momentum. Good agreement is found for the efficiency as a
function of the $N$th jet for different jet multiplicities with the same
threshold (e.g.\ \trig{L1_6J15}, \trig{L1_4J15} and \trig{4j45}, \trig{5j45}) and between data and simulation for the HLT.

\begin{figure}[htbp]
\centering
\subfloat[]{\includegraphics[width=0.5\textwidth]{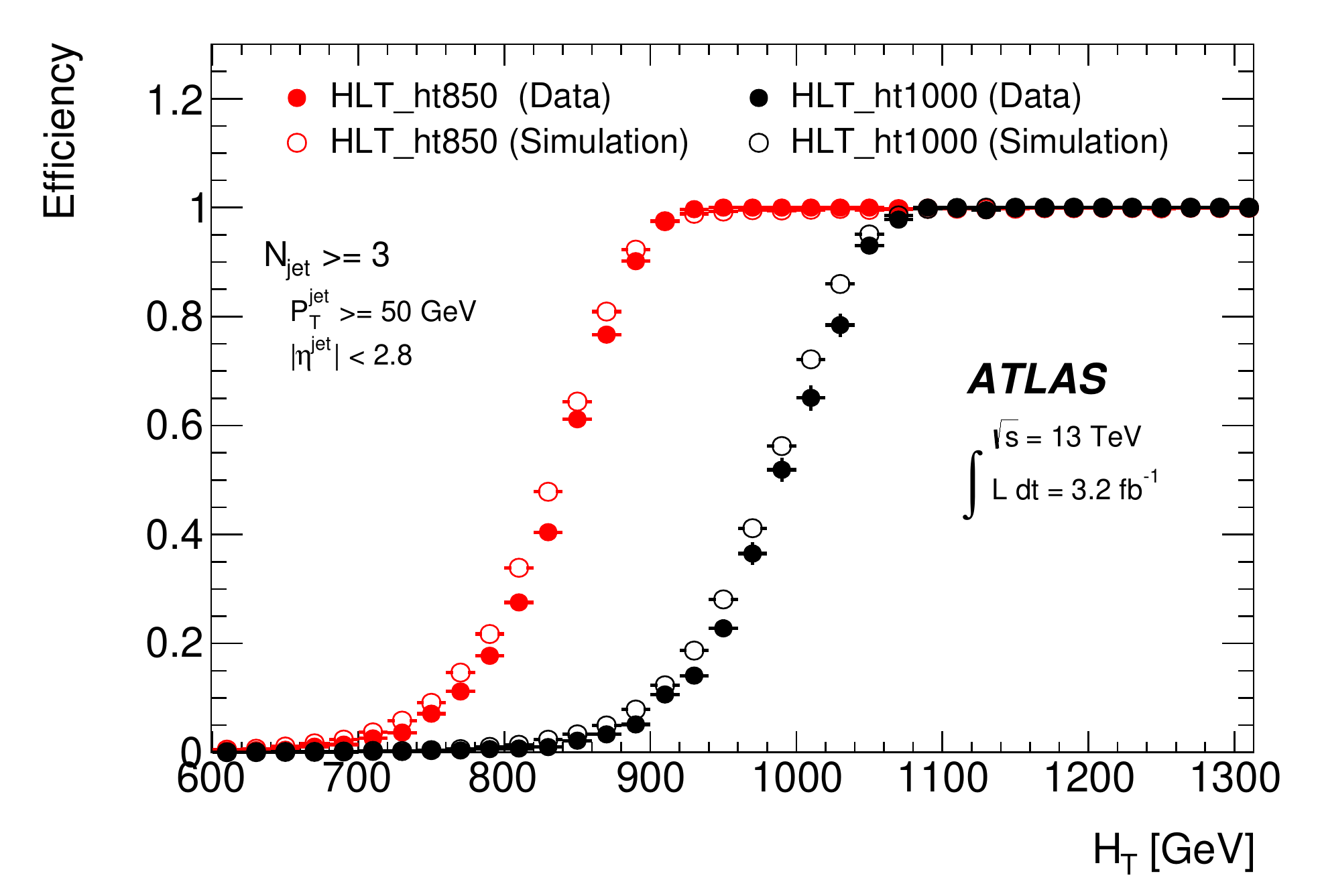}}
\subfloat[]{\includegraphics[width=0.5\textwidth]{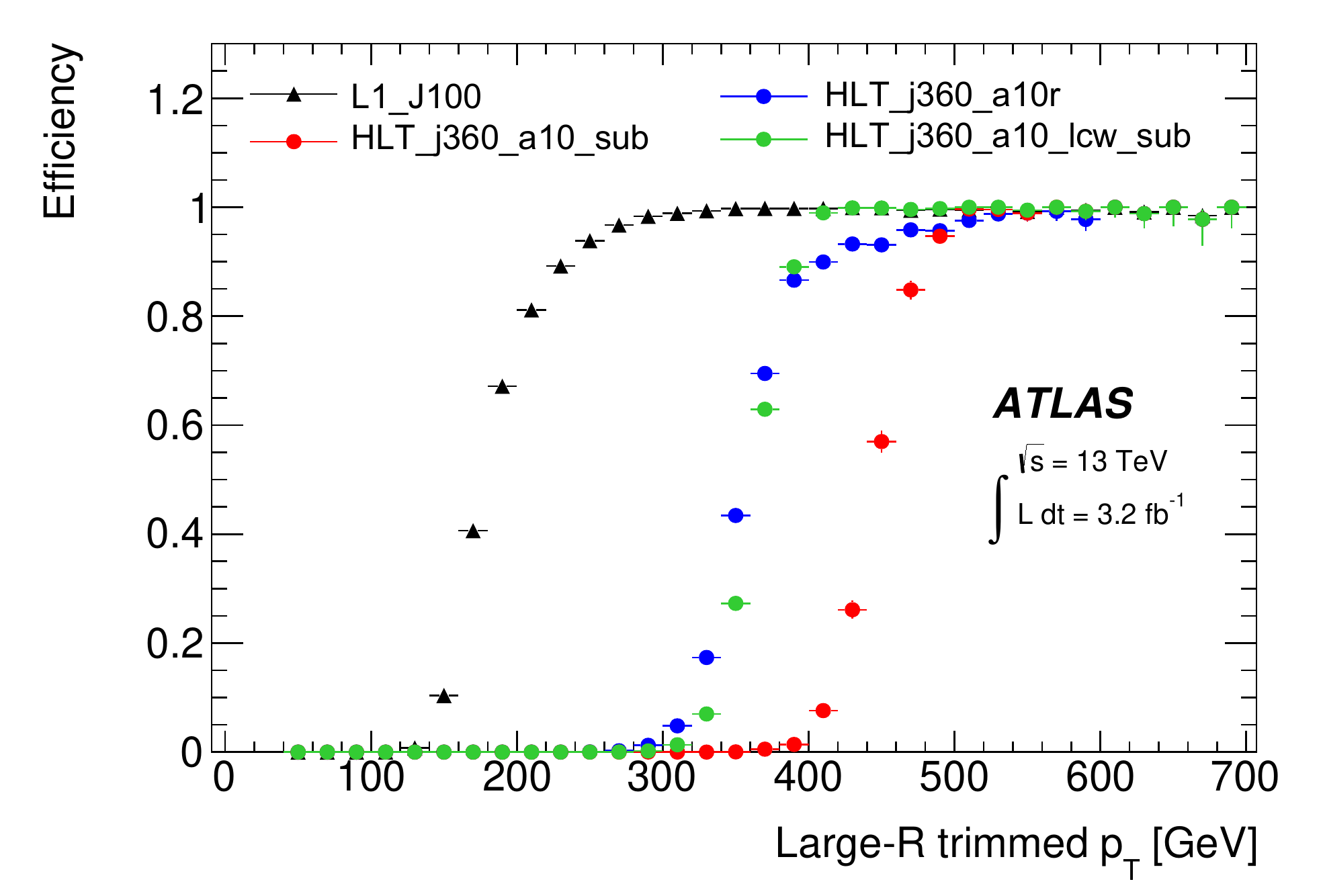}}
\caption{Efficiency of (a) \HT\ triggers as a function of offline \HT\
  and (b) large-$R$ ($R=1.0$) single-jet triggers as a function of offline \pT{}.
  \HT\ is defined as the summed transverse energy of all jets that are
  reconstructed above a transverse energy threshold of \SI{50}{\GeV}.}
\label{fig:jet_efficiency_otherjet}
\end{figure}

Finally, the efficiency of the \HT\ and large-$R$ ($R=1.0$) triggers are shown in
Figure~\ref{fig:jet_efficiency_otherjet}. The \HT\ trigger efficiencies are measured with
respect to the \trig{HLT_j150_L1J40} trigger. There is a small
offset in the efficiency curves for data and simulation for both thresholds. For the
large-$R$ triggers, the HLT threshold is set to \SI{360}{\GeV} and the efficiency curves are shown
for three different calibrations and jet input options: jets built from topo-clusters at the EM scale with
a pile-up subtraction applied (\trig{a10_sub}), jets built from topo-clusters with local calibration weights
and pile-up subtraction applied (\trig{a10_lcw_sub}) and reclustered jets built from $R=0.4$ jets using both
pile-up subtraction and local calibration weights (\trig{a10r}).

\subsubsection{Jets and Trigger-Level Analysis}
\label{sec:TLA}

\begin{figure}[htbp]
\centering
\includegraphics[width=0.5\textwidth]{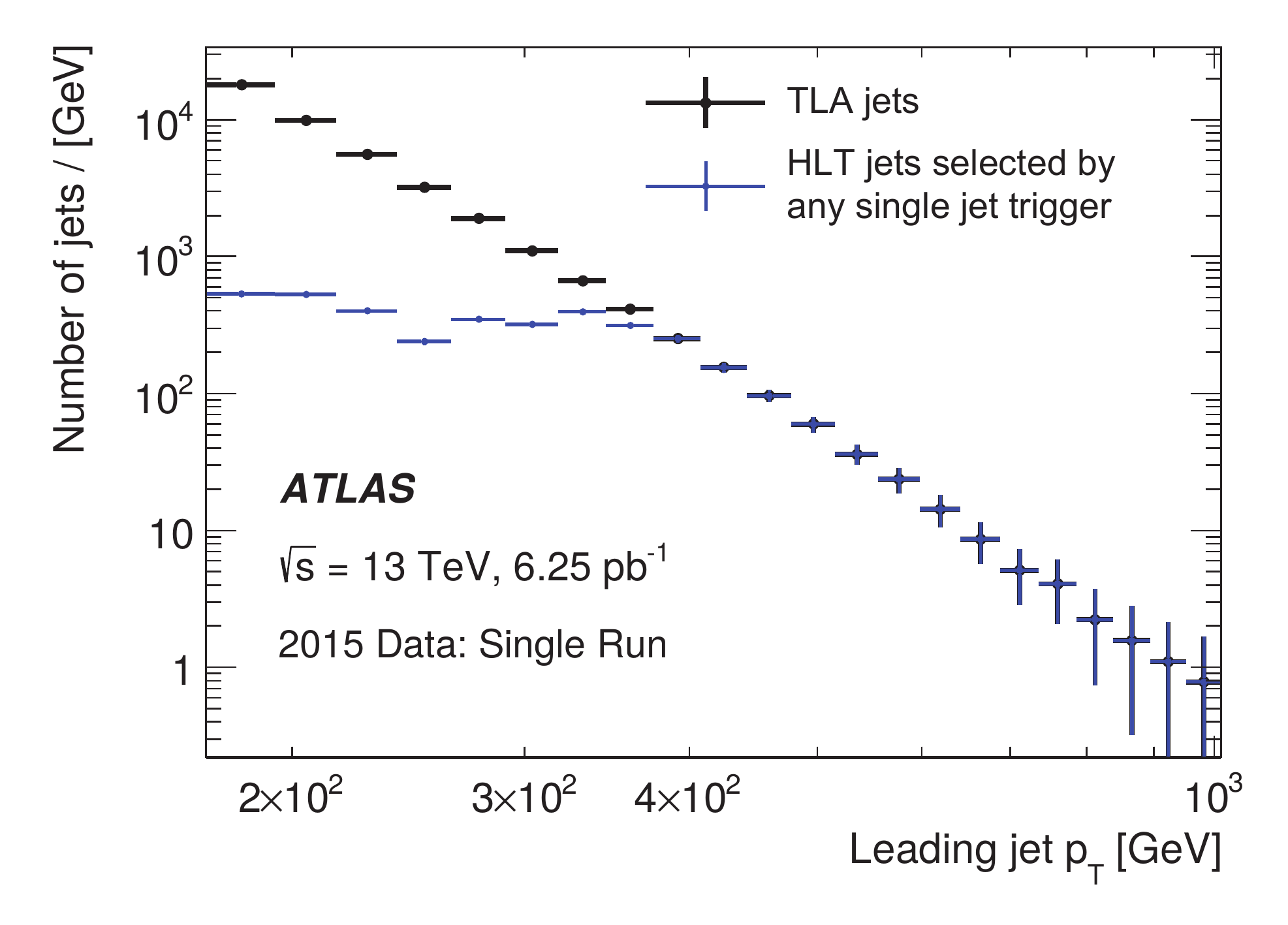}
\caption{Jet \pt\ spectrum after the basic kinematic selection for the TLA trigger
  jets (black) compared to trigger jets recorded by all single-jet triggers (blue).}
\label{fig:run281317_rawJets_JetPt}
\end{figure}

Searches for dijet resonances with sub-\si{\TeV} masses are statistically
limited by the bandwidth allocated to inclusive single-jet triggers. Due to
large SM multi-jet backgrounds, these triggers must be prescaled in
order to fit within the total physics trigger output rate of \SI{1}{\kHz}.
However, as the properties of
jets reconstructed at the HLT are comparable to that of jets reconstructed
offline, one can avoid this rate limitation by using Trigger-Level Analysis (TLA) triggers that record partial events,
containing only relevant HLT jet objects needed for the search, to a dedicated
stream. Using Trigger-Level Analysis triggers allows a factor of 100 increase in the event recording rates, and
results in a significant increase in the
number of low-\pt jets as shown in
Figure~\ref{fig:run281317_rawJets_JetPt}. 
Dedicated calibration and jet identification procedures are applied to these partially built events, 
accounting for differences between offline jets and trigger jets as well as for the lack
of detector data other than from the calorimeters. These procedures are described in detail in Ref.~\cite{ATLAS-CONF-2016-030}.

\FloatBarrier
\newcommand{\tauhad}{\tau_{\mathrm{had}}}

\subsection{Tau leptons}
\label{sec:tau}

Tau leptons are a key signature in many SM measurements and searches for new physics.
The decay into tau lepton pairs provides the strongest signal for measurements of the SM
Higgs boson coupling to fermions. Final states containing tau leptons are also often
favoured by heavier Higgs bosons or other new resonances in many scenarios beyond the SM.
Most (about 65\%) of tau leptons decay hadronically. Hence an efficient trigger
on hadronic tau decays is crucial for many analyses using tau leptons.

Dedicated tau trigger
algorithms were designed and implemented based on the main features of hadronic
tau decays: narrow calorimeter energy deposits and a small number of associated
tracks. Due to the high production rate of jets with features very similar to
hadronic tau decays, keeping the rate of tau triggers under control is
particularly challenging.

\subsubsection{Tau reconstruction and selection}
\label{sec:tau:rec}

At L1 the tau trigger uses the algorithms described in
Section~\ref{sec:l1calo}. The isolation requirement was tuned with
\SI{13}{\TeV} simulation to yield an efficiency of 98\% and is not applied for tau
candidates with a transverse energy above \SI{60}{\GeV}.

At the HLT three sequential selections are made. First, a minimum requirement is applied
to the transverse energy of the tau candidate. The energy is calculated using the locally
calibrated topo-clusters of calorimeter cells contained in a cone of size
$\Delta R=0.2$ around the L1 tau RoI direction taken from the L1 cluster.
A dedicated tau energy calibration scheme is used. 
Second, two-stage fast tracking (Section~\ref{sec:MultipleStageTracking}) is used
to select tau candidates with 
low track multiplicity. A leading track is sought within a narrow cone
($\Delta R=0.1$) around the tau direction followed by a second fast tracking
step using a larger cone ($\Delta R=0.4$) but with the tracks required to
originate from within a fixed interval along the beam line around the leading track.
Tracks with $\pt>\SI{1}{\GeV}$ are counted in the core cone region
$\Delta R<0.2$ and in the isolation annulus $0.2<\Delta R<0.4$ around the tau
candidate direction. A track multiplicity requirement selects tau candidates
with $1\le N^{\mathrm{trk}}_{\Delta R<0.2}\le 3$ and $N^{\mathrm{trk}}_{0.2<\Delta R<0.4}\le 1$. 
Finally, the HLT precision tracking is run, and a collection of variables built from
calorimeter and track variables are input to a Boosted Decision Tree (BDT), which produces
a score used for the final tau identification. The implementation of those variables
follows closely their offline counterparts as described in Ref.~\cite{ATL-PHYS-PUB-2015-045}.
In addition, the same
BDT training is used offline and online to ensure a maximal correlation between
online and offline identification criteria. The performance of the offline training was
found to be comparable to a dedicated online training. To ensure a robust response under
differing pile-up conditions, corrections as a function of the average number of
interactions per bunch-crossing are applied to the discriminating variables. 
Working points of the BDT are tuned separately for 1-prong and 3-prong candidates. The baseline
{\em medium} working point operates with an efficiency of 95\% (70\%) for
true 1-prong (3-prong) taus.

\subsubsection{Tau trigger menu and rates}
The primary tau triggers consist of triggers for single high transverse momentum
taus, and combined
$\tau+X$ triggers, where $X$ stands for an electron, muon, a second tau or \met. The
transverse momentum thresholds used
in the single-tau and ditau triggers in 2015 are indicated in Table~\ref{T:AtlasTriggerMenu}. 
For all tau triggers the L1 isolation, HLT track multiplicity and online \emph{medium}
identification requirements are applied to the tau candidates.

\begin{figure}[htbp]
  \centering
   \subfloat[]{
\includegraphics[width=0.5\textwidth]{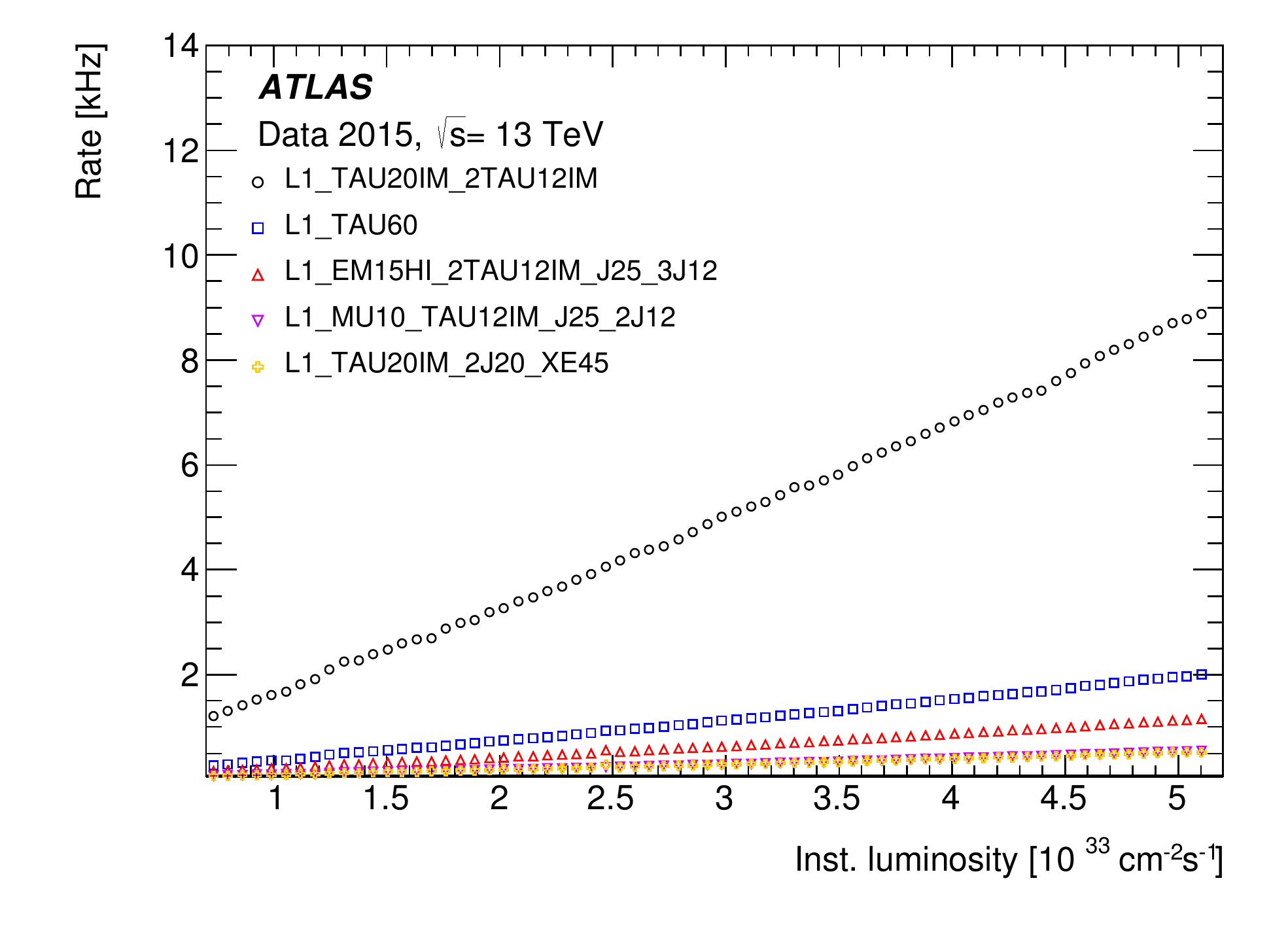}\label{fig:tau:L1rates}
   }
   \subfloat[]{
\includegraphics[width=0.5\textwidth]{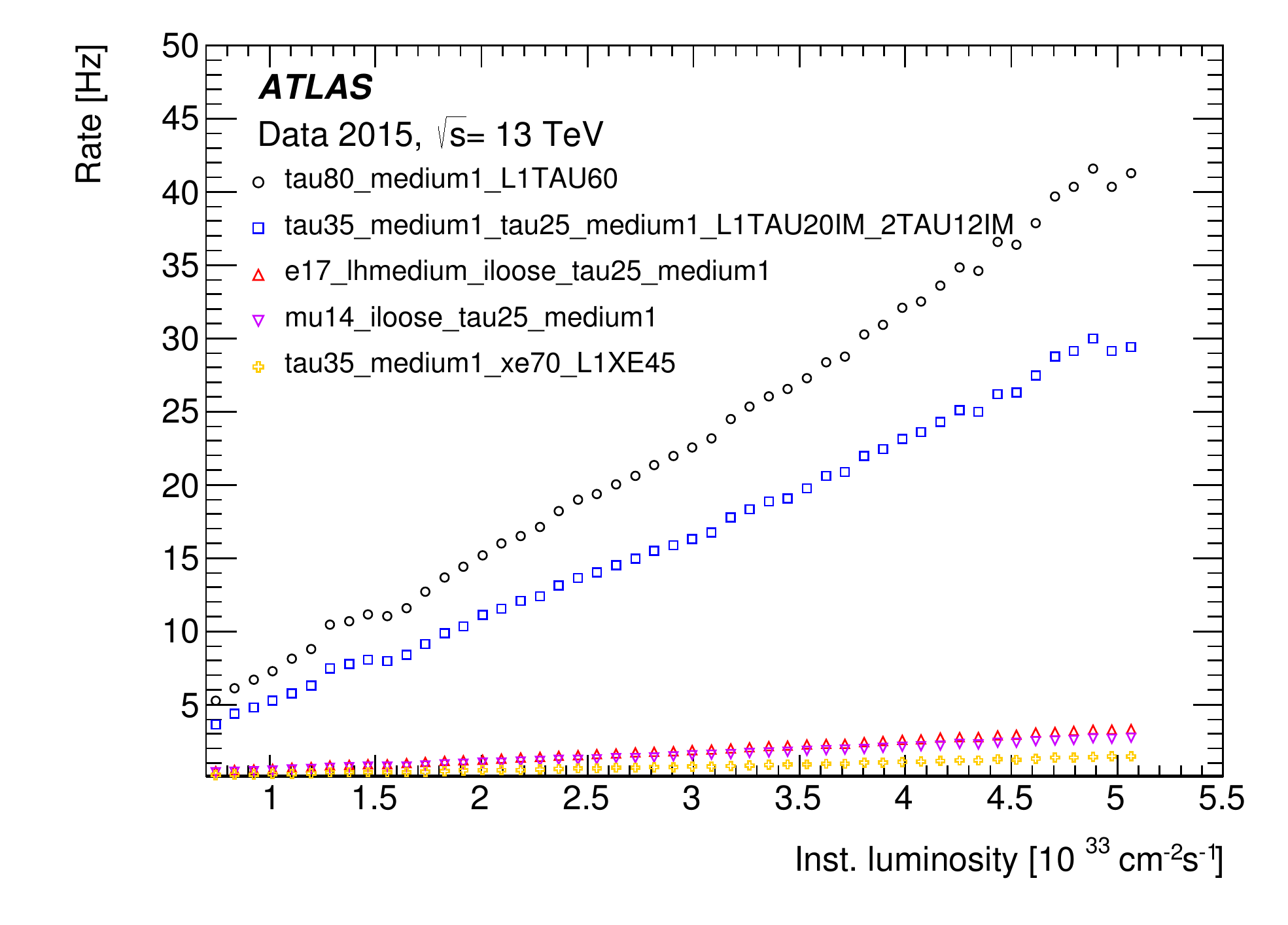}\label{fig:tau:HLTrates}
}
   \caption{Trigger rates as a function of instantaneous luminosity for several (a) L1
     and (b) HLT tau triggers. }
    \label{fig:tau:rates}
\end{figure}

Due to L1 rate limitations, the combined triggers $\tau+(e,\mu)$ and $\tau+$\met
require the presence of an additional 
jet candidate at L1 with transverse momentum above \SIlist{25;20}{\GeV}, respectively.  
Variants of these triggers with higher thresholds for the tau transverse momentum and
without the L1 
jet requirement are also included in the trigger menu.
Figure~\ref{fig:tau:rates} shows the L1 and HLT output rates as function of the
instantaneous luminosity for the primary single-tau, ditau, $\tau+e$, $\tau+\mu$ and
$\tau+$\met\ triggers.   

\subsubsection{Tau trigger efficiencies}

The efficiency of the tau trigger was measured using a tag-and-probe (T\&P) method in an enriched  
sample of $Z\to\tau_\mu\tauhad\to\mu+2\nu+\tauhad$ events, where $\tau_\mu$ is a tau lepton
decaying to $\mu \nu \nu$ and $\tauhad$ is a tau lepton decaying hadronically.
Events are selected by the lowest unprescaled single-muon trigger and are tagged
by an offline reconstructed and isolated muon with transverse momentum above \SI{22}{\GeV}. 
The presence of an offline reconstructed tau candidate with transverse momentum above \SI{25}{\GeV}, 
one or three tracks, fulfilling the \emph{medium} identification criteria and with electric 
charge opposite to the muon charge is also required. 
This reconstructed tau candidate is the probe with respect to which the tau trigger
  efficiency is measured.
  The event selection used to enhance the sample with $Z\to\tau_\mu\tauhad$
  events and 
  therefore the purity of the probe tau candidate is similar to the one described in
  Ref.~\cite{ATL-PHYS-PUB-2015-045}: 
to reject $Z(\to\mu\mu)+\text{jets}$ and $W(\to\mu\nu)+\text{jets}$ events, the invariant mass of the muon and the 
offline tau candidate is required to be between \SIlist{45;80}{\GeV}, 
the transverse mass, $m_{\mathrm{T}}$, composed of the muon \pt\ and \met\ ($m_{\mathrm{T}}^2=2p_{\mathrm{T}}^{\mu}\met(1-\cos\Delta\phi(\mu,\met))$) 
is required to be smaller than \SI{50}{\GeV}, and the variable built from the difference
in azimuth between
the muon and \met\ and between the offline tau candidate and \met\ ($\cos\Delta\phi(\mu,\met)+\cos\Delta\phi(\tau,\met)$) 
is required to be above~$-0.5$.
The dominant sources of background events in the resulting sample are $W(\to\mu\nu)+\text{jets}$ and
multi-jet events and 
their contributions are determined in data as described in Ref.~\cite{ATL-PHYS-PUB-2015-045}.
The multi-jet contribution is estimated from events where the offline tau candidate and the
muon have the same electric charge. 
The $W(\to\mu\nu)+\text{jets}$ contribution is estimated from events with high $m_{\mathrm{T}}$.

\begin{figure}[htbp]
  \centering
  \subfloat[]{
    \includegraphics[width=0.5\textwidth]{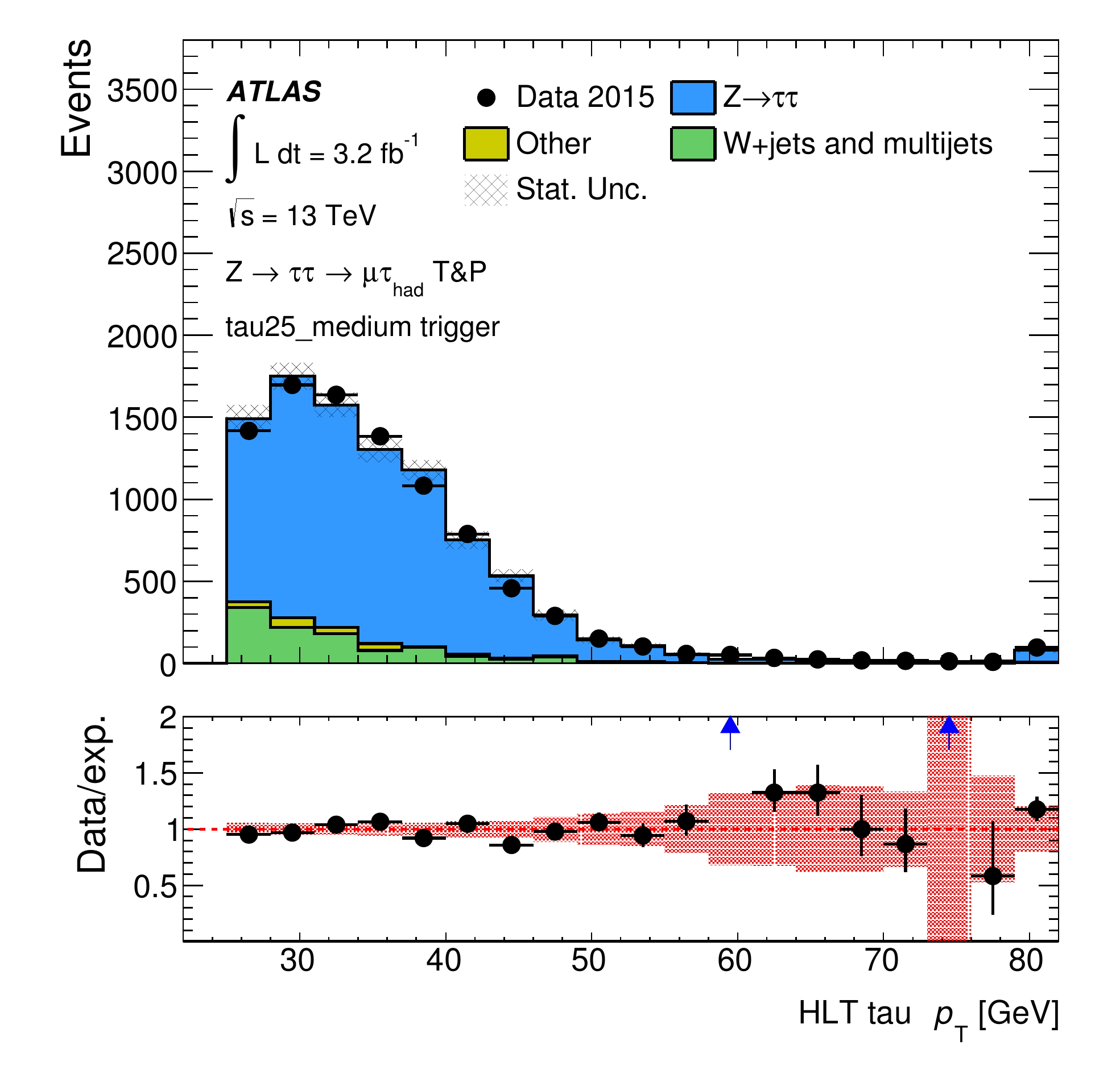}
    \label{fig:tau:HLTpt}
  }
  \subfloat[]{
    \includegraphics[width=0.5\textwidth]{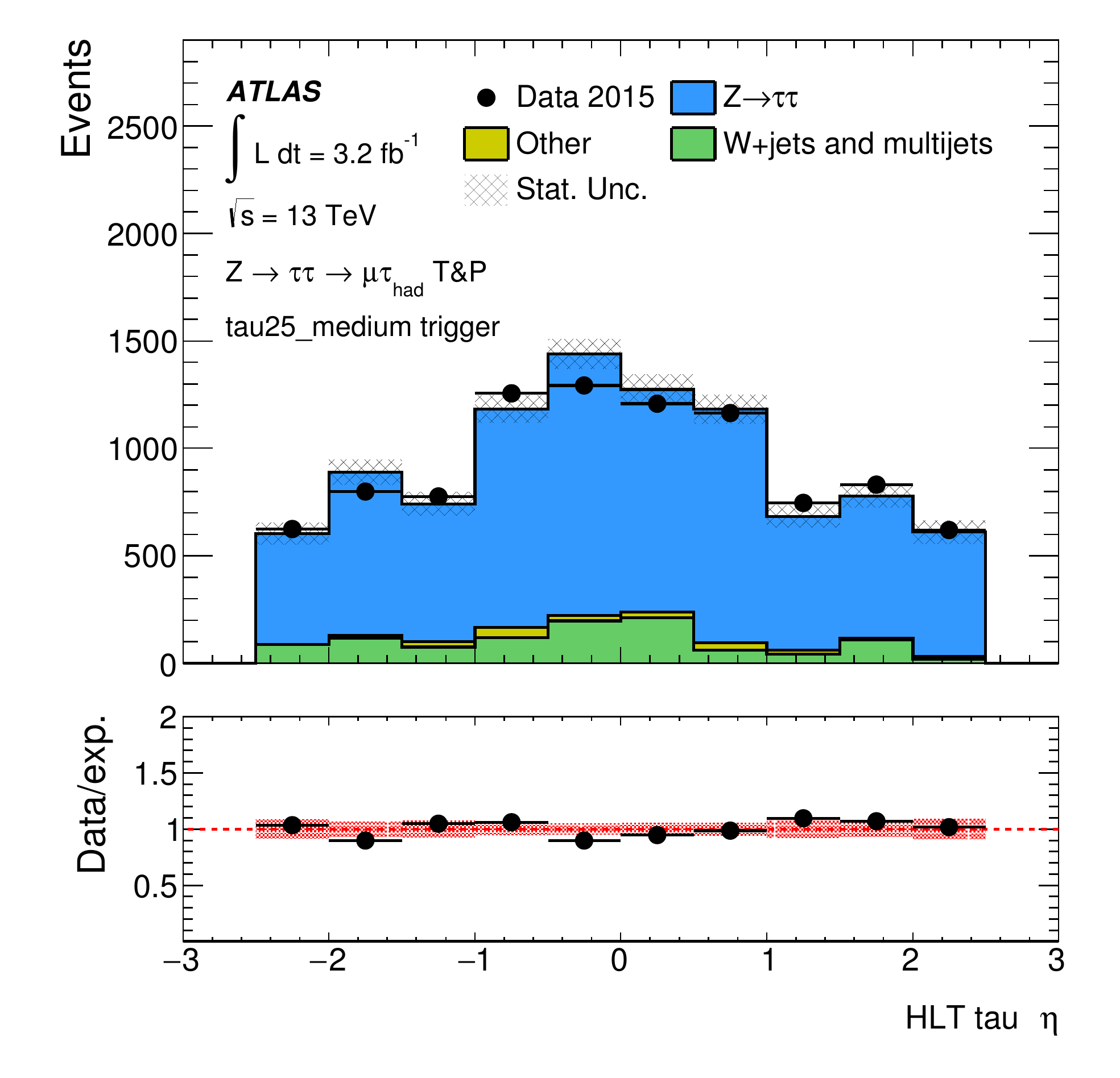}
    \label{fig:tau:HLTeta}
  }\\
  \subfloat[]{
    \includegraphics[width=0.5\textwidth]{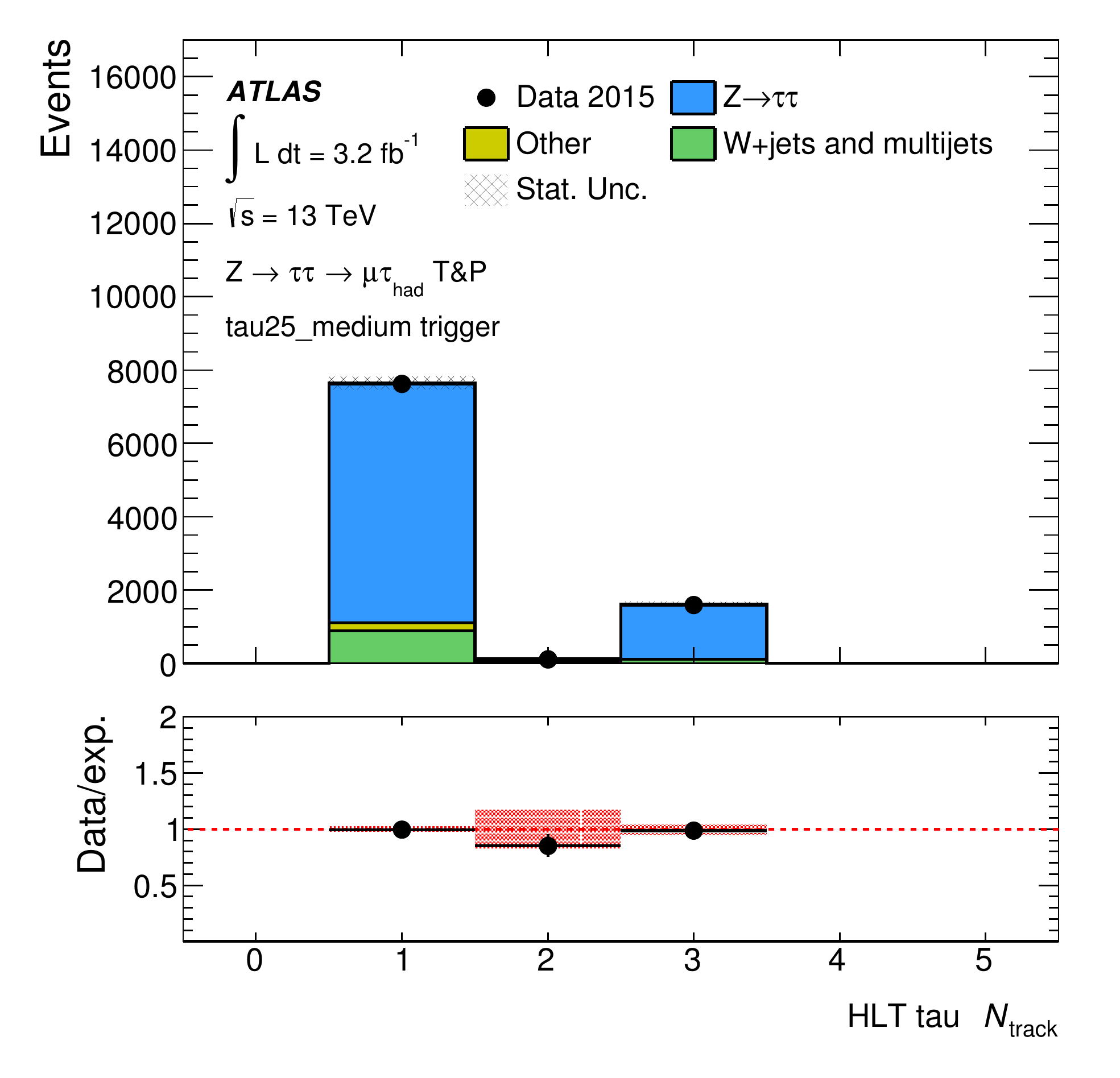}
    \label{fig:tau:HLTNCore}
  }
  \subfloat[]{
    \includegraphics[width=0.5\textwidth]{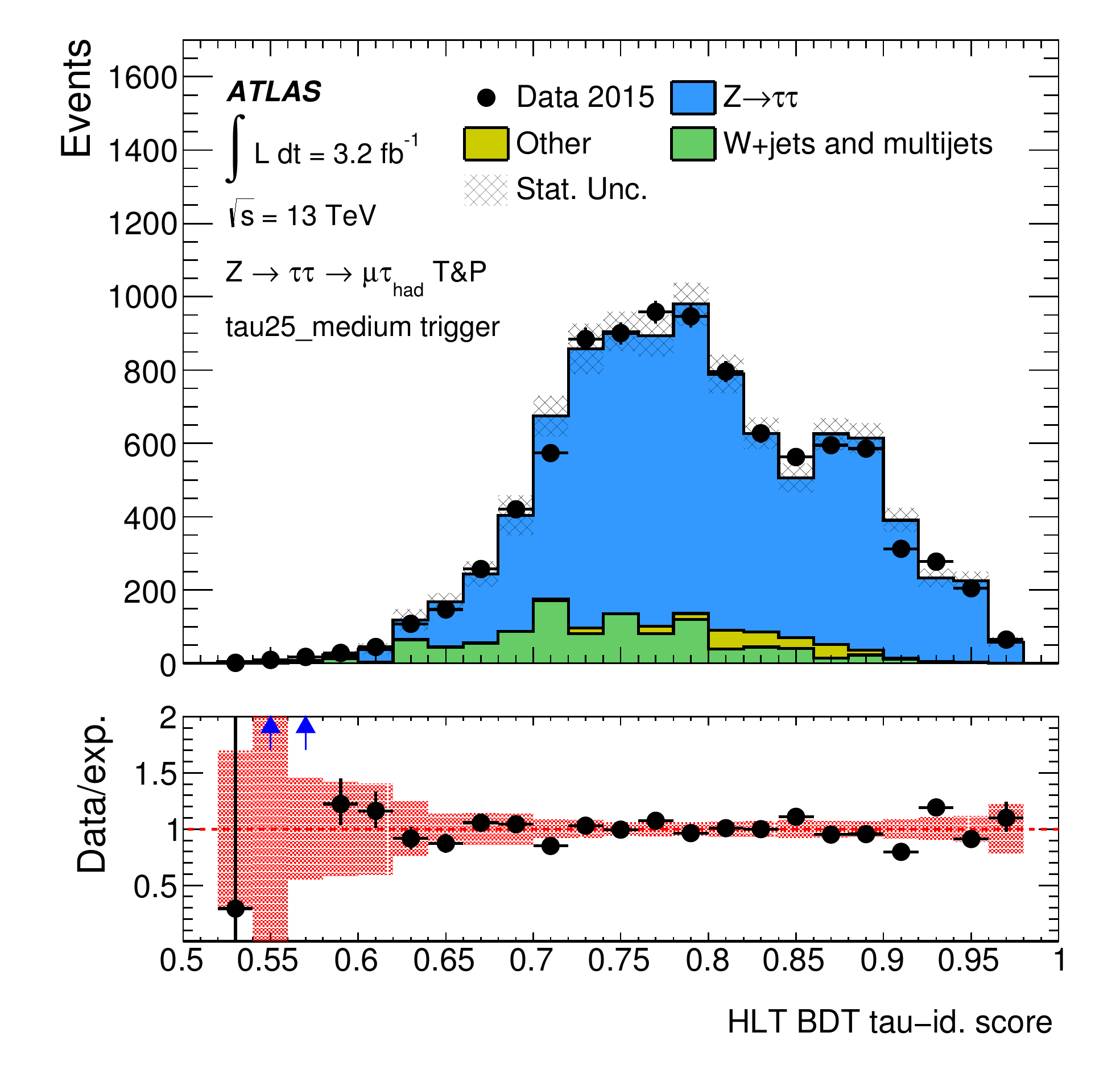}
    \label{fig:tau:HLTBDTScore}
  }
  \caption{Distributions of the HLT tau candidates passing the
    \trig{tau25_medium} trigger: (a) transverse momentum, (b) pseudorapidity,
    (c) track multiplicity distributions of the core tracks $\Delta R<0.2$ of
    the tau-axis and (d) online BDT identification score. The HLT tau
    candidates are matched to offline tau candidates with transverse momentum
    above \SI{25}{\GeV}, with one or three tracks and satisfying the offline
    \emph{medium} tau identification criterion. Only statistical uncertainties
    are shown, and the last bin in (a) contains overflow events. The ratio of the
    observed data to the expected signal and background events is also shown,
    where the red band shows the statistical uncertainty of the total
    prediction.}
  \label{fig:tau:kin}
\end{figure}

Distributions of the transverse momentum, pseudorapidity, track multiplicity and BDT discriminant
score for the HLT tau candidates matched to the offline probe tau candidates are shown
in Figure~\ref{fig:tau:kin}.
The HLT tau candidates pass the \trig{tau25_medium} trigger, which requires an isolated L1
RoI with transverse momentum
above \SI{12}{\GeV} and a tau candidate at the HLT with transverse momentum above \SI{25}{\GeV} satisfying
the track multiplicity 
and the online \emph{medium} identification criteria.   
The observed distributions in data are in good agreement with simulation.

\begin{figure}[htbp]
  \centering
  \subfloat[]{
    \includegraphics[width=0.5\textwidth]{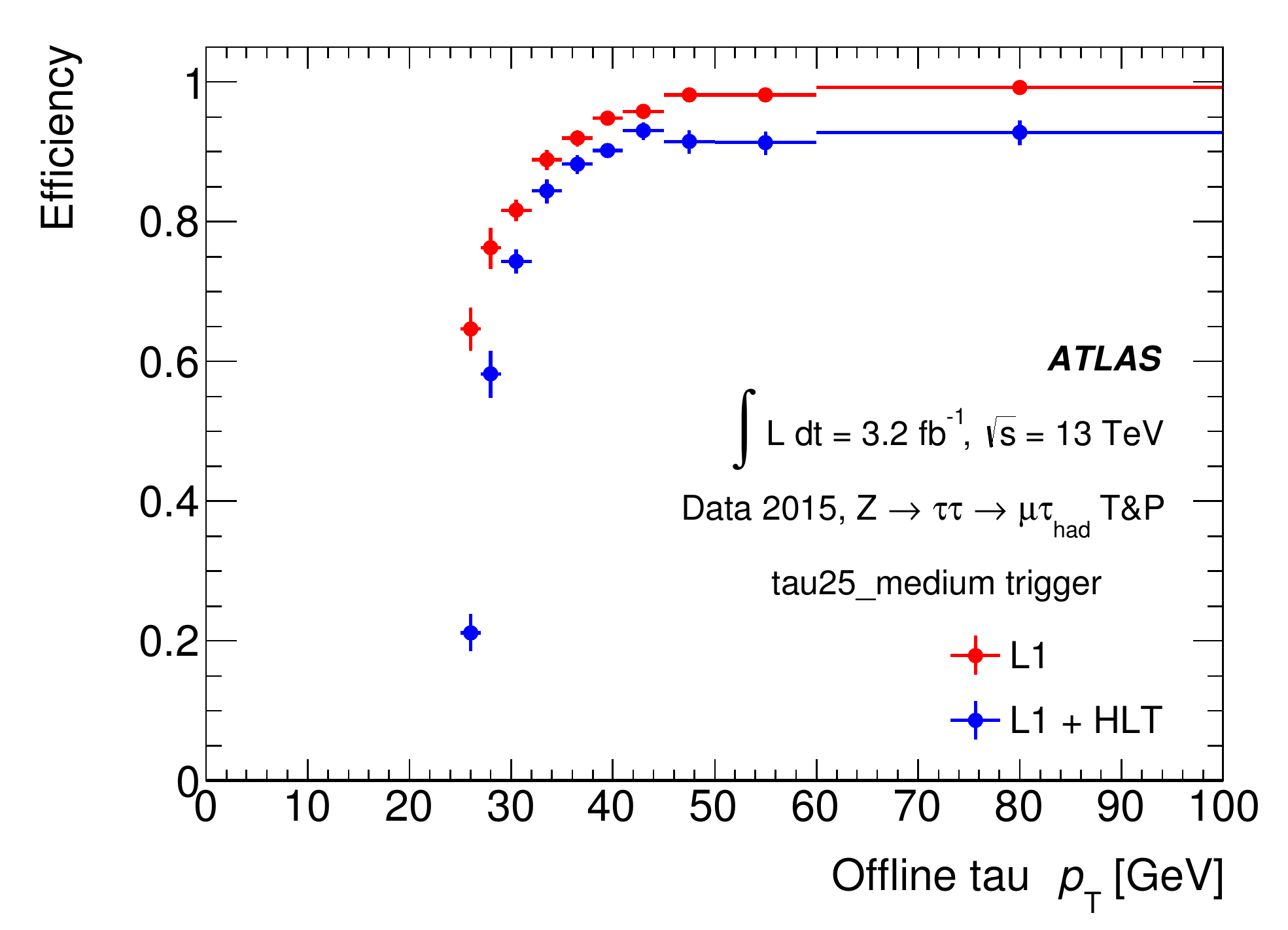}
    \label{fig:tau:pteff}
  }
  \subfloat[]{
    \includegraphics[width=0.5\textwidth]{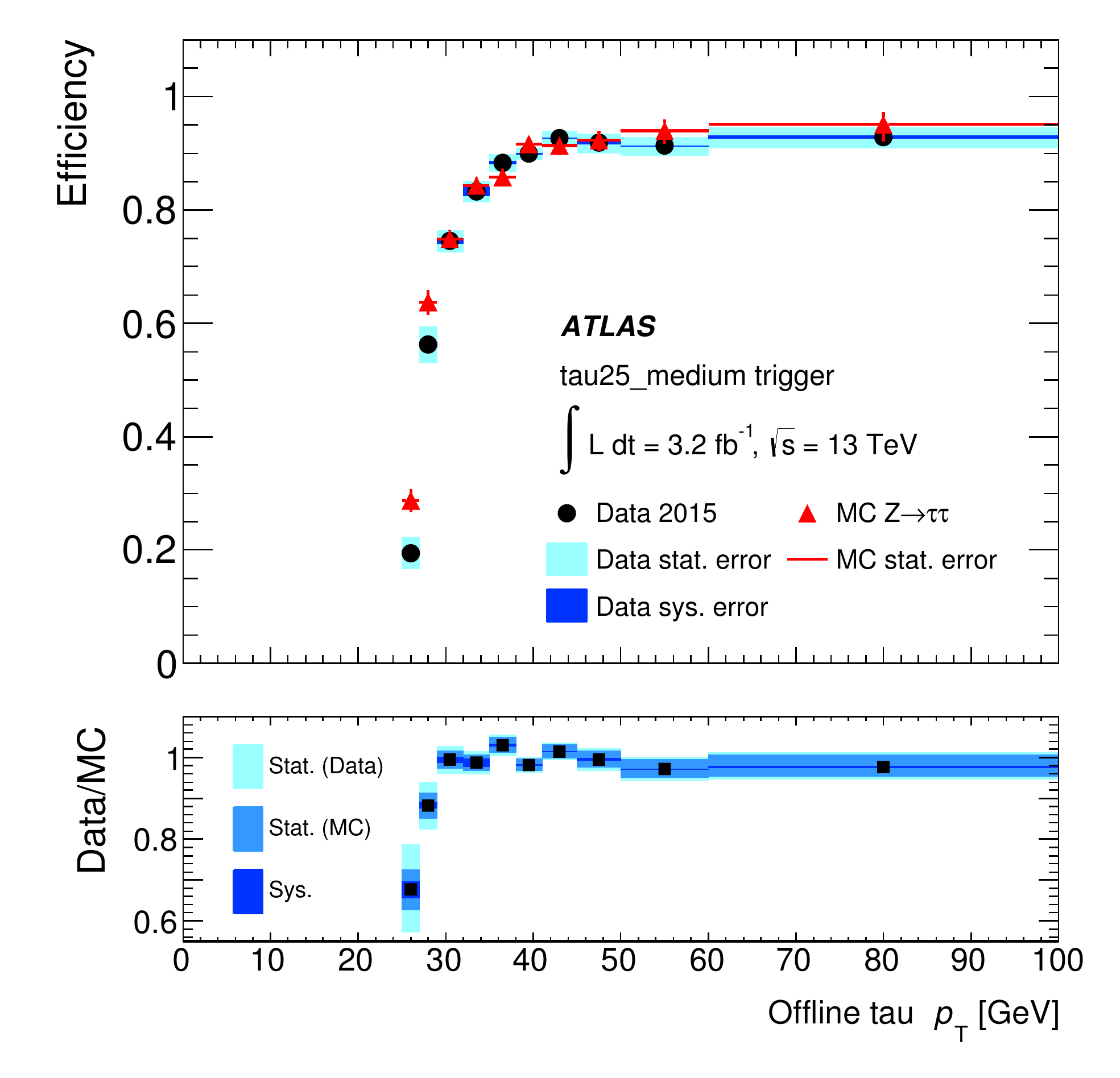}
    \label{fig:tau:efftp}
  }
  \caption{Efficiency of the \trig{tau25_medium} trigger measured in data as a function of the
    offline tau \pt\ for offline tau candidates with \pt\ above
    \SI{25}{\GeV}, one or three tracks 
    and satisfying the offline \emph{medium} identification requirement. The expected background
    contribution has been subtracted from the data. (a) Efficiencies after the L1 (red) and
    L1+HLT (blue) selections are shown separately with only statistical uncertainties. (b) Comparison
    of the measured efficiency after L1+HLT to simulation. Statistical uncertainties associated
    with data and simulation and the systematic uncertainty associated with the background
    subtraction procedure in data are shown.}
  \label{fig:tau:eff}
\end{figure}

The estimated background is subtracted from data and the uncertainty in this subtraction is
considered as a systematic uncertainty in the measured efficiency. This systematic uncertainty
includes uncertainties in the background contributions estimated from both simulation and data.
Figure~\ref{fig:tau:pteff} shows the measured efficiency for the \trig{tau25_medium} trigger
as a function of the transverse momentum of the offline tau candidate.
The efficiency loss of the HLT with respect to L1 is mainly due to the HLT's track multiplicity
selection and its BDT selection, which uses slightly different input variables online and
offline. In Figure~\ref{fig:tau:efftp} this efficiency is compared 
with simulation. The statistical uncertainties in data and simulation are shown
together with the systematic uncertainties associated with the background subtraction procedure
in data.

\FloatBarrier
\subsection{Missing transverse momentum}
\label{sec:met}

\newcommand{\ptmissvec}{\ensuremath{{\vec{p}_{\mathrm{T}}^{\mathrm{\,miss}}}}}
\newcommand{\pxmiss}{\ensuremath{{\vec{p}_y^{\mathrm{\,miss}}}}}
\newcommand{\pymiss}{\ensuremath{{\vec{p}_x^{\mathrm{\,miss}}}}}

\newcommand{\metLOne} {L1}
\newcommand{\metcell} {\trig{xe}}
\newcommand{\mettc}   {\trig{xe_tc_lcw}}
\newcommand{\metpufit}{\trig{xe_tc_pufit}}
\newcommand{\metpueta}{\trig{xe_tc_pueta}}
\newcommand{\metmht}  {\trig{xe_tc_mht}}
\newcommand{\nvtx}{\ensuremath{N_{\mathrm{vtx}}}}

The \met{} trigger is used in searches where the final state
contains only jets and large \met. The
\met trigger can also be the most efficient trigger for selecting final states
that contain highly energetic muons. An example is searches for
supersymmetric particle production where jets, leptons and invisible particles
are produced.  Another major use is for multi-particle final states where the
combination of \met\ with other trigger objects such as jets, electrons, or
photons enables lower thresholds to be used for these other objects than would
otherwise be possible.  Finally, the \met{} trigger collects data samples used
for detector performance studies.  For example, the data set used for electron
efficiency calculations in events consistent with a $W$ boson is selected with
an \met{} trigger.

\subsubsection{\met reconstruction and selection}

The very large rate of hadronic jet production means that, even with reasonably
good calorimeter resolution, jet energy mismeasurement can lead to an
unaffordably large \met{} trigger rate. The difficulty is exacerbated by pile-up
collisions that add energy to the calorimeter and hence degrade the \met{}
resolution. Controlling the rate via increased trigger thresholds usually
reduces the efficiency for analyses.

The improvements in the L1 \met{} determination, including the L1 dynamic
pedestal correction described in Section~\ref{sec:l1calo}, have been important
in maintaining L1 performance. In particular they have permitted the
\trig{L1_XE50} trigger to be used without prescale throughout 2015.

To fulfil the desired broad \met-based physics programme, different HLT
algorithmic strategies based on cells, jets or topo-clusters in addition to two
methods for correcting the effects of pile-up were developed during LS1 and
deployed during 2015 data-taking. While the offline algorithms do often include reconstructed muons in the \met\
calculation, the trigger algorithms described herein use only energy
measurements in the calorimeter. Five different algorithms, involving different
levels of complexity (and thus different CPU requirements) were commissioned and
evaluated with data during 2015. Since the time-consuming (topo-)clustering is shared
between the different algorithms, running them all in parallel does only require a small
amount of extra CPU time.
The algorithms are as follows:
\begin{itemize}
\item[$\bullet$] \textbf{Cell algorithm} (\metcell): The measured energy in each LAr and Tile calorimeter cell, labelled $i$,
  and the position of the cell in the detector are used to obtain the components of the cell measured momentum in the massless
  approximation, i.e.\ $p_{x,i}=E_i \sin\theta_i\cos\phi_i$ and $p_{y,i}= E_i\sin\theta_i\sin\phi_i$.
  To suppress noise and cells with large negative energy, only those cells with energy
  satisfying $|E_i| > 2\sigma_i$ and $E_i > -5\sigma_i$,        
        are considered further, 
        where $\sigma_i$ is the noise in the cell energy measurement, including the noise-like effects from 
        pile-up.\footnote{A one-sided $2\sigma$ noise cut was used during \runi, which resulted in a bias 
towards higher \met-values.}
        Non-functioning calorimeter cells are masked out and do not contribute to the calculation.
        The total missing transverse momentum two-vector $\ptmissvec = -\sum_i (p_{x,i}, p_{y,i})$ is found from the 
        remaining contributing cells, and the \met calculated from its norm $\met=|\ptmissvec|$.
	
      \item[$\bullet$] \textbf{Jet-based algorithm} (\metmht): \met\ is calculated directly from the negative of the transverse
        momentum vector sum of all jets reconstructed by the jet trigger algorithm
	presented in Section~\ref{sec:jet}, which have been corrected for the energy contribution from pile-up.

      \item[$\bullet$] \textbf{Topo-cluster algorithm} (\mettc): Topo-clusters (described in Section~\ref{sec:calo:algo}) are built for the entire
        calorimeter and used for the \met reconstruction.
        For each topo-cluster $j$, the momentum components $(p_{x,j}, p_{y,j})$ are calculated in the approximation that the particles contributing energy to the cluster are massless, 
        and, in a manner similar to the cell algorithm, the missing transverse momentum is calculated from the negative vector sum
        of these components.

      \item[$\bullet$] \textbf{Pile-up suppression algorithm} (\metpueta): This algorithm is based on the topo-cluster \met algorithm
        described above, but includes a further pile-up suppression method that is intended to limit the degradation of
	the \met resolution at very high pile-up. The method starts by calculating the average topo-cluster energy and standard deviation
        in ten regions of pseudorapidity covering,
	in equal steps, $-5.0<\eta<5.0$ in the calorimeter. In each pseudorapidity region, known as a ring, the topo-clusters of energy
        above $2\sigma$ are omitted and the average energy of the 
	residual topo-clusters is calculated. This average represents an estimate of the energy contribution from pile-up in that ring.
        The pile-up energy density in each ring is
	obtained by dividing the average energy by the solid angle of the ring. This energy density is then multiplied by the solid
        angle of each topo-cluster
	and then subtracted from the energy of that topo-cluster to obtain a topo-cluster energy measurement corrected for pile-up. The \met
        is recalculated as described above using the $(p_{x,j}, p_{y,j})$ of topo-clusters after the pile-up subtraction.

      \item[$\bullet$] \textbf{Pile-up fit algorithm} (\metpufit): Starting again from the topo-cluster \met described above, a
        different pile-up suppression method is used in this algorithm. 
        The calorimeter is partitioned into 112 towers each of size $\eta\times\phi\approx 0.71\times0.79$.
        For each tower, the $p_x$ and $p_y$ components of all the topo-clusters with centres in that tower are summed to obtain
        the transverse momentum $\vec{p}_{{\mathrm T},k}$ of that $k$th tower. The transverse energy sum of the
        tower $E_{\mathrm{T},k}$ is also calculated from the scalar sum of the \pt\ of the individual clusters.
        If $E_{\mathrm{T},k} < \SI{45}{\GeV}$, the tower is determined to be below threshold
        and its energy assumed to be due to pile-up.
        The average pile-up \et\ density is calculated from $\sum_k E_{\mathrm{T},k}/\sum_k A_k$ of all the towers below threshold, where $A_k$
        is the total area in $(\eta$,$\phi)$ coordinates of those towers.       
        A fit estimates the \et\ contributed by pile-up in each tower above
        threshold using the average pile-up \et\ density and constraining the
        event-wide \met\ from pile-up to be zero within resolution.
        These estimated pile-up contributions are subtracted from the
        corresponding \et\ measurements for towers above threshold, and these
        corrected \et\ values are used to calculate \met.
\end{itemize}

\begin{figure}[htbp]
\centering
\includegraphics[width=0.5\textwidth]{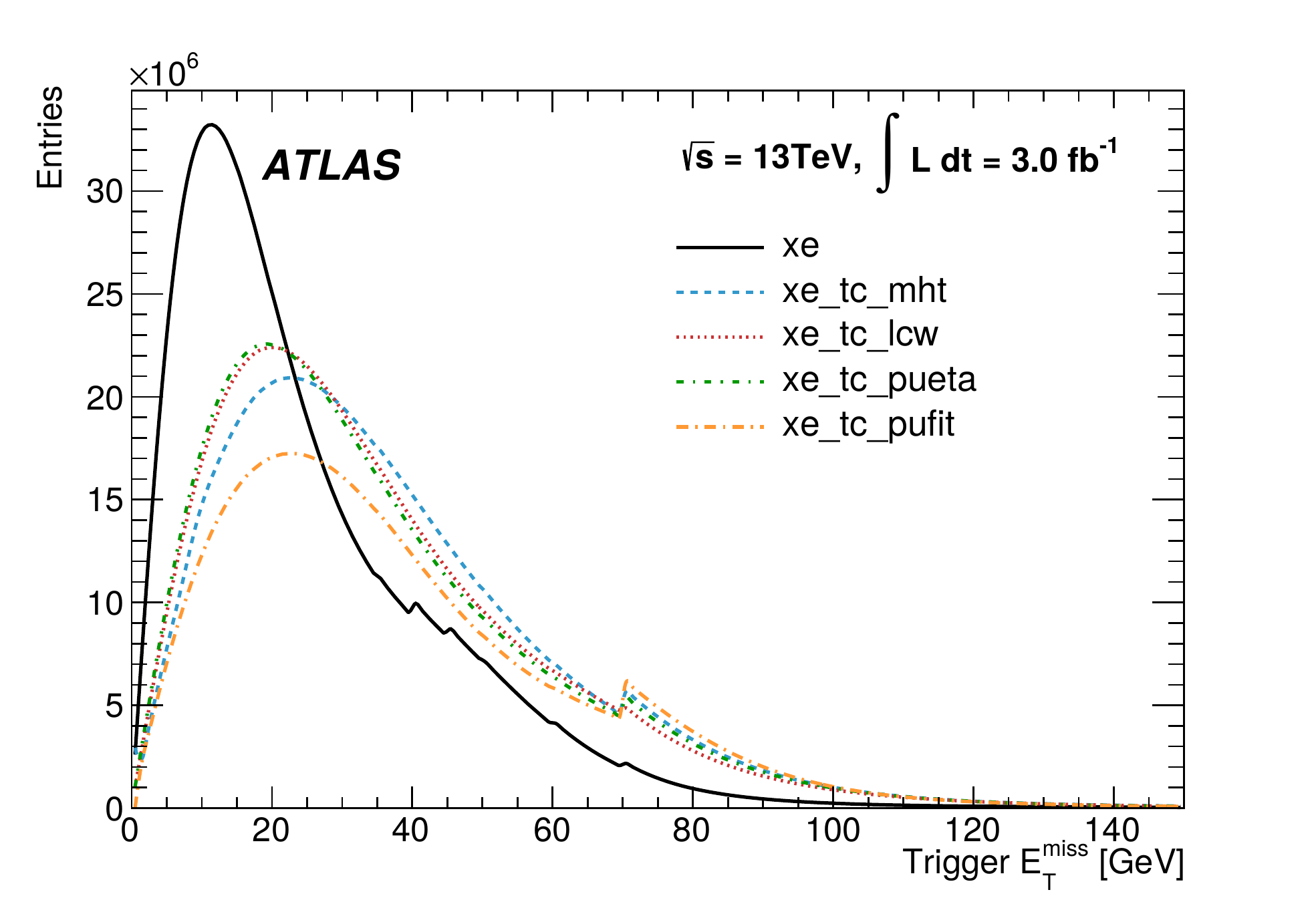}
\caption{Comparison of the different \met distributions for events accepted by
  the HLT into the \emph{Main} physics stream. The algorithms consist of a
  cell-based \met (\metcell) and different topo-cluster-based algorithms
  described in the text. The zero entries of the \metpufit algorithm, which
  occur when no tower is above threshold, have been suppressed. The steps in
  the distributions are caused by the various trigger thresholds.}
\label{fig:met:metdist}
\end{figure}

Figure~\ref{fig:met:metdist} shows the \met\ distribution of the various HLT algorithms for events accepted into
the \emph{Main} physics stream.
The differences observed between the cell-based and the topo-cluster-based \met distributions
are caused in part by different calibration; the cell-based algorithm is calibrated at the EM scale, 
while algorithms based on topo-clusters generally have larger values of \met as they include a 
correction for the calorimeter response to hadrons (hadronic scale).
Differences between the \met\ distributions for the various pile-up
correction schemes are small, since these algorithms were optimised to improve the resolution
at large pile-up values of 80 overlapping interactions that will only be achieved in future LHC runs.

\subsubsection{\met trigger menu and rates}

\begin{figure}[htbp]
\centering
\subfloat[]{
  \includegraphics[width=0.5\textwidth]{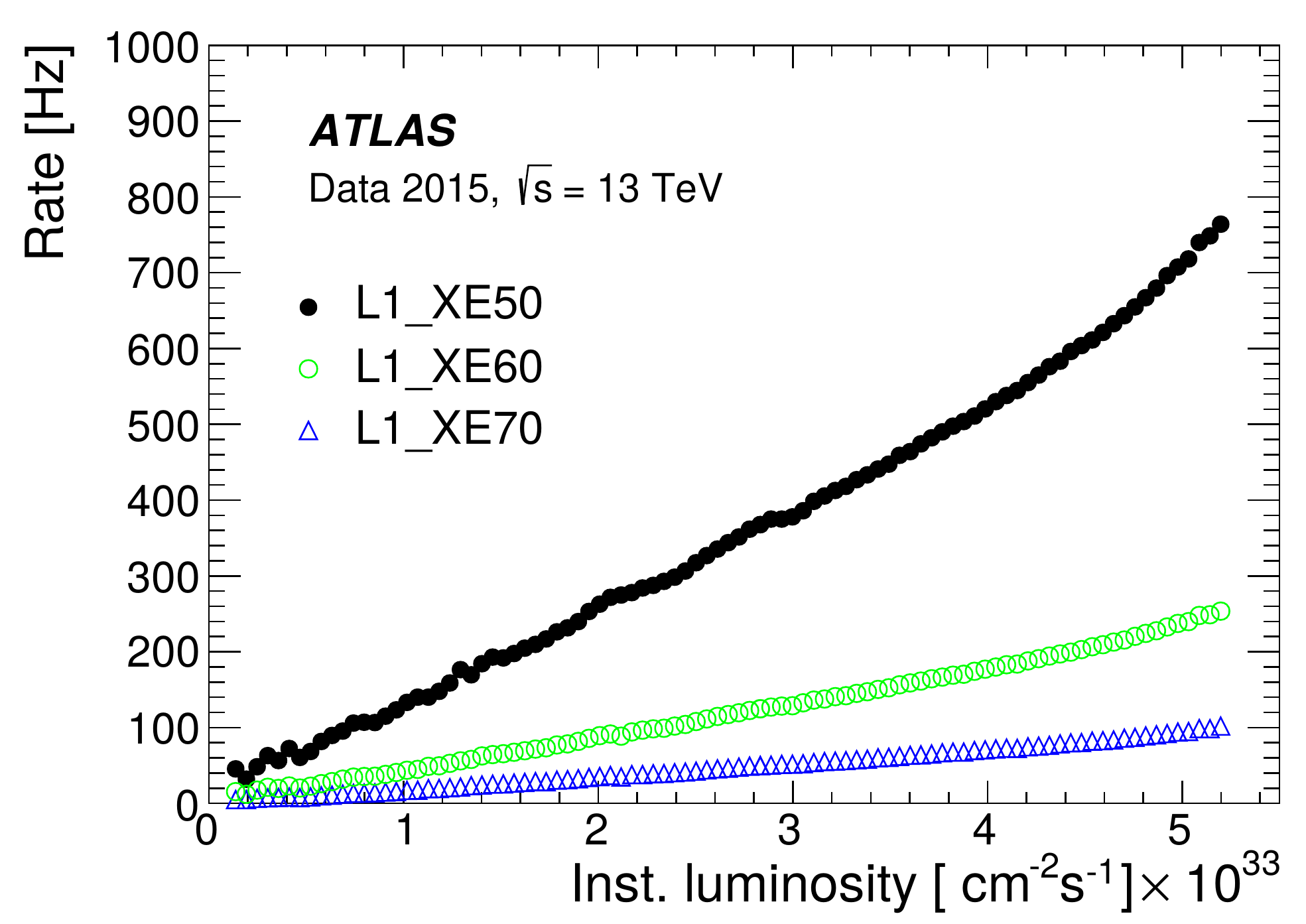}
  \label{fig:met:l1rate}
}
\subfloat[]{
  \includegraphics[width=0.5\textwidth]{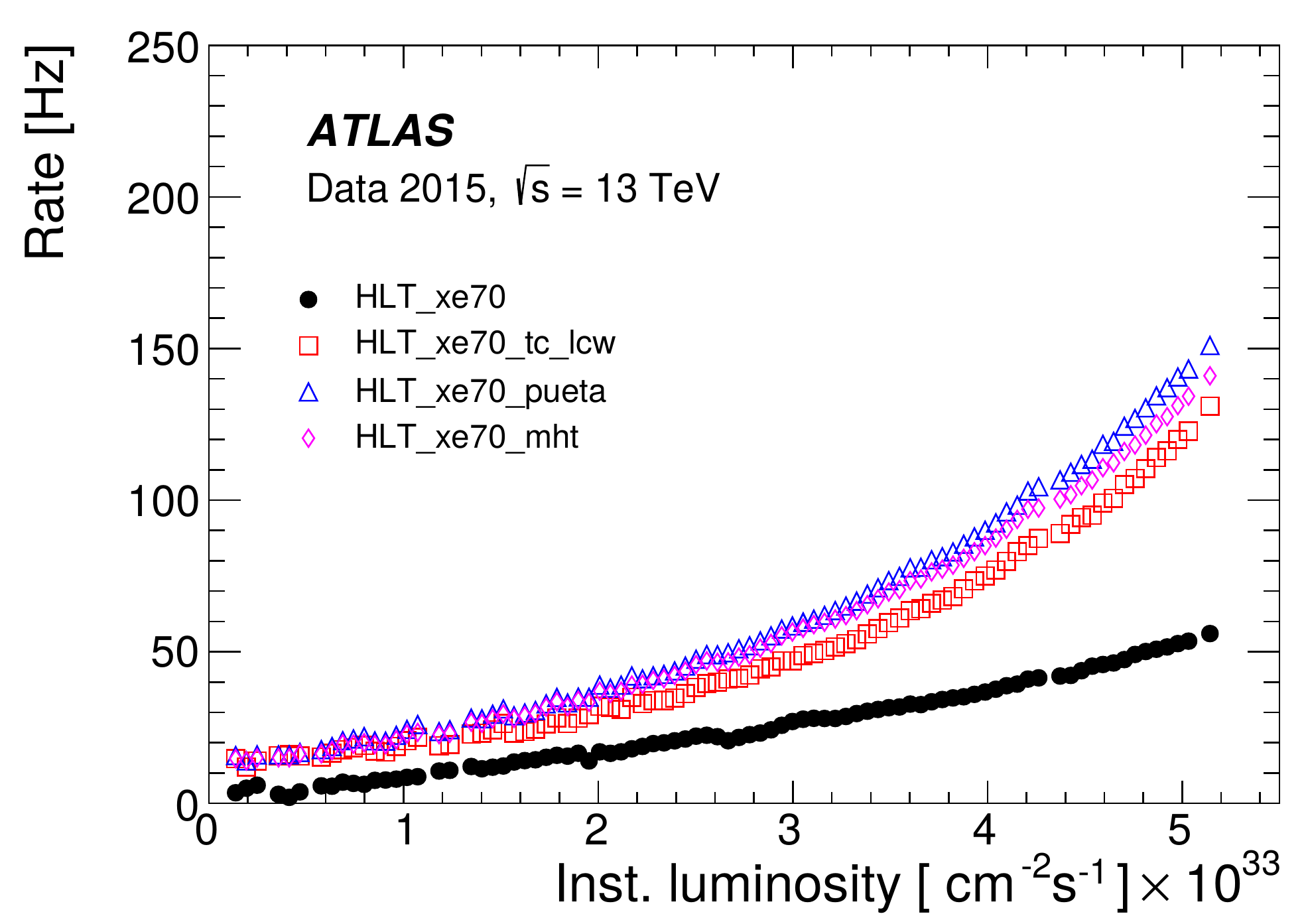}
  \label{fig:met:hltrate}
}
\caption{\met{} trigger rates (a) at L1 and (b) 
  for various HLT algorithms operating with nominal thresholds of \SI{70}{\GeV}.
  The HLT algorithms are each seeded by \trig{L1_XE50}.
  Rates are shown as a function of instantaneous luminosity from various runs
  taken in 2015 excluding periods with atypically
  high or low rates arising from different pile-up conditions for the same instantaneous luminosity.}
\end{figure}

All the primary HLT \met{} algorithms used in 2015 were seeded by the \trig{L1_XE50} trigger with 
a nominal threshold, calibrated at the EM scale, of \SI{50}{\GeV}.
The \trig{L1_XE50} output rate was approximately \SI{700}{\Hz} at an instantaneous 
luminosity of $5\times\lumi{e33}$ as shown in Figure~\ref{fig:met:l1rate}.
The HLT \metcell{} trigger with a threshold of \SI{70}{\GeV} remained unprescaled throughout the
2015 data-taking period. The typical output rate for this trigger was approximately \SI{50}{\Hz}
at the same luminosity as seen in Figure~\ref{fig:met:hltrate}.
The topo-cluster-based algorithms, all of which are calibrated at the hadronic scale,
had rates of approximately \SI{110}{\Hz} at the equivalent nominal threshold of \SI{70}{\GeV}.
The output rate from these algorithms is larger for the same nominal threshold
due in part to the different calibration methods.
Prescaled triggers at a set of lower L1 and HLT thresholds, with HLT output rates of order \SI{1}{\Hz} each,
were included in the menu to record a sample of data from which the efficiency of the unprescaled, primary physics triggers could be calculated.
Further triggers based on the significance of the observed \met{}, known as \trig{xs} triggers~\cite{ATLAS-CONF-2014-002} were used to select
$W\to e\nu$ events for electron reconstruction performance studies. Triggers used during \runi for selecting events based on the scalar sum of
the transverse energy of all calorimeter cells $\Sigma\ET$ were found to have a high sensitivity to pile-up~\cite{ATLAS-CONF-2014-002}, and so
were not used during the proton--proton run in 2016.\footnote{A $\Sigma\ET$ trigger was used during heavy-ion collisions at L1.}

\subsubsection{\met trigger efficiencies}

\begin{figure}[htbp]
\centering
\subfloat[]{\includegraphics[width=0.5\textwidth]{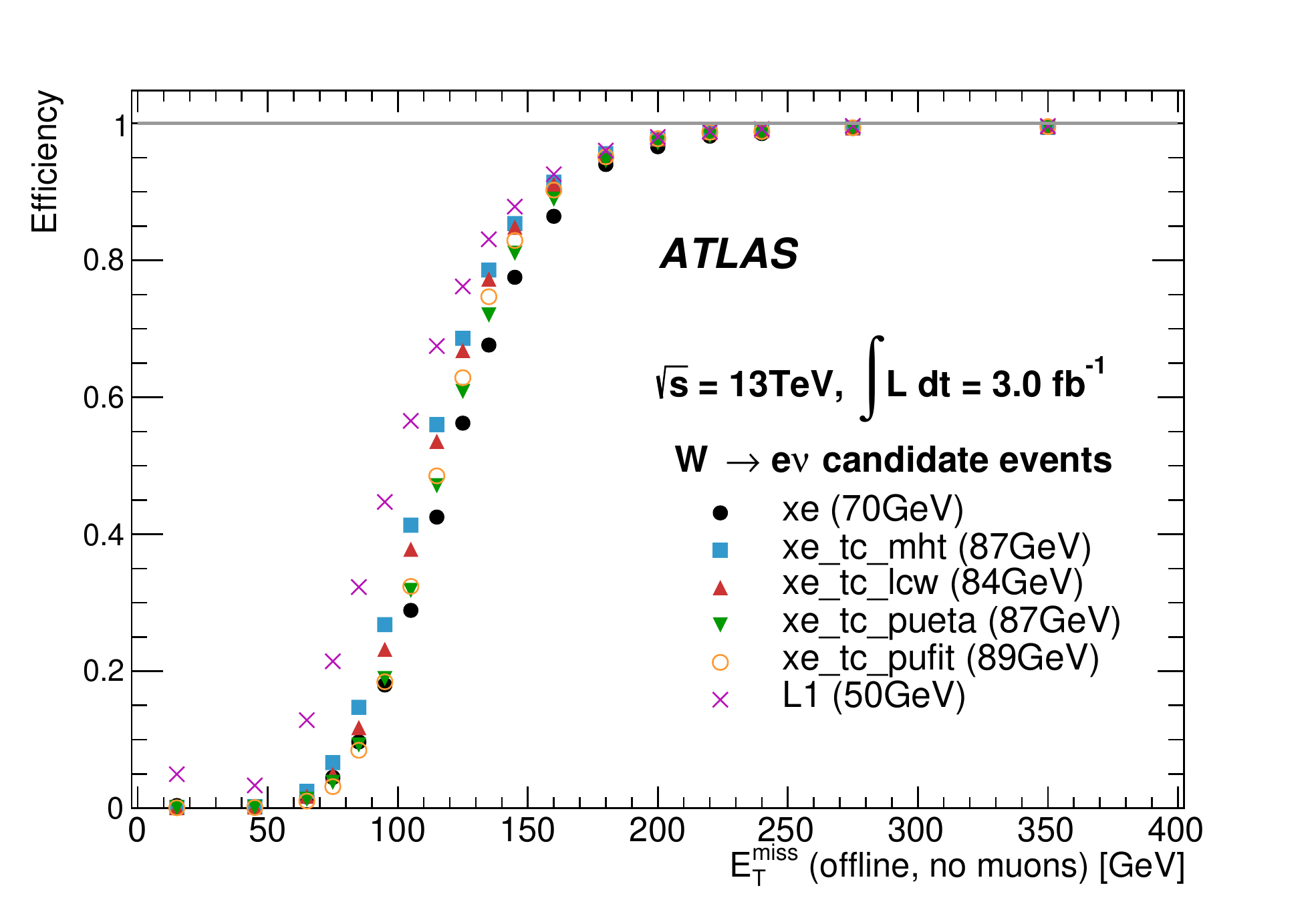}}
\subfloat[]{\includegraphics[width=0.5\textwidth]{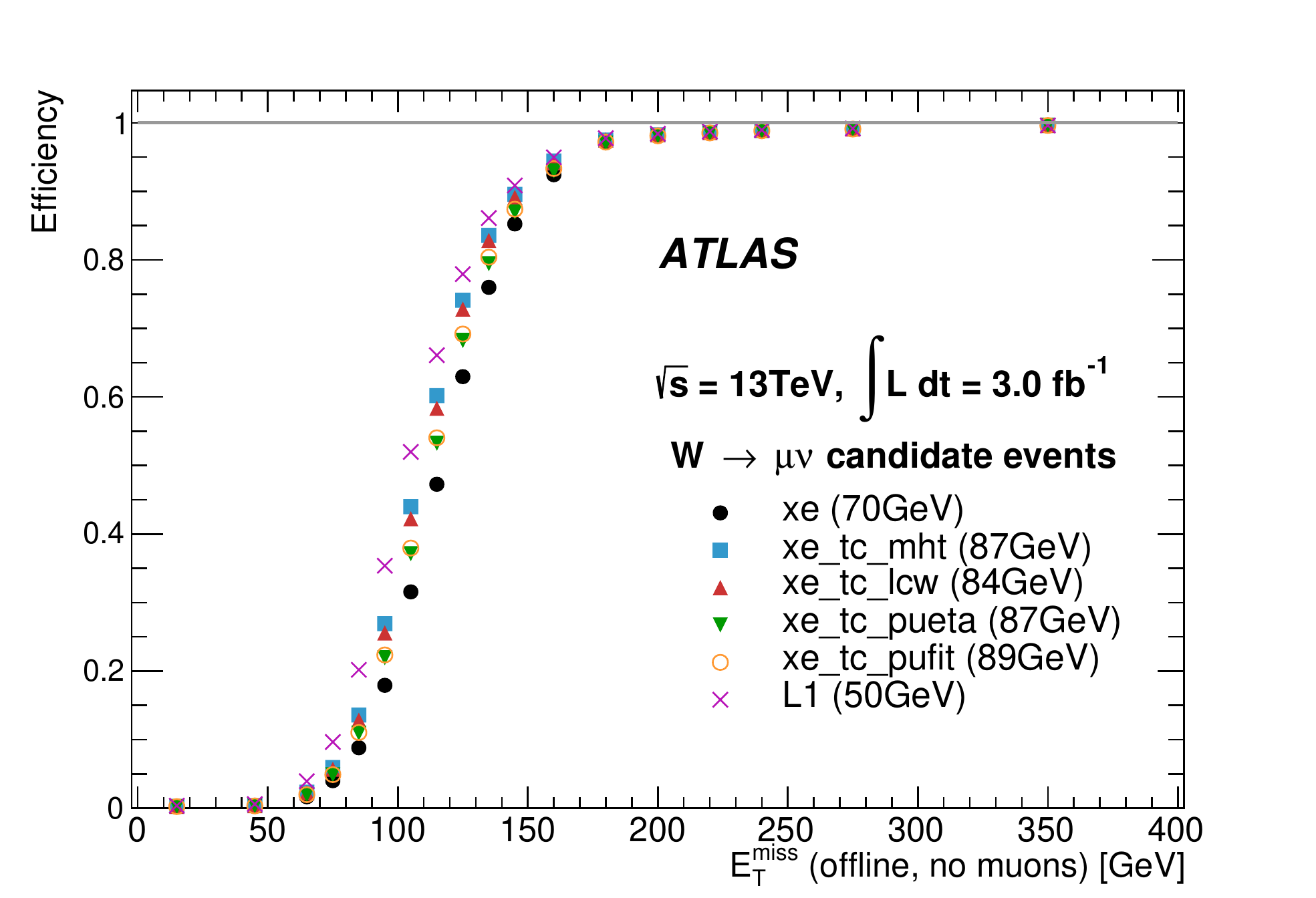}}
\caption{\met{} trigger efficiency curves with respect to the \met{}
  reconstructed offline without muon corrections for all events passing the
  (a) $W\to e\nu$ or (b)
  $W\to \mu\nu$ selections. The different efficiencies were
  measured for \metLOne, and for the combination of \metLOne\ with each of the
  HLT \met algorithms. The thresholds for the different algorithms correspond to
  an approximately equal trigger rate.}
\label{fig:met:metturnon}
\end{figure}

Since \met{} is a global observable calculated from many contributions, 
each of which has its own detector resolution,
the efficiency of the \met{} trigger for any particular analysis inevitably depends on the event selection used in that analysis.
The efficiency turn-on curves of the various \met{} trigger algorithms are shown
in Figure~\ref{fig:met:metturnon}, for $W\to e\nu$ and $W\to
\mu\nu$ selections.  
The selection is similar to that of the 
$W$ boson cross-section measurement~\cite{STDM-2015-03}, requiring
exactly one lepton (electron or muon)
with $\pT>\SI{25}{\GeV}$, transverse mass $m_{\mathrm{T}}>\SI{50}{\GeV}$,
and a single lepton trigger (\SI{24}{\GeV} single-electron or \SI{20}{\GeV} single-muon).
The efficiencies are shown as a function of a modified
offline \met{} calculation with no muon correction, emulating the calorimeter-only
\met{} calculation used in the trigger.
The event kinematics for the same \met{} are very different for the decays into electron and muon,
since the energy of the electron for $W\to e\nu$ is included in both the online 
and offline calculations of \met,
whereas this is not the case for the muon in $W\to\mu\nu$. Events with high \pt muons
are recorded by the muon triggers.

The turn-on curves are shown for different nominal HLT \met{} thresholds, selected such that
they give rates close to that of the \metcell{} algorithm at its lowest
unprescaled (\SI{70}{\GeV}) threshold.  All the HLT algorithms, 
with their stated thresholds, 
are close to fully efficient with respect to the offline \met{} for values of 
$\met>\SI{200}{\GeV}$. At that
value of \met, the \trig{L1_XE50} trigger itself has an efficiency in the range
of 95--99\%, depending on the exact event selection required. The topo-cluster-based
algorithms, and in particular \metmht{} have higher efficiency in the
turn-on region than the cell-based algorithm.

\begin{figure}[htbp]
\centering
\subfloat[]{
  \includegraphics[width=0.5\textwidth]{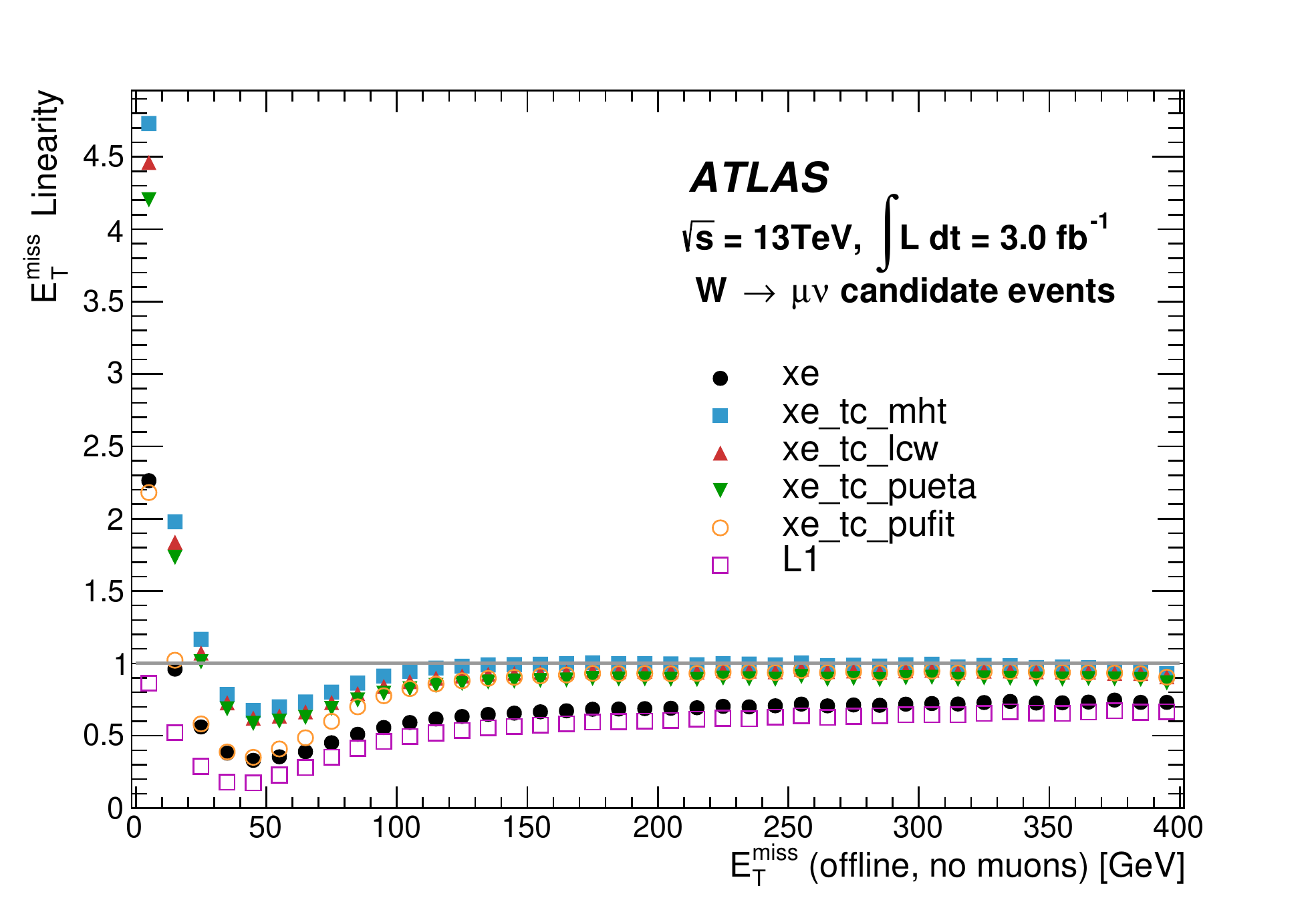}
  \label{fig:met:lin}
}
\subfloat[]{
  \includegraphics[width=0.5\textwidth]{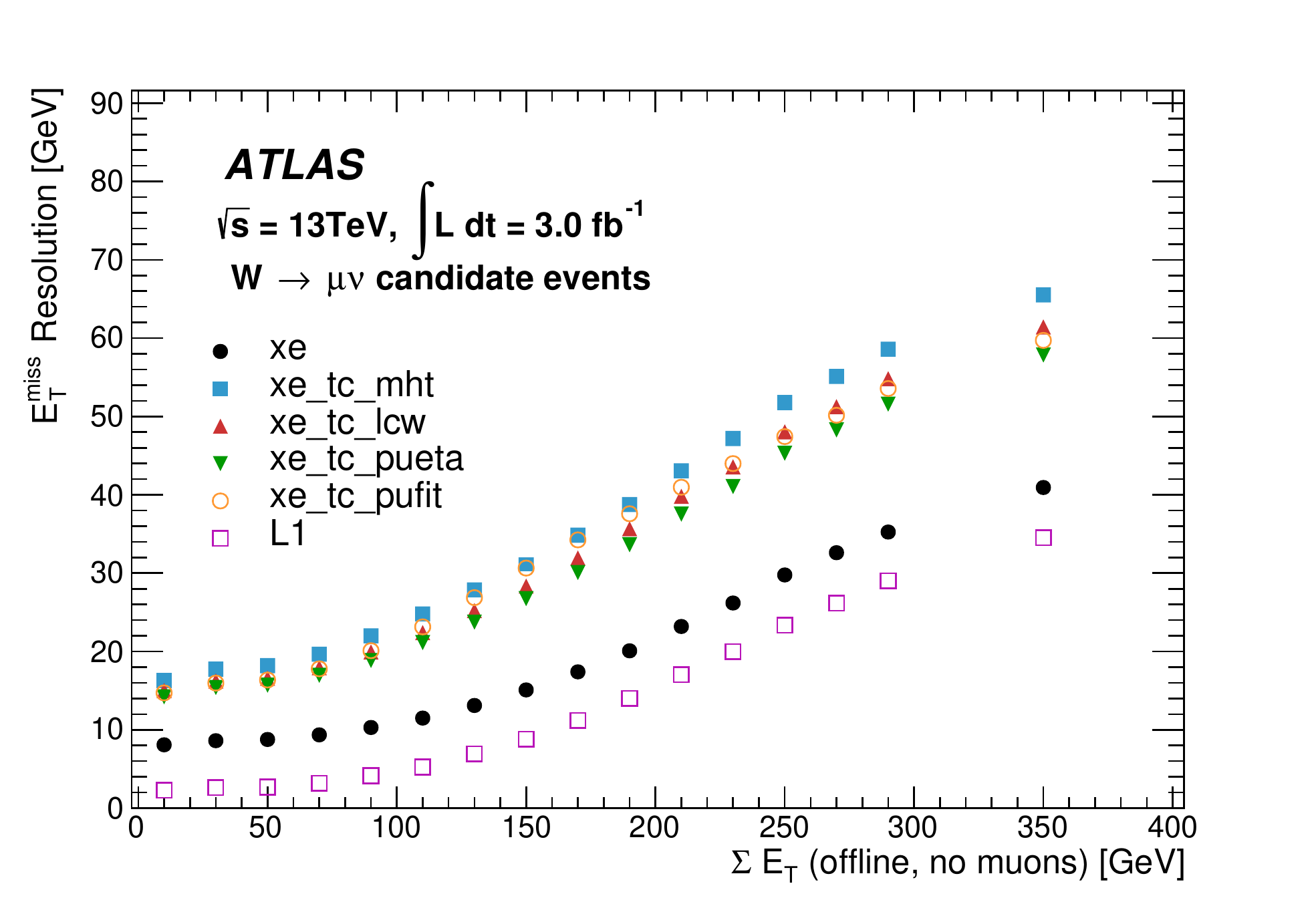}
  \label{fig:met:res}
}
\caption{(a) \met\ trigger linearity with respect to the \met\
  reconstructed offline without muon corrections and
  (b) \met\ trigger resolution with respect to the $\Sigma\ET$
  reconstructed offline without muon corrections, for all events passing $W\to \mu\nu$
  selections for \metLOne\ and for each HLT \met\ algorithm. Linearity and resolution are defined in the text.}

\label{fig:met:linresol}
\end{figure}

The linearity of the \met{} trigger is defined as the average ratio 
of the trigger \met{} to the offline \met{}.
The linearity of the \metLOne{} algorithm and the various HLT algorithms is shown in 
Figure~\ref{fig:met:lin}. 
For the larger values of offline \met{} where the triggers approach full efficiency,
the topo-cluster-based HLT algorithms
show good linearity at values close to unity.
The \metLOne{} and the \metcell{} HLT algorithms
also show stable linearity in the trigger efficiency plateau, but at a lower value,
reflecting their calibration at the EM scale rather than the hadronic scale. 

The \met{} resolution is defined as the RMS of the
$x$-component of the core of the $\ptmissvec$ distribution.  Since the
resolution is dominated by the stochastic fluctuations in calorimeter energy
measurements, it is shown in Figure~\ref{fig:met:res} as a function of the offline value of $\Sigma\ET$
(reconstructed offline without muon corrections). The expected approximate
scaling of \met{} with $\sqrt{\Sigma\ET}$ can be observed. The stochastic
contribution to the resolution can be seen to be accompanied by an offset that
varies from algorithm to algorithm and that is lower in the cell-based,
electromagnetically calibrated \metLOne\ and \metcell\ algorithms.  Such
differences are expected because different noise suppression schemes are used to
define calorimeter cells and topological clusters.

\begin{figure}[htbp]
\centering
\includegraphics[width=0.5\textwidth]{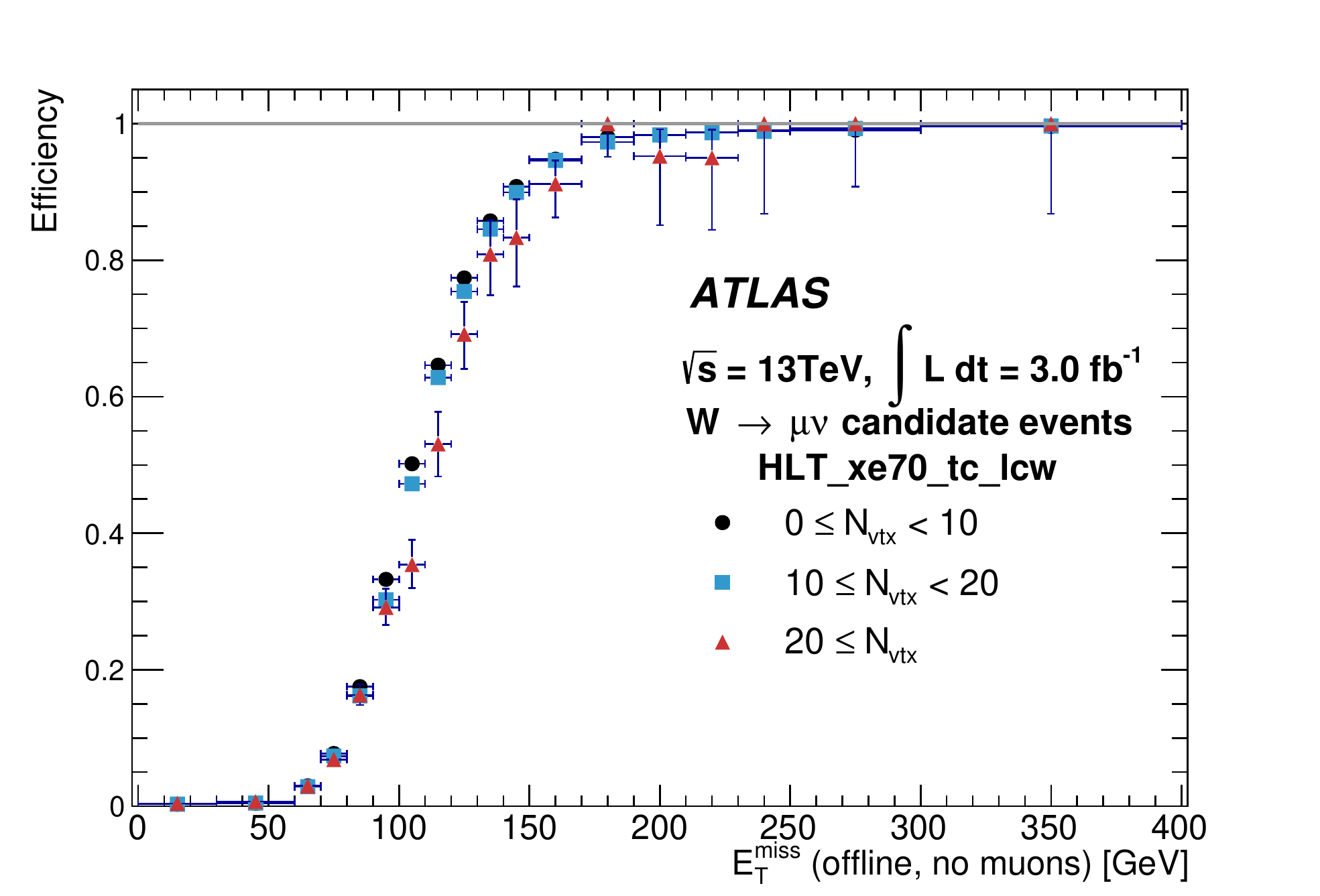}
\caption{\met trigger efficiency curves with respect to the \met
  reconstructed offline without muon corrections for 
  the $W\to \mu\nu$
  selection. The different efficiencies were obtained for different
  pile-up conditions expressed in terms of various ranges of the average number
  of reconstructed vertices per bunch-crossing (denoted here as \nvtx). The
  efficiency of the \metLOne{} algorithm is included.}
\label{fig:met:metpileup}
\end{figure}

Figure~\ref{fig:met:metpileup} shows the efficiency of the trigger-level \met{} algorithm for $W\to\mu\nu$ events 
for several ranges of the number of reconstructed vertices. 
The effect of pile-up on the \met{} turn-on curves can be seen in this figure
for the topo-cluster algorithm (\mettc), which does not employ any pile-up correction methods. 
Some degradation of efficiency is observed for larger numbers of proton--proton vertices~\nvtx.
The larger pile-up both increases the trigger rate, through increasing the probability to 
pass the trigger at lower \met{}, and degrades the efficiency in the turn-on region.


\FloatBarrier
\subsection{$b$-Jets}
\label{sec:bjet}

Bottom-quark-initiated jet (`$b$-jet') triggers are designed to identify heavy-flavour content
in real time and provide the means to efficiently record events with fully hadronic final states
containing $b$-jets.
Various signatures from the Higgs boson or physics beyond the SM rely on triggering on $b$-jets. 
These include the SM processes $t\tbar H(H\to b\bbar)$ and vector-boson fusion
production with $H\to b\bbar$, the supersymmetric decay $bA\to bb\bbar$, search for 
di-$b$-jet resonances, and resonant and non-resonant Higgs boson pair production $HH\to b\bbar b\bbar$.

\subsubsection{$b$-Jet reconstruction and selection}

Several $b$-hadron properties are exploited to identify (tag) $b$-jets. 
The $b$-hadrons have a mean lifetime of $\sim$\SI{1.5}{\ps} and often travel several millimetres before decaying. 
Consequently, a secondary vertex (SV) displaced from a primary interaction point
characterises the decay. 
Reconstructed tracks associated with this SV have large transverse and
longitudinal ($z_0$) impact parameters with respect to the primary vertex. 
In addition, $b$-hadrons go through hard fragmentation and have a relatively high mass
of about \SI{5}{\GeV}.
Thus, in addition to the decay length, b-jets can be distinguished from 
light-quark jets by having a large invariant mass, a large fraction of 
jet energy carried by tracks and a large track multiplicity.

As track and vertex reconstruction are crucial for the identification of $b$-jets,
the $b$-jet trigger relies heavily on the performance of the ID tracking
described in Section~\ref{sec:id}.
Several improvements in the ID tracking made for \runii have directly benefited the $b$-jet trigger. 
The new IBL improves the impact parameter resolution of reconstructed tracks, leading
to better $b$-jet identification and overall performance of the $b$-jet triggers~\cite{B-layerRef}.
Another improvement for \runii is the multiple-stage tracking described in
Section~\ref{sec:MultipleStageTracking}.
This new approach provides improved primary vertex finding and mitigates CPU requirements
in the face of increased pile-up.

\begin{figure}[htbp]
\centering
\subfloat[]{\includegraphics[width=0.5\textwidth]{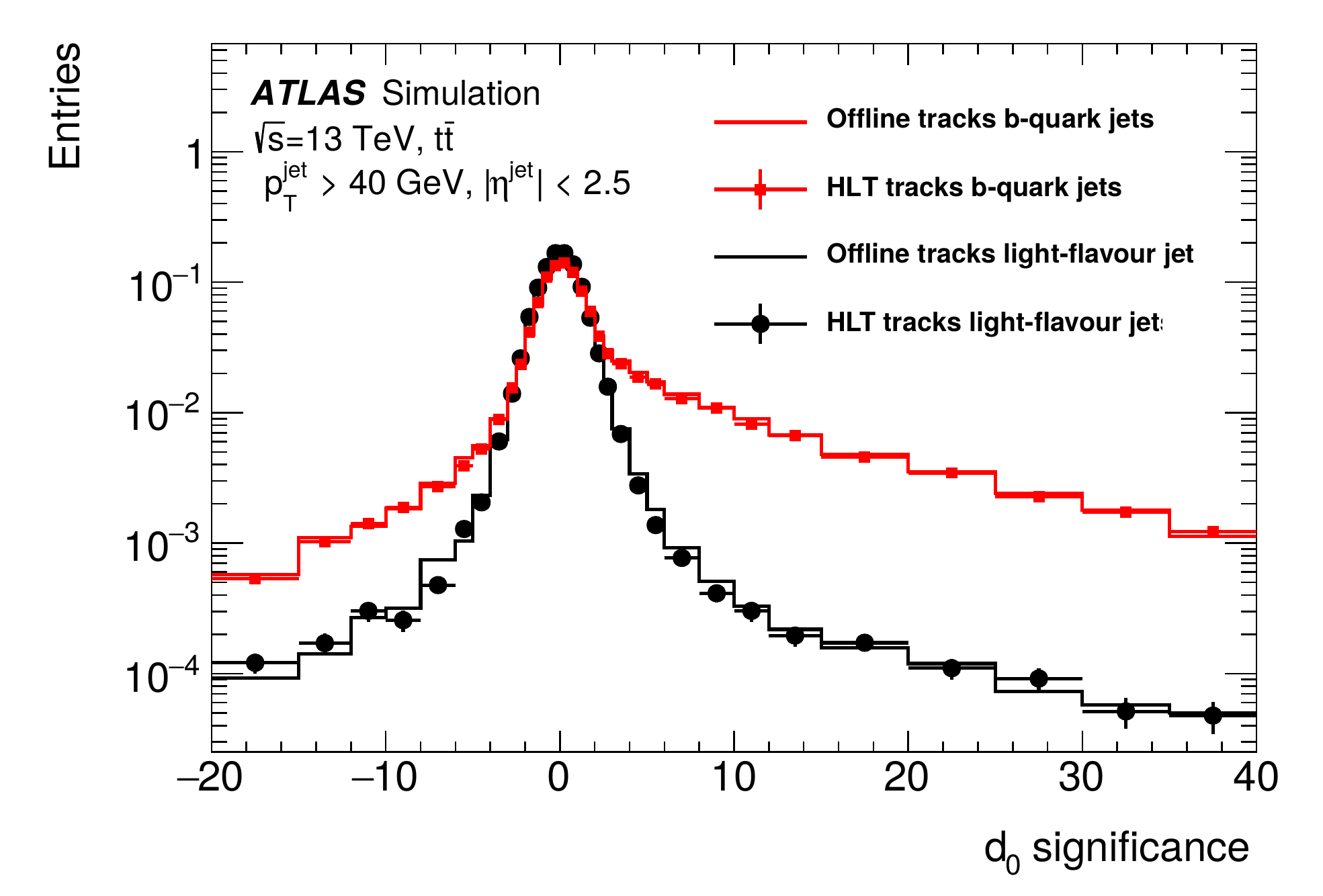}}
\subfloat[]{\includegraphics[width=0.5\textwidth]{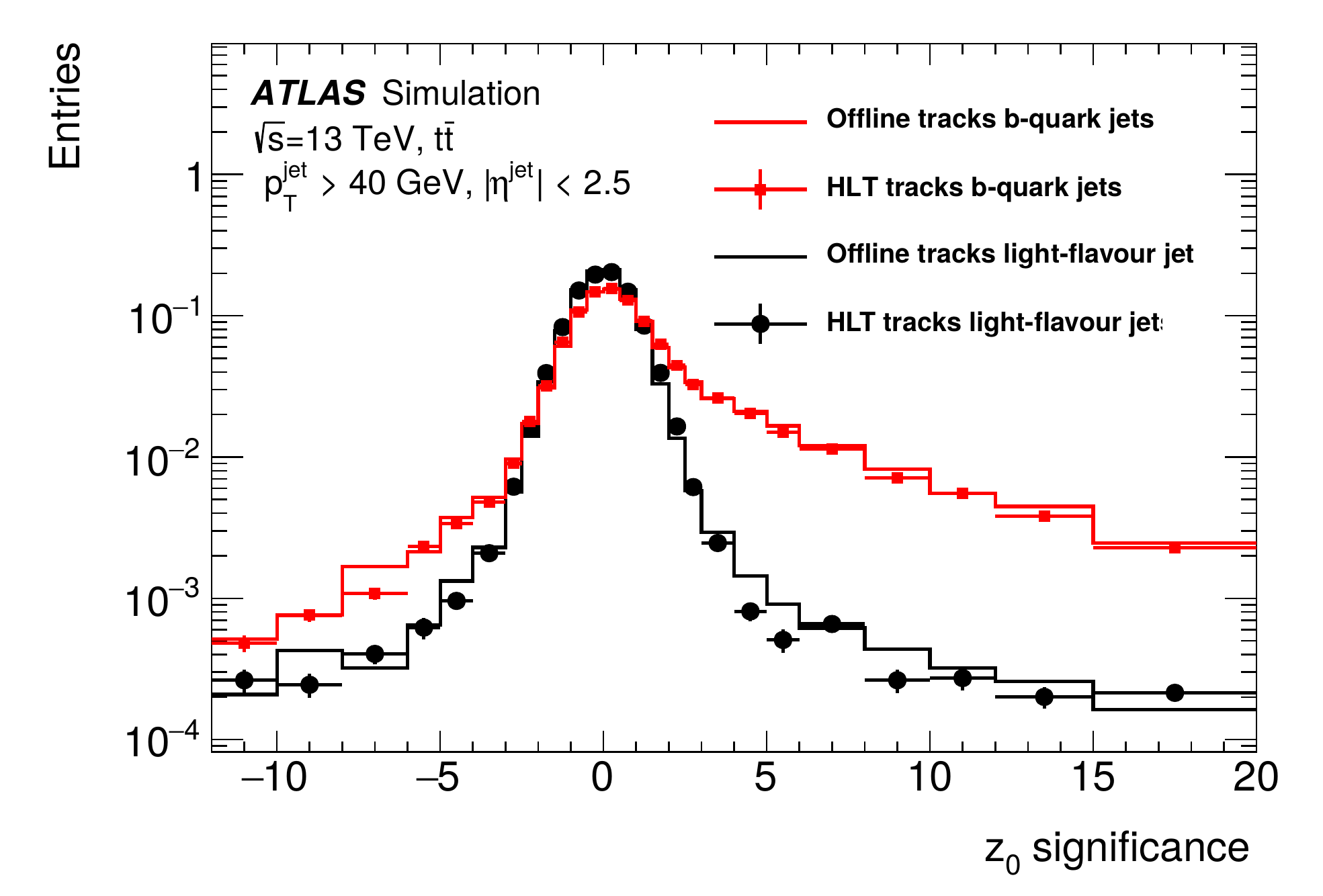}}
\caption{(a) Transverse and (b) longitudinal impact parameter significance for
  tracks associated with light-flavour (black) and $b$-quark (red) jets measured
  in a sample of simulated \ttbar\ events.  The solid lines show the
  distribution for the offline tracks.  The points show the corresponding
  distribution for tracks used in the $b$-jet trigger.  The impact parameter
  significance is defined as the impact parameter divided by the associated
  uncertainty.  The impact parameters are signed such that track displacements
  in the direction of the jet have positive values, while tracks with
  displacements opposite of the jet direction are negative.}
\label{fig:impactParameterSignificances}
\end{figure}

%
%
The basic inputs to $b$-tagging are reconstructed jets, reconstructed tracks and the
position of the primary vertex. 
The jet reconstruction used in the trigger is described in Section~\ref{sec:jet_reco}.
The $b$-jet trigger uses tracks from the precision stage of the ID trigger reconstruction.
The  beam-spot location is used for the position of the primary vertex in the plane
transverse to the beam line. 
Dedicated algorithms are run online to reconstruct and monitor the position of the
beam spot in real time. 
The position of the primary vertex along the beam line is taken from the $z$ position
of the primary vertex reconstructed as described in Section~\ref{sec:MultipleStageTracking}.
Distributions of the transverse and longitudinal impact parameter significances for
light-flavour and $b$-quark jets are shown in Figure~\ref{fig:impactParameterSignificances} for 
a sample of simulated \ttbar\ events.
Tracks used in the online $b$-tagging are compared to the corresponding tracks used offline.

%
%
During \runi, the $b$-jet triggers used a combination of two likelihood-based algorithms,
IP3D and SV1 \cite{PERF-2012-04}. 
The IP3D algorithm discriminates between $b$- and light-jets using the two-dimensional distribution of
the longitudinal and transverse impact parameter significances.
The SV1 algorithm exploits properties of the secondary vertex such as the invariant mass of
tracks matched to the vertex, the fraction of
the jet energy associated with the
secondary vertex and the number of two-track vertices. 
These \runi algorithms, optimised for \runii conditions, were used during 2015 data-taking. 
Three operating points, \emph{loose}, \emph{medium} and \emph{tight}, are defined to
correspond to $b$-jet identification efficiencies obtained from simulated \ttbar\ events of 79\%, 72\% and 62\%, respectively.

\begin{figure}[htbp]
\centering
\includegraphics[width=0.45\textwidth]{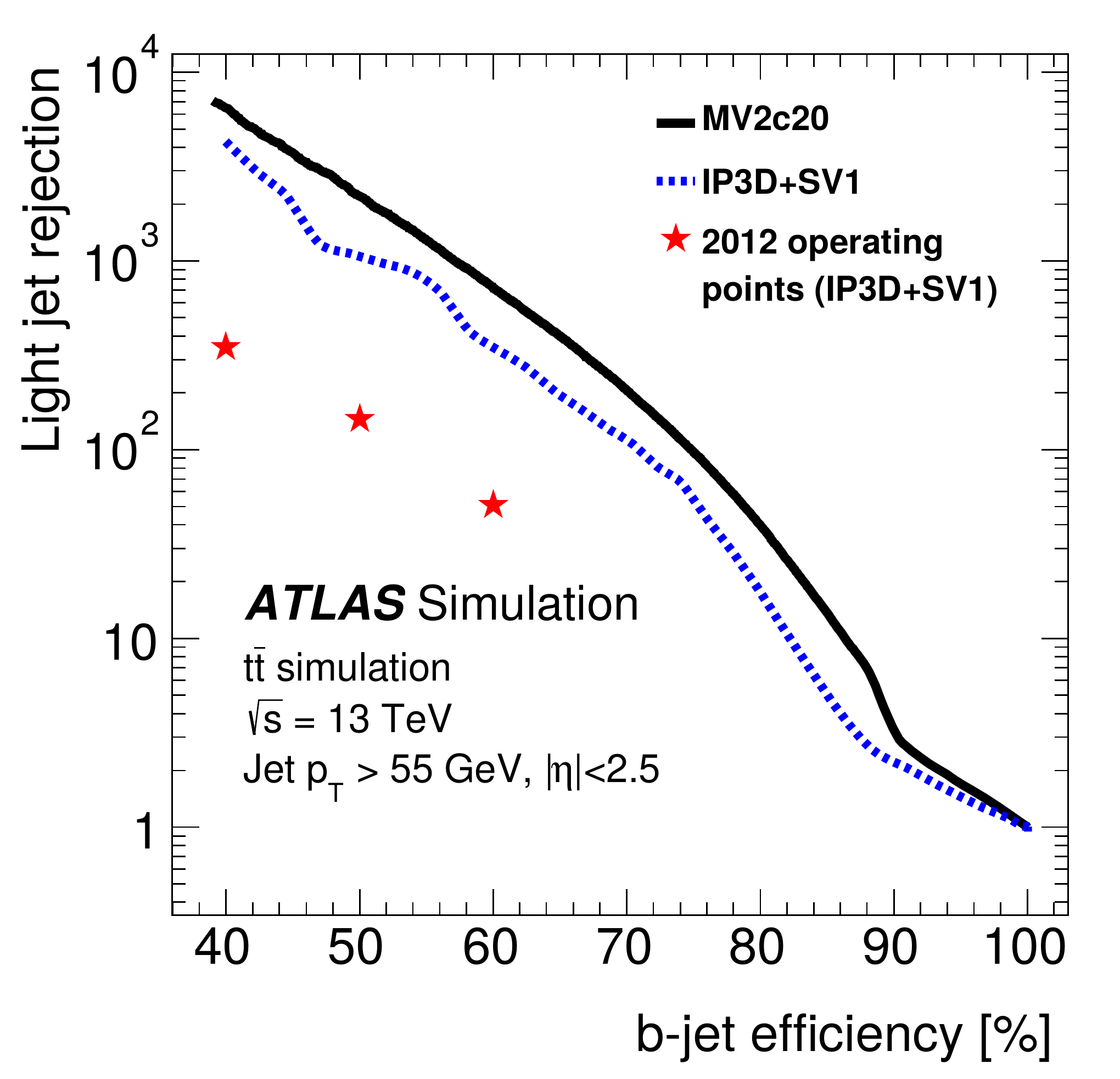}
\caption{The expected performance of the MV2c20 trigger tagger (solid black line) in terms of light-jet 
rejection is shown together with the expected
  performance of the IP3D+SV1 trigger tagger in \runii (dashed blue line) and its actual
  performance achieved during \runi (red stars).}
\label{fig:bjetROC}
\end{figure}

%
%
Another major development in the $b$-jet trigger for \runii is the adaptation of the offline $b$-tagging 
algorithms~\cite{ATL-PHYS-PUB-2015-022} for use in the trigger.
The use of the offline MV2 multivariate $b$-tagging algorithm provides better
 online $b$-jet identification and leads to a higher level of coherence between the online
and offline $b$-tagging decisions.
The MV2 algorithm uses inputs from the IP3D, SV1 and JetFitter algorithms.
The JetFitter algorithm exploits the topological structure of weak $b$- and $c$-hadron
decays inside the jet. 
The MV2 algorithm used in the trigger was optimised to identify $b$-jets using a
training sample with a background composition of 80\% (20\%) light- ($c$-) jets and is
referred to as MV2c20.
Operating points analogous to \emph{loose}, \emph{medium} and \emph{tight} were defined for MV2c20
and give light-flavour rejections similar to the corresponding operating points of
the \runi $b$-tagging algorithm.
Triggers utilising the MV2c20 $b$-tagging algorithm were run in 2015 for commissioning
purposes.  
MV2c20 is the baseline $b$-tagging algorithm for 2016.
Figure~\ref{fig:bjetROC} shows the expected performance of the MV2c20 and
the IP3D+SV1 trigger taggers in \runii compared to the actual performance of the 
IP3D+SV1 tagger that was achieved during \runi.

\begin{figure}[htbp]
\centering
\includegraphics[width=0.5\textwidth]{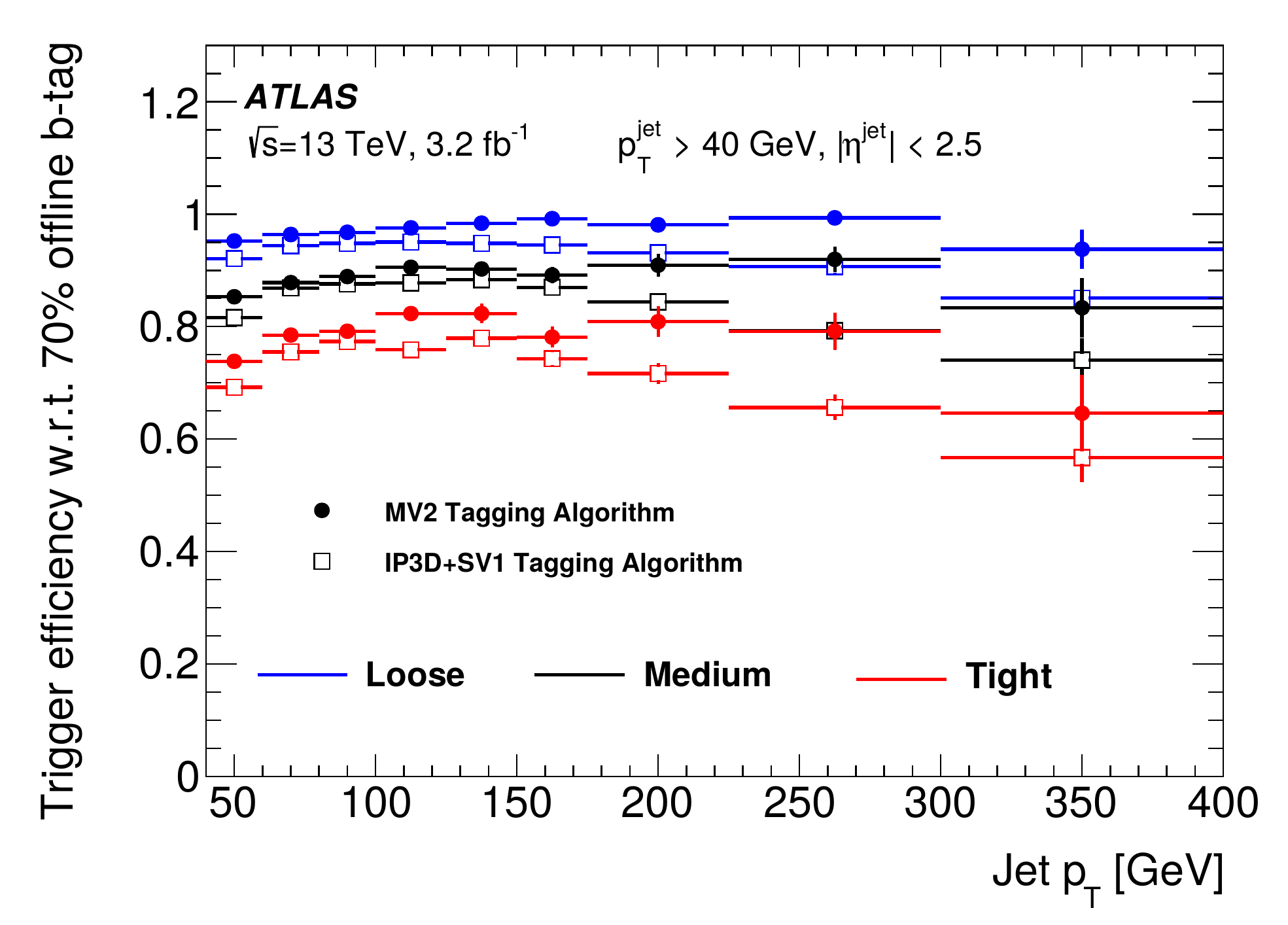}
\caption{$b$-jet trigger efficiency as a function of jet \pt\ for the loose
  (blue), medium (black) and tight (red) operating points.  The open squares
  show the trigger efficiency using the IP3D+SV1 $b$-tagging algorithm.  The
  closed circles show the trigger efficiency for the corresponding MV2
  $b$-tagging algorithm working points.  The efficiencies are measured in a pure
  sample of $b$-jets selected in \ttbar\ events and are computed with respect to
  jets identified by the 70\% working point of the offline MV2c20 $b$-tagging
  algorithm.}
\label{fig:bjeteff}
\end{figure}

Figure~\ref{fig:bjeteff} shows the efficiency of the online $b$-tagging as a
function of jet \pt\ for the three operating points.
The efficiencies are calculated in a pure sample of $b$-jets
from fully leptonic \ttbar\ decays and are computed with respect to jets identified
by the 70\% working point of the MV2c20 algorithm. Events used in the efficiency calculation
require an online jet with \pt\ greater than \SI{40}{\GeV}. A
significant gain in trigger efficiency is seen when moving to the MV2
$b$-tagging algorithms.

\subsubsection{$b$-Jet trigger menu and rates}

\begin{figure}[htbp]
\centering
\includegraphics[width=0.5\textwidth]{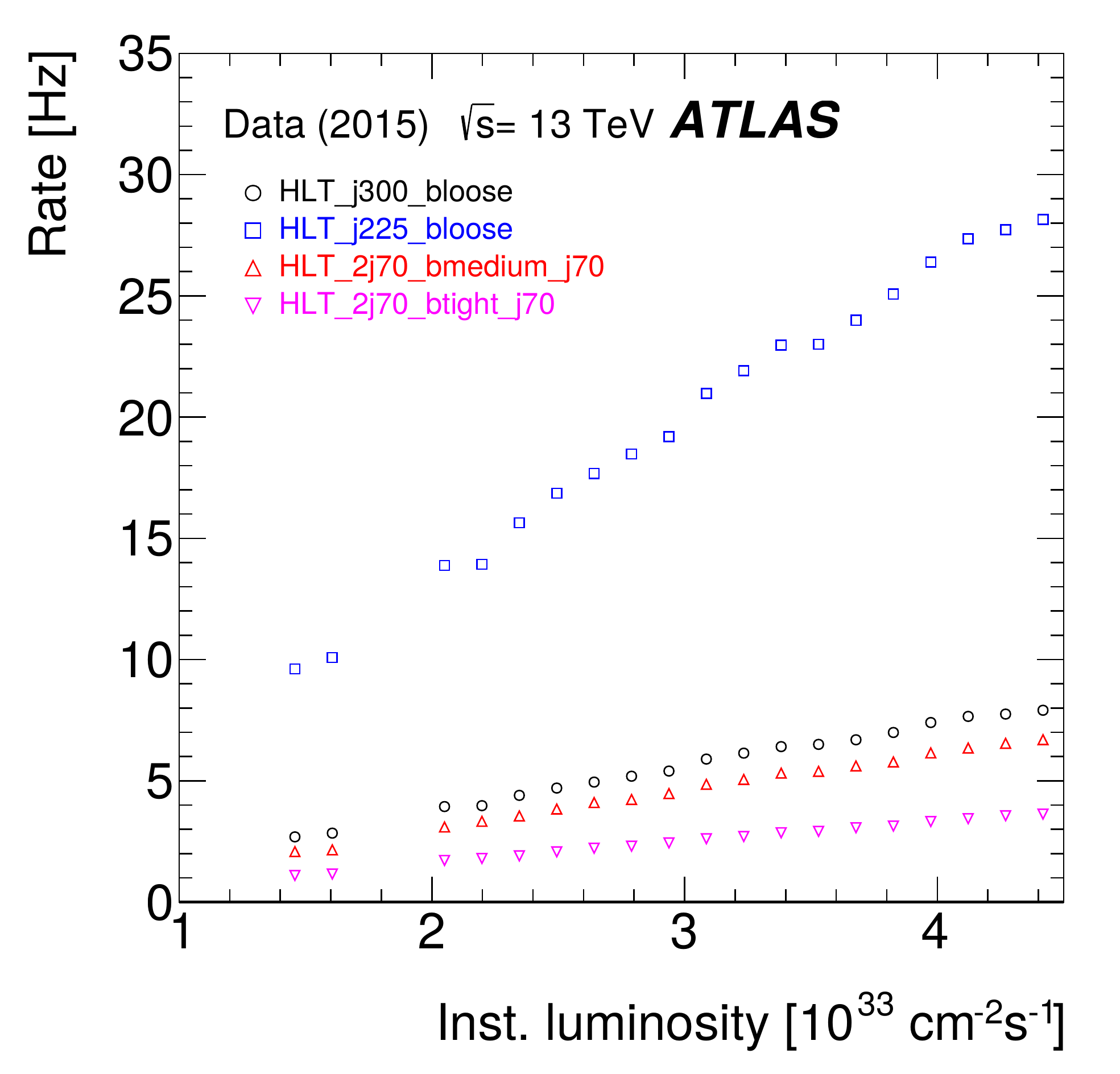}
\caption{Rates of $b$-jet triggers as a function of the instantaneous luminosity.}
\label{fig:bjetdist}
\end{figure}

Several $b$-jet triggers have been implemented with different combinations of jets
and $b$-tagged jets, using different \pt\ thresholds and $b$-tagging operating points.
The operating points, thresholds and multiplicities, for several of the primary
$b$-jet triggers are listed in Table~\ref{T:AtlasTriggerMenu}.
The jet multiplicities vary between one and four, with up to two $b$-tagged jets. 
The $b$-jet triggers are typically seeded at L1 using either a single jet
with $\ET>\SI{100}{\GeV}$ or three jets with $\ET>\SI{25}{\GeV}$ and
pseudorapidity $|\eta| < 2.5$.
Rates of various $b$-jet triggers as a function of luminosity are shown in
Figure~\ref{fig:bjetdist}.

The benefit of exploiting $b$-tagging in the HLT can be seen by comparing the
thresholds used in jet triggers with and without $b$-tagging.
The threshold for the lowest unprescaled single-jet trigger without $b$-tagging is \SI{360}{\GeV}.
A \emph{loose} requirement in the trigger allows this threshold to be lowered to \SI{225}{\GeV}.
For the four-jet trigger, \SI{85}{\GeV} thresholds are used when no $b$-tagging is applied.
Requiring two jets to satisfy the \emph{tight} $b$-tagging requirement allows the four-jet
threshold to be lowered to \SI{35}{\GeV}.

\FloatBarrier
\subsection{$B$-physics}
\label{sec:bphys}

The trigger selection of events for $B$-physics analyses is primarily based on the identification
of $b$-hadrons through decays including a muon pair in the final state. Examples
are decays with 
charmonium, $B\to\jpsi(\to\mu\mu)X$, rare decays $B^0_{(s)}\to\mu\mu$, and  semileptonic $B\to\mu\mu X$.
Decays of prompt charmonium and bottomonium are also identified through their dimuon decays,
and are therefore similar to $b$-hadron decays, apart from the lack of measurable displacement
from the $pp$ interaction point. 

\subsubsection{$B$-physics reconstruction and selection}

The primary suite of triggers require two muons at L1.
Their rate is substantially reduced compared to single-muon L1 triggers. However, this results in
inefficiencies at high transverse momentum, where the opening angle of the two muons
becomes small for low-mass resonances, and the granularity at L1 is not sufficient to form separate RoIs.
At the HLT, muons are reconstructed using the same algorithms as described in Section~\ref{sec:muonrec}
with the additional requirement that the two muons should have opposite charges and form a good vertex (where the fit
is performed using the ID track parameters) within a certain invariant mass window.
The primary triggers use three dimuon mass windows: \SIrange{2.5}{4.3}{\GeV} intended for the selection of
\jpsi and $\psi(2\mathrm{S})$ decays into muon pairs (including charmonia produced in $b$-hadron
decays), \SIrange{4.0}{8.5}{\GeV} for $B^0_{(s)}\to\mu\mu$ decays, and \SIrange{8}{12}{\GeV} for
$\Upsilon(1,2,3\mathrm{S})\to\mu\mu$ decays. These invariant mass selections are indicated by
the \trig{bJpsimumu}, \trig{bBmumu} and \trig{bUpsimumu} suffixes in the trigger names, respectively.

Additional primary and supporting triggers are also implemented. Triggers using a single L1 muon
RoI with an additional track found at the HLT do not have similar opening angle issues, but
suffer from high rates and run with high prescale factors. These combined muon triggers
are, however, essential components in
data-driven estimates of the dimuon trigger efficiencies.
Triggers requiring three muons at L1 help to maintain the lowest muon $\pT$ thresholds for
certain event signatures with a likely presence of a third muon. Finally, for selecting
semileptonic decays, such as $B^0\to\mu\mu K^{*0}(\to K^+\pi^-)$, searches for additional ID
tracks and a combined vertex fit are performed assuming a few exclusive decay hypotheses. This
reduces the rate with respect to a simple dimuon vertex selection thus allowing the
dimuon mass window to be widened to the full kinematically allowed range. The corresponding trigger names
use the \trig{bBmumuxv2} suffix.

\subsubsection{$B$-physics trigger menu and rates}

\begin{figure}[htbp]
\begin{center}
\subfloat[]{\includegraphics[width=0.5\textwidth]{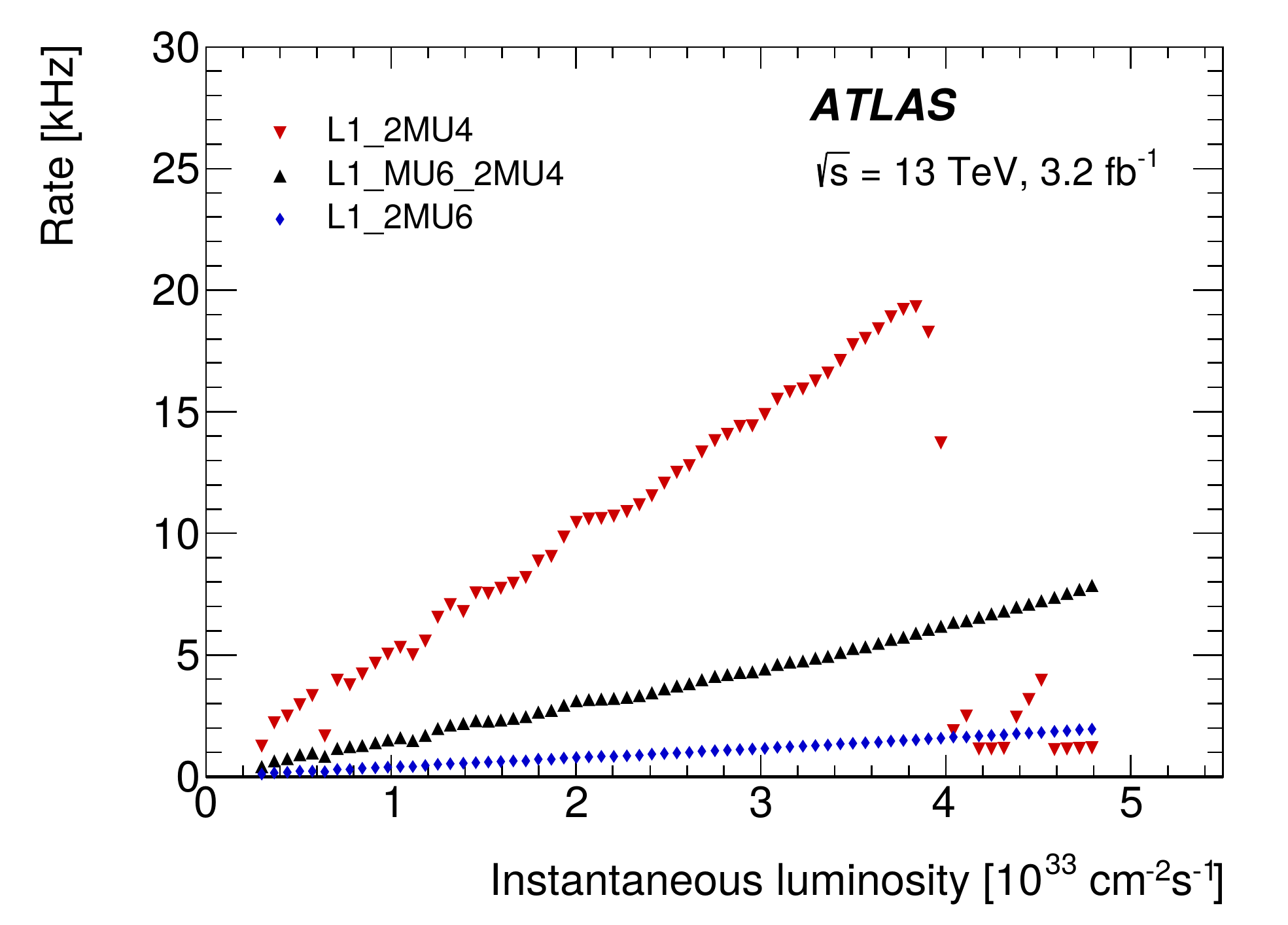}}
\subfloat[]{\includegraphics[width=0.5\textwidth]{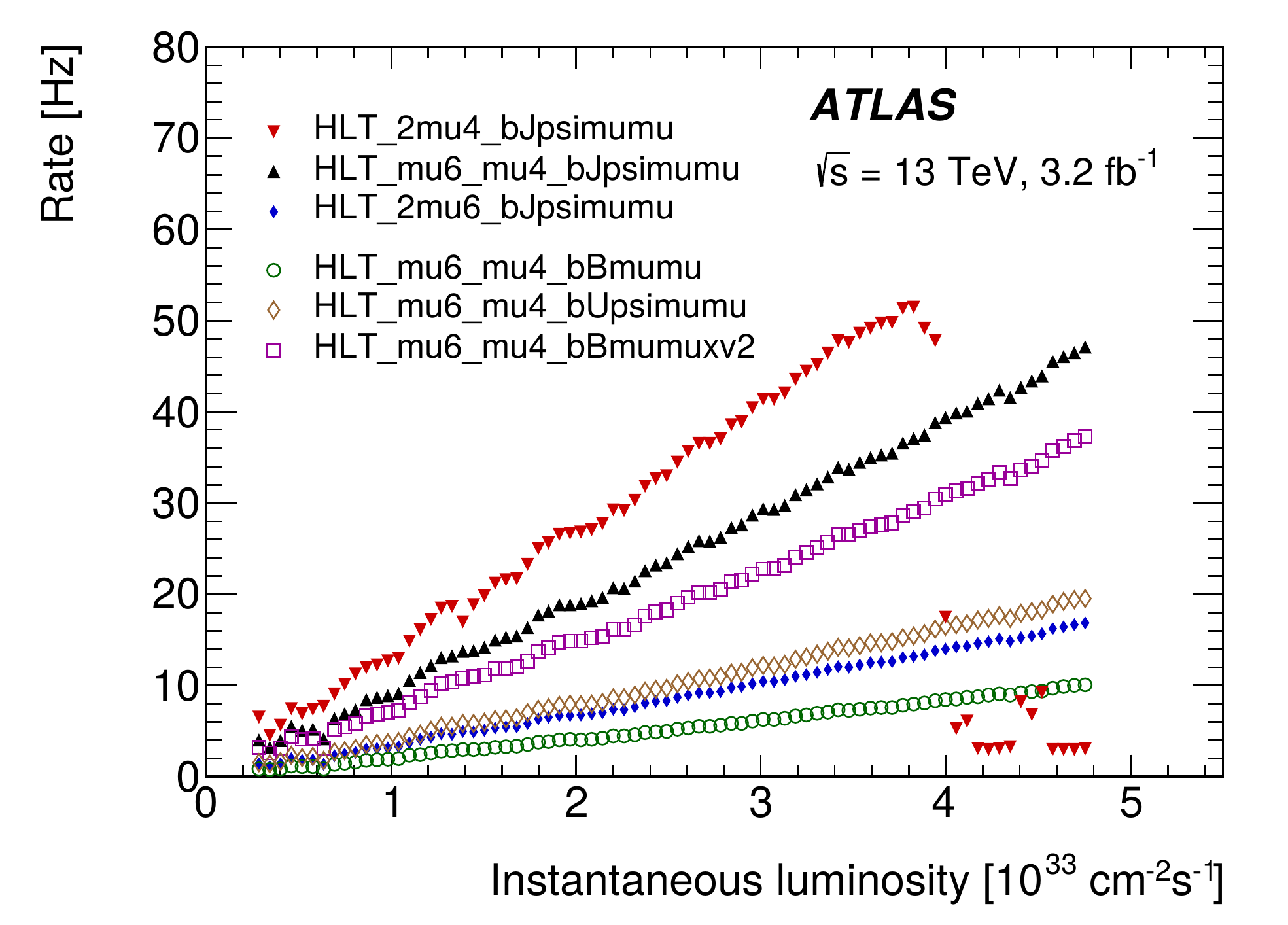}}
\caption{Trigger rates for (a) low-$\pT$ dimuon L1 triggers with various muon \pT thresholds 
  and (b) primary HLT $B$-physics triggers as a function of instantaneous luminosity. 
  (b) shows triggers requiring two muons to pass various \pT
  thresholds, to have an invariant mass within the $\jpsi$ mass window, and to form a good vertex (full markers); also shown are 
  triggers requiring
  two muons with $\pT > 6$ and \SI{4}{\GeV} and either having an invariant mass in a different
  window ($B^0_{(s)}$, $\Upsilon(1,2,3\mathrm{S})$) or forming a $B\to\mu\mu X$
  candidate after combination with additional tracks found in ID (open markers).
  As \trig{L1_2MU4} was
  prescaled at luminosities above $4\times 10^{33} {\mathrm cm^{-2}s^{-1}}$, the rate of \trig{2mu4_bJpsimumu} seeded
  from this L1 trigger drops above that luminosity.
\label{fig:bphys:rates}}
\end{center}
\end{figure}

Dimuon trigger rate restrictions at L1 define the lowest muon transverse momentum thresholds for
primary $B$-physics triggers in 2015 data-taking.
HLT triggers using \trig{L1_2MU4} were unprescaled up to a luminosity of $4\times\lumi{e33}$.
Above this, triggers seeded from \trig{L1_MU6_2MU4},\footnote{L1 muon thresholds are inclusive, i.e.\ \trig{L1_MU6_2MU4} is a dimuon trigger.} which requires two muons with \pt\ above 4 and 
\SI{6}{\GeV}, were unprescaled. The overall loss of
events collected with the former amounts to 15\%. Higher-threshold triggers seeded
from \trig{L1_2MU6} and \trig{L1_2MU10} were also active. Figure~\ref{fig:bphys:rates}
shows the L1 rates for low-$\pT$ dimuon triggers as well as the HLT rates for various primary triggers seeded from them, as a function of the instantaneous luminosity.

\begin{figure}[htbp]
\begin{center}
\includegraphics[width=0.6\textwidth]{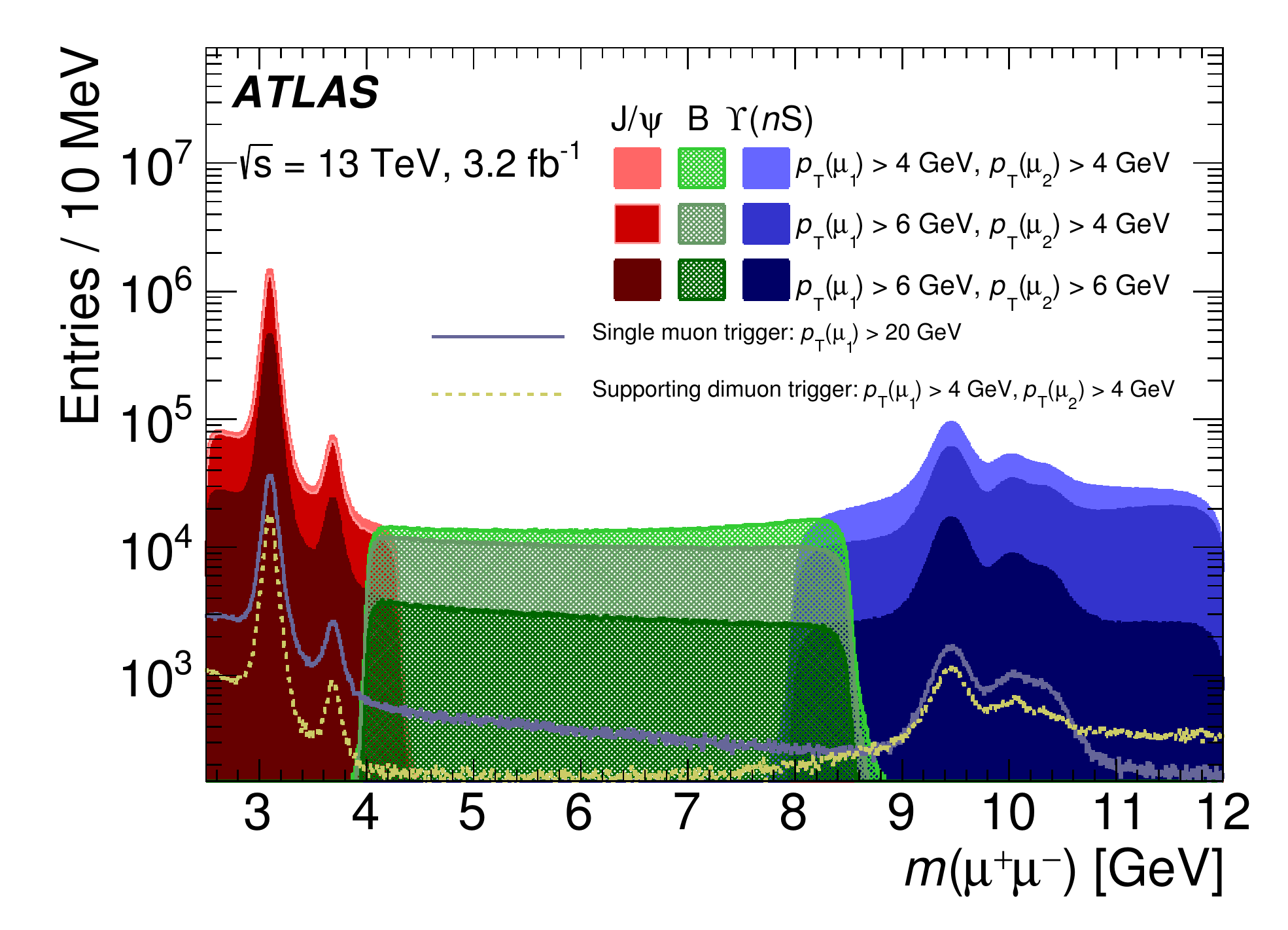}
\caption{Invariant mass distribution of offline-selected dimuon candidates passing the lowest
  thresholds of dimuon $B$-physics triggers. Triggers targeting different invariant mass
  ranges are illustrated with different colours, and the differing thresholds are shown with
  different shadings. No accounting for overlaps between triggers is made, and the distributions
  are shown overlaid, and not stacked. 
  For comparison, the number of candidates passing the lowest unprescaled
  single-muon trigger and
  supporting dimuon trigger is also shown.
  \label{fig:bphys:invmass}}
\end{center}
\end{figure}

The invariant mass distribution of offline reconstructed dimuon candidates passing the suite
of primary triggers is shown in 
Figure~\ref{fig:bphys:invmass}. For comparison, the number of candidates passing the lowest unprescaled single-muon
trigger is also shown, as well as the supporting dimuon trigger with wide invariant mass range.

\subsubsection{$B$-physics trigger efficiencies}
To evaluate the efficiency of the $B$-physics selection at the HLT, two supporting triggers
with and without the opposite-sign and vertex criteria are used.
The first trigger requires that the events contain two opposite-sign muons and form a 
good fit to a common vertex, using the ID track parameters of the identified muons
with a $\chi^{2} < 20$ for the one degree-of-freedom. This selection is the same as used in 
primary dimuon triggers but has a wider invariant mass window.
The second trigger differs by the absence of the muon charge selection and vertex fit.
The efficiency is calculated using a sample collected by these triggers.

\begin{figure}[htbp]
\begin{center}
\includegraphics[width=0.55\textwidth]{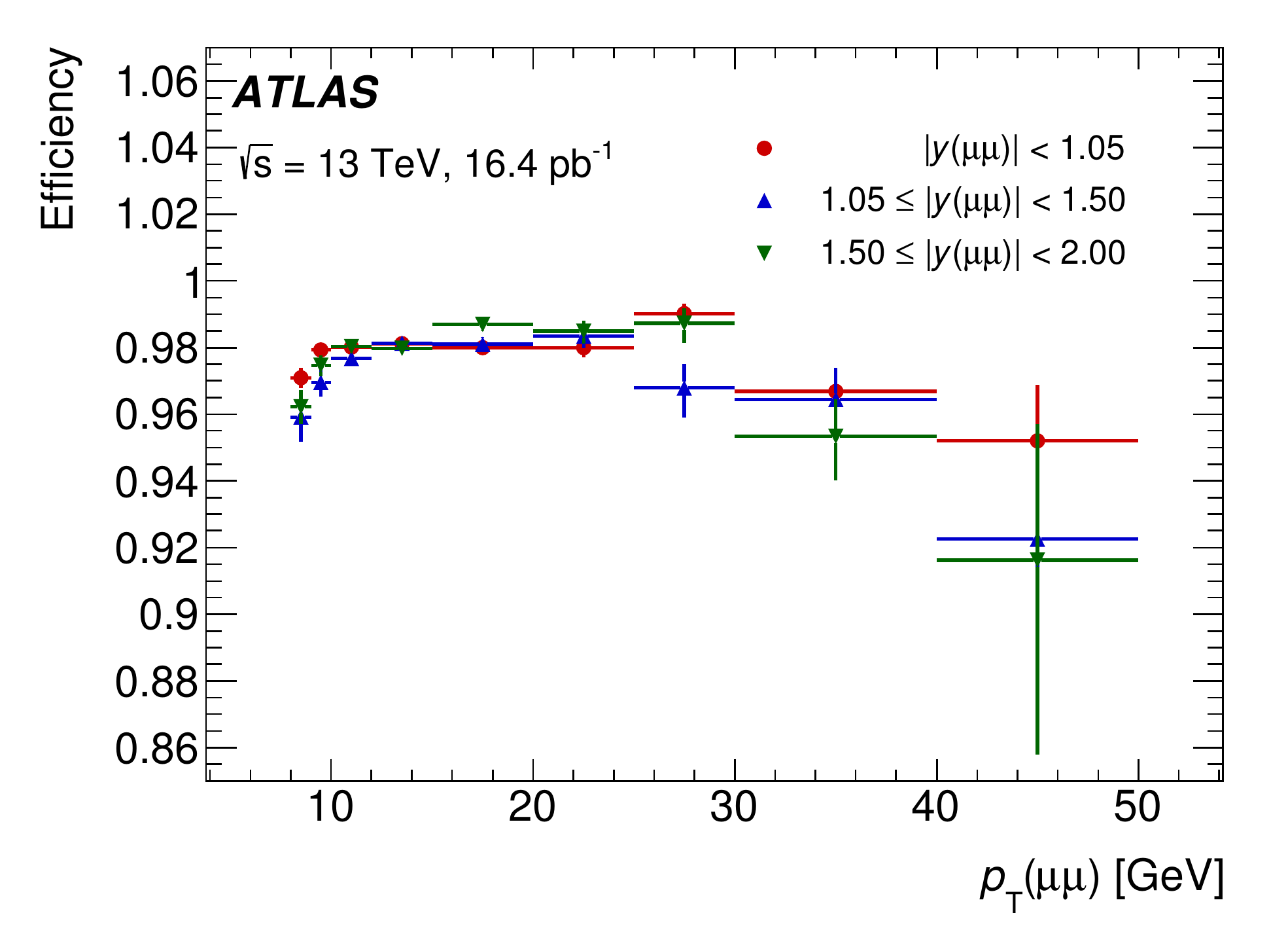}
\caption{The efficiency of the opposite-sign muon requirement and vertex quality selection 
  applied for dimuon $B$-physics triggers
  as a function of $\pt(\mu\mu)$ for three rapidity regions. 
  Supporting dimuon triggers
  with and without the selection criteria applied
  are used to determine the efficiency. 
  The integrated luminosity shown takes into account
  the high prescale factors applied to the supporting triggers.
\label{fig:bphys:dimuvtxeff}}
\end{center} 
\end{figure}

For the efficiency measurement, events are selected by requiring two offline reconstructed 
combined muons satisfying the 
\emph{tight} quality selection criteria and $\pt(\mu) > \SI{4}{\GeV}$, $|\eta(\mu)|<2.3$. The offline
muons are fit to a common vertex, using their ID track parameters, with a fit
quality of $\chi^2/\mathrm{dof}<10$ and invariant mass $|m(\mu\mu) - m_{J/\psi}| < \SI{0.3}{\GeV}$.
The number of $\jpsi$ candidates is determined from a fit to the offline 
dimuon invariant mass distribution.
The efficiency of the opposite-sign muon requirement and vertex quality selection
is shown in Figure~\ref{fig:bphys:dimuvtxeff} as a function of the offline 
dimuon transverse momentum $\pt(\mu\mu)$ calculated using the track parameters extracted after the 
vertex fit, for three slices of $\jpsi$ rapidity. The observed small drop in efficiency at high $\pt(\mu\mu)$
is due to the increasing collinearity of the two muons.

\FloatBarrier


\section{Conclusion}
\label{sec:conclusion}
\label{sec:conclusions}
A large number of trigger upgrades and developments for the ATLAS experiment were made during
the first long shutdown of the LHC in preparation for the \runii data-taking. A summary of
the various updates as well as the first \runii performance studies can be found
in this paper.

Many improvements in the L1 trigger were implemented including the addition of
completely new systems. Upgrades in the L1
calorimeter trigger included the implementation of a dynamic pedestal correction
to mitigate pile-up effects. In the L1 muon trigger, a new coincidence
logic between the muon end-cap trigger and the innermost muon chamber has been
used since 2015, and it is being extended with the hadronic calorimeter, to suppress the
fake-muon rate. New chambers were also installed to increase the trigger
coverage. In addition, the new central trigger processor doubles the number of
L1 trigger thresholds and the L1 output rate limit has increased from
\SIrange{70}{100}{\kHz}. Furthermore, a new topological processor was installed and is
being commissioned. A new HLT architecture was developed to unify the
Level-2 and Event Filter scheme used in \runi, improving the flexibility of the
system. The HLT software was also upgraded, making the algorithms and
selections closer to the offline reconstruction to maximise the efficiency, and
making use of the newly installed systems such as the innermost pixel layer IBL.

The trigger menu was revisited and redesigned to cope with the greater
rates due to the higher centre-of-mass energy and increasing instantaneous
luminosity. The different trigger signatures were set up according to the
physics needs, considering different luminosity scenarios. The ATLAS trigger
system was successfully commissioned with the first data acquired at
\SI{13}{\TeV}. First performance studies of the different trigger signatures and
trigger efficiencies with respect to the offline quantities are presented
using the 13 TeV proton--proton collision data with a \SI{25}{\ns} bunch
separation collected during 2015.



\section*{Acknowledgements}

We thank CERN for the very successful operation of the LHC, as well as the
support staff from our institutions without whom ATLAS could not be
operated efficiently.

We acknowledge the support of ANPCyT, Argentina; YerPhI, Armenia; ARC, Australia; BMWFW and FWF, Austria; ANAS, Azerbaijan; SSTC, Belarus; CNPq and FAPESP, Brazil; NSERC, NRC and CFI, Canada; CERN; CONICYT, Chile; CAS, MOST and NSFC, China; COLCIENCIAS, Colombia; MSMT CR, MPO CR and VSC CR, Czech Republic; DNRF and DNSRC, Denmark; IN2P3-CNRS, CEA-DSM/IRFU, France; SRNSF, Georgia; BMBF, HGF, and MPG, Germany; GSRT, Greece; RGC, Hong Kong SAR, China; ISF, I-CORE and Benoziyo Center, Israel; INFN, Italy; MEXT and JSPS, Japan; CNRST, Morocco; NWO, Netherlands; RCN, Norway; MNiSW and NCN, Poland; FCT, Portugal; MNE/IFA, Romania; MES of Russia and NRC KI, Russian Federation; JINR; MESTD, Serbia; MSSR, Slovakia; ARRS and MIZ\v{S}, Slovenia; DST/NRF, South Africa; MINECO, Spain; SRC and Wallenberg Foundation, Sweden; SERI, SNSF and Cantons of Bern and Geneva, Switzerland; MOST, Taiwan; TAEK, Turkey; STFC, United Kingdom; DOE and NSF, United States of America. In addition, individual groups and members have received support from BCKDF, the Canada Council, CANARIE, CRC, Compute Canada, FQRNT, and the Ontario Innovation Trust, Canada; EPLANET, ERC, ERDF, FP7, Horizon 2020 and Marie Sk{\l}odowska-Curie Actions, European Union; Investissements d'Avenir Labex and Idex, ANR, R{\'e}gion Auvergne and Fondation Partager le Savoir, France; DFG and AvH Foundation, Germany; Herakleitos, Thales and Aristeia programmes co-financed by EU-ESF and the Greek NSRF; BSF, GIF and Minerva, Israel; BRF, Norway; CERCA Programme Generalitat de Catalunya, Generalitat Valenciana, Spain; the Royal Society and Leverhulme Trust, United Kingdom.

The crucial computing support from all WLCG partners is acknowledged gratefully, in particular from CERN, the ATLAS Tier-1 facilities at TRIUMF (Canada), NDGF (Denmark, Norway, Sweden), CC-IN2P3 (France), KIT/GridKA (Germany), INFN-CNAF (Italy), NL-T1 (Netherlands), PIC (Spain), ASGC (Taiwan), RAL (UK) and BNL (USA), the Tier-2 facilities worldwide and large non-WLCG resource providers. Major contributors of computing resources are listed in Ref.~\cite{ATL-GEN-PUB-2016-002}.


\printbibliography

\newpage 
\begin{flushleft}
{\Large The ATLAS Collaboration}

\bigskip

M.~Aaboud$^\textrm{\scriptsize 137d}$,
G.~Aad$^\textrm{\scriptsize 88}$,
B.~Abbott$^\textrm{\scriptsize 115}$,
J.~Abdallah$^\textrm{\scriptsize 8}$,
O.~Abdinov$^\textrm{\scriptsize 12}$,
B.~Abeloos$^\textrm{\scriptsize 119}$,
R.~Aben$^\textrm{\scriptsize 109}$,
O.S.~AbouZeid$^\textrm{\scriptsize 139}$,
N.L.~Abraham$^\textrm{\scriptsize 151}$,
H.~Abramowicz$^\textrm{\scriptsize 155}$
% ...
\bigskip
\\
% ...
$^{8}$ Department of Physics, The University of Texas at Arlington, Arlington TX, United States of America\\
% ...
$^{12}$ Institute of Physics, Azerbaijan Academy of Sciences, Baku, Azerbaijan\\
% ...
$^{88}$ CPPM, Aix-Marseille Universit{\'e} and CNRS/IN2P3, Marseille, France\\
% ...
$^{109}$ Nikhef National Institute for Subatomic Physics and University of Amsterdam, Amsterdam, Netherlands\\
% ...
$^{115}$ Homer L. Dodge Department of Physics and Astronomy, University of Oklahoma, Norman OK, United States of America\\
% ...
$^{119}$ LAL, Univ. Paris-Sud, CNRS/IN2P3, Universit{\'e} Paris-Saclay, Orsay, France\\
% ...
$^{137}$ $^{(a)}$ Facult{\'e} des Sciences Ain Chock, R{\'e}seau Universitaire de Physique des Hautes Energies - Universit{\'e} Hassan II, Casablanca; $^{(b)}$ Centre National de l'Energie des Sciences Techniques Nucleaires, Rabat; $^{(c)}$ Facult{\'e} des Sciences Semlalia, Universit{\'e} Cadi Ayyad, LPHEA-Marrakech; $^{(d)}$ Facult{\'e} des Sciences, Universit{\'e} Mohamed Premier and LPTPM, Oujda; $^{(e)}$ Facult{\'e} des sciences, Universit{\'e} Mohammed V, Rabat, Morocco\\
% ...
$^{139}$ Santa Cruz Institute for Particle Physics, University of California Santa Cruz, Santa Cruz CA, United States of America\\
% ...
$^{151}$ Department of Physics and Astronomy, University of Sussex, Brighton, United Kingdom\\
% ...
$^{155}$ Raymond and Beverly Sackler School of Physics and Astronomy, Tel Aviv University, Tel Aviv, Israel\\
% ...
$^{a}$ Also at Department of Physics, King's College London, London, United Kingdom\\
$^{b}$ Also at Institute of Physics, Azerbaijan Academy of Sciences, Baku, Azerbaijan\\
$^{c}$ Also at Novosibirsk State University, Novosibirsk, Russia\\
$^{d}$ Also at TRIUMF, Vancouver BC, Canada\\
$^{e}$ Also at Department of Physics {\&} Astronomy, University of Louisville, Louisville, KY, United States of America\\
% ...
\end{flushleft}




\end{document}
